# TAX POLICY HANDBOOK FOR CRYPTO ASSETS

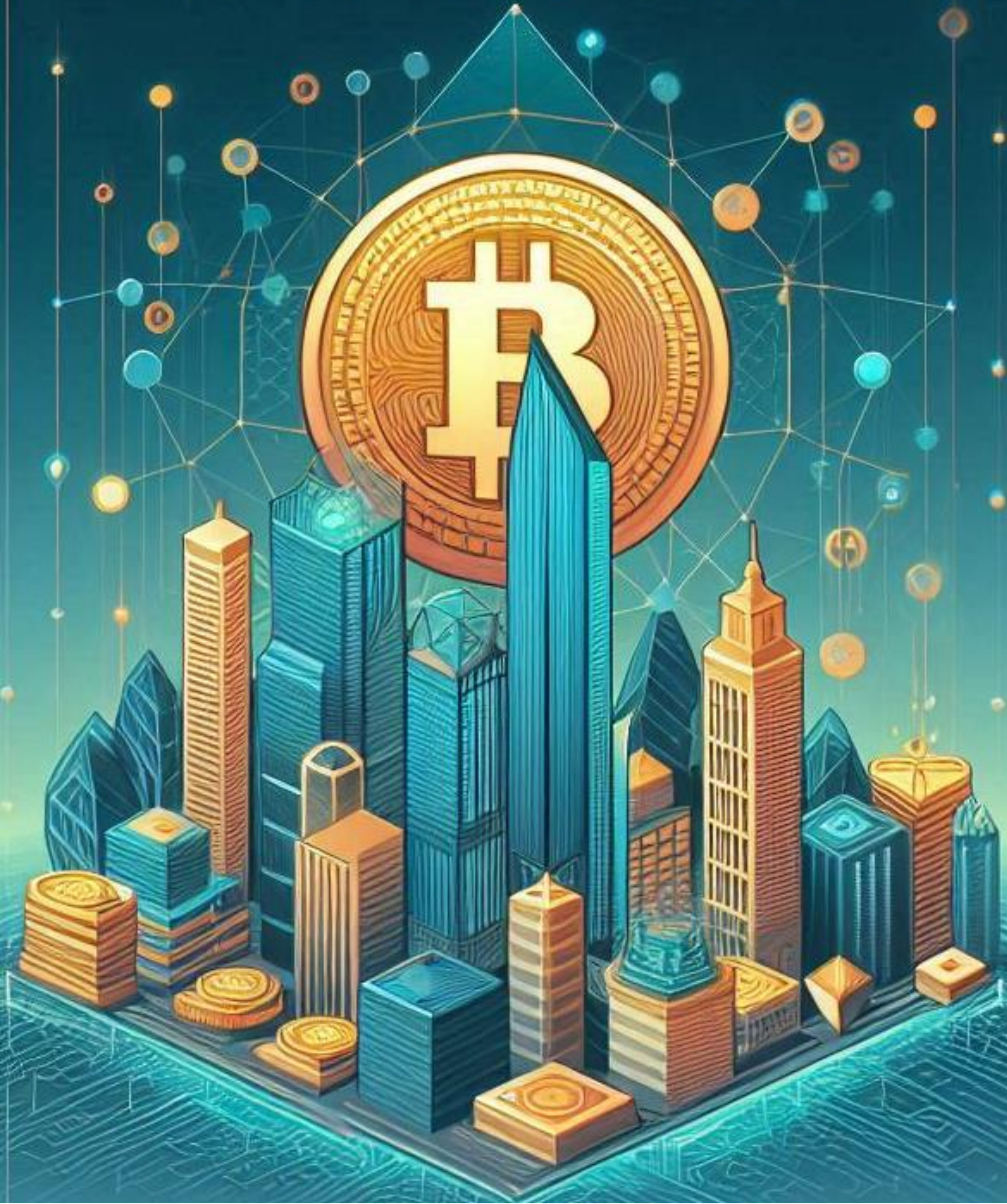

**TAX POLICY HANDBOOK FOR CRYPTO ASSETS**

*Arindam Misra*

Abstract

The Financial system has witnessed rapid technological changes. The rise of Bitcoin and other crypto assets based on Distributed Ledger Technology mark a fundamental change in the way people transact and transmit value over a decentralized network, spread across geographies. This has created regulatory and tax policy blind spots, as governments and tax administrations take time to understand and provide policy responses to this innovative, revolutionary, and fast-paced technology.

Due to the breakneck speed of innovation in blockchain technology and advent of Decentralized Finance, Decentralized Autonomous Organizations and the Metaverse, it is unlikely that the policy interventions and guidance by regulatory authorities or tax administrations would be ahead or in sync with the pace of innovation. This paper tries to explain the principles on which crypto assets function, their underlying technology and relates them to the tax issues and taxable events which arise within this ecosystem.

It also provides instances of tax and regulatory policy responses already in effect in various jurisdictions, including the recent changes in reporting standards by the FATF and the OECD. This paper tries to explain the rationale behind existing laws and policies and the challenges in their implementation. It also attempts to present a ballpark estimate of tax potential of this asset class and suggests creation of global public digital infrastructure that can address issues related to pseudonymity and extra-territoriality. The paper analyses both direct and indirect taxation issues related to crypto assets and discusses more recent aspects like proof-of-stake and maximal extractable value in greater detail. The overall objective of the paper is to enable a tax policymaker, an auditor, or an investigator to obtain a reasonable understanding of this technologically challenging realm of economic activity, and formulate tax policy and laws according to the specific requirements of their individual jurisdictions.

# Index

## 1. Introduction

On 3rd January 2009, Satoshi Nakamoto mined the first block of Bitcoin (the genesis block) and created a revolutionary system of storing and transmitting value in a trustless[1] manner over an open network. The genesis block also insinuated at the philosophy behind the newly created network. The transaction which generated the first 50 Bitcoins contained the phrase "The Times 03/Jan/2009 Chancellor on brink of second bailout for banks." which alluded to the problems of the contemporary monetary and financial system. However, Bitcoin was and remains largely a network for storing and transmitting value through the blockchain.

Few years later, on 30th July 2015 Vitalik Buterin started Ethereum, which had more features as compared to Bitcoin and worked like a global general purpose computing virtual machine. The ability to run customized code on blockchain started an entire ecosystem of applications which provide innovative financial products from lending and borrowing, to insurance through smart contracts[2]. With blockchain networks like Solana the transaction speeds can outnumber traditional networks like Visa[3]. Entities like Decentralized Autonomous Organizations, Decentralised Applications and the Metaverse challenge the traditional notions of source and residence-based taxation and pose new challenges to taxing this form of digital economy.

This has given rise to a new asset class which has unique features like openness, decentralization, pseudonymity, and transparency. This asset class also aims at improving the efficiencies of the current financial and monetary system where transfer of securities can take a couple of days, or sending/receiving huge sums of money across nations can take a few days. It enables creating a financial system where users with lower capital do not get differentially treated than those with huge capital[4]. Although there are studies that indicate that entities with large capital can get certain advantages in the crypto asset ecosystem (Aramonte et. al 2021).

This has also created new challenges for tax policy and administration, which is more accustomed to function in a centralized banking ecosystem where identities of the participants are known or possible to be known. The beneficial ownership of entities can be found and the source and origins of transactions can be traced in the centralized system. However, with crypto assets, especially those held and transacted using non-custodial wallets[5] it is difficult, if not impossible, to identify the natural or juridical person behind the transaction. The nature of these assets also blurs regulatory boundaries as they possess the characteristics of money, equity, commodity, financial instruments, and property

---

[1] A trustless system is one in which the participants do not need to trust a third party or each other for the system to function as intended.

*The author belongs to the Indian Revenue Service and is currently working as a Joint Commissioner at the Tax Policy Research Unit, Department of Revenue, Government of India. Email: misra.arindam@gov.in

**This paper is not a legal or investment guide or advice and is purely academic in nature. Readers should draw their own legal conclusions and investment decisions regarding Crypto Assets and their Taxation.

*** The findings, interpretations, and conclusions expressed herein are those of the author and do not necessarily reflect the views of the Government of India. This paper represents the personal views of the author only, not the views of the Government of India. The author accepts sole responsibility for any errors.

**** Readers are advised not to send any Crypto Assets to any public addresses given in the paper, the author is not responsible for loss of any Crypto Assets due to any such actions.

[2] Smart Contracts are analogous to automated agreements stored on the Blockchain which can accept or release crypto assets or perform certain actions when certain conditions specified in the form of code are met

[3] Solana: Better Than Your Credit Card?

[4] https://hbr.org/2022/05/how-digital-currencies-can-help-small-businesses

[5] Non-Custodial wallets allow users to have full ownership of their crypto assets without relying on any central custodian of the secret keys corresponding to the crypto assets

(Auer, R., & Claessens, S. (2018)). Consequently, many countries have no regulatory or taxation framework for this asset class, and there is wide variety in regulation and taxation of this asset class across jurisdictions which have formulated policies and guidance for crypto assets.

The increasing volumes of transactions in crypto assets make it a sizeable tax base, which if not taxed appropriately, can lead to potential revenue loss. Thus, it becomes important for policymakers to have a conceptual understanding of this asset class and its financial ecosystem, to effectively formulate and administer realistic tax policies and laws regarding crypto assets, or to use the existing legal and policy framework to collect due taxes. It is also imperative for investigators and law enforcement agencies to acquaint and train themselves with this technology, as they are likely to have more frequent encounters with it while performing audit and investigating tax fraud cases.

There is abundant literature and guidance available to the tax authorities, practitioners, and taxpayers regarding the regulatory and taxation regimes for crypto assets in different jurisdictions. However, it is imperative to have a grasp of the underlying technology and dynamics of these assets and not just the legal provisions, to be able to effectively regulate and tax them. For example, notions like the site of storage of Bitcoins and crypto assets, and the services provided by various members of this ecosystem and their place of supply have a bearing on multiple issues related to direct and indirect taxes. It is important to understand the mining process using proof-of-work and proof-of-stake mechanisms, and the nexus between a miner or validator and crypto asset owner whose transaction is included on the blockchain by the miner or validator (service provider) for the purpose of indirect taxation.

Also, questions like tax treatment of Non-Fungible Tokens (NFTs) and what exactly the owner of an NFT gets legally, are central to their taxation. This paper tries to relate the technological underpinnings of this asset class to the tax events they trigger and the source of economic value creation in this ecosystem, which can enable the tax administrations to tax them intuitively based on the existing tax provisions or any other specific provisions or guidance for crypto assets. It also tries to quantify their tax potential from publicly available data on crypto asset transactions.

This paper first explains the basic concepts of blockchain technology using Bitcoin as a classic example. It also explains the vulnerabilities of the Bitcoin Blockchain to emerging technologies like quantum computing. The later sections deal with Ethereum and its recent change to the proof-of-stake consensus mechanism. The discussion tries to link the underlying technology to the regulatory and tax implications. It also tries to make the reader understand why a specific regulatory or enforcement action which is effective in the current centralized system, may not be practical in case of crypto assets.

The paper also includes the regulatory and tax policy responses by multiple jurisdictions to specific areas in crypto assets to enable the tax practitioners to understand the various approaches used to tax this asset class. For example, a law enforcement agency's request to a Decentralized Application to identify the beneficial owner of a particular crypto asset address might be impossible for the application to process for purely technological reasons. The understanding of underlying technology is also important to formulate practical and realistic policies to ring fence this technology from its potential misuse for tax evasion, money laundering, terror financing, proliferation financing and other illegal activities and at the same time, not to stifle the innovative spirits of the torch bearers of this technology, which also has huge potential benefits (Marian, 2013), (Kapsis, 2023), (Shin & Rice, 2022).

# 2. Bitcoin and the Blockchain

## 2.1 The Origins

The Oxford English dictionary defines Bitcoin as "a system of electronic money, used for buying and selling online and without the need for a central bank" and "a unit of the bitcoin electronic system of money" which essentially captures the notion we attribute to Bitcoin, when we refer to the Bitcoin network for storing and transmitting value, whose unit of accounting is unsurprisingly known as, Bitcoin. It also refers to the protocol or software that various nodes in the Bitcoin network run to transmit and include the transactions on the blockchain in the form of blocks.

At a very basic level, the Bitcoin Blockchain records transactions between entities on a public ledger structured in the form of blocks by using cryptographic techniques without involving any central authority like a bank. However, to prevent fraud and double-spend transactions[6] the blockchain relies on a network of nodes which maintain the integrity of the chain of transactions cryptographically and are in-turn rewarded for that.

## 2.2 Bitcoin in a nutshell

Bitcoin has an open permissionless blockchain where anyone who generates a correct cryptography-based pseudonym (Bitcoin address) and its corresponding private key, can start transacting. Also, anyone with appropriate hardware and network capabilities can join the network and get rewarded (depending on their computational power) for adding transactions bundled into blocks on the blockchain (mining), securing and maintaining the blockchain. Unlike a banking system where users must undergo a Know Your Customer (KYC) procedure and risk profiling before opening an account, one can instantaneously generate Bitcoin pseudonym(s) and start transacting on the blockchain.

In a common banking transaction between Bob and Alice the account balance of Bob is reduced by a certain amount and that of Alice is increased by that amount by a central authority. Bob may also pay some transaction fee to the bank for facilitating the transaction. The central authority essentially plays the role of minimizing the counter party risk involved in the transaction.

Similarly in the Bitcoin Blockchain, users transmit amount denominated in Bitcoins to each other. However, instead of relying on the bank for processing the transaction they use the Bitcoin Blockchain which approves this transaction with the help of various nodes(miners) which cryptographically record the set of transactions called a block in a public blockchain using a process called *mining.* Instead of paying the bank fee the users pay fees to the miners who maintain the integrity of the Bitcoin Blockchain and get this fee along with an in-built economic incentive in the Bitcoin Blockchain as a reward.

Transmitting value through the Bitcoin Blockchain involves no settlement intermediaries and the blockchain is oblivious and agnostic to the geographical location of the owner of the Bitcoin pseudonymous address. This creates and maintains an immutable global public ledger in which all the BTC value ever exchanged by a Bitcoin pseudonymous address is readily and publicly available, and is used to verify and validate future transactions.

[6] A double-spend transaction is one in which a Bitcoin owner in the ecosystem, Alice, owns a certain BTC value and transmits it to Bob for buying a good or a service. However, the flaw in the network does not prevent her from resending the same BTC value to Carol, a third user. This allows Alice to spend the same Bitcoins twice.

## 2.3 Basic Concepts

The underlying technology of Bitcoin tries to solve the problem of transmitting value from one user to another in a trustless manner, using a distributed public ledger without involving any central authority. This problem requires technological solutions which ensure trust in a trustless environment. Just as a banking system needs to develop mechanisms to verify identity and maintain the security of the bank accounts, the Bitcoin system also solves these problems by combining cryptography, game-theory, computer networking and economics.

Cryptography is extensively used in the Bitcoin ecosystem. Hash functions and digital signatures form the bedrock on top of which the entire technological stack of Bitcoin and other crypto assets is built. To understand the mechanics of any transactions in this ecosystem, it is imperative to understand how these fundamental building blocks like hash functions and digital signatures along with economic incentive mechanisms make this system work effectively in a fault tolerant manner. The next section explains the basic concepts of hash functions and digital signatures which will be used repeatedly in the subsequent sections.

### 2.3.1 Hash Functions

In a blockchain like Bitcoin the copy of the public ledger is stored by many participants of the Bitcoin Blockchain called nodes. As each individual node can store, transmit, and validate transactions, it is important that the transactions are verifiable and temper-evident, as in an open and permissionless system like Bitcoin it is impossible to prevent tempering by malicious nodes, but possible to make the tempering evident through cryptography. In solving this problem in the blockchain, hash functions have a critical role to play. Many of us might have come across hash of a particular file while downloading it from a website. The purpose of the hash is to enable the user downloading the file to confirm that the file has not been tempered and is same as present on the website.

Hash functions are mathematical functions which map a string of any length to a string of fixed length based on certain algorithm that scrambles the input string to generate a seemingly unrelated output string of fixed length. It is like a unique fingerprint of given data string. It can be represented as shown in Fig.1

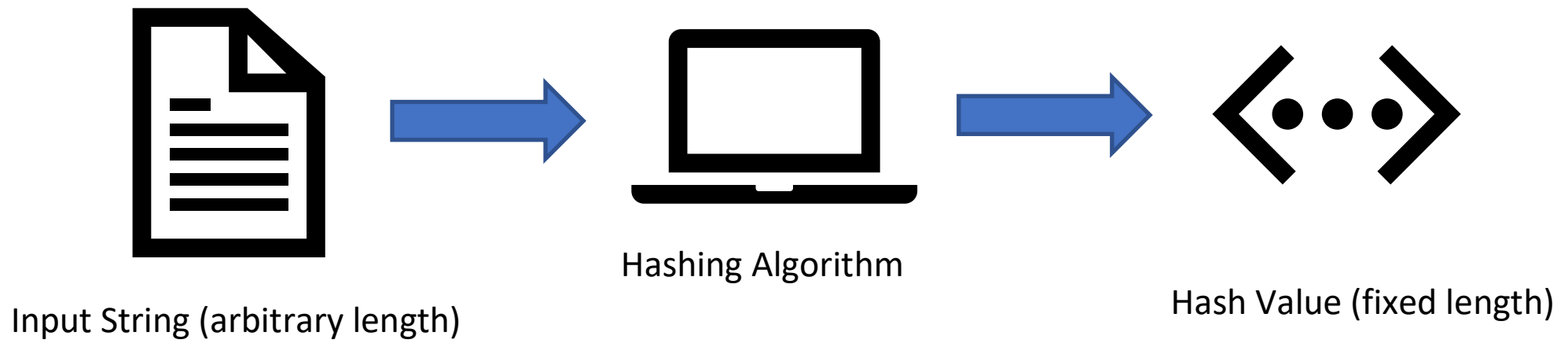

Fig.1 Hash Functions

For any hash function to be cryptographically secure it must have the following properties:

a) Computational Efficiency: For an input string of certain size, the time taken by the algorithm to calculate the hash or digest of the input should be a linear function of the size of the input string.

b) Collision Resistance: The hash value function maps the infinite set of strings of variable length to a finite set of hash values of fixed length. From the pigeonhole principle[7] it can be inferred that there exists a string $y$ for a string $x$ and hash function $H()$, such that $H(x) = H(y) = z$ when $x \neq y$ . Such a case is called a collision, when two non-equal input strings have the same hash value. However, for cryptographically secure hash functions, even though the collisions exist, it should be computationally infeasible to find two strings that have the same hash.

c) Hiding: It should be computationally infeasible for an attacker to guess the input string correctly by looking at the hash. And even if a slight change is made in the input string, the cryptographic hash changes completely.

This can be seen with the SHA256 hash of the string "Crypto Asset taxation is interesting" concatenated with various special characters as shown in Table 1.

| String | SHA256 hash |
|---|---|
| Crypto Asset taxation is interesting~ | 94c68e0e2fe8e816b7290352dd689c54cca15d4c18717586c84e37c6a81c1356 |
| Crypto Asset taxation is interesting@ | a233387a164f7ef4d4c630e5f1c73cdc1c87a3e68e437a917d3cdc971b0d603c |
| Crypto Asset taxation is interesting# | 5e47cbeb3a3926059ab42cef858386caa05ac6776fc8bfeba0294bd2b8a4266d |
| Crypto Asset taxation is interesting$ | b2a85c2e59a7e480950c885b8e28730addbe40fc713e5268d7e46a1d2f5c9397 |
| Crypto Asset taxation is interesting% | 9050a1d69e3d4cff17cbcfd73166d78f827d125e040641341e36f9a8d5bd6aa2 |
| Crypto Asset taxation is interesting& | cc7375d79bab0cac25f831dfb9fa05458423d37cfeaedfbdb61f577f94a6a601 |
| Crypto Asset taxation is interesting* | 7bd72477ad02701a977019bba8d1c919b2775358f9c25817cb442bc83bed459d |

Table 1. Changing hash Values with changes in input string

From the table, it can be inferred that changing a single character generates a totally different hash value. This characteristic of the SHA256 Cryptographic hash function is important for making the blockchain temper-evident, which means that every participant in the blockchain would be able to identify any tempering by a malicious actor with transactions or other block fields in the blocks that are already a part of the blockchain.

### 2.3.2 SHA256 security

Bitcoin uses the SHA256 algorithm for hashing. It is a secure hashing algorithm designed by the United States National Security Agency[8]. It is widely believed to be collision resistant as for a given string $x$ an attacker trying to find an input string $y$ for a hash function $H()$ such that $H(x) = H(y) = z$ would have to calculate at least $2^{128}$ hashes, which is an extremely large number. The highest hash rate generated by the Bitcoin network till March 2024 is ~600 Exa[9] hashes per second. Even at that hash rate it would take around 17,983,805,117[10] years to find a collision by brute force calculation, which is more than the estimated age of our universe. However, there are certain hash functions which do have efficient collision detection methods.

In case of SHA256 we do not have knowledge of any such efficient collision detection methods as of now, and it is largely believed to be collision resistant, as despite best efforts, collisions could not be found[11]. However, there is no way to prove that a hash function is collision resistant. In the past, hash functions have been broken and phased out of security systems. A famous case is that of MD5 hashing algorithm which was widely used but eventually phased out for use as a cryptographic hash function.

---

[7] If n+1 pigeons occupy n holes, then some hole must have at least 2 pigeons

[8] National Institute of Standards and Technology (NIST). Secure Hash Standard (SHS). Federal Information Processing Standards Publication 180-4.

[9] $10^{18}$

[10] ((2^128) ÷ (600 × (10^18) ×365×86400))

[11] https://steviecellis.medium.com/the-beautiful-hash-algorithm-f18d9d2b84fb

Wang & Yu (2005) found and published ways to attack MD5 hashing algorithm[12] and Magnus & Stefan (2005) produced two Post Script files which had the same MD5 hash[13].

Hash functions are so critical for blockchains as they are the primary mechanisms of creating trust about the integrity of information being exchanged between the nodes of the network. If an underlying algorithm like SHA256 breaks down, the trust in the Bitcoin blockchain would also crumble. Quantum computers can also significantly affect the cryptography used in Bitcoin. It is estimated that in the 2030s quantum computers would be able to crack SHA256[14] unless Bitcoin Blockchain migrates to quantum safe cryptography. As the blocks and transactions in the Bitcoin Blockchain are identified using the hash value of their header, a well-orchestrated attack by an attacker having a reasonably sized quantum computer can potentially cause significant damage to trust in the Bitcoin Blockchain. However, just as the traditional banks continuously improve the security of their credit cards and internet banking solutions, Bitcoin Blockchain can also find potential solutions to circumvent the likely threat from quantum computers.

### 2.3.3 Digital Signatures

The Bitcoin Blockchain needs to identify the originators of transactions and owners of funds who can spend their Bitcoins on the Bitcoin Blockchain using valid transactions. This is analogous to a banking system where the bank assigns a customer identification number to each customer and each customer has their respective username and passwords for online banking. It is a mechanism to establish the identity, to verify the originator and authorizer of the transactions. However, in centralized banking systems, additional checks are possible to verify the identity of the originator of the transaction in circumstances of suspicion and there is a rigorous process of onboarding which includes mapping the customer identification number to a natural person using Social Security Number, Passport etc. The beneficial ownership of the account is also captured while opening the bank account. As the Bitcoin Blockchain does not have any centralized authority to verify identity, it relies on cryptographic techniques like digital signatures to validate the identity of the participants in the system.

A digital signature is like a stamp of authentication of the origin of a message, transaction or document which establishes that the information did indeed originate from the person who has signed the message, transaction or document and it has not been altered. An excellent overview of how digital signatures work can be found in security tip (ST04-018) of the US Cybersecurity and Infrastructure Security Agency[15]. We often come across digital signatures in our daily lives while receiving digital documents from the government. Various PDF documents contain digital signatures which are verified as soon as the PDF file is opened. One such example is the AADHAAR document (Indian social security number) downloaded in PDF format.

The user downloading the document can ensure that the document has indeed been issued by the issuing authority i.e., the Unique Identification Authority of India (UIDAI) and has not been tempered with. Fig. 2 and Fig. 3 show how a user can verify the identity of the issuer and absence of any tempering in the document.

---

[12] Wang, X., & Yu, H. (2005, May). How to break MD5 and other hash functions. In Annual international conference on the theory and applications of cryptographic techniques (pp. 19-35). Springer, Berlin, Heidelberg.

[13] Daum, Magnus & Lucks, Stefan. (2005). Attacking Hash Functions by Poisoned Messages "The Story of Alice and her Boss.

[14] In the 2030s, quantum computers will be able to crack the SHA-256 algorithm used by Bitcoin

[15] https://www.cisa.gov/uscert/ncas/tips/ST04-018

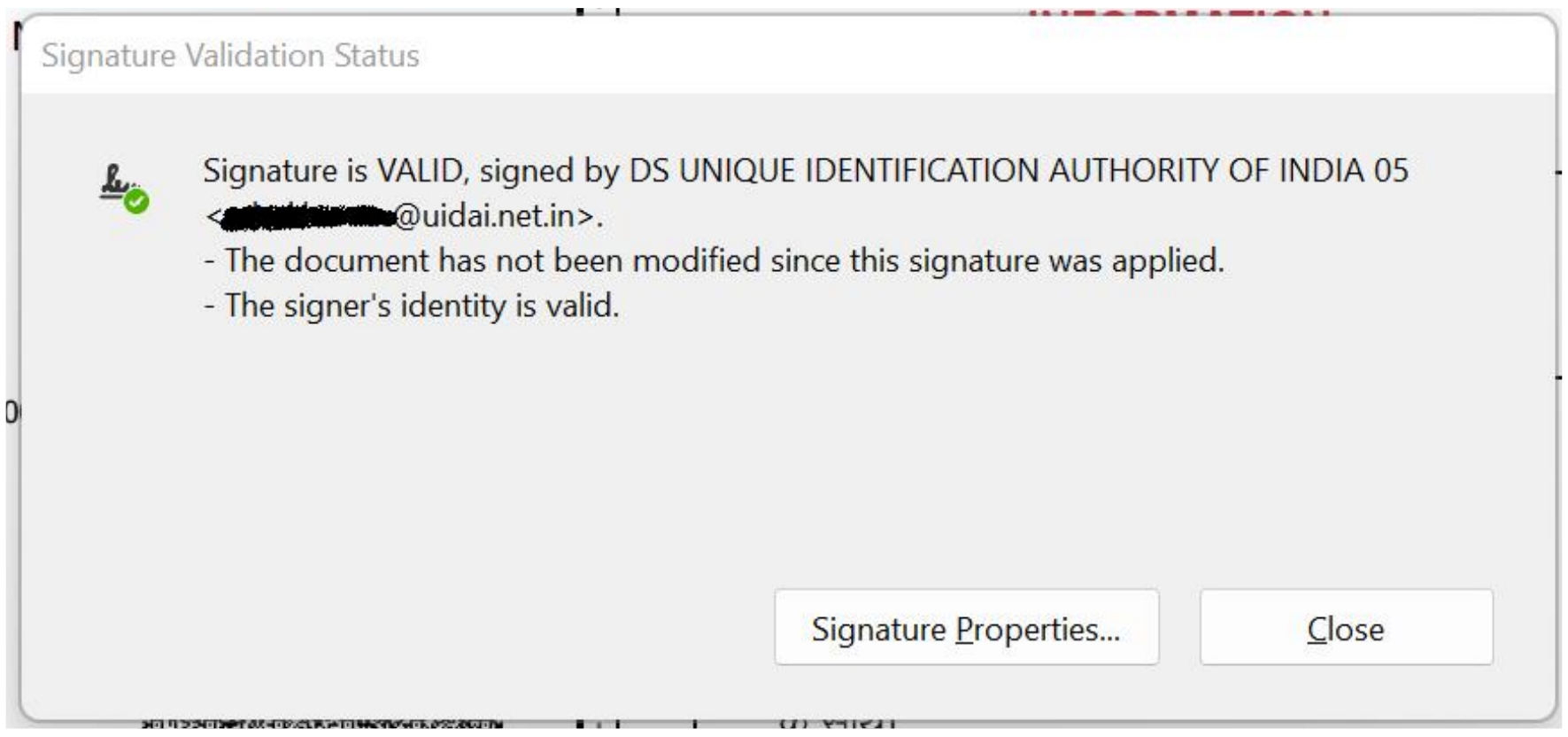

Fig. 2 Digital Signature being used to ensure that the document has not been modified

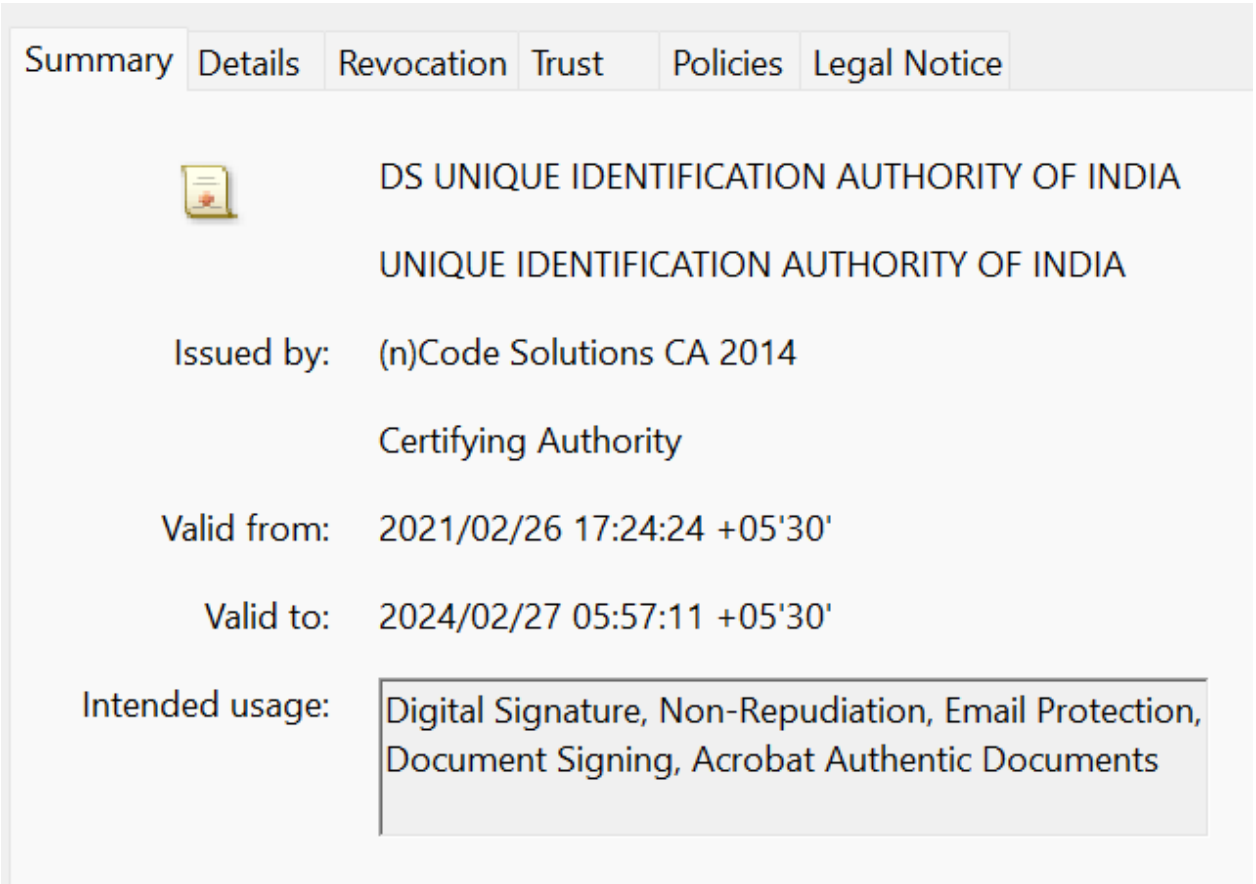

Fig. 3 Properties of digital signature

From the above figure it is clear that the said digital signature is issued by a Certifying Authority which contains the public key and the identity of the Unique Identification Authority of India (UIDAI). It is a kind of authorised directory of public keys which maps them to a natural person or an organization. As shown in Fig. 3 the digital signature can also be used for non-repudiation.

The signature of the Unique Identification Authority of India (UIDAI) on the AADHAAR document makes it impossible for the UIDAI to claim that it has not issued and signed the said document. The understanding of non-repudiation property of digital signatures is crucial for an investigator or tax auditor to prove that the funds in a crypto asset wallet are indeed controlled and beneficially owned by a natural or juridical person.

Digital signatures are based on public key cryptography. A public key cryptographic system uses a pair of keys with each pair consisting of a public and a private key. The public key is publicly known and can be shared with various users and the private key is kept secret. To prove identity, the transaction is signed(encrypted) using the private key and the signed as well as the original transaction is transmitted to the receiver by the sender. The signed(encrypted) message is then decrypted by the receiver using the public key of the sender, which is known to the receiver, and is matched with the original (unencrypted) transaction. If the two match, then it can be inferred that the message did indeed originate from the owner of the private key corresponding to the public key used to decrypt the signed message. For example,

Consider the private key
1ee283aa3024205e69c14b11b65685ac367213aba131e654d97020cfe5aa5818

and its corresponding public key
fc1a62a566ce576431cc2c96dac6766fb8fcea90b8da28307b1af1647a361f874c9604c3c98497a11538 7b0382df141d4cdda97bc8502a7cf7af267ce91d3f45

Alice has to send the message "Crypto Taxation is interesting" to Bob, and Bob needs to verify that this message is indeed from Alice, and that it has not been tempered. Thus, Alice uses her private key to encrypt the message "Crypto Taxation is interesting" and generates the signed (encrypted) message. She transmits the signed message along with the original message to Bob. Cryptographically the signed message would be

02df346b572d370d5bb19cd5a6de115335284d4c6de7683078eef9d6df11e14c236af35e817a303d3d ed6b3e63531e5ad26f6db006fbbf97ca6c96de60db3326.

Bob receives the signed message as well as the original message "Crypto Taxation is interesting". He decrypts the signed message using the known public key of Alice and compares it with the original message "Crypto Taxation is interesting". If the two messages match, then Bob can say with certainty that the message is indeed sent by Alice and has not been tempered. If a malicious actor tempers even a single bit of the original or signed message sent by Alice, Bob would come to know that the message has been tempered or was not sent by Alice. This use of digital signatures in blockchains makes them temper-evident. This can be seen pictorially in Fig. 4.

Public key cryptography is central to the addresses used by Bitcoin to send and receive Bitcoins. The public key is analogous to the bank account number in a banking system and private keys are analogous to the passwords used to authorize transactions. The public key is derived from the private key using a one-way function[16], which has the nature of making the computation possible only from the private key to the public key. These one-way functions are based on properties of large prime numbers. For example, it is very easy to find the product of two very large prime numbers, but it is extremely difficult, given the product, to factorize it into the two large prime numbers.

The public key can be easily derived from the private key using the one-way function, but it is practically impossible to derive the private key from a public key. Bitcoin uses the Elliptic Curve Digital Signature Algorithm (ECDSA) which has 256-bit private keys. This means that a Bitcoin private key is a random 256-bit number. 256-bit keys are highly secure and currently it is practically impossible for an attacker to derive the private key from a public key using brute force.

In the best case the attacker would have to guess $2^{128}$ keys, and even if the attacker uses the fastest supercomputer in the world with capacity of $1.1 \times 10^{18}$ floating point calculations per second, with 10 floating point operations to be carried to try a key, it would take around 1 trillion years to derive the private key. However, with the advent of quantum computers it is believed that algorithms like the Shor's algorithm can break public key cryptography in reasonable time by adopting more efficient methods of factorizing large numbers (Fahmy, 1997). This can pose a significant threat to public key cryptography in general and crypto assets in particular.

### 2.3.4 ECDSA and Bitcoin addresses

Bitcoin uses the Elliptic Curve Digital Signature Algorithm (ECDSA) for its digital signature scheme. There are multiple elliptic curves in ECDSA. Bitcoin uses the Secp256k1 elliptic curve to derive the public keys from private keys. However, even though public and private keys are required for bitcoin transactions, in order to reduce the size of transactions, instead of using the public key, Bitcoin uses the public address which is a modified version of the public key.

[16] In computer science, a one-way function is a function which is easy to compute for every input, but given a computed output, it is very hard to invert.

This transformation creates various types of addresses through the algorithm depicted in Fig. 5. Besides allowing for a greater number of transactions in a block due to lower transaction size, this transformation has some positive security implications, as it provides quantum resistance to existing public keys[17] which hold Bitcoin and reduces the possibility of errors that users might make in specifying the public addresses while sending the Bitcoins, due to the presence of a checksum.

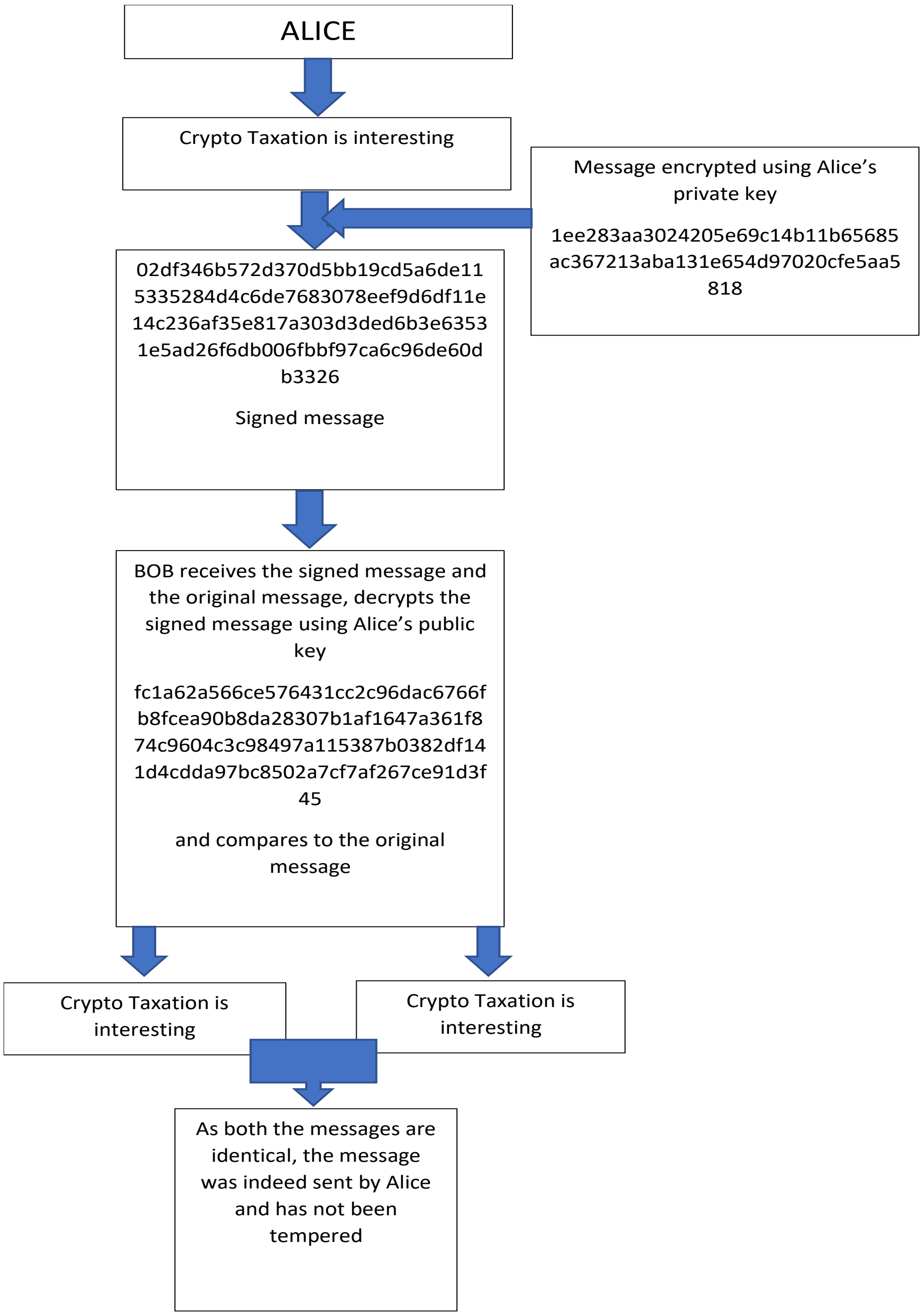

Fig.4 Signature verification in public key cryptography

[17] As discussed in 2.3.3 using Shor's algorithm on quantum computers it would be possible to obtain the private key corresponding to a given public key efficiently. However, if the attacker has a hashed version of the public key and not the public key itself, he/she need to find the public keys first using Grover's Algorithm

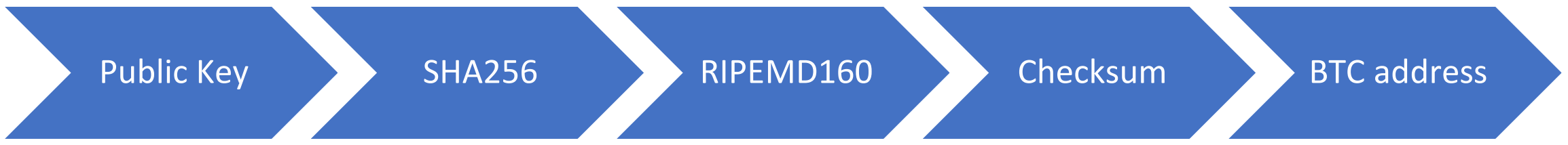

Fig. 5 Deriving a Bech32 Bitcoin address from a public key

For getting an identity on the Bitcoin Blockchain, the user must generate a random number between 0 and $2^{256}$, which becomes the private key of the user. The user derives the corresponding public key of this private key using the Secp256k1 elliptic curve of ECDSA and it is transformed into a Bitcoin address using the process shown in Fig. 5. In this process it is critical to have a good source of randomness as the key generation algorithm uses a random seed to generate the key pair.

There are multiple websites which help users to generate such key pairs, they use various methods to generate a truly random number by using techniques like moving a mouse pointer or entering random text as input. One such website is www.bitaddress.org . Fig. 6 is a screenshot from the website, which shows how it uses the mouse movement to generate a truly random number which is used to generate Bitcoin key pairs.

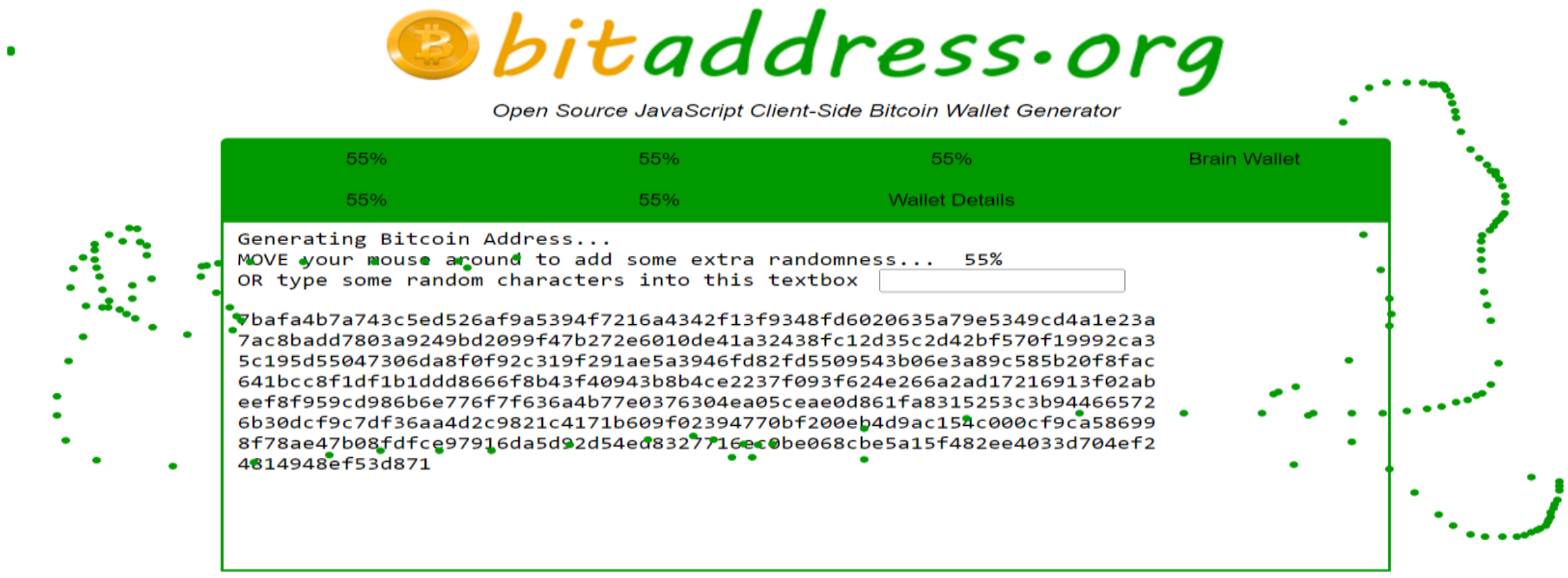

Fig. 6 Random number generation using mouse pointer for generating Bitcoin private key

The generated Bitcoin address is shown in Fig. 7. It is important to note that in this process of generating a Bitcoin address and its corresponding private or secret key, there is no requirement to provide any identity related information. No Know Your Customer (KYC) procedure is required to generate this valid Bitcoin address which can be used for transaction on the Bitcoin Blockchain. The validity of the bitcoin address can be easily verified on the Bitcoin Blockchain Explorer as given in Fig. 8.

As we have freshly created this address there are no transactions on this address. Similarly, a user can create any number of addresses and make transactions on the Bitcoin Blockchain without undergoing any Identity verification or KYC. Unlike the digital signatures which had their public keys verified by a central authority like in case of UIDAI, the Bitcoin Blockchain does not require attaching any real-world identity to the digital signature.

This has implications for tax enforcement and other law enforcement agencies, as this identity on the Bitcoin Blockchain is not mapped to a real-world identity. Unlike details like Swift Code and a name or possibly a social security number attached with a bank account, there is no such personal or geographical information associated with a Bitcoin address. However, there are some mechanisms to do so on a probabilistic basis. This makes it pseudonymous.

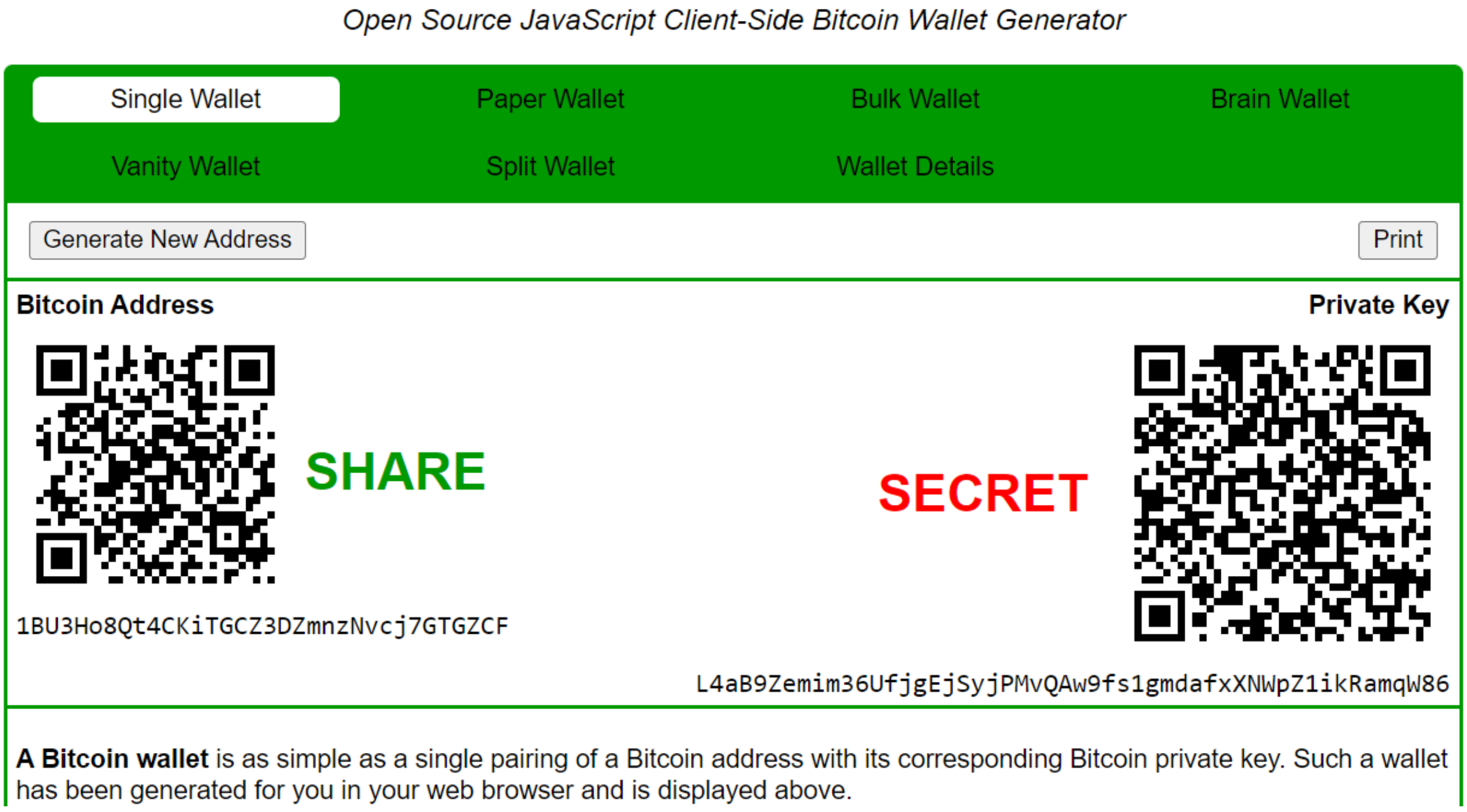

Fig. 7 Bitcoin address generated using the Random Seed

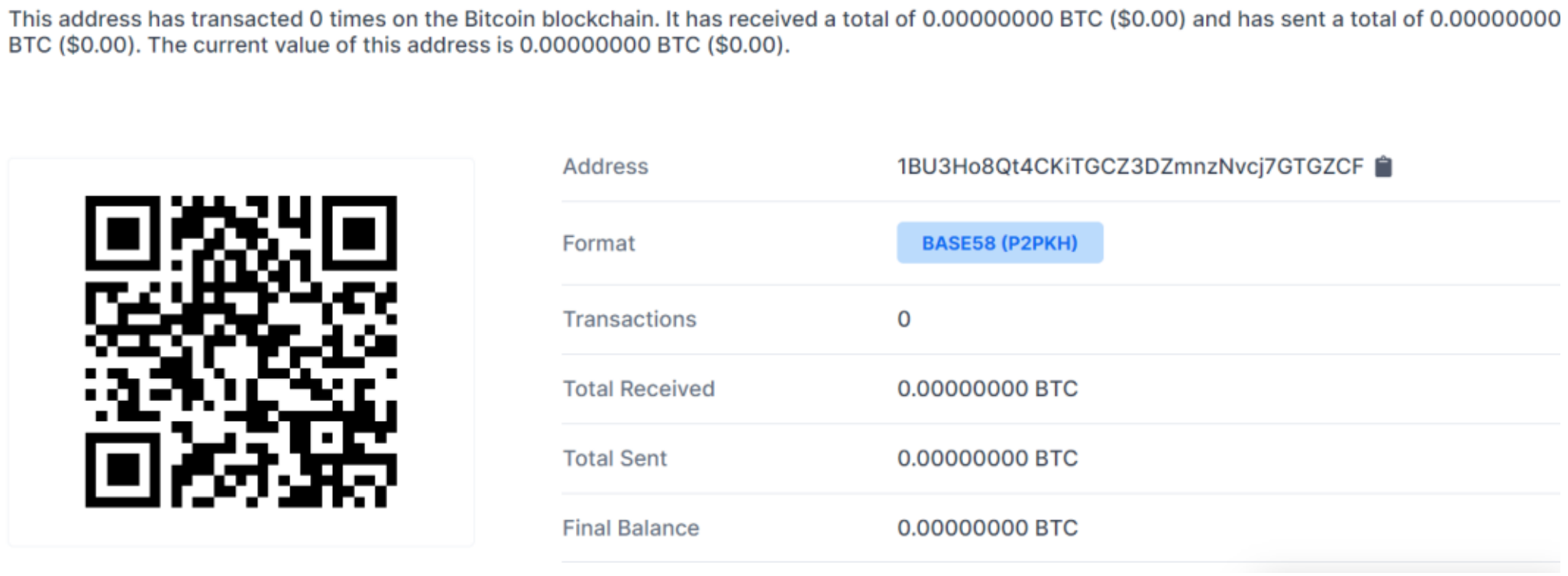

Fig. 8. The newly created Bitcoin address displaying zero BTC amount on Bitcoin Blockchain

In this case, methods like subpoena and summons cannot be issued to a central authority like a bank or a stock exchange to obtain information about the beneficial owner of this Bitcoin address. Methods to map a real-world identity to this Bitcoin address are usually based on blockchain analytics and inference. In case of a centralized exchanges where users do not generate their own key pairs and the exchange does this for them, the KYC information is captured before on-boarding a customer and can be obtained after satisfying the legal requirements, through their law enforcement portal. However, such centralized exchanges might primarily rely on the self-declaration of tax residency by the user.

The Bitcoin address is basically a 160-bit number. Therefore, a maximum of $2^{160}$ unique addresses can be generated. This might raise concerns for some users about collisions in Bitcoin address generation, i.e., two users independently generating same Bitcoin addresses and their corresponding private keys. However, as the number of addresses is too large, the probability of a collision is almost non-existent. For a detailed mathematical analysis of this problem, you may refer to an article by andytoshi on The Bitcoin Birthday Paradox[18]. It concludes that there would be a 0.1% chance of a collision if $5.4x10^{22}$ addresses are in existence. Still there is no reason to worry about collisions as:

[18] https://download.wpsoftware.net/bitcoin-birthday.pdf

i) The probability of collision is extremely low (1 in $10^{48}$) for each independent collision
ii) It would take trillions of years to reach a stage where so many Bitcoin addresses are generated that the newly generated addresses have a significant chance of colliding with the old ones.
iii) Since only 21 million Bitcoins can ever be generated and the smallest possible unit of Bitcoin is 1 Satoshi or $10^{-8}$ Bitcoin, only $2.1 \times 10^{14}$ addresses can possibly have any amount of Bitcoin in them and an attacker would have to generate trillions of Bitcoin addresses before he/she can find a collision address which has Bitcoin balance.

Thus, it is practically impossible for two Bitcoin users to generate the same address independently unless they have used the same or predictable source of randomness. So, if a taxpayer presents the argument of his/her Bitcoin being stolen by someone who maliciously generated their public and private key pair, the taxman should know that it is practically impossible for anyone to do so, and the private key might have been compromised in some other way. In the next section we would see how the addresses and their corresponding private keys are used to transact on the Bitcoin Blockchain.

From the point of view of tax investigation and enforcement as well as for other law enforcement agencies it is important to have a general understanding about the characteristics of the public and private keys as well as the public addresses of Bitcoin and other crypto assets. For investigators involved in enforcement actions as well as the forensic analysis of evidence, it is important to be able to identify crypto assets address and keys to find the quantum and flow of funds.

It would also be important to use regular expressions to identify certain crypto asset addresses and keys located in mobile devices or hard drives. Although, nowadays various forensic investigation tools provide this functionality. However, knowledge about the various types of addresses of crypto assets might help the investigators and the taxman to prima facie identify the crypto assets involved in the investigation.

The algorithm shown in Fig. 5 generates 3 types of Bitcoin addresses which can enable an investigator to identify/suspect a string of characters representing a Bitcoin address. Awareness about Bitcoin address types can also aid an investigator in tracking the flow of funds in cases of theft, fraud or establishing the ownership or control of a Bitcoin address. The three types of Bitcoin addresses generally used in transactions are:

i) P2PKH addresses: These addresses are the first version of Bitcoin addresses and are also known as legacy addresses. They start with the number '1' and have 26 to 36 characters. The transactions using these addresses are larger in size and hance require more transaction fee making this address type inefficient. For example, 1KfVFNtxkknugXhA9uYkohMWxG8f78nyax
ii) P2SHaddresses: These are like P2PKH addresses and provide more complex functionality like multiple signatures for spending the Bitcoins. They start with the number '3'. For example, 3FMFtFAJxKmvpumn3sGjA6xjXw1M4Bw97v
iii) P2WPKH or Bech32: These transactions are advance transactions which help to reduce transaction size and make them faster and more efficient. These are used in SegWit transactions, a new format for Bitcoin transactions in which to accommodate more transactions in a block, certain information (regarding witnesses) was removed from the input field of the block. These addresses start with 'bc1'. For example, bc1q9jayxqvah5gynukddmmvs7jc9xwjc0c6emulpp

Similarly the Bitcoin private keys start with 'K','L' or '5' like

```
Kx4cBkAHgD9CrYNhTM12P5cNgVfwTeG5nN2R4KxcZjPPLx7DfrEr
L5WNtTRjhqisXpnkCbr2mRu6oR7k34axzggh4nyXKuQU6aozLnkQ
5J5PZqvCe1uThJ3FZeUUFLCh2FuK9pZhtEK4MzhNmugqTmxCdwE
```

These patterns can help investigators to identify the use of Bitcoin blockchain by the taxpayer or suspect. The regular expressions[19] for Bitcoin and other crypto assets' addresses and keys can be used to look for use of such assets by the taxpayer or suspect on a computer or disk using specialized software.

In Bitcoin, mnemonics-based keys, often referred to as mnemonic seeds or recovery phrases, as shown in Fig. 8A, are a human-readable and memorable representation of the private key or seed used to generate a deterministic wallet. These phrases typically consist of a sequence of 12, 18, or 24 words chosen from a predefined list of words. Mnemonics serve as a convenient way for users to back up and restore their cryptocurrency wallets without needing to store or remember the raw cryptographic keys themselves. They are crucial for wallet recovery in case of loss, theft, or hardware failure, as users can simply input their mnemonic seed into a compatible wallet software to regain access to their funds. Mnemonics-based keys adhere to the BIP-39 standard[20], ensuring compatibility across different Bitcoin wallets and implementations. Thus, whenever an auditor or investigator comes across such mnemonics it should be identified as a seed phrase for a crypto asset wallet.

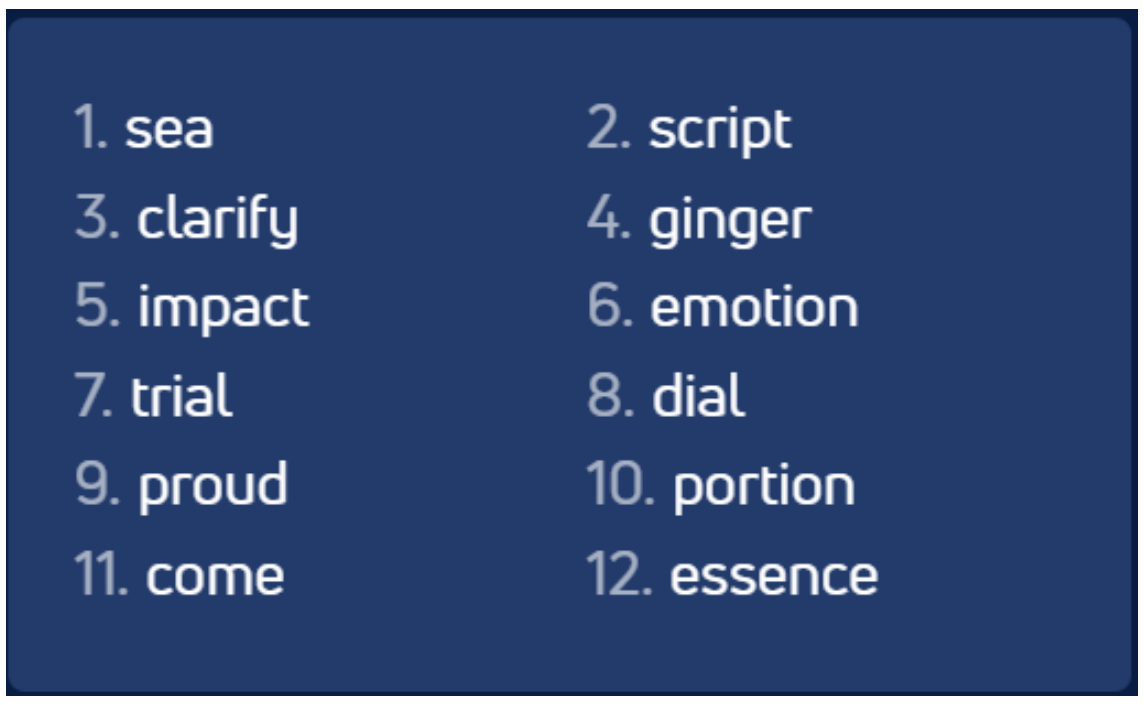

Fig. 8A mnemonics-based keys

### 2.3.5 Bitcoin Transactions

Bitcoin is primarily a means for transmitting value from one user to another through an open, permissionless and trustless network. Bitcoin Blockchain has multiple mechanisms to ensure the validity of transactions which make sure that it is usually impossible to carry out fraudulent transactions. A transaction is usually of the form of Alice paying Bob a certain Bitcoin value. The various aspects of ensuring security and integrity of transactions are:

i) The users must own the Bitcoin value that they wish to transfer and should be able to prove the ownership unambiguously on the network. For example, if Alice has 2.5 BTC the Bitcoin Blockchain based on a consensus, should agree that Alice owns 2.5 BTC
ii) Appropriate funds should be available with the user. It should be impossible to transfer 3 BTC when a user owns only 2.5 BTC[21].
iii) The originator and authorizer of the transaction must be verifiable as Alice
iv) The transaction should not have been tempered

One of the most important ideas in a transaction on Bitcoin Blockchain is that of an Unspent Transaction Output (UTXO). UTXOs are quantities of Bitcoin sent to various public addresses which have not yet been spent by them. They represent the units of Bitcoin which can be referenced by the owners of the Bitcoin addresses to which the UTXOs are assigned, and can be spent on the Bitcoin Blockchain.

[19] Regular Expressions are patterns which are used for matching strings by computers
[20] https://en.bitcoin.it/wiki/BIP_0039
[21] There is no concept of loans or any overdrafts in Bitcoin transactions.

For every valid transaction Alice must reference a UTXO which is a result of previous transaction by Alice herself or some other address in the Bitcoin Blockchain which transferred Bitcoins to Alice's Bitcoin address. The only exception to this is the first transaction in every block, called the Coinbase transaction, which results in newly created bitcoins being assigned to the miner of the block as reward. Coinbase transactions do not require any input UTXO, but only the pay-out Bitcoin address.

UTXOs are analogous to cheques which are written in the name of a specific Bitcoin address owner and can be spent only by them. It also important to note that while Bitcoins are fungible, UTXOs are not fungible. A user must use the entire value of the input UTXO either by transferring Bitcoin value to other Bitcoin addresses, their own address or as transaction fee for miners. The difference in the value of the input and output UTXOs is deemed to be the transaction fee for miners.

UTXOs can be imagined as piggybanks assigned to specific users and every time the user intends to spend the amounts kept in the piggybank, they must break the entire piggybank assigned in their name, and create new piggybank(s) in the name of the intended transferees or the transferor himself/herself which can only be spent by them. In the transaction depicted in Fig. 9 these new output UTXOs (piggybanks) are for the Bitcoin addresses of Bob, Carol, and Alice. Any change that is not assigned to either Bob and other transferees or Alice's new piggybank, is by default the miner fee given by Alice for adding her transaction on the Bitcoin Blockchain.

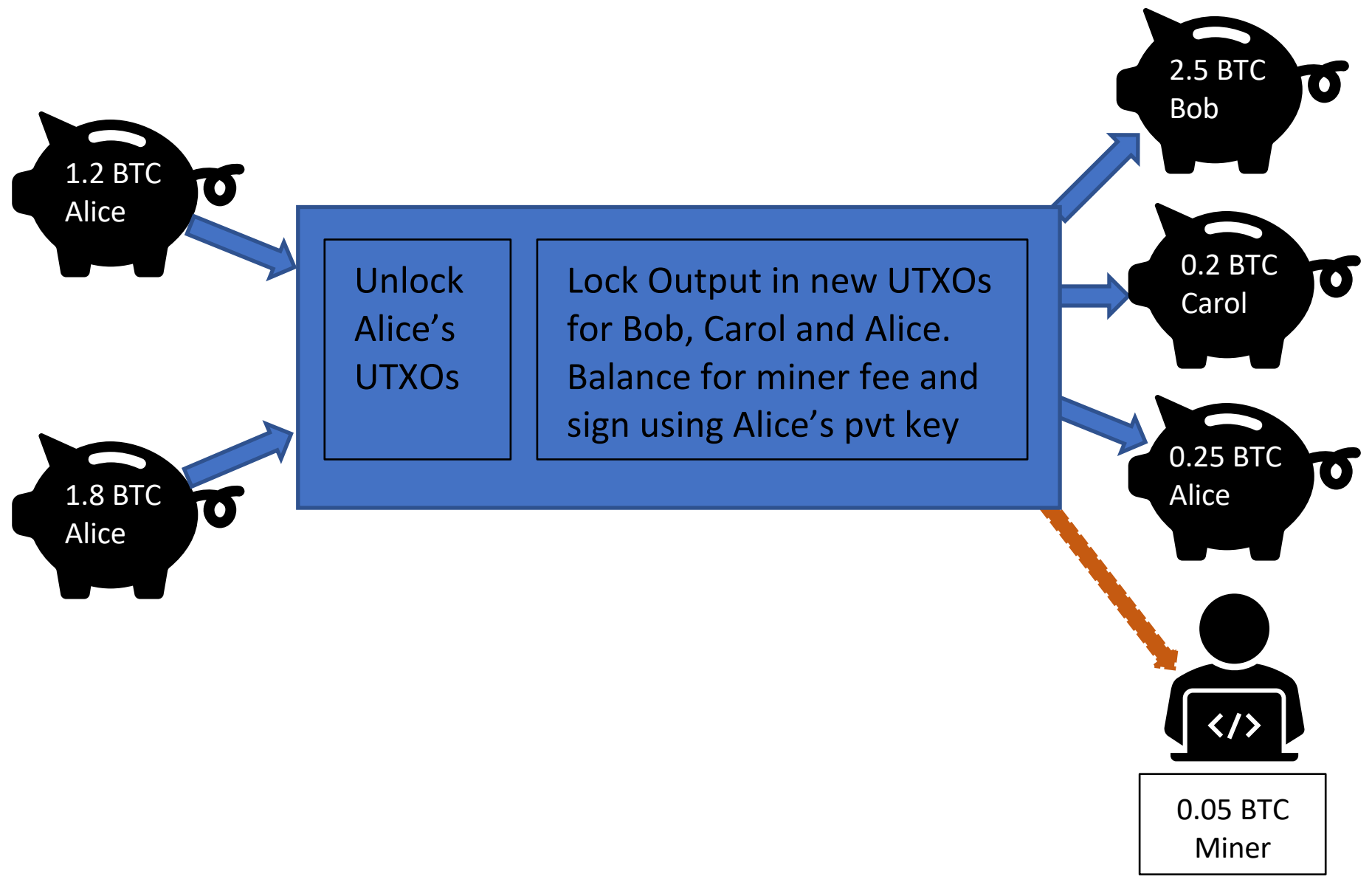

Fig. 9 A Bitcoin Transaction

In the above transaction Alice sends 2.5 BTC to Bob, 0.2 BTC to Carol and 0.25 BTC to herself (to a new Bitcoin address controlled by her or her existing Bitcoin address) the balance amount of 0.05 BTC is the miner fee. The UTXOs are in the form of a locked script which can only be spent by a user who has a public address, same as that specified in the locking script. It is a kind of cheque written for the user whose public address is same as that specified in the script.

The input UTXOs of 1.2 and 1.8 BTC have been locked using a script by the sender for Alice's Bitcoin address. So, Alice refers to the input UTXOs in the input side of her transaction and specifies the output Bitcoin addresses of Bob, Carol, and herself along with locking scripts and signs the transaction using her private key. She then broadcasts the transaction signature along with her public key which allows the nodes of the Bitcoin Blockchain to make sure that the transaction has indeed originated from Alice and that she is the one entitled to spend the input UTXOs of 1.2 and 1.8 BTC.

The nodes compare Bitcoin address corresponding to the public key of Alice with the Bitcoin address specified in the input UTXO, if they match, then they infer that indeed Alice is entitled to spend the Bitcoin amount contained in the input UTXO. Then the nodes verify the signature of the transaction broadcasted by Alice. If the signatures are verified, then it can be inferred that indeed Alice has authorized the spending of the input UTXOs.

This ensures that only Alice can spend the input UTXOs owned by her Bitcoin address. Suppose an eavesdropper Eve knows Alice's public key somehow and tries to spend her UTXOs. She cannot do it as Eve does not know the private key of Alice and would have to sign the transaction using her own private key or any other private key. The nodes of the Bitcoin network would fail to verify the signatures and would come to know that it is not Alice but someone else who has signed the transaction or the transaction has been tempered.

In case Eve uses her own private and public key pair to create a transaction having UTXOs meant for Alice as input, the Bitcoin address corresponding to the public key of Eve will not be same as that of Alice's public key and the nodes will be unable to successfully run the unlocking script of UTXOs meant for Bitcoin address of Alice, making Eve unable to spend Alice's UTXOs. This combination of locking and unlocking scripts along with digital signatures makes sure that UTXOs are not spent fraudulently by malicious actors.

While broadcasting the transaction, Alice transmits the signature of the transaction along with her public key. This has potential implications for security threats from quantum computers, as until only the Bitcoin address is known to the attacker with a reasonably resourceful quantum computer, he/she will have to first find out the public key itself using Grover's Algorithm and then use Shor's Algorithm to find out Alice's private key in reasonable time. However, with the disclosure of public key while spending the UTXOs, it might become easier for an attacker to find the private key of Alice before the transaction is included in the next few blocks, allowing him/her to spend Alice's UTXOs fraudulently.

It can also be seen that a Bitcoin transaction[22] on the blockchain has no narration, purpose code or identity or IP address information of any device and natural or juridical person attached with it. Thus, by merely looking at the transaction it is not possible to know the device or IP address that originated the transaction as well as the real natural or juridical persons behind it or the purpose of the transaction. However, some blockchain analytics and inferences can help in such identification.

#### 2.3.5.1 Tax Implications of UTXO Based Transactions

The UTXO based design of Bitcoin and other similar crypto assets like Bitcoin Cash and Monero can have implications for determining the cost basis for calculation of the capital gains liability. As the UTXOs are not fungible, each UTXO(s) belonging to a public key(s) of a taxpayer can be uniquely identified. While transacting on the Bitcoin Blockchain, the taxpayer might choose an optimal combination of the input UTXOs which minimizes the transaction fee, but this might give rise to different tax liabilities due to possibly different cost basis of each UTXO. This can lead to significant complexity for owners of crypto assets while reporting the basis of transactions and calculating their tax liability from crypto asset transactions.

[22] https://www.blockchain.com/explorer/transactions/btc/e9412cfda50be55c0a1989fb0ac7a2514a2f47852804d4d7cfdebb33f5e780f3

To simplify the calculation of cost basis and the tax liability many tax administrations allow various accounting methods to determine the cost basis of crypto assets. The methods are as follows:

i) First In First Out (FIFO): The earliest brought crypto asset is sold first
ii) Last In First Out (LIFO): The most recently brought crypto asset is sold first
iii) Highest In First Out (HIFO): The most expensive crypto asset is sold first
iv) Specific Identification (Spec ID): The specifically identified asset in record is sold. This is the method most consistent with the UTXO based design of Bitcoin and other similar crypto assets.
v) Average Cost Basis (Known as Share Pooling in the UK): A method wherein identical assets are pooled and the average cost basis is taken for each of these pools. Depending on the average basis being taken for the entire financial year or for each disposal the methods are also known as the total average and moving average methods respectively.
vi) Periodic Basis: The cost basis in some jurisdictions is determined by the value of the crypto asset at the beginning of each tax year. This method is used by the Belastingdienst
vii) PVCT (Plus-value à court terme) method: While disposing a crypto asset, the cost basis is calculated by the fraction of the acquisition cost of the entire crypto holdings relative to the sales proceeds. This method is used by the Direction Générale des Finances Publiques

As one cannot transact with oneself, the transfer of crypto assets from on wallet address, hosted or un-hosted, owned by the same person or entity is not taxed. However, such transfers would require maintaining records of the original incoming crypto asset for an accurate determination of the cost basis for determining capital gains or losses at the time of disposal of the crypto asset.

Choice of different methods can have an impact on the tax liabilities of individuals as well as the revenue collected through taxation on crypto assets. As crypto assets resemble securities in many aspects and some have even been classified as securities by regulators like the SEC, they are also susceptible to "wash sales". However, as in some jurisdictions, wash sale rules apply only to securities, and crypto assets are classified into different asset classes in multiple jurisdictions, investors can use the regulatory arbitrage to lower their tax liability. There is also evidence of Tax Loss Harvesting (Cong, Landsman, Maydew, & Rabetti, 2023) in jurisdictions like the US to lower the tax liability.

### 2.3.6 The Bitcoin Blockchain

Wikipedia[23] provides a comprehensive definition of a blockchain as "*A blockchain is a decentralized, distributed, and often public, digital ledger consisting of records called blocks that are used to record transactions across many computers so that any involved block cannot be altered retroactively, without the alteration of all subsequent blocks. They are authenticated by mass collaboration powered by collective self-interests. This allows the participants to verify and audit transactions independently and relatively inexpensively. A blockchain database is managed autonomously using a peer-to-peer network*"

The distributed public ledger of Bitcoin is based on the blockchain. The transactions in a blockchain are grouped together in the form of blocks and the subsequent blocks are cryptographically linked to the previous blocks, thus forming a chain. A complex interplay of cryptography, distributed consensus regarding the state of the blockchain and a system of economic incentives keep the Bitcoin Blockchain secure. It leads to only valid blocks being added to the blockchain and prevents malicious actors from compromising the integrity of the blockchain. Blockchain is not the only way to structure data and blocks for crypto assets. Some crypto assets like IOTA, Nano and Obyte use Directed Acyclic Graph framework to store data[24].

---

[23] https://en.wikipedia.org/wiki/Blockchain
[24] https://finance.yahoo.com/news/cryptocurrencies-dag-based-framework-why-081019399.html

Unlike the banking system, the copies of ledgers in a public blockchain are maintained in the form of a distributed database which can be openly accessed by anyone. The transactions broadcasted by the users of the blockchain are verified by the nodes of the Bitcoin network and are included into a block only when they are found to be valid. Addition of each block to the top of the existing blockchain requires solving a rigorous mathematical puzzle by nodes called miners which add a new block to the blockchain. The promptest miner to solve the puzzle gets to add the block created by it on top of the existing blockchain and extend it. It also earns him/her the mining rewards in the form of newly created Bitcoins through the Coinbase transaction.

This difficulty in adding another block to the existing blockchain makes it practically impossible for any malicious actor to temper the earlier blocks or transactions. Any effort at tempering becomes evident to the participating nodes and due to the enormous computational resources required for mining, it is extremely difficult for a malicious actor to create a blockchain which is a few blocks different from the original blockchain. A system of economic incentives promotes honest behaviour by participants, as they own or hope to own Bitcoin in the form of miner's reward, and have an incentive to keep the trust in the blockchain by maintaining its integrity.

The structure of the blockchain is shown in Fig 10. Each block contains a header and several valid transactions. The header of the block consists of the following:

i) Merkle Root: It is the hash of the merkle tree[25] which contains all the transactions in the block. Its main objective is to ensure that no transaction has been modified in the block. It is also used to quickly verify if a transaction is a part of a block.

ii) Nonce: It is a value which is the solution of the mathematical problem that the miner must find to mine the block. A nonce value once found by a miner can be easily verified by anyone.

iii) Previous Block Hash: Every Bitcoin block contains the hash of the previous block. This makes the blocks connect in the form of a chain as shown in Fig. 10, hence the name blockchain. This property makes the blockchain temper evident as any change in a block changes the block hash and breaks the blockchain as shown in Fig. 11

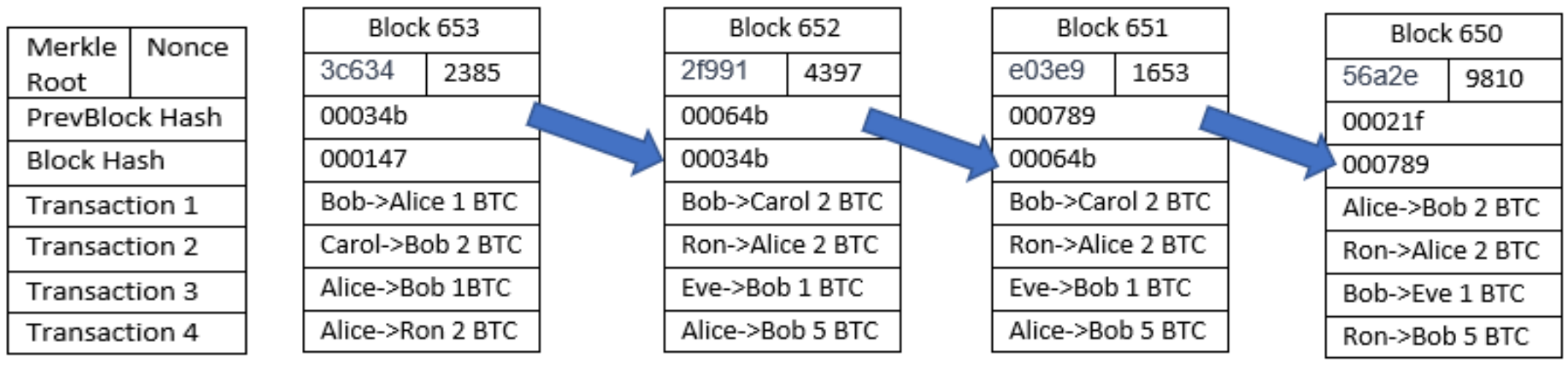

Fig. 10 Blockchain Structure (representative)

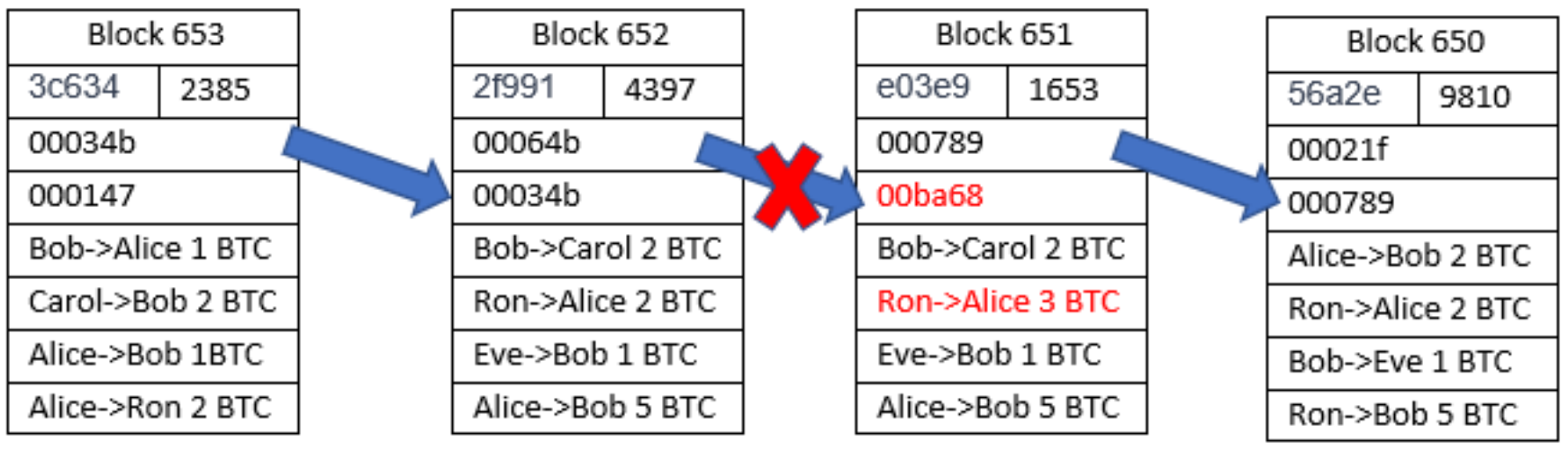

Fig. 11 Maliciously modified blockchain

[25] In Computer Science, A merkle tree is a data structure which is used for verification of data in a large distributed dataset. It allows huge amount of data to be efficiently mapped in the form of a tree. If it is required to check if any changes have been made to the data in the merkle tree it can be done by checking if the data is consistent with the root hash instead of traversing the entire merkle tree.

If a malicious actor tries to alter the transaction on Block 651 where Ron sends 3 BTC to Alice instead of 2 BTC (Assuming that the added transaction is indeed authorized by Alice). As one of the transactions in the merkle tree has been changed, the merkle root will change and consequently the block hash would change. This would make the actual hash of the Block 651 inconsistent with the hash of the Block 651 recorded in the header of Block 652. The nodes on the blockchain would take cognizance of this and rely on the longest consistent chain instead and continue adding new blocks on top of that chain.

Bitcoin blocks are identified using the unique hash of their header. This relies on the collision resistance of SHA256 hashing algorithm. This is the reason why if an attacker, equipped with a quantum computer can find an input $x$ such that $H(x) = H(Header\ of\ Block\ 651) = 00064b$ then the integrity of the blockchain can be seriously compromised.

### 2.3.7 Bitcoin Mining

The Bitcoin Blockchain needs to achieve a distributed consensus in which the participants of the network agree on which blocks form a part of the blockchain. As various nodes broadcast new transactions to be included in blocks and added to the existing blockchain, there must be a mechanism to decide which transactions will be added to the blockchain and which node adds the new block. The new blocks added to the blockchain must be verified and the nodes must be in sync about the current state of the blockchain. This common view of the blockchain is called a consensus.

We come across the term miners and mining very often in the context of crypto assets. To effectively tax mining it is important to understand what exactly does it mean to mine Bitcoin. This facilitates the understanding of the direct and indirect tax implications of this service of adding blocks to the blockchain which creates economic value in the Bitcoin ecosystem. Many tax administrators might want to have an idea about the trends and extent of mining in their jurisdiction. It can be estimated using trends in network traffic and some services like Bitnodes[26]. However, many Bitcoin nodes use the ToR network and are very hard to trace using their IP addresses.

Also, as the probability of earing mining rewards is based on the percentage of total hashing rate (computational power) available with the miner, most individual miners join a mining pool. A mining pool significantly reduces the variance of rewards received by a miner. For a fee, the mining pool makes the rewards almost certain if the individual miner has a hashing rate above a certain threshold. Mining activities can result in direct as well as indirect tax events which are discussed later.

The Bitcoin Blockchain needs to have a robust mechanism to ensure that only valid transactions are included in the blockchain and a system of incentives is in place to encourage honest actors and disincentivize malicious actors in the system. It is important to note that not all nodes of the Bitcoin network are involved in mining, as mining is a very resource intensive activity. Some nodes only relay the transactions received by them to broadcast those transactions to the entire network effectively. Only nodes involved in mining can add another block to the blockchain.

One of the important problems to be solved by a miner node is to select the transactions out of the broadcasted transactions to be included in the block which the miner would try to mine. When a Bitcoin user submits a transaction to the Bitcoin network, it is broadcasted using a flooding algorithm to the entire Bitcoin network and the miner nodes come to know about its existence. The broadcasted transactions form what is known as a mempool, which is a pool of unverified transactions being broadcast on the Bitcoin network[27]. The transaction fee and the block reward are the incentives for the miner nodes to verify transactions and mine new blocks to be added to the blockchain.

[26] https://bitnodes.io/nodes/live-map/

To gain maximum reward in Bitcoins, the miners solve an optimization problem like the knapsack problem[28] in Computer Science (Fig. 12) while picking up the transactions from the mempool for inclusion in Blocks to maximize their block rewards. The miners check if the transactions are valid and the senders have the funds and the authorization to spend the UTXOs and construct the merkle tree of the transactions to form a block. The miner also puts its own Bitcoin address in the Coinbase transaction for receiving the block reward. It is usually believed that miners pick up the transaction with the highest transaction fees, but it is not entirely correct. Bitcoin transactions have multiple inputs and outputs which are measured in bytes. However, for the purpose of inclusion in the blocks, another criterion called weight units is used, as all bytes in a Bitcoin transaction do not have the same weight.

The miner uses the criterion of "fee per weight unit" for inclusion in the mining block and maximizes his/her payoff. For example, if a miner must choose between a transaction which has 4 input UTXOs and 4 output addresses and the other transaction having 2 input UTXOs and 1 output address, and both transactions have the same transaction fee. Then the miners would pick up the second transaction over the first one as it is smaller and has more transaction fee per weight unit. This would enable the miner to fit more transactions into the block as the limit for Bitcoin transactions in a block is 4 million weight units[29].

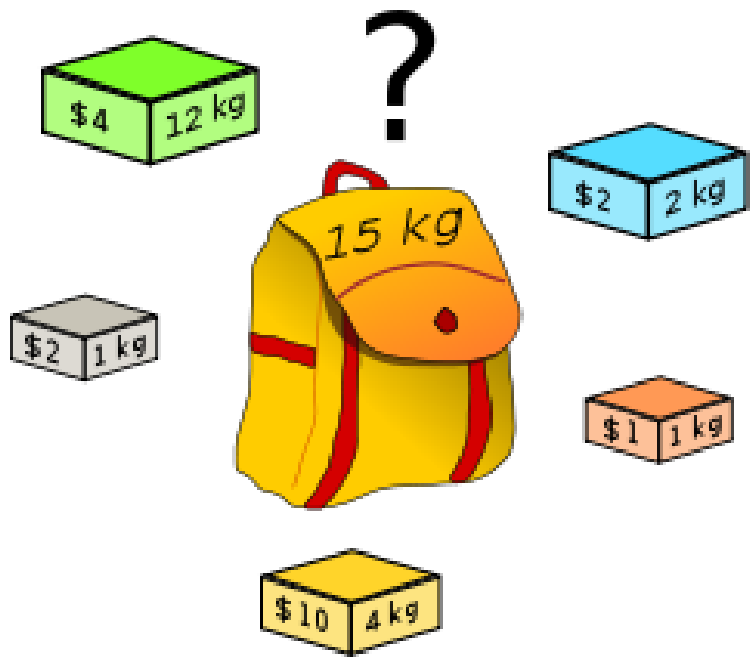

Fig. 12 The Knapsack Problem[30]

The next problem in mining is to select the node which would add a valid block to the blockchain. One possible solution for it could be to randomly select a miner node and allow it to add a valid block to the blockchain. This solution will rely on the premise that most of the times an honest miner node will be selected and it would add a valid block to extend the blockchain. However, this solution has a potential problem of the network getting flooded by numerous malicious miner nodes largely as the marginal cost of adding another malicious miner node would not be much. Thus, there needs to be some barriers to entry for participation in the validation process of transactions, adding new blocks and gaining from the consequent rewards. Computational intensity of mining acts as such a barrier.

Bitcoin solves this problem using proof-of-work mechanism. In the proof-of-work mechanism, each node involved in mining must perform intensive computation to find the solution to a mathematical problem to '*mine*' a block and add it to the blockchain. Once the solution has been found, a node can broadcast the solved block to the entire Bitcoin network and claim the reward for validation. When a node involved in mining solves the mathematical puzzle and adds a block to the blockchain, it is said to have *mined* a block.

[27] This pool of transactions getting arranged into blocks and being mined, along with other parameters of Bitcoin network can be visualized at https://mempool.space/ and https://mempool.space/mining
[28] https://en.wikipedia.org/wiki/Knapsack_problem
[29] https://en.bitcoin.it/wiki/Weight_units
[30] Source: https://upload.wikimedia.org/wikipedia/commons/thumb/f/fd/Knapsack.svg/375px-Knapsack.svg.png

To understand the mathematical puzzle that the miners solve to mine a Bitcoin block, it is pertinent to look the block hash of some of the actual Bitcoin blocks mined recently and added to the blockchain. One can go to any Blockchain Explorer[31] and find out the block hash of the last 10 recently mined blocks of Bitcoin Blockchain. They are shown in Table 2.

| Block Height[32] | Block Hash |
|---|---|
| 832072 | 00000000000000000000fd8369848fa6bcb3bef188d76fc7a1d38992a5e881d1 |
| 832071 | 000000000000000000023f411f7ebea61731f0d7990a4da1f7106375501d100e |
| 832070 | 00000000000000000002a7473666215bd65b39e74ffbd7b9d2eccb8336430f4a |
| 832069 | 00000000000000000001907e66f40f2bd56e267d73dd37a7541161efea891808 |
| 832068 | 000000000000000000004df71f0b4419e9c3355dd2c399b19bd56f65ad58c113 |
| 832067 | 000000000000000000004ff4ea42aa1a467363d22e524913e6913eaf5d24e8fb |
| 832066 | 000000000000000000009de6eee5f331dfdbb1ef59a15ea095552a00983bdbc0 |
| 832065 | 00000000000000000002782005e1e33e276d47763cc86822983cd258e04d1614 |
| 832064 | 000000000000000000021a6cf69f6deabb2641d4afb5aede607584f9c02e6566 |
| 832063 | 0000000000000000000211c4647e2610ae16307851e38b7953c399b501b788e5 |

Table 2 Block Hash of 10 Bitcoin Blocks

A striking feature of the block hash values of these blocks are that the first 19 digits of the hexadecimal hash are 0s. This is not a coincidence, and is related to the mining puzzle which each miner tries to solve to '*mine*' a block and receive the block reward. The Bitcoin block header is 80 bytes in size and consists of multiple fields as shown in Fig. 13

To add a block to the blockchain the miner must find a nonce, which is a number when entered in the block header and subsequently the SHA256 hash of the block header is calculated, the resulting hash value is less than the target hash. Or mathematically:

$$H(version||PrevBlockHash||MerkleRootHash||time||nBits||nonce) < target\ hash$$

This is the reason why in Table 2 the hash values of blocks have their first 19 digits as zero. Finding such a nonce is an extremely challenging computational problem and this is the reason why massive hardware is deployed in various mining farms to try as many nonces as possible in the least possible time, so that the miner can find the right nonce and broadcast the '*mined*' block to the rest of the network and get the miner's fee as well as the block reward. Once a valid nonce is found, the other nodes can very easily verify that the hash value in the proposed block header is indeed lower than the hash value target.

To ensure that the mining difficulty keeps pace with the rapidly evolving hashing capacity of the network, it is in-built in the Bitcoin Core software[33] to recalibrate the difficulty level of the problem after every 2016 Blocks (~ two weeks) such that a Bitcoin block is mined every 10 minutes on an average. This can be practically seen in the hash values of blocks mined at various points of time in

[31] A Blockchain Explorer is a website that enables anyone to search any information regarding a Transaction, Address or Block in the Blockchain. Different Crypto Assets have different Blockchain Explorer websites.
[32] Block Height is the number of blocks mined since the Genesis block of Bitcoin was mined in January 2009
[33] https://github.com/bitcoin/bitcoin/blob/master/src/pow.cpp

## Block Headers

Block headers are serialized in the 80-byte format described below and then hashed as part of Bitcoin's proof-of-work algorithm, making the serialized header format part of the consensus rules.

| Bytes | Name | Data Type | Description |
|---|---|---|---|
| 4 | version | int32_t | The block version number indicates which set of block validation rules to follow. See the list of block versions below. |
| 32 | previous block header hash | char[32] | A SHA256(SHA256()) hash in internal byte order of the previous block's header. This ensures no previous block can be changed without also changing this block's header. |
| 32 | merkle root hash | char[32] | A SHA256(SHA256()) hash in internal byte order. The merkle root is derived from the hashes of all transactions included in this block, ensuring that none of those transactions can be modified without modifying the header. See the merkle trees section below. |
| 4 | time | uint32_t | The block time is a Unix epoch time when the miner started hashing the header (according to the miner). Must be strictly greater than the median time of the previous 11 blocks. Full nodes will not accept blocks with headers more than two hours in the future according to their clock. |
| 4 | nBits | uint32_t | An encoded version of the target threshold this block's header hash must be less than or equal to. See the nBits format described below. |
| 4 | nonce | uint32_t | An arbitrary number miners change to modify the header hash in order to produce a hash less than or equal to the target threshold. If all 32-bit values are tested, the time can be updated or the coinbase transaction can be changed and the merkle root updated. |

The hashes are in internal byte order; the other values are all in little-endian order.

Fig. 13 A Bitcoin Header[34]

| Block Height | Date | Block Hash |
|---|---|---|
| 50 | 2009-01-11 | 0000000026f34d197f653c5e80cb805e40612eadb0f45d00d7ea4164a20faa33 |
| 75000 | 2010-08-18 | 00000000000ace2adaabf1baf9dc0ec54434db11e9fd63c1819d8d77df40afda |
| 150000 | 2011-10-20 | 0000000000000a3290f20e75860d505ce0e948a1d1d846bec7e39015d242884b |
| 225000 | 2013-03-09 | 000000000000013d8781110987bf0e9f230e3cc85127d1ee752d5dd014f8a8e1 |
| 300000 | 2014-05-10 | 000000000000000082ccf8f1557c5d40b21edabb18d2d691cfbf87118bac7254 |
| 375000 | 2015-09-18 | 000000000000000009733ff8f11fbb9575af7412df3fae97f382376709c965dc |
| 450000 | 2017-01-26 | 0000000000000000014083723ed311a461c648068af8cef8a19dcd620c07a20b |
| 525000 | 2018-05-30 | 0000000000000000002ffaa108b110ff7fd64475841b6ab65abd9bb43f8ede1d |
| 600000 | 2019-10-19 | 00000000000000000007316856900e76b4f7a9139cfbfba89842c8d196cd5f91 |
| 675000 | 2021-03-17 | 000000000000000000057b6df3f61f96fbdd4b9c4d76fcb975cb0e8a56577d51 |
| 750000 | 2022-08-18 | 0000000000000000000592a974b1b9f087cb77628bb4a097d5c2c11b3476a58e |
| 825000 | 2024-01-09 | 00000000000000000001432b1ea8b3b710c3fb7e628d605cf4a42c25d7822431 |

Table 3. Block Hash values after every 75000 blocks

the past. Table 3 shows the block hash values recorded after every 75000 mined blocks. The number of zeros in the hash is has been increasing, indicating lower and lower target hash values. This can also be seen with the graphs of difficulty and hash rates of the Bitcoin network given in Fig. 14.

The Bitcoin mining difficulty and Bitcoin network's hash rate follow a close trajectory. This is also a mechanism through which the Bitcoin protocol ensures that there are always enough participants in the Bitcoin network which validate the transactions and keep the blockchain secure. When the number of miners goes down, the hash rate of the Bitcoin network goes down and it takes longer to mine new Blocks at the existing difficulty level. Consequently, the mining difficulty is lowered and this attracts more miners till a new equilibrium is achieved. The sharp fall in the hash rate around mid-2021 can be attributed to the crypto ban by China. Subsequently the difficulty of mining reduced and hash rate started having an upward trend with mining activity increasing in USA and Kazakhstan, consequently increasing the mining difficulty again. This can be seen in Fig. 15 showing the country-wise share of hash rate.

[34] Source: https://developer.bitcoin.org/reference/block_chain.html#block-headers

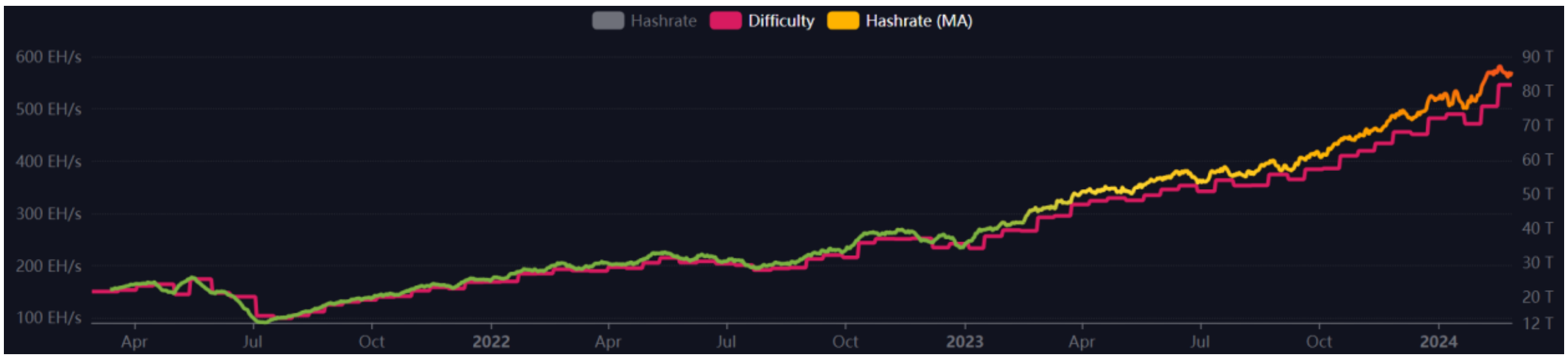

Fig. 14 Bitcoin mining difficulty and Bitcoin network hash rate[35]

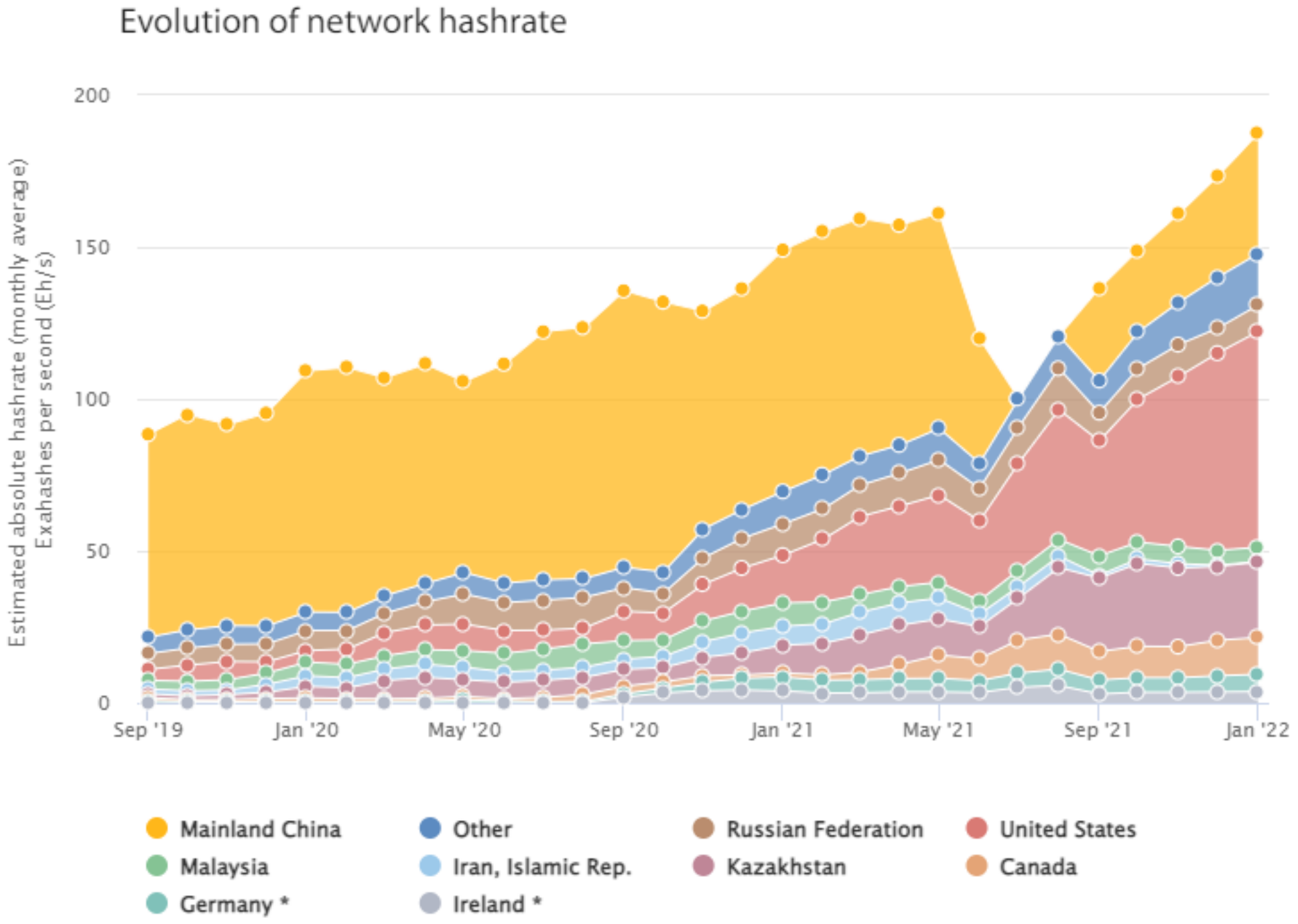

Fig. 15 Country-wise share of Bitcoin hash rate[36]

Mining is a highly resource intensive activity and miners deploy significant computational capacity in the form of Application-Specific Integrated Circuits (ASICs) which are optimized for producing trillions of SHA256 hashes per second. This also results in consumption of massive amounts of electricity in operating the hardware as well as cooling it. It is estimated that if Bitcoin mining consumes more electricity than 167 countries[37].

The probability of a miner successfully mining a block is directly proportional to the percentage of the total hash rate of the Bitcoin network that the miner can produce. Thus, if a miner produces a hash rate which is 0.001% of the total hash rate of the Bitcoin network, the miner can expect to mine one in 100000 blocks. As we know that approximately 2016 blocks are mined in two weeks, this means that it would take around 100 weeks or two years on an average for a miner to successfully mine a block. Till then he/she would have to make the capital investment and incur the costs of depreciation of hardware, the Internet and electricity.

[35] Source: https://mempool.space/graphs/mining/hashrate-difficulty#3y
[36] Source: https://ccaf.io/cbeci/mining_map
[37] https://www.aa.com.tr/en/economy/bitcoin-mining-consumes-as-much-energy-as-167-countries/3109915

### 2.3.7.1 Mining Pools

As discussed above, the possibility of a standalone miner successfully mining a block is miniscule and it involves high degree of risk, as even after spending substantial amount on electricity and IT hardware, it is possible that a standalone miner might not successfully mine a block for several years. Thus, to hedge this risk, miners join mining pools. A mining pool is a group of crypto asset miners who pool their computational resources over a network to increase the probability of success in mining crypto assets, the rewards are divided amongst the participants of the pool rather than a winner-takes-all case in standalone mining. The mining pools are operated by a centralized entity which divides the task of computation between the pool members – the 'worker' miners. A comparison of the leading mining pools can be found at Bitcoin Wiki[38].

Joining a mining pool significantly reduces the risk and uncertainty associated with mining. The mining pools have pool managers which manage the pay outs of mining rewards to the members of the pool depending on the computation done by them. In Bitcoin mining, the mining pool puts its own Bitcoin address in the Coinbase transaction and shares a work template with the 'worker' miner who returns 'shares' to the mining pool. A share is not exactly a solved block, but one in which the block hash is only a few times higher than the target hash. For example, a mining pool taking share which are nearly 10 times of the target hash can be mathematically expressed as:

$$H(version||PrevBlockHash||MerkleRootHash||time||nbits||nonce_{share}) < 10 \times target\ hash$$

This is a proof that the 'worker' miner has indeed put in a lot of computational effort as getting a hash which is close to the target hash is also computationally intensive. This results in a pay out from the mining pool depending on the pay-out policy. Some pools pay 'worker' miners for every share whereas some pools pay out only when the pool finds a block. When a 'worker' miner joins a mining pool even though he/she finds solves a block, he/she cannot take the entire block reward as the Coinbase transaction has the Bitcoin address of the mining pool. This is an important aspect from the perspective of taxation. The shares made to a mining pool in lieu of pay-outs by the mining pools can have direct and indirect tax consequences which have been discussed later. There are also multiple issues related to the identification of miners and determining their tax residency for the purpose of direct and indirect taxation. These issues will also be discussed subsequently.

### 2.3.7.2 Estimating the extent of Mining

Mining is an important part of any blockchain which ensures trust and security in the blockchain. It produces economic value in the blockchain ecosystem and leads to generation of income which might be liable to tax. However, for mining Bitcoin or any other crypto asset mined using the proof-of-work mechanism, only hardware that can produce high hash rates, the Internet and electricity are required. Usually, a 'worker' miner joins a mining pool based on self-certification, the pool usually does not carry out any Know Your Customer verification or Anti-Money Laundering measures to link the pseudonymous 'worker' miner to a natural or juridical person.

Tax administrations would also be interested to know the incidence of mining in their jurisdiction. As mining activity is mostly pseudonymous and only estimates and trends can be drawn from data either obtained from the mining pools or through open-source network analytics. (Sun et al., 2022) studies the spatial distribution of Bitcoin mining through bottom-up tracking and geospatial statistics. The Cambridge Centre for Alternative Finance has developed a mining map[39] which shows the geographical distribution of the total hash rate of Bitcoin over a period. This is based on the data obtained from various mining pools and extrapolating it to provide the monthly percentage share of the country in providing the global hash rate of the Bitcoin network. It provides a snapshot of the extent of mining in each country from September 2019 to January 2022. The methodology adopted for this estimation has some important assumptions like inability to specify the geolocation of miners using VPN or proxy services[40]. Which indicates that it can be very difficult to map various miners with certainty.

[38] https://en.bitcoin.it/wiki/Comparison_of_mining_pools

Mining pool networks have specific characteristics which can be used to detect the miners who are members of the mining pool. Using public network analysis tools like Shodan it is possible to identify the IP addresses of some miners who use specific port numbers to connect to the pools. Shodan[41] is a search engine for devices connected across the Internet. It can help to find IP addresses and number of users using a specific port number in a country or region. By having a look at the number of ports and port numbers used by a particular IP it can be inferred if it is a miner or a general computer. For example, some popular pools use port number 3333 for Transmission Control Protocol (TCP) connection with the participant miners. The result of such a Shodan query for finding IP addresses who are using port 3333 is given in Fig. 16

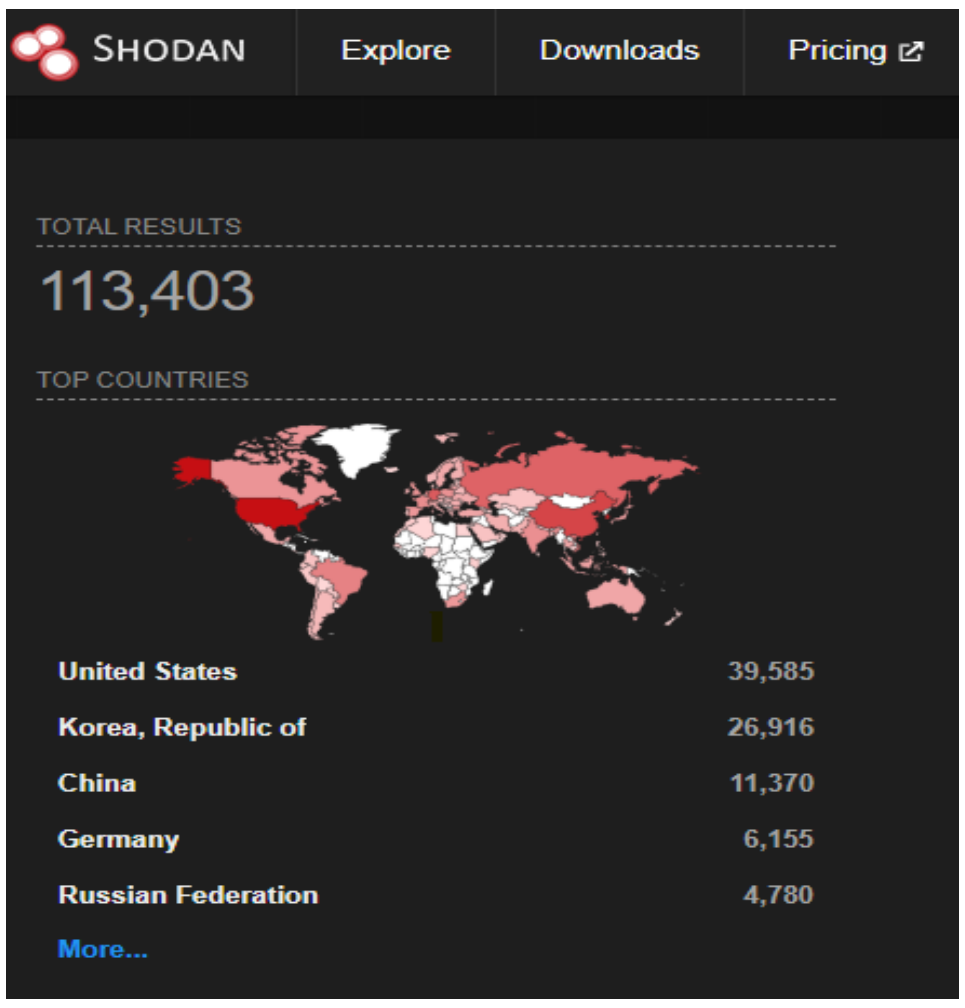

Fig. 16 Query for port number 3333 on Shodan

Various IP addresses which have port 3333 open are displayed. The information about two such IP address from South Korea and Poland is shown in Fig. 17 and Fig. 18 respectively. This potential miner has only few ports open as compared to a general computer which might need to have various other open ports. This makes it highly probable that these computers/ASICs are involved in mining. Due to the imperfection of this method in finding an accurate number of miners, it may only enable users to see the trends in mining in a country, region, or city.

39 https://ccaf.io/cbeci/mining_map
40 https://ccaf.io/cbeci/mining_map/methodology
41 https://www.shodan.io

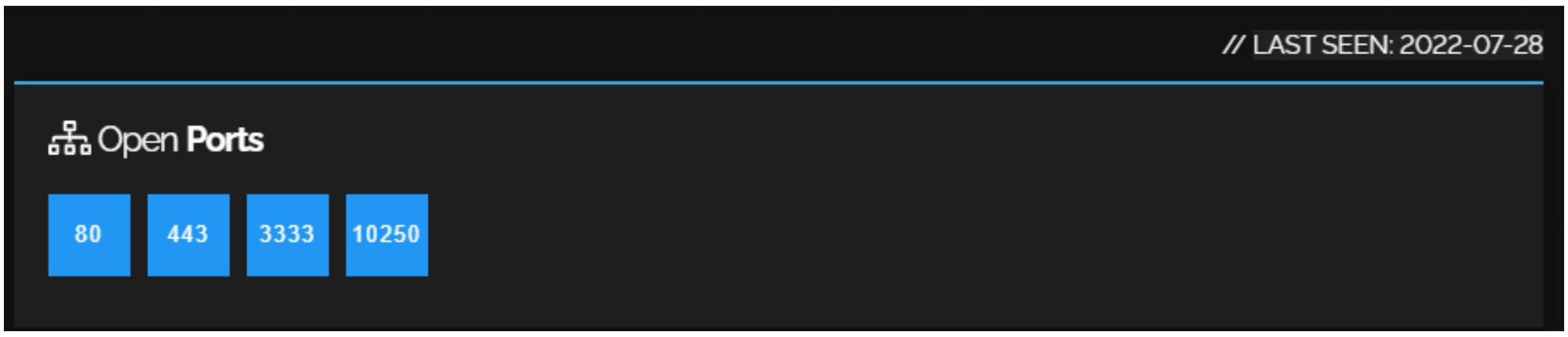

Fig. 17 Open ports of a potential miner in South Korea as seen on Shodan

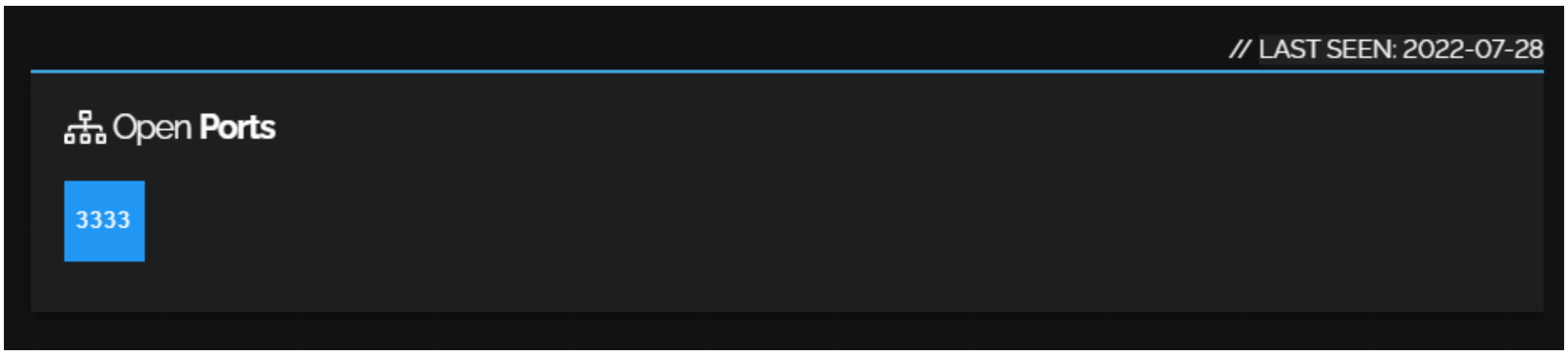

Fig. 18 Open ports of a potential miner in Poland as seen on Shodan

#### 2.3.7.3 Bitcoin Mining and Game Theory

The participants in the Bitcoin Blockchain are rational and they try to maximize payoffs for their actions in the Bitcoin ecosystem. As Bitcoin is based on a decentralized network, malicious nodes/actors cannot be prevented from joining the network. Bitcoin solves this problem effectively by using incentives and cryptographic checks which can detect and counter the actions of malicious actors. The behaviour of the participants in the Bitcoin ecosystem can be understood and predicted using game theory. The application of game theory to crypto asset mining is an active area of research.

Various participants try to maximize their payoffs and adopt strategies accordingly. Mining can be considered a repeated zero-sum game between various rational miners who do not know each other's strategy. Thus, to gain maximum out of the game, the miners tend to follow the strategy of honesty and submit only verified blocks to the blockchain. The miners have a disincentive in duping the system as the validity of the fraudulent transaction can be verified and the other honest nodes would discard the proposed block. As mining is a resource intensive activity, any such efforts would cost the miner dearly even if he/she has a substantial part of the total hash rate of the network.

Some mining pools can adopt a strategy to infiltrate other mining pools and make their infiltrator workers send the shares to the pool but withhold the shares whenever they 'solve' a block. This can enable other pools to bleed profits out of their competitors. This strategy can harm the profits of rival pools but the workers cannot divert the profits to the original pool as the work templates shared by the victim pools would have their Bitcoin address in the Coinbase transaction. A detailed study on such pool attacks has been done by Li et al., (2020).

As discussed earlier, the percentage of hash-power of miners/mining pools determines their probability of mining the next block. The hash-rate share of various mining pools in the Bitcoin network in the past one year is shown in Fig. 19[42]

[42] Source: https://mempool.space/graphs/mining/pools#1y

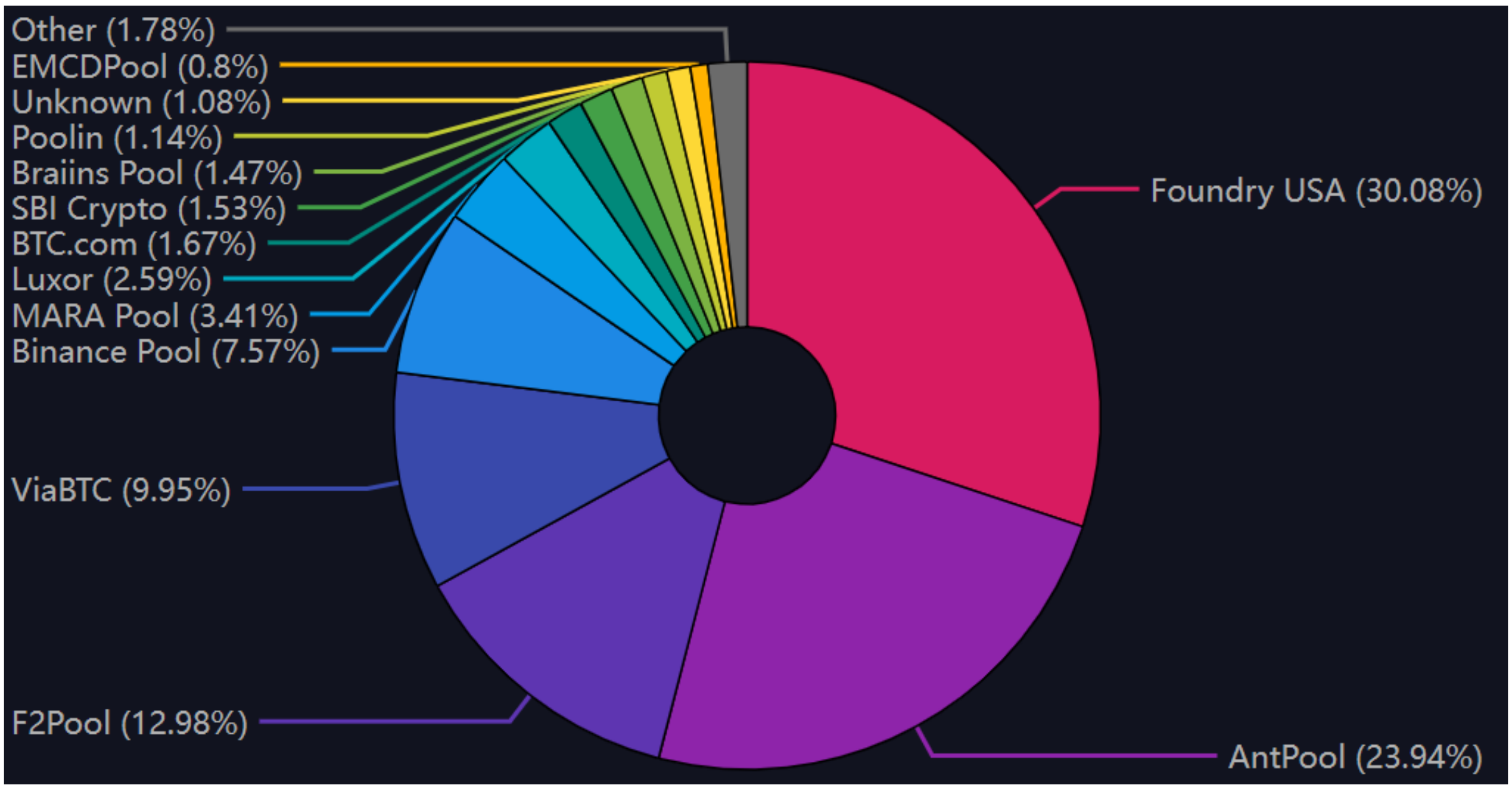

Fig. 19 hash-rate share of various mining pools in the Bitcoin network in the past one year[43]

There can be a scenario in which a single pool has more than 51% of the hash-power of the network. This can enable the pool to delay the transactions of some Bitcoin addresses and carry out double spends. The delay in service can be caused if the pool excludes the transactions received from a set of Bitcoin addresses which it wants to exclude from the network. However, since all nodes would not have this discriminatory behaviour, the pool can at best delay the transactions of a subset of Bitcoin addresses. Double spending can be done by such a pool as it can simultaneously send two transactions in the network that spend the same Bitcoins. The first transaction would pay the provider of the good or service purchased with some UTXOs and the second transaction would use the same UTXOs for payment to an address controlled by the pool. As the pool controls more than 51% of hash-rate of the network it can get the second transaction included in the longest blockchain resulting in orphaning of the first transaction and loss to the supplier. This would significantly reduce the confidence in the Bitcoin network and adversely affect its price.

As the pools also have an incentive to maintain or enhance the value of Bitcoin which they get as reward, it disincentivizes any pool to grow to such a large size that it affects Bitcoin and they cause losses to everyone including themselves. However, such an attack can be launched by an entity which does not have much stake in the Bitcoin ecosystem or which is facing a larger loss in value terms in real world.

It is sometimes misunderstood that a mining pool with 51% of hash rate can steal Bitcoins out of the addresses of users. Due to the digital signature-based mechanism of transactions it is impossible for a mining pool to maliciously steal the Bitcoins out of any Bitcoin addresses. The pool can at best delay certain transactions or carry out double spends as described above.

### 2.3.8 Taxation of Mining Activity and Rewards

As discussed above, mining is an important source of value creation in the blockchain as it adds new transactions to the blockchain and provides trust and security. It creates an incentive for the miners due to which they participate in validating transactions and add blocks to the blockchain. The miner essentially provides a service to the Bitcoin owners by verifying the validity of transactions, arranging them in the form of a block and solving for the nonce to mine the block, in lieu of the Bitcoin block rewards and user fees. Mining activity can have direct and indirect tax consequences for miners as well as mining pool operators. The following sections first discuss the direct tax issues in mining related to the acquisition and 'disposal' of crypto assets obtained through mining. The subsequent section discusses the indirect tax issues related to mining. In the discussion to follow, mining mainly refers to mining using the proof-of-work mechanism. The proof-of-stake mechanism and its taxation is discussed subsequently.

[43] https://mempool.space/graphs/mining/pools#1y

#### 2.3.8.1 Direct Taxes on Mining

Mining provides economic value to the miners in lieu of the services provided by them to blockchain users as well as the blockchain protocol. The income of miners arises either from the protocol rewards for 'mining' a new block, the user fee paid by the users transacting on the blockchain or pay-outs from a mining pool for 'shares' by the 'worker' miners. As this constitutes income or accrual of assets/value, a tax liability may arise, depending on the tax jurisdiction of the miner, as different tax administrations tax mining activity differently.

Moreover, the disposal/sale or any further transactions with the mined crypto assets are also taxable events in various jurisdictions. The tax treatment of the income or accrual of assets/value and disposal/sale or any further transactions of mined crypto assets depends on the nature and scale of the activity carried out by miners. The two potential taxable events in the mining process are:

i) Acquisition of crypto assets as mining rewards
ii) Disposal/transfer of crypto assets acquired as mining rewards

##### 2.3.8.1.1 Taxation of crypto asset acquired as mining rewards

A miner carries out mining activity using IT hardware that generates trillions/billions of hashes per second. As soon as a valid block is found by the miner, it is transmitted to other nodes in the Bitcoin network. If the 'worker' miner is a part of a mining pool, it sends 'shares' to the mining pool over the Internet and the pool, in turn, transmits the 'mined' block over the Bitcoin network. The rewards are received by the miner when it successfully mines a block or when the 'worker' miner in a mining pool cashes out his/her accrued rewards.

The acquisition of crypto assets as mining rewards is subject to income tax in most jurisdictions. The miners are supposed to report the mining incomes separately as 'miscellaneous income' in many jurisdictions and the tax liability arises on the fair market value of the acquired crypto assets at the time of acquisition, with allowable deductions for the costs of mining. Many tax jurisdictions carve out an exception for 'hobby miners' who undertake mining activities out of interest or as a pastime on a small scale, with an intention to accumulate the mined crypto assets instead of trading them to earn a profit.

In some jurisdictions, hobby miners are not liable to pay income tax on the acquired crypto assets at the time of acquisition of such assets. However, the expenses incurred by the hobby miners are treated as the basis of acquisition. On the other hand, many jurisdictions take the basis of acquisition of crypto assets of hobby miners as zero and allow no deductions for expenses.

Besides this, in some jurisdictions like the US, crypto mining is considered as a self-employment activity and miners are subject to payment of employment taxes as well as social security contributions for their income from mining. However, as highlighted above, as it is difficult to find the tax jurisdictions of miners, the payment of taxes is largely subject to self-compliance by the individual miners. Since the mining of first block by Satoshi Nakamoto in 2009 till February 2024 mining rewards and user fee of approximately 60 billion USD has accrued to miners[44]. Similarly, the mining and transaction rewards accrued to Ethereum miners till 15th September 2022, when Ethereum switched to proof-of-stake based consensus mechanism are approximately 35 billion USD[45]. These are significant amounts for tax administrations as the direct and indirect taxes on it might run into billions of dollars after deducting the relevant expenses. Moreover, as the market capitalization of Bitcoin reaches an all-time high of ~1.3 trillion USD[46] and that of Ethereum nears 450 billion USD[47], the amount of taxes due on these mined crypto assets might be in hundreds of billions of dollars.

###### 2.3.8.1.2 Disposal/transfer of crypto assets acquired as mining rewards

In most jurisdictions, the tax treatment of disposal of the crypto assets acquired as mining rewards depends upon the nature and scale of activity by the miners. The income/gain from the disposal of crypto assets acquired as mining rewards is classified as business income or capital gains. If the miners deploy large number of equipment to carry out mining on a commercial scale or frequently engage in trading of crypto assets acquired as mining rewards instead of holding and accumulating them, they are classified as commercial miners and the profits from such activities are taxed as business income/commercial profits, after deducting allowable expenses.

In other cases, the disposal of crypto assets acquired as mining rewards is considered to give rise to capital gains which are taxed at the time of 'disposal' of the crypto assets. In some jurisdictions which do not have a capital gains tax, such disposals may not result in a capital gains tax liability. However, for calculating capital gains, issues related to determination of the basis, as discussed earlier, might arise, and may require clear guidance by the tax administrations. Also, unlike crypto assets acquired as mining rewards which can be clearly ascertained on the blockchain through the Coinbase transactions, the accrual of capital gains cannot be easily ascertained through blockchain analysis alone as many of the apparent 'disposals' of the crypto assets acquired as mining rewards might either be payouts by the mining pool to the 'worker' miners or transfers to other addresses controlled by the same taxable entity/person, resulting in no tax liability for the transaction.

##### 2.3.8.2 Indirect Taxes on Mining

As described above, Bitcoin mining activity involves various actors like:

i) Bitcoin users who broadcast transactions
ii) Individual miners who combine the transactions in the mempool into a block and mine the blocks to include in the Bitcoin blockchain.
iii) Mining pools which create blocks of transactions and divide the computation task between various 'worker' miners
iv) 'Worker' miners which try to find the nonce for a block given by a mining pool and get paid based on 'shares' sent to the mining pool

The Bitcoin network and the users provide economic incentive to the miners to perform mining and record their transactions on the blockchain. These activities are akin to providing a digital service to the users who transmit transactions and pay the transaction fee and the Bitcoin protocol which provides the block reward in lieu of this service. However, the high degree of uncertainty associated with mining rewards leads to many miners joining mining pools as 'worker' miners, this complicates the indirect tax treatment of the services provided. This gives rise to two category of service providers namely:

---

[44] https://explorer.btc.com/btc/blocks
[45] https://etherscan.io/chart/blockreward
[46] https://coinmarketcap.com/currencies/bitcoin/
[47] https://coinmarketcap.com/currencies/ethereum/

A) Individual miners and mining pools which create blocks of transactions broadcasted by the users of Bitcoin network and add new blocks to the blockchain through mining.
B) 'Worker' miners which get a pre formed block from a mining pool, which contains the payout address of the mining pool in the Coinbase transaction. They provide the mining pool with 'shares' and get rewarded for the same.

There are also the mining pools which provide services of reducing the variance of payout rewards of 'worker' miners which join the mining pools instead of mining Bitcoin as individual miners. These service providers in the Bitcoin ecosystem are depicted in Fig. 20

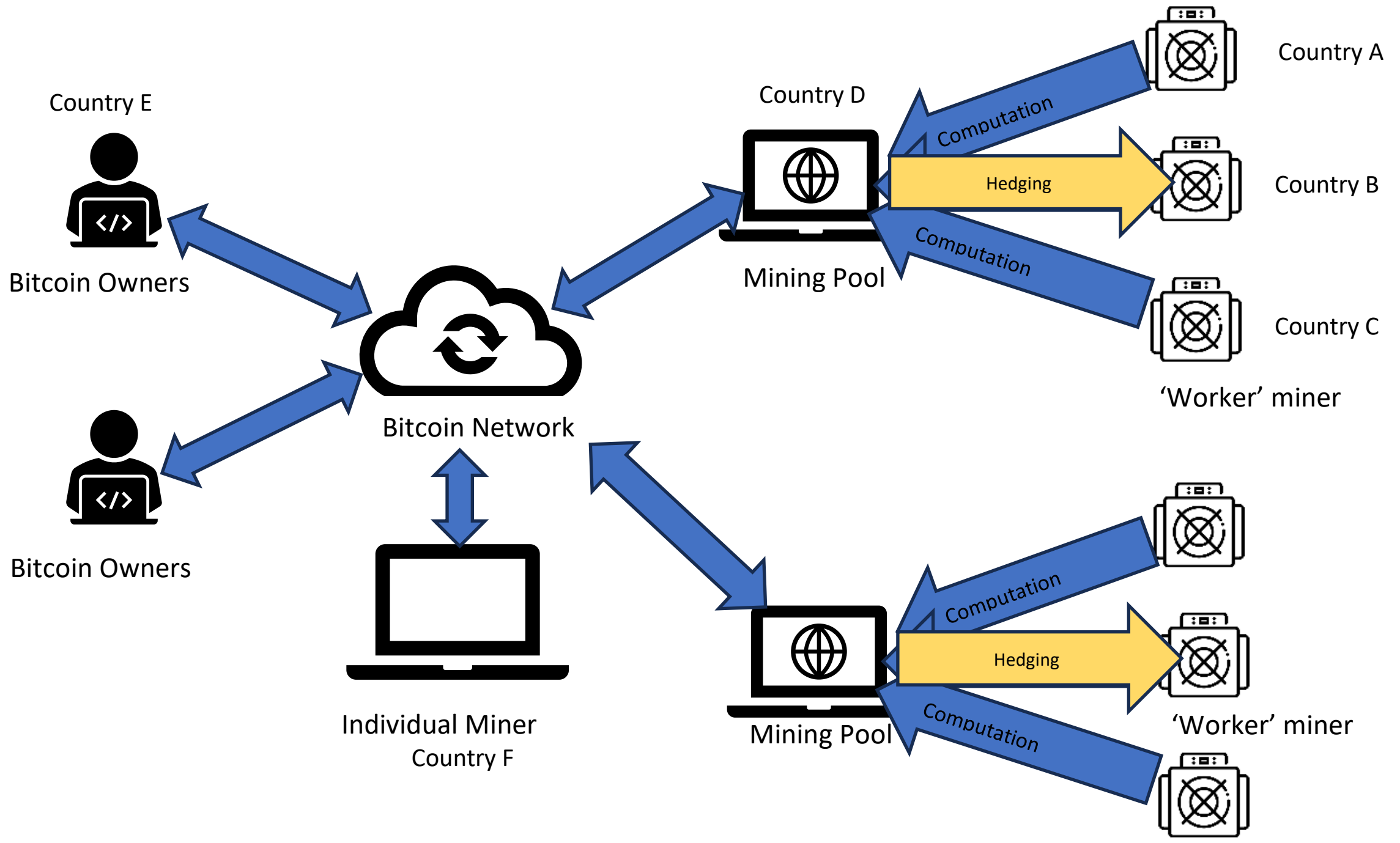

Fig. 20 Mining services by various service providers

The tax treatment of the services offered by these two categories of miners are based on various questions directly related to the service offered by the two categories of miners to Bitcoin users (which is rewarded in the form of transaction fee) and the Bitcoin protocol (which is rewarded in the form of new Bitcoins issued through the Coinbase transaction). Some of the important questions that are critical in determining the indirect tax treatment of mining activities are:

i) Does there exist a direct nexus or a legal synallagmatic relationship between the user broadcasting a transaction and the miner adding the transaction on the blockchain?
ii) Does the service offered by miners to the blockchain network, for which they are rewarded with the block reward, constitute a taxable service for a consideration?
iii) Does the service provided by the 'worker' miners to the mining pool have a bearing on the nature of service provided and its tax treatment?
iv) What is the place of supply of such services?

These questions have a critical role in determining the levy of indirect taxes on mining activities. Multiple jurisdictions have different legal provisions for taxability and exemptions to services like Bitcoin mining, with most jurisdictions treating mining activity as exempt for indirect tax purposes. Two important issues related to the questions raised above are those of existence of a nexus/link/contract between the miner and the users transacting on the blockchain, which also determines the place of supply of the services, and if such nexus is established or deemed to exist, the issue of exemption of services of miners as financial services, which is contingent upon the treatment of Bitcoin and other crypto assets as a negotiable instrument or any other financial instrument which is exempt. However, the broad principles and considerations while determining the levy of GST/VAT on mining activities remain common and provide the answers to the questions above.

As laws and jurisprudence in this area seem to be most developed in the EU the following discussion analyses these issues from the perspective of the EU Law, although the broad principles and considerations would apply to almost all jurisdictions.

The primary consideration for levying an indirect tax on a miner is that he/she should be engaged in an 'economic activity.' The miner should be performing his/her activities for an income that is linked to its activity. The Article 9 of the EU Directive 2006/112/EC of 28 November 2006 on the common system of value-added tax defines a 'taxable person' and 'economic activity' as:

*'Taxable person' shall mean any person who, independently, carries out in any place any economic activity, whatever the purpose or results of that activity.*

*Any activity of producers, traders or persons supplying services, including mining and agricultural activities and activities of the professions, shall be regarded as economic activity. The exploitation of tangible or intangible property for the purposes of obtaining income therefrom on a continuing basis shall in particular be regarded as an economic activity.*

As per the Council Implementing Regulation (EU) No 282/2011

*"Electronically supplied services as referred to in Directive 2006/112/EC shall include services which are delivered over the Internet or an electronic network and the nature of which renders their supply essentially automated and involving minimal human intervention, and impossible to ensure in the absence of information technology"*

Also, mining activities could also be seen as falling within the definition of an electronically supplied service as per section 3.1.5 of the Working paper No 854 of the EU VAT Committee which states:

*"When the platform is run in an automated manner with minimal human intervention and the provision of the service is impossible without information technology, then the access to such platform supplied for consideration should be seen as covered by the definition of electronically supplied service"*

Also, in a recent case before the Dutch Court of The Hague the claimant performed mining activities and raised a question that does the incentive/remuneration received by her pertain to her mining activities and would the mining activity be seen as an 'economic activity'?

The court adjudicated that the transaction fee as well as the block reward were considerations for validating transactions on the Bitcoin Blockchain and that the fact that the petitioner does not always receive the transaction fee was irrelevant[48]. The court compared this to a home buyer reaching out to multiple realtors to buy a home but choosing only one of them to buy the house and pay commission. The other realtors would still be considered to be engaged in economic activity with only one being remunerated.

Further the court argued that validating the transactions was important for blocks to be created and added to the blockchain and this made the validation, verification and mining of coins inseparably intertwined. However, the mining activity was exempted from VAT by the court by virtue of Article 135(d) of the EU VAT Directive.

[48] https://www.lexology.com/library/detail.aspx?g=e4cfac02-9134-4629-85cb-cffaa9859dc0

From a theoretical standpoint also, the presence of incentives to the miners is an existential factor for crypto assets like Bitcoin, as the fundamental problem in the Bitcoin Blockchain of making multiple nodes agree on transactions added to the blockchain, when it is known that some of the nodes can be faulty or malicious, is theoretically impossible to achieve without incentives. This is a widely studied problem known as the Byzantine Agreement problem in Computer Science, which is named after the Byzantine Generals' Problem, in which generals of the Byzantine army must agree on a coordinated attack or retreat. There are no incentive mechanisms in this problem and it is proven that the problem cannot be solved if more than 1/3rd of the generals are traitors[49]. Also, another theoretical result related to distributed systems - the FLP (Fischer-Lynch-Paterson) impossibility theorem states that no deterministic protocol solves the byzantine agreement problem in the asynchronous model, with even a single faulty node (Fischer, Lynch, & Paterson, 1985).

However, in the Bitcoin Blockchain we observe that with the presence of incentives in the form of block rewards and transaction fees for the miners, the Bitcoin distributed consensus protocol for the ledger has worked well for years. This makes incentives the cornerstone of Bitcoin as a network and a crypto asset. It leads to the alignment of the interest of the miners with the holders of Bitcoin transmitting transactions on the Bitcoin Blockchain to be included in the public ledger. This makes the activities performed by the miner in expectation of incentives an economic activity by design.

As discussed earlier, the probability of an individual miner mining a Bitcoin block is directly proportional to the fraction of the hash rate of the miner to the total hash rate of all the miners combined. This results in high variance and uncertainty in rewards received by the solo miners who deploy resources for mining. To reduce the uncertainty in mining rewards, solo miners often join a mining pool.

The 'worker' miners share hashes with the pool and based on different criteria like the number of hashes shared below a target range (not the same as the Bitcoin hash target range), or some other pay-out agreement, pay-outs are made by the mining pool. This provides the 'worker' miners frequent pay-outs depending on the number of 'shares' made with the mining pool. The mining pools charge a fee for this service (usually 1-3%)[50] and deduct the same from the pay-outs given to the 'worker' miners.

The mining pool creates a block of transactions and distributes the mining computation between the various 'worker' miners which join the mining pool. The template shared by the mining pool with the 'worker' miners contains the transactions picked by the mining pool and the pay-out address in the Coinbase transaction is that of the mining pool. In case a valid nonce is found by a 'worker' miner which mines the current block, the entire block reward goes to the mining pool and the 'worker' miner cannot earn the full block reward, but is entitled to the reward as per the pay-out scheme agreed between the 'worker' miner and the mining pool.

This makes the operation of the mining pool an 'economic activity' which provides a service of hedging the risks by significantly lowering the variance of reward payments for mining activity. The consideration received for this service (usually 1-3%) is in the form of crypto assets being mined, which is deducted from the pay-out due to the 'worker' miner. This is an important aspect from the perspective of indirect taxation as the 'worker miners' and the mining pools provide two services to each other in a symbiotic relationship. The mining pools help to reduce the uncertainty and variance of the mining rewards of the 'worker miners,' who in turn, provide computational power in the form of billions/trillions of hashes per second (shares) to the mining pool, which may or may not result in 'mining' a block by the mining pool and can never result in direct block rewards to the 'worker' miner. There are multiple case laws like C-2/95 SDC[51] and C-350/10 Nordea[52] which provide that for a service to be VAT exempt, the service must be an exempt service itself. The argument that it is an input service to an exempt service does not suffice for VAT exemption to a service. Considering the nature of these services, neither of these services may classify as a financial service which can exempt the service providers from VAT/GST. However, the service provided by the mining pool or individual miners to the blockchain and its users might be exempt from the perspective of VAT/GST as financial services, depending on the classification of the crypto assets involved as negotiable instrument or some other instrument which qualify the service as financial services in the domestic law of a jurisdiction.

[49] https://en.wikipedia.org/wiki/Byzantine_fault
[50] https://fastercapital.com/topics/understanding-pool-rewards-and-payouts.html

Now that we have a case where VAT/GST might be leviable on the services provided by the mining pool and the 'worker' miners to each other, it is important to determine the place of supply of such services as it may be zero rated in most jurisdictions, if the services are supplied to a non-resident. The taxability will also be affected by whether an individual or enterprise is providing and/or availing the services and whether such entities have obligations for VAT/GST registration. It may also lead to implications for providing input tax credit and refunds for suppliers of zero-rated mining related services.

This problem is like the problem of taxation of cross-border services, using crypto assets-based payments. Brondolo, (2021) provides a detailed discussion on potential solutions to this issue. Under the vendor collection model, the mining pools, if they are non-resident suppliers of hedging services to the 'worker' miners, would be required to register for VAT/GST in the jurisdictions of the 'worker' miners and charge and collect VAT/GST and pay the collected amounts to the respective jurisdictions. For example, in India these hedging services would qualify as an Online Information Data Base Access and Retrieval (OIDAR) service and the mining pool is obligated to pay 18% GST on it, if the service is availed by an Indian 'worker' miner, especially if it is not registered under GST. In case of Business-to-Business supplies the recipient may be obligated to collect VAT/GST on a reverse charge basis.
Also, the services exported by the mining pool to the worker miner can be zero-rated and entitle the mining pool for a refund of the input tax credit in its country of registration, with the 'worker' miner paying taxes for import of services in its own jurisdiction.

This might be a humongous compliance burden for mining pools which might be required to collect the tax residency and VAT/GST information from 'worker' miners. However, this may be simplified if the tax jurisdictions create crypto asset wallets for depositing the aggregated VAT/GST collections along with filing of returns by the mining pools to the respective jurisdictions. A similar arrangement may also be required for the computational services of 'worker' miners used by the mining pools in return of pay-out for 'shares' by them. Alternatively, the mining pool may also collect and pay VAT/GST net of the fee charged to 'worker' miners in its own jurisdiction on a reverse charge basis.

In various tax jurisdictions the transaction fee by the users in lieu of the service of including the user's transaction in the blockchain through 'mining' might also be taxable as a service[53]. In that case, in line with the destination principle of VAT/GST the blockchain user/mining pool (in a vendor collection model) might be liable to pay VAT/GST in the jurisdiction of the user, depending on turnover threshold

---

[51] https://eur-lex.europa.eu/legal-content/EN/ALL/?uri=CELEX%3A61995CJ0002

[52] https://curia.europa.eu/juris/document/document.jsf;jsessionid=CD1328EA651A7F81A11CBDAB3C8E3AC9?text=&docid=108324&pageIndex=0&doclang=EN&mode=lst&dir=&occ=first&part=1&cid=1203786

[53] The service provided by the mining pool to the users might not be exempt from the perspective of VAT/GST as financial service, depending on the classification of the Crypto Assets in the domestic law of a jurisdiction

and other requirements for VAT/GST collection as per the domestic laws of the user's jurisdiction. However, the determination of tax jurisdiction of crypto asset users who transmit transactions on the blockchain without any information about the tax residency of the user, which has a bearing on the place of supply of the service, remains the biggest challenge for this. A more complicated issue will be the taxation of services provided by the mining pools or individual miners to the blockchain itself, for which they receive the fixed block reward programmed in the blockchain. The place of supply of such service cannot be determined and the payment for the service may amount to billions of dollars each year.

If we consider the scenario depicted in Fig. 20, A mining pool registered in country D might be liable to pay VAT/GST on the services provided to the 'worker' miners of country A, B and C as per the domestic laws of countries A, B and C, and might be required to register for VAT/GST in countries A, B and C especially if the 'worker' miners are not registered under GST. The GST/VAT amounts would be paid by the mining pool to the countries A, B and C for providing hedging services to minimize the variance and uncertainty of block rewards in mining. Also, the individual 'worker' miners in the countries A, B and C would also have an indirect tax liability for providing computational services for Bitcoin mining to the mining pool registered in country D. Alternatively, as discussed earlier, the mining pool registered in country D may also collect and pay VAT/GST net of the fee charged to 'worker' miners in countries A, B and C, in its own jurisdiction country D on a reverse charge basis. The mining pool might also have a VAT/GST liability for providing transaction validation services to a user in country E, where the final consumption of the service of the 'worker' miners in countries A, B and C takes place, as the Bitcoin user in country E pays a transaction fee to the mining pool for mining his/her transaction.

The individual miner in country F might also be liable to pay VAT/GST in country E on the transaction fee paid by the Bitcoin user if it validates the transaction of the Bitcoin user in country E and includes it in a block mined by it. Depending upon the turnover thresholds of various jurisdictions, the individual miner might not be liable to pay VAT/GST on the transaction fee.

Some mining pools like Terra Pool[54] implement KYC and provide real-time AML measures. Other top mining pools also specify in their privacy policy that they collect information like username, email addresses, wallet addresses, IP addresses, unique device IDs, information related to amount of computing power provided to the pool, the rewards, and processed payouts. The privacy policy of many mining pools makes the users aware that their personal data can be used by the mining pool to comply with domestic and international legal obligations. This can be used by the mining pools to determine residency of 'worker' miners and enable tax administrations or other law enforcement agencies to seek information regarding their tax residents using the services of such pools for both direct and indirect tax purposes. However, the determination of place of supply for the ordinary Bitcoin or other crypto asset users will remain a challenge.

#### 2.3.8.2.1 Indirect Taxes on Supply of crypto assets

In various jurisdictions there are specific provisions in the indirect tax law which exempt certain financial transactions, loans deposits etc. from VAT/GST. For example, in India the Notification No. 12/2017- Central Tax (Rate) exempts loans, deposits purchase of foreign currency etc. from GST. However, in many jurisdictions there are no such explicit exemptions for crypto assets. Thus, it is possible to interpret the indirect tax law to consider the crypto asset transactions as a taxable supply and levy GST on such supplies. However, many jurisdictions do not treat crypto asset transactions as taxable supply, but one such exception is Singapore where prior to 1st January 2020, the supply of virtual currencies (including cryptocurrencies such as Bitcoin) was treated as a taxable supply of services.

[54] https://terrapool.io/

#### 2.3.8.2.2 Indirect Taxes on use of crypto assets for payments for goods and services

In most jurisdictions, the ‘disposal’ of crypto assets in lieu of any goods or services attracts GST/VAT like fiat currency and the ‘disposal’ event might also attract income tax or capital gains liability based on the domestic provisions for taxation of crypto assets which are mainly based upon the nature and scale of engagement of the individual or entity in crypto assets transactions. Baer et al., (2023) highlight the profound risks that crypto assets might pose for collection of VAT/Sales Tax on final sales of goods and services.

#### 2.3.8.2.3 Taxing the externalities of mining using the proof-of-work mechanism

As highlighted in earlier sections, crypto asset mining using proof-of-work consensus mechanism is a resource intensive activity with a huge carbon footprint of both the electricity consumption and the specialized IT hardware that is used to find nonces to ‘mine’ a block of Bitcoin. This produces externalities which calls for imposition of taxes like the excise tax which discourage mining or make up for its associated costs on the environment.

The Biden administration has proposed a 30% excise tax on electricity used for Bitcoin mining. It is called the Digital Asset Mining Energy (DAME) tax. Similarly other options like a graded tax on electricity consumption by miners as being used by other jurisdictions like Kazakhstan can be used. Moreover, measures to incentivise miners to switch from proof-of-work based mining to staking and forging, which is based on the proof-of-stake protocol (as discussed below) and has a much less carbon footprint, can be considered. Alternatively, if the excise taxes are found difficult to administer or impose, miners can be denied to claim deductions for electricity consumption and depreciation for IT hardware from their revenues to calculate taxable profits.

### 2.3.9 Proof-of-Stake and Forging

Proof-of-work and proof-of-stake are two different consensus mechanisms in blockchain technology. A consensus mechanism is a method for maintaining the integrity of the blockchain in which nodes of the blockchain network develop a consensus about the blocks and length of a blockchain. The proof-of-stake consensus mechanism requires much less computation and resources to add blocks to the blockchain than proof-of-work. In this method, nodes do not compete against each other to mine the block, and have to stake their own minimum amount of crypto asset to be a part of the consensus mechanism[55] Nguyen et al. (2019). These assets act as a collateral and are locked and cannot be moved before the locked period. If a node wants to stop being a forger, its staked crypto assets are released after a certain period, once it has been verified that it has not been involved in any malicious behaviour in adding blocks and verifying transactions. By staking crypto assets in a wallet the validator nodes make themselves available to propose new blocks and validate the blocks proposed by other forgers. The proof-of-stake mechanism is depicted in Fig 21.

A node is selected by the blockchain to be the proposer of the next block using a pseudo-random process. The proposed block is then verified by other nodes and if it is verified by more than 2/3rd of the nodes, the block is added to the blockchain. The proposer node gets the reward in the form of transaction fee. The proof-of-stake method usually leads to generation of new crypto assets and the process of adding new blocks to the blockchain is called forging instead of mining.

[55] A detailed analysis of proof-of-stake consensus mechanism has been done by Nguyen et al. (2019) Nguyen, C. T., Hoang, D. T., Nguyen, D. N., Niyato, D., Nguyen, H. T., & Dutkiewicz, E. (2019). Proof-of-stake consensus mechanisms for future blockchain networks: fundamentals, applications and opportunities. *IEEE Access*, *7*, 85727-85745.

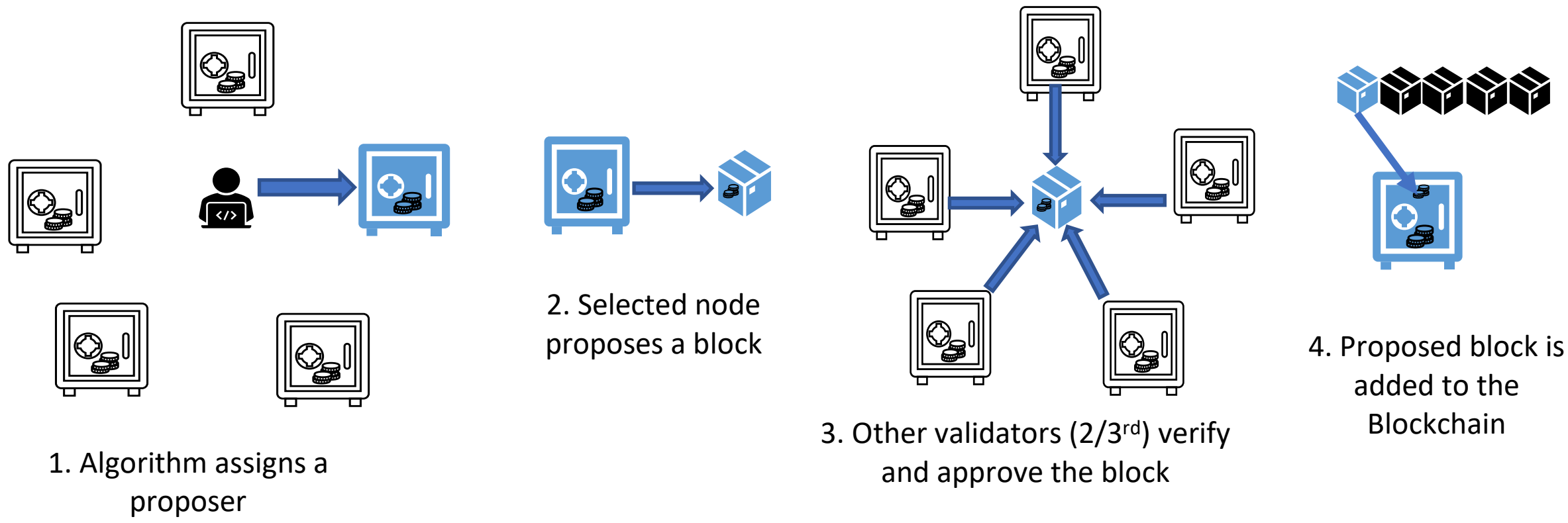

Fig.21 proof-of-stake consensus mechanism

Crypto assets frequently begin by selling pre-mined coins or they begin with the proof-of-work consensus mechanism and then move to the proof-of-stake consensus mechanism (like in case of Ethereum). Various blockchains use different methods to choose the node for proposing the next block in proof-of-stake. The two most popular methods are randomized block selection and coin age selection. In randomized block selection, the node with the lowest hash value and the highest stake is chosen. As stake sizes are public, the next forger can typically be predicted by other nodes. The coin age selection method selects nodes depending on the length of time that their crypto assets have been staked, the age of a coin is determined by multiplying the number of days the coins have been staked by the number of coins staked. After a forging block a node's currency age is reset to zero and they must wait a specific amount of time before forging another block, this prevents large stake nodes from controlling the consensus mechanism.

The proof-of-stake consensus also has a penalty mechanism which discourages forger misbehaviour which discourages malicious activity and promotes honest nodes. If a node fails to propose a block on its turn or proposes more than one block then a slashing penalty is imposed on the node by taking away a portion or entire staked crypto asset. The malicious node can even be banned for the current epoch or permanently. A part of the slashed funds is usually given to the node which reports the malicious behaviour and the rest are burnt[56]. On one hand this provides the blockchains with the best and most effective forgers, on the other hand the inability of a node to propose a block due to factors like server downtime, buggy algorithms and network attacks can penalize forgers heavily.

Although the proof-of-stake consensus described above is broadly the same as one used by the Ethereum blockchain, however there are marked differences between the two. The proof-of-stake mechanism currently used by the Ethereum Blockchain is described in detail in the section on Ethereum. The taxable events in this consensus mechanisms with the Ethereum Blockchain as the example are also discussed in the subsequent sections.

[56] Crypto Assets are burnt by removing them from circulation permanently by sending them to a special address, called a burn address, that cannot send or receive any Crypto Assets, making them inaccessible and effectively destroying them.

### 2.3.10 The Bitcoin Network

The Bitcoin nodes are connected to each other through a decentralized network where there is no central server. It is essentially a peer-to-peer network which relays and validates Bitcoin transactions and secures and maintains the Bitcoin Blockchain. Anyone is free to join the network if they have access to the Internet. The collection of such peer-to-peer nodes can be called as the Bitcoin network. All the nodes in the network perform basic functions like discovering peer nodes and maintaining a connection with them, as well as propagation and validation of transactions and blocks.

The Bitcoin network consists of various types of nodes as listed below:

a) Mining nodes: These nodes run full copy of the blockchain and produce valid blocks for being added to the blockchain. They get rewards in the form of newly created Bitcoins.
b) Full Nodes: These nodes also maintain a full copy of the blockchain and can independently verify transactions.
c) Super Nodes: These nodes have many connections with the full nodes and act as redistribution relays to ensure that every node in the network has the latest copy of the blockchain.
d) Light/Simple Payment Verification (SPV) Nodes: These nodes do not maintain a full copy of the Bitcoin Blockchain and operate on small devices like smartphones or tablets. They verify transactions by querying their peers to retrieve the subsets of the blockchain they require to validate a transaction.

The Bitcoin network scheme is shown in Fig. 22

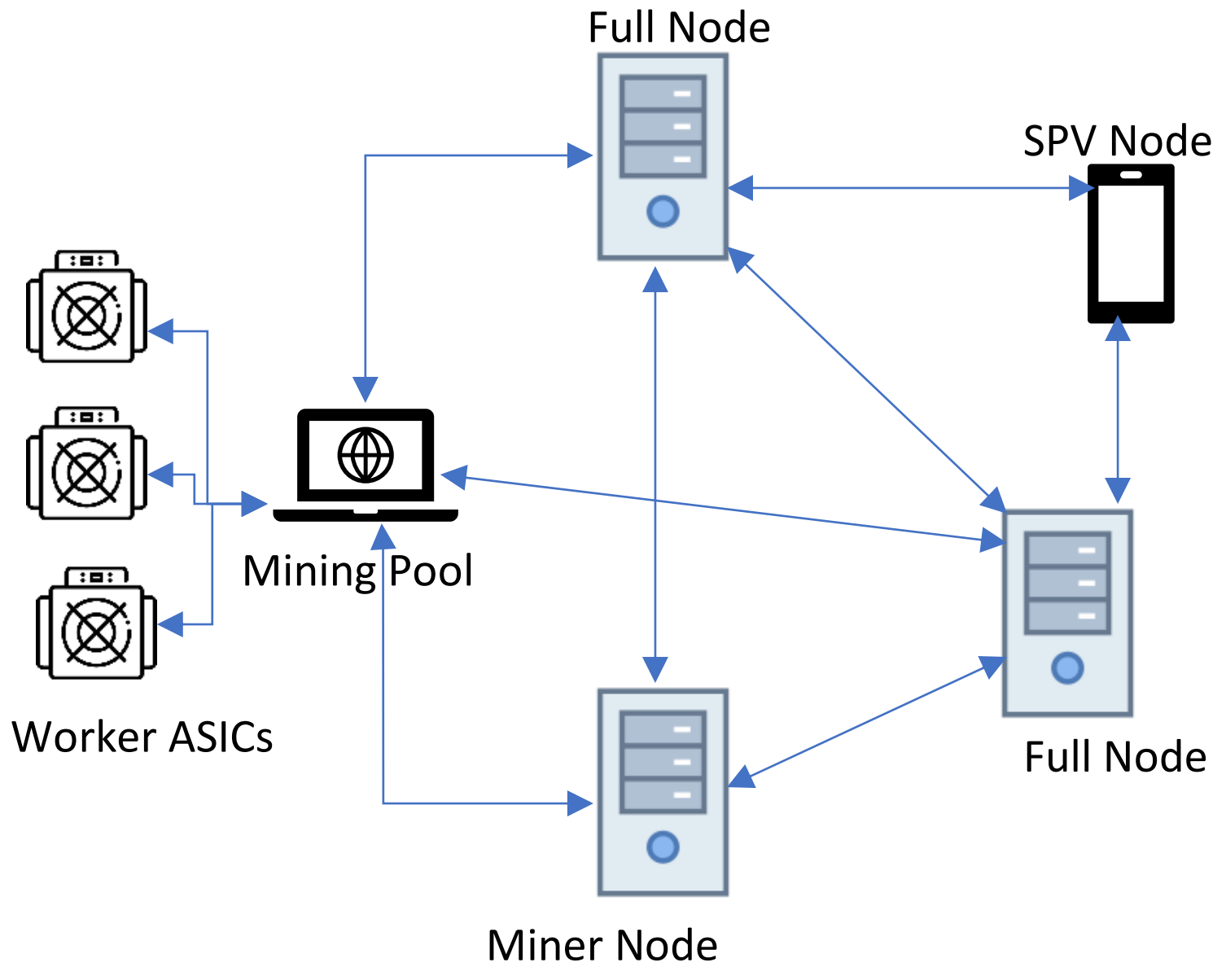

Fig. 22 The Bitcoin network

When a node broadcasts a transaction to its peers, they further broadcast it and flood the network. Thus, the propagation of transactions in the network is not instantaneous and every node does not have the same view of the mempool. When a transaction is broadcasted over the network, as discussed earlier, there is no IP address related information associated with the transaction. Thus, it is not usually feasible to find out the IP address of the node that broadcasted a specific transaction. This makes it difficult to attach IP addresses to Bitcoin addresses by listening to transactions on the Bitcoin network.

Juhász, Stéger, Kondor, & Vattay, (2018) have tried a Bayesian approach to identify Bitcoin users where they installed more than 100 Bitcoin clients and analysed the propagation of these messages in the Bitcoin network. They claim to have identified several thousand Bitcoin clients and bind their transactions to geographical locations. However, such an approach would provide an estimate which might not enable a tax administration to assign an IP address and subsequently an IP address owner/user to a Bitcoin transaction beyond reasonable doubt.

As the Simple Payment Verification/Light nodes depend on other nodes for verification of transactions, they can be used to identify Bitcoin addresses in their wallet, and enable associating IP addresses to Bitcoin addresses. To prevent this, the SPV Nodes use bloom filters[57]. However, Gervais, Capkun, Karame, & Gruber, (2014) found that SPV clients with bloom filters also leak considerable information regarding Bitcoin addresses in their wallet and make associating IP addresses to Bitcoin addresses possible.

Another feature of the Bitcoin network is that the participating nodes usually connect to each other using the port number 8333[58]. This can create a potential problem for the network in case of a widespread disruption in traffic on this port, which can seriously affect the mining and validation of blocks and consequently, the price of Bitcoin significantly. However, this issue has been addressed in Bitcoin Core 23.0[59] .

### 2.3.11 Forking

Just like a fork in a road, forking in a blockchain refers to a situation where the blockchain diverges into two potential paths forward. This can arise due to a temporary lack of consensus in the network regarding the block(s) to be added to the main blockchain, resulting in a temporary fork. This can also be due to changes in the software run by various nodes which validate transactions and add blocks to the blockchain. If the blockchain software undergoes a major change, for example due to a major security threat or a necessary improvement, and the new blockchain is incompatible with the previous version, such a change is called a *Hard Fork*. If the software updates are backward compatible it results in a *Soft Fork*.

As the Bitcoin Blockchain is decentralized with each node maintaining its own copy of the blockchain, it is possible that at some instance divergence might arise between the nodes about the blockchain. Temporary forks can emerge in cases like when two blocks are mined within a short interval by two different miners, and they reach various nodes almost simultaneously. This results in some nodes adding the first block and some nodes adding the second block to the blockchain, resulting in the nodes having two different perspectives of the blockchain at that moment. The blockchain eventually reconverges as more blocks are added to one of the forks. The blocks added to the other fork get orphaned and are discarded, the transactions in orphaned blocks remain unconfirmed. This is the main reason why it is recommended to wait for 6 confirmations of the transaction i.e., waiting for six subsequent blocks to be added to the blockchain after the transaction for confirming the validity of the transaction. This is graphically depicted in Fig. 23

#### 2.3.11.1 Hard Fork

A hard fork occurs in a blockchain when a major change happens in the blockchain software/protocol and these are incompatible with the previous version of the blockchain. This is a way through which new features can be incorporated in the blockchain protocol or security loopholes in the software can be plugged. However, all nodes of the Bitcoin network might not accept the change and upgrade. As the changes introduced make the nodes that upgrade incompatible with the old protocol, a split occurs in the blockchain. The blocks created by the nodes running the old version of the protocol would be invalid for the upgraded nodes. As a result, two crypto assets are created and the owners of the crypto assets prior to the point of hard forking get equal number of crypto assets in the both the blockchains right after the hard forking.

---

[57] https://bitcoinops.org/en/topics/transaction-bloom-filtering/
[58] https://bitcoin.org/en/full-node#network-configuration
[59] https://github.com/bitcoin/bitcoin/blob/e88a52e9a2fda971d34425bb80e42ad2d6623d68/doc/release-notes/release-notes-23.0.md#p2p-and-network-changes

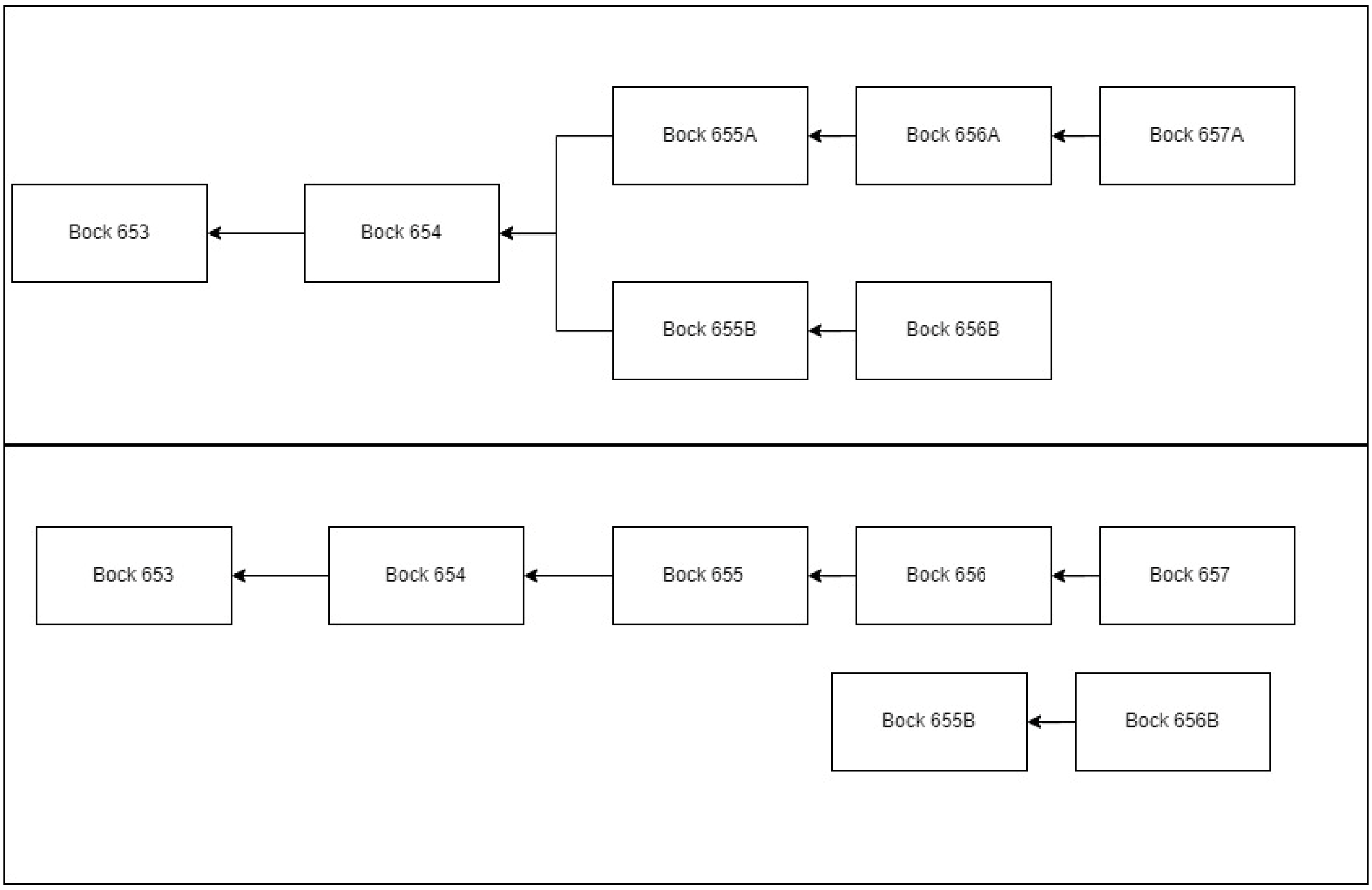

Fig. 23 Temporary Blockchain Fork

A prominent example of hard forking is the creation of Bitcoin Cash Blockchain from the Bitcoin Blockchain. To increase the number of transactions that can be included in a block in Bitcoin, it was proposed to segregate the signatures from the main transaction data to accommodate more transactions in the 1MB block size limit. To avoid the update, some Bitcoin developers and users decided to initiate a hard fork and created the Bitcoin Cash which has a block size of 8MB and does not segregate signatures. Hard forks are important events from tax perspective as it results in the ownership of crypto assets in both the forks of the blockchain after the fork has taken effect, for those who owned any Bitcoin amount right before the fork. For example, the hard fork in Bitcoin took effect from block number 478558 of Bitcoin blockchain. If we look at the address 3FGs7JfaoAZTT6Sda73XrJ6i5Gwsuw9GUC which received 8.0 BTC in a transaction included in the no. 478557[60] right before the hard fork happened and query the Blockchain Explorer[61] at present, we find a result as shown in Fig 24.

This shows that this Bitcoin address also became a Bitcoin Cash address and the owner of this address received 8.0 Bitcoin Cash by virtue of the hard fork. This is analogous to splitting of shares in equity market and is an important taxable event.

60 https://www.blockchain.com/btc/tx/98ecc4189d3f33eee96afbff948219ba0ba2342c263241602b95437050c1130b

61 https://www.blockchain.com/search?search=3FGs7JfaoAZTT6Sda73XrJ6i5Gwsuw9GUC

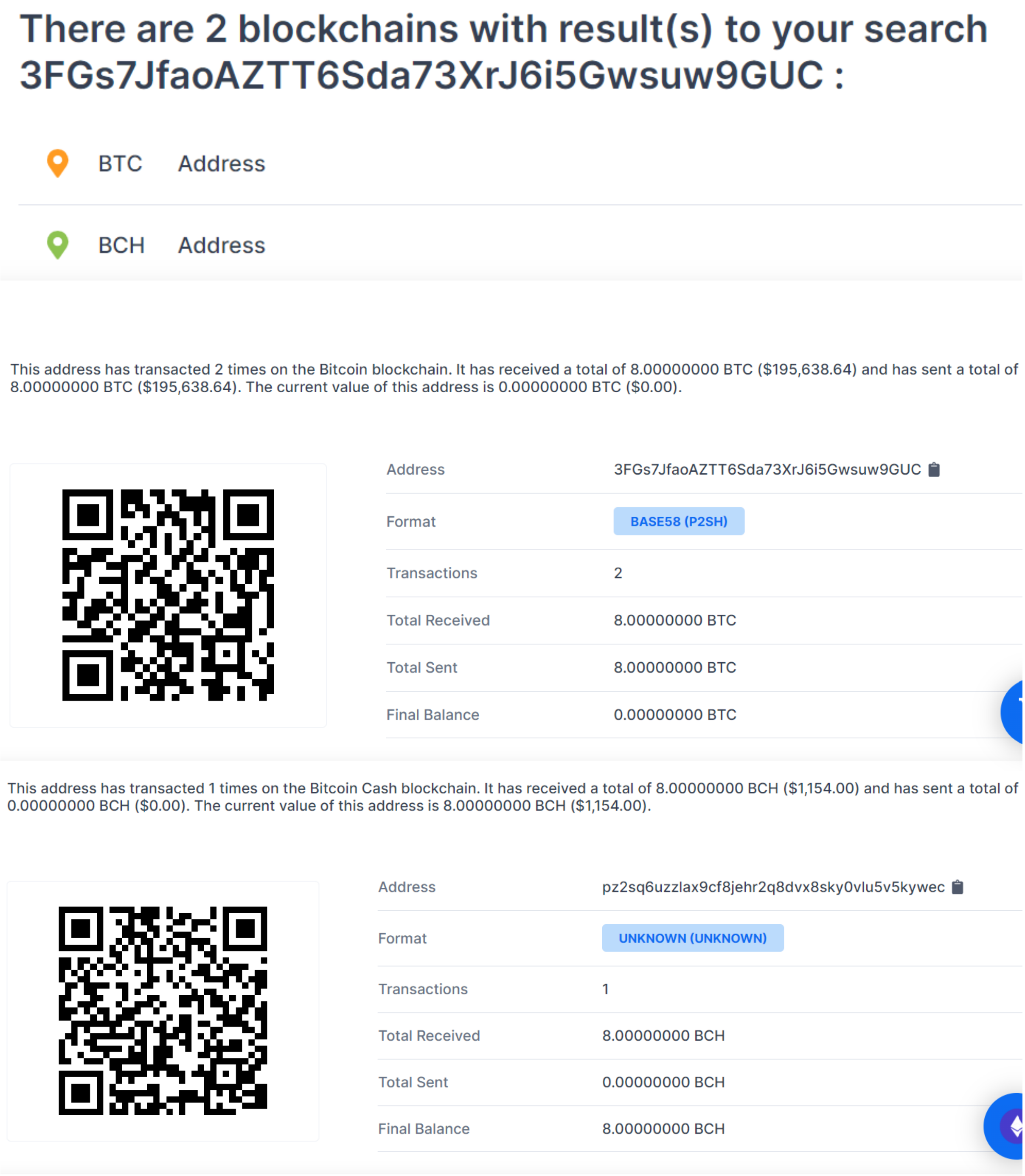

Fig. 24 Receipt of Bitcoin Cash by owner of Bitcoin due to a hard fork

### 2.3.11.2 Soft Fork

Soft Forks are minor changes in the blockchain protocol/software that are backward compatible. Thus, a new blockchain is not created and the blocks created by nodes running the updated software are treated as valid by the nodes running the old version of the software. However, the blocks created by old nodes would be treated as invalid by the nodes running the upgraded software. If a significant number of nodes upgrade their software, this results in no disruption in the functioning of the blockchain. Soft forks usually do not have any tax implications unless they trigger a hard fork like in case of Bitcoin.

### 2.3.11.3 Taxation of forking events

A) Hard Forks: As explained earlier for those who own any crypto asset amount right before the fork, hard forks result in the ownership of crypto assets in both the forks of the blockchain after the fork has taken effect. This results in acquisition of crypto assets by the users of one blockchain on another blockchain without any consideration. The two taxable events that arise out of a hard fork are:

   i) Acquisition of crypto assets in the other blockchain
   ii) Disposal of the crypto assets acquired as result of the hard fork

   The tax treatment of acquisition of crypto assets in the other blockchain due to a hard fork differs across jurisdictions. While some jurisdictions like the US and Japan consider crypto assets received because of hard forks as taxable income, with the value of the newly acquired crypto assets taken as the fair market value of the crypto assets at the time of the hard fork.

   Other jurisdictions like the UK, Finland, Sweden etc. do not tax the crypto assets received because of hard forks at the time of their receipt, and their cost of acquisition is taken as zero. However, upon the disposal of the crypto assets received because of hard forks, a capital gain or income tax may be charged depending upon the local law (presence or absence of capital gains) with the tax liability depending upon the basis/cost of acquisition. In Australia, the Australian Tax Office has issued guidance for hard forks where individuals/entities running crypto asset businesses need to apply trading stock tax rules instead of Crypto Tax Rules in hard fork events.

B) Soft Forks: As soft forks are minor software changes and do not result in a new blockchain, they are not taxable events and result in no tax liability.

### 2.3.12 Bitcoin Wallets

Bitcoin Wallets are software that store the address(es) of a Bitcoin user (Fig. 25). A wallet is a software that keeps a track of the crypto assets owned by various addresses in the wallet, generates and manages new addresses for new transactions (as it is recommended to generate new addresses for every Bitcoin transaction). The wallet software also provides a consolidated interface enabling the owner to find the aggregate sum of all UTXOs owned by him/her in the form of a Bitcoin balance. Recommendations of transaction fees are also provided by wallet software after analysing the network congestion and transaction fee trends.

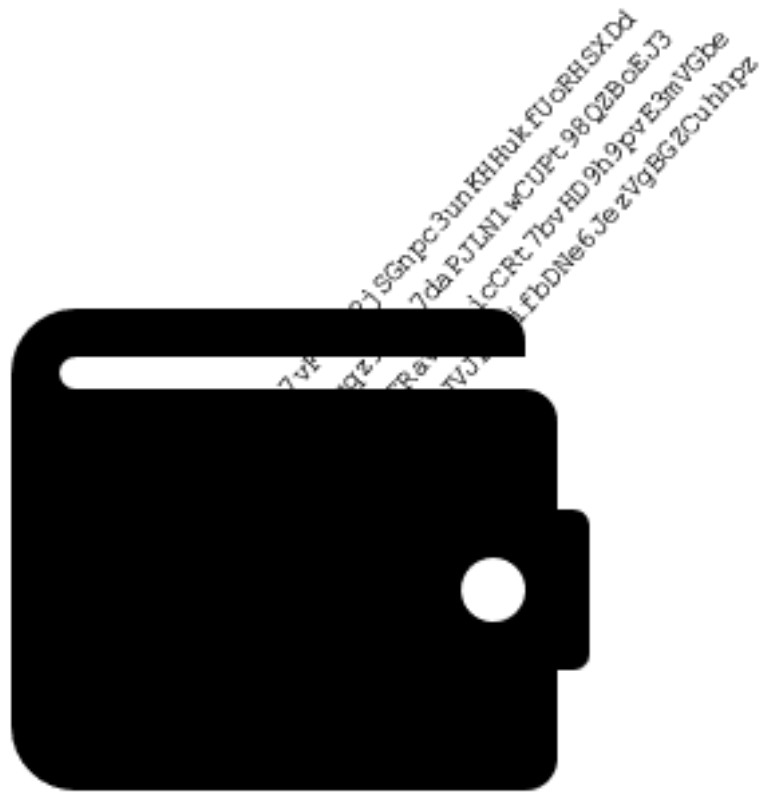

Fig. 25 A Bitcoin wallet (representation)

There are mainly two types of wallets used by Bitcoin and other crypto assets. Cold wallets are the wallets stored on a paper or some specialized hardware wallets that are not connected to any network or the Internet. A collection of Bitcoin addresses as shown in Fig. 7 is known as a paper wallet. There are multiple hardware wallets like the Trezor hardware wallet and the Ledger Nano hardware wallet.

These USB drive like wallets do not communicate with any other device unless physically plugged while accessing the private keys. It is like the digital signature USB drives which need to be plugged-in to sign a document.

Hot wallets are connected to the Internet and the blockchain network. They are often in the form of smartphone apps like the Trust wallet or computer applications and browser extensions like the Coinbase wallet and Jaxx Liberty wallet respectively. Since, hot wallets are connected to the Internet, they are usually recommended to be used only for small transactions. For privacy, the wallets create a new address for every Bitcoin transaction. Many wallets also use a mnemonic seed phrase along with a password and using this seed key they can generate many public and private key pairs and use a new key for every transaction. Such wallets are known as Hierarchical Deterministic wallets. One such wallet is shown in Fig. 26.

The practice of generating a new Bitcoin address for every transaction also protects Bitcoin users from risk of quantum computer enabled attacks. As explained earlier, while broadcasting a signed transaction, a user Alice needs to send her public key along with the signature. Thus, even if a quantum computer enabled attacker Qua finds Alice's private key using Shor's algorithm, he can derive no benefit, as Alice's address would have no UTXOs to be spent once the transaction is recorded on the blockchain.

Such wallets make it difficult to track the transactions on the blockchain, as a wallet can have numerous Bitcoin addresses and it can be difficult to associate them to a single user by analysing transactions on the blockchain.  Also, the Bitcoin balance shown by the wallet on the dashboard is an aggregate of all the UTXOs that can be spent by all the Bitcoin addresses in the wallet. Thus, there is nothing like One Bitcoin which resides in a wallet, a balance of 1.2 BTC on the wallet means that the Bitcoin Blockchain agrees that the Bitcoin addresses in the wallet have ownership of UTXOs which aggregate to 1.2 BTC.

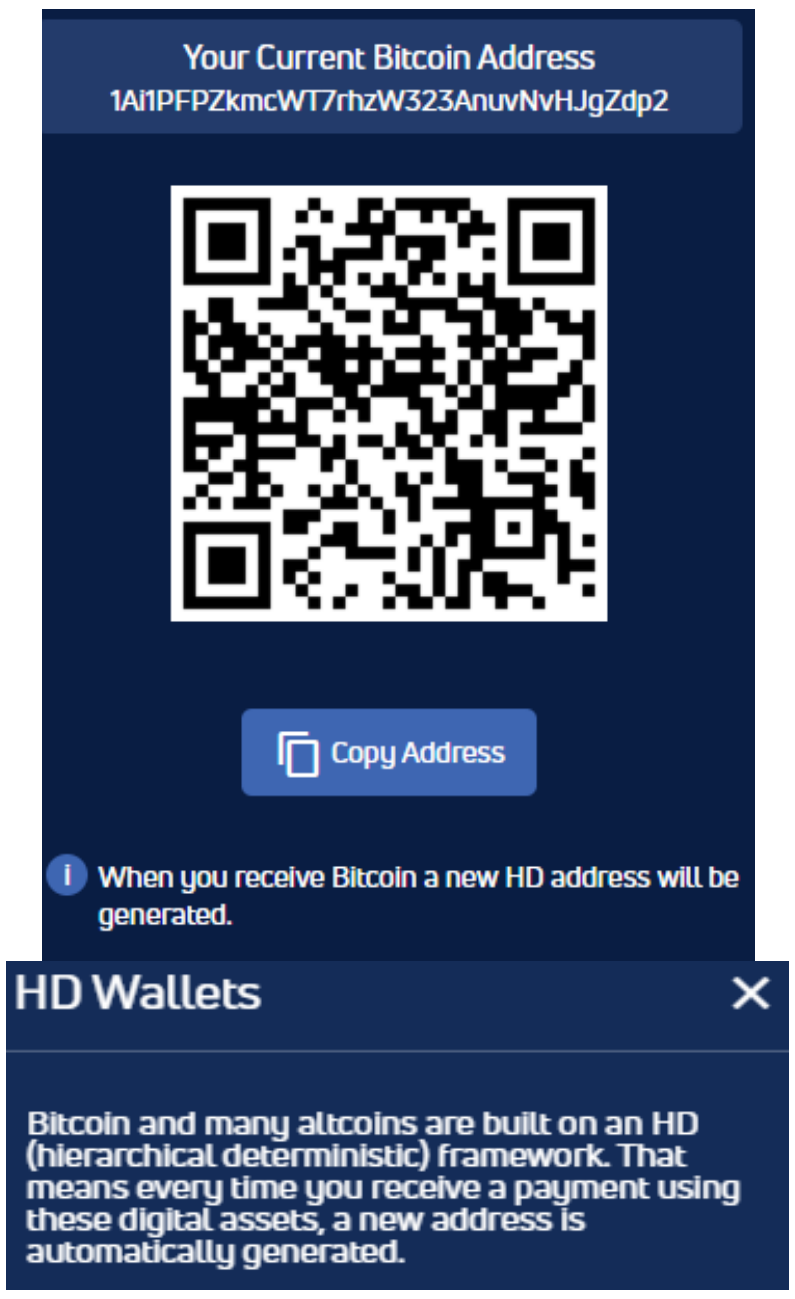

Fig. 26. A Hierarchical Deterministic wallet that generates a new address for every transaction.

These wallet services generate and manage addresses using ECDSA, as described above and provide services like creating transactions with optimum number and amounts of UTXOs, signing them and broadcasting on the Bitcoin network. Most wallets do not charge any fees for it. However, if the wallet charges a fee, it might be subject to service tax/GST in various jurisdictions. In many jurisdictions the wallet services charging fee might not be exempted from VAT/GST as for a service to be VAT Exempt the service must itself be an exempt service, and the argument that it is an input service to an exempt service does not suffice for VAT exemption to a service (C-2/95 SDC[62] and C-350/10 Nordea[63])

It is also important to note that many such hot and cold wallets do not capture any identity information regarding the user or carry out any KYC compliance procedures. Anyone with access to the Internet and a smartphone or PC can create an account or buy a hardware wallet and start transacting in crypto assets. This kind of wallets where the keys are owned and directly controlled by the user of the wallet are known as non-custodial wallets.

## 3. Crypto Exchanges

The Bitcoin Blockchain allows users to transact 24x7 using wallet software which facilitate signing and broadcasting transactions to be included in blocks that get added to the Bitcoin Blockchain. However, Bitcoin owners also have two other fundamental requirements of being able to convert fiat currency into crypto assets and vice versa, as well as exchanging one crypto asset into another. The crypto exchanges provide these facilities and many more to the owners of crypto assets. In this ecosystem the exchanges can be divided into two categories, centralized and decentralized exchanges. The major differences between them are:

| **Centralized Exchanges** | **Decentralized Exchanges** |
|---|---|
| Exchange is the custodian of private keys | Owner is the custodian of private keys |
| KYC is required | Usually, No KYC is required |
| Easy to use | Relatively difficult to use |
| Products like derivatives available | Offer only basic products |
| Higher liquidity | Lower liquidity |
| Can convert fiat currency to Crypto assets | Cannot convert fiat to crypto assets directly |

The wallets provided by crypto asset exchanges are known as custodial wallets as the keys contained the wallet are in the custody of the crypto exchanges. The users access their accounts and authorize transactions using a PIN or a password which the exchange executes locally or on the blockchain, on behalf of the user, charging a fee.

Some of the other facilities provided by the exchanges are:

i) Liquidity: Due to the participation of market makers, the crypto exchanges can provide liquidity to the markets. It enables facilitation of trades as enough number of buyers/sellers are present at any given point through facilities like margin trading.

ii) Derivatives: Products like Futures and Options based on the underlying crypto assets are offered by exchanges.

iii) Passive Income: By providing services like liquidity to exchanges, staking crypto assets in proof-of-stake consensus mechanism, loans, and yield farming[64], many crypto exchanges allow users to earn passive income on their crypto assets. Many centralized and decentralized exchanges offer such services and in the latter case they form a part of

[62] https://eur-lex.europa.eu/legal-content/EN/ALL/?uri=CELEX%3A61995CJ0002

[63] https://curia.europa.eu/juris/document/document.jsf;jsessionid=CD1328EA651A7F81A11CBDAB3C8E3AC9?text=&docid=108324&pageIndex=0&doclang=EN&mode=lst&dir=&occ=first&part=1&cid=1203786

[64] A form of platform arbitrage where users staking or lending their Crypto Assets move their assets between platforms that give the highest yield/return.

Decentralized Finance (DeFi) which is an emerging area in crypto asset ecosystem and warrants an in-depth discussion as a part of a separate paper.

Besides centralized and decentralized exchanges there are also some peer-to-peer trading platforms like LocalBitcoins and Paxful which list various individual suppliers of crypto assets who provide quotes and acceptable volumes of trade. The suppliers specify methods like credit card payments, bank transfers, gift cards etc. for paying for crypto assets like Tether, Bitcoin or Ethereum. These platforms provide escrow service which minimizes the counter party risk and the seller transfers the crypto assets to the buyer once the receipt of the payment by the seller is confirmed.

As the sellers on such platforms are persons who might be neither registered as crypto asset traders nor regulated, they can be potentially used by taxpayers to evade their due taxes on their crypto asset transactions. However, to comply to AML/CFT laws and regulations such services have verification levels depending on the types of transactions being undertaken by the user. For example, LocalBitcoins has 4 verification tiers in which the most basic level captures information regarding Full Name, Country of Residence, Email address and Phone number[65] of the user. This can enable tax administrations and other law enforcement agencies to gather information regarding such transactions in their jurisdiction.

Many centralized crypto exchanges execute trades based on order matching and get their liquidity from market makers and margin trading facility. As the transaction costs of recording the transactions on the blockchain are high, the centralized exchanges execute transactions by passing accounting entries in their own accounts of crypto assets or accounts of other users maintained with them. This off-chain nature of transactions enables faster processing times and greater scalability compared to on-chain transactions, which must be validated and confirmed by the blockchain network. For example, if Alice buys 0.1 BTC from a centralized exchange, the exchange might debit its own BTC account by 0.1 and credit 0.1 BTC to Alice's account in the form of an accounting entry. The exchange may also match the sell order of Bob against the buy order of Alice and debit Bob's BTC account by 0.1 BTC and credit 0.1 BTC to Alice's account. The exchange would also credit the fiat currency value of 0.1 BTC to Bob's fiat currency funds and debit the equivalent amount from Alice's fiat currency funds kept with the exchange.

Usually, the transactions undertaken on a centralized exchange cannot be traced on the blockchain except when the funds are transferred in the form of crypto assets from one exchange to another by the user, or to/from any other non-custodial wallet addresses outside the exchange. Similarly, some exchanges may utilize on-chain settlement mechanisms for certain types of trades or transactions to provide greater transparency and security. However, most trading activities on centralized exchanges are conducted off-chain to optimize performance and scalability, albeit at the expense of decentralization and trustless nature inherent to blockchain technology.

The management of user keys also differs between centralized exchanges and decentralized exchanges. In centralized exchanges, users typically do not have direct control over their private keys, which are used to sign transactions and authorize the transfer of funds. Instead, users rely on the exchange to manage their keys on their behalf, with the exchange holding custody of the keys associated with user accounts. This centralized custody model introduces counterparty risk, as users must trust the exchange to safeguard their assets and manage their keys securely. However, some exchanges offer additional security features such as two-factor authentication and cold storage to mitigate this risk.

---

[65] https://localbitcoins.com/guides/verification-guide

The exchange charges a service fee for this transaction and is liable to pay VAT/Service Tax/GST on it. It is also the case that the service fee can be paid in the form of another crypto asset (which is native to the exchange or freely tradable) at a discount. This makes the transaction equivalent of buying a service in exchange of crypto assets and may be taxed accordingly.

## 4. Source of Value of crypto assets and Bootstrapping

There are multiple examples of assets which have a value without an underlying asset. For example, the valuation of shares of many start-ups are based on the perception of the company in the minds of the shareholders and the expectation of higher profits in the future. Tulip Mania, the Dutch speculative bubble in the 17th century which led to prices of some Tulip bulbs reaching unprecedented levels, is another such case where the Tulip bulbs did not have any inherent value or utility, but derived their value from the value buyers ascribed to them.

Bitcoin and other crypto assets have some characteristics of other conventional assets which ascribe value to them. Scarcity is one feature which makes it rare, just as Gold and Diamond derive their value from scarcity, the fact that only 21 million Bitcoins can ever be issued, makes it scarce. There is also some evidence to suggest that the lack of centralized control over Bitcoin makes it trustworthy and a hedge against inflation, as no central bank can ever impose an 'inflation tax' on Bitcoin owners. Various instances have come to light where citizens of a country who have reduced faith in their Central Banks use Bitcoin as a safe store of value[66]. Also, the secure nature of immutable blockchain which is almost impossible to modify along with cryptographic safeguards guarantee the owners of Bitcoin the ownership and the right to transfer the assets securely. If the users believed that an attacker can maliciously steal Bitcoins from them, Bitcoin would not have much value as an asset. The fact that the users can transfer a part or whole of their Bitcoins to another user who is willing to transact, irrespective of his/her location establishes the acceptance and portability of Bitcoin.

This trust in the value of Bitcoin also has positive feedback on the nodes which add blocks to the blockchain (miners). It creates an incentive for them to secure the blockchain and maintain its integrity. The incentive mechanism promotes honest behaviour by nodes and miners which maintains the integrity and trust in the Bitcoin Blockchain which has not suffered outages. As seen earlier, if the number of miners decreases resulting in lower hash rate, the mining difficulty decreases resulting in an increase in the number of miners. The value ascribed to Bitcoin is a complex interplay between the factors depicted in Fig. 27

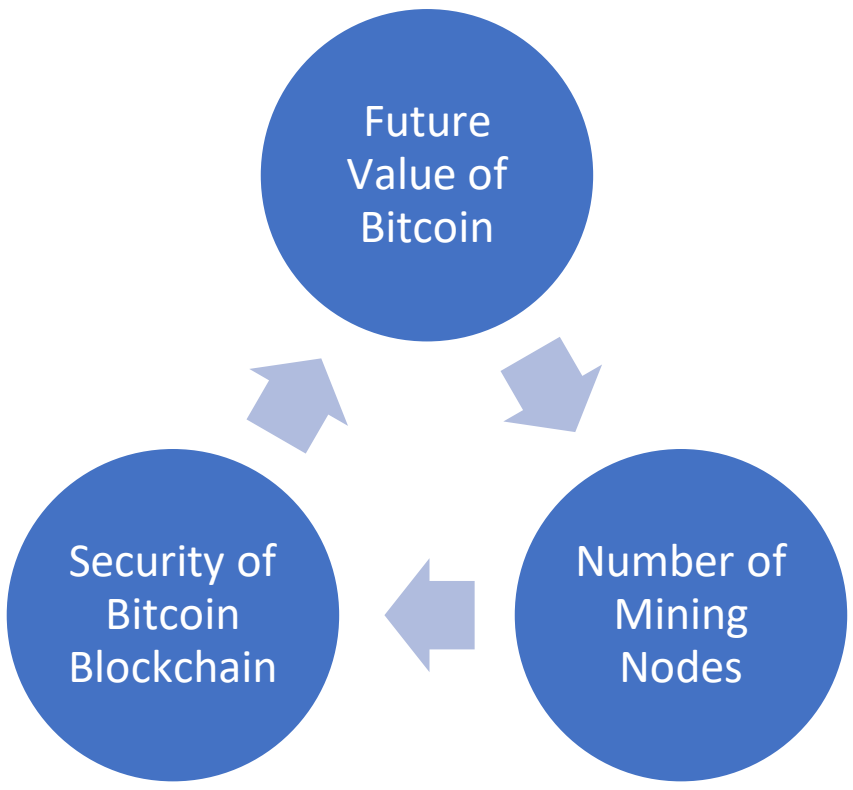

Fig. 27 Source of Bitcoin Value

[66] https://www.mariblock.com/africans-should-embrace-stablecoins-safeguard-savings-against-inflation-currency-devaluation/

In the above figure, the perceived future value of Bitcoin is an incentive for a sizeable number of nodes to validate the transactions and mine blocks, which enhances or maintains the security of the blockchain and thus, in turn keeps Bitcoin valuable.

Other crypto assets like Ethereum, Tether and Solana also derive their value from properties and mechanisms of Bitcoin with a few modifications. Ethereum with is smart-contract[67] ecosystem provides a wide platform on top which various services like lending and borrowing can be offered. Another token, Solana claims highest transaction speed of 65,000 transactions per second with average transaction fees of $0.00025[68]. It is also growing its stack of services which can offered on top of Solana Blockchain. Such USPs and Network effects also contribute to the value of crypto assets.

The security provided by high hash rates and large number of miners can explain the value ascribed to crypto assets in stable blockchains like Bitcoin. However, to start and establish a blockchain or a crypto asset on top of an existing blockchain like Ethereum, the developers need to bootstrap it to create a virtuous cycle and enhance its prospects and investment. This is done through various incentives like higher initial block rewards, initial coin offerings and airdrops along with aggressive marketing. In case of Bitcoin when Satoshi Nakamoto mined the first block of Bitcoin the Bitcoin Blockchain was not secured by many nodes producing trillions of hashes per second. To incentivise more nodes and miners to join the Bitcoin network, the block reward was 50 BTC per block. The idea of a new decentralized system of exchanging value in a trustless manner attracted users and miners who also believed that the price of Bitcoin would increase once it gains wider acceptance, which was indeed the case. This system of incentives has also resulted in honest behaviour being the dominant strategy for miners and users if bitcoin mining is considered a repeated zero-sum game.

## 5. Initial Coin Offerings

Many blockchain developers undertake projects to launch their own blockchain or another token on top of an existing blockchain like Ethereum. This is somewhat like a start-up launching its product. For this purpose, developers need funds for development as well as a user base for the token or blockchain who participate in the network and contribute in terms of nodes, value in tokens and transactions over the blockchain. Thus, many developers come out with initial coin offering and sell/promise to sell the new tokens/crypto assets to investors at a price, just like in the IPO of a company, to raise capital for the project. To enable investors/buyers to understand the proposed token/blockchain the developers usually come out with a whitepaper which contains:

- The long terms goals of the developers for the token/blockchain
- The detailed architecture and code of the token/blockchain
- Plan for marketing and advertisement
- The USP of the token/blockchain
- Details like developer fee, the initial distribution of assets and rights and permissions on the blockchain/token for user and developers
- Details of reward and incentive mechanism in mining and transactions, if any

This makes the participants and investors stakeholders in the success of the project and spreads the risk involved in launching a new token/blockchain. However, ICOs are not regulated in many jurisdictions or their regulation is not clear. This results in investors not having rights and safeguards similar to the equity markets. As many crypto assets are very similar to securities in nature there is a possibility of intervention of securities regulators which aim at enforcing investor protection measures similar to securities markets. There are numerous examples of various successful ICOs. Ethereum was funded through an ICO in 2014 where buyers exchanged Bitcoin for Ether. In the Ethereum ICO some amount was set aside for the developers and the Ethereum Foundation. Nowadays Ethereum through its smart contract execution enables the creation of new tokens and platforms which are interoperable on the Ethereum Blockchain.

[67] A self-executing contract with the terms of the agreement between buyer and seller being directly written into lines of code
[68] Digital Assets Primer: Only the first inning- Bank of America

ICOs are important taxable events as the fair market value of new crypto assets at the time of their acquisition would be the basis for calculating the capital gains on the 'disposal' of such assets in lieu of fiat currency or another crypto asset (subject to domestic tax laws). Also, as most crypto assets are highly centralized in the initial phases and its only after the crypto asset gains some traction that their control is transferred to a completely decentralized organization, such crypto assets may be classified as securities and subject to taxes due on securities and regulations of the securities regulator.

## 6. Airdrops

Another method used by developers of new tokens or blockchains is to give out crypto tokens for free to incentivise its usage and enhance market presence and exposure. These free tokens are known as airdrops. They are also a kind of reward for the early adopters of the new crypto asset. For example, Uniswap, a decentralized exchange gave 400 Uniswap token to all the Ethereum accounts that had interacted with its smart contract prior to September 1, 2020. This was aimed at rewarding users for using the Uniswap exchange and make them stakeholders in the success of Uniswap.

To receive airdrops users are required to perform actions like using the platform or service, signing up for newsletters and mailing lists, hold tokens for a specified period, contribute to the development of the project etc. Non-Fungible Tokens or NFTs can also be airdropped for using an NFT platform or buying/selling NFTs.

Airdrops are also important taxable events and in most jurisdictions an income tax is chargeable upon the receipt of an airdrop. The receipt of an airdrop is not taxed but is considered acquisition of a crypto asset with zero basis in some jurisdictions. Also, the subsequent sale, swap, spend or gift transactions are also subject to capital gains. In some jurisdictions, depending upon the domestic tax laws and guidance, gifts might not be taxed.

## 7. Ethereum

Bitcoin is mainly used to transfer value from one address to another in a secure and efficient manner. However, it does not offer many opportunities to build applications on top of the Bitcoin Blockchain as the Bitcoin script is not Turing complete[69] and does not allow user defined logic and customized functions to execute in transactions. As seen earlier, Bitcoin transactions also require the entire UTXO amount to be spent and do not provide control over the amount that can be withdrawn from a UTXO. Ethereum is different from Bitcoin as it provides above and beyond the capacity to transfer value from one address to another. Ethereum is mistakenly considered a crypto asset whereas it is a decentralized platform that is designed to run smart contracts, with Ether as its native asset which gives basis to its value.

As Ethereum's language is Turing complete, it can act like a giant decentralized general-purpose computer which is censorship resistant and minimizes third party risks in transactions. To understand Ethereum it is important to imagine a blockchain as a state-machine where transactions change the state of the blockchain. A blockchain is essentially a "cryptographically secure transactional singleton machine with shared-state."[70]. This is depicted pictorially in Fig.28 [71] . The consensus mechanism essentially aims to make all the nodes/validators of the Ethereum network agree on the current state of the blockchain.

[69] In Computer Science a Turing Complete system is one that can mimic a Turing Machine. A Turing Machine is a theoretical machine with a memory tape of infinite length which can calculate or compute anything for which an algorithm exists.

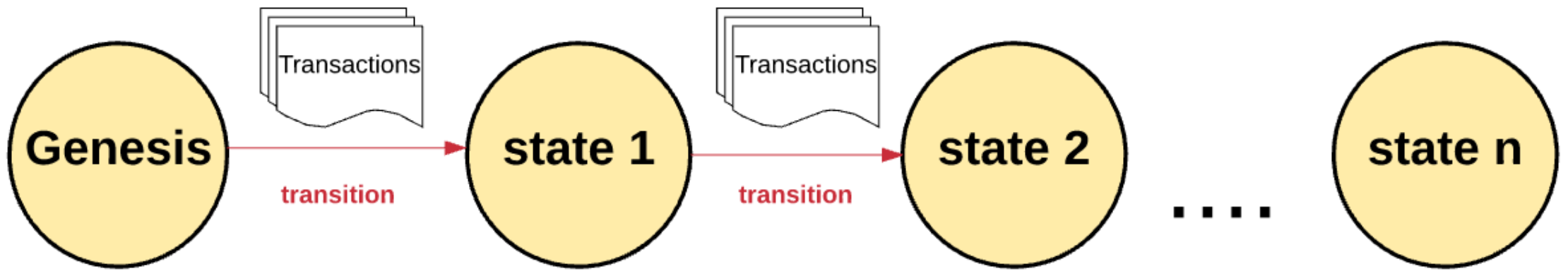

Fig. 28 Blockchain as a State Machine[72]

Ethereum is an account based blockchain unlike Bitcoin which is UTXO based. For a Bitcoin user the 'Balance' of his/her account is the sum of all the UTXOs for which he/she owns the private keys. In Ethereum, the accounts keep a track of the balance automatically. There are two types of accounts in Ethereum a) Externally Owned Accounts which have private keys and no code associated with them and b) Contract Accounts which do not have private keys and have code associated with them. Both these types of accounts have a Balance field associated with them, which shows the amount of Ether owned by these accounts. The state of the Ethereum Blockchain consists of the balance, data, code, and all other fields of all the accounts. The fields in a transaction on the Ethereum Blockchain are given in Table 4.

A contract account contains the code of the smart contract and it is controlled by the smart contract code. In an Ethereum transaction when the destination address is a contract address, the smart contract code is executed on the Ethereum Virtual Machine. The smart contract function specified in the data payload is called, if no function is specified, a fallback function is called.

| | |
|---|---|
| From | Address of externally owned account initiating the transaction |
| To | The address of the receiver. 0x00000 in case of creation of a smart contract |
| Value | Amount in Wei ($10^{-18}$Ether) |
| Data | Bytecode sent as input for contract creation or execution |
| Gas Limit | The maximum amount of gas user is willing to pay for the transaction |
| Gas Price | The amount of Ether the user is willing to pay for each unit of gas. In gwei(=$10^{-9}$ Ether) |
| Nonce | Sequence number of transaction for the EOA |
| Signature | Proves the origin of the transaction, not required in transactions where one contract invokes a function in another contract. |

Table. 4 fields in a transaction on the Ethereum Blockchain

An Ethereum block consists of transactions which i) change the balance associated with EOA and smart contract accounts ii) change data stored in a smart contract or iii) create, destroy(deprecated) or change the code of a smart contract. Each transaction affects the data stored in contracts in every single full node on the Ethereum network. Thus, Ethereum can be imagined as a giant decentralized computer where users can run their code and store data after paying the fee for the transaction. However, these computations are mainly used for basic functions like changing the balance and ownership of tokens held by the smart contract and validating signatures instead of storing large files or executing complex programs.

[70] https://docs.ethhub.io/ethereum-roadmap/ethereum-2.0/stateless-clients/
[71]https://www.preethikasireddy.com/post/how-does-ethereum-work-anyway
[72]Source: https://uploads-ssl.webflow.com/5ddd80927946cdaa0e71d607/5ddd80927946cdd1dd71d6f1_how-does-ethereum-work-anyway-2.png

Ethereum also has a much smaller block time of 12 seconds as opposed to ~10 Minutes in Bitcoin. It also has much lower fees as compared to the Bitcoin Blockchain. The fee to be paid for any transaction in Ethereum is called gas. To complete a transaction, the user needs to provide appropriate amount of gas depending on the size and nature of the transaction. Simple transactions to transfer Ether from one account to another do not cost as much gas as breeding CryptoKitties[73] using a smart contract. The user must provide adequate gas for the entire transaction to go through, if the entire gas is not used for the transaction, the excess gas is refunded. However, if the gas amount is not enough for the entire transaction to take place, the entire gas amount is consumed and forfeited.

Gas also protects the Ethereum Blockchain from infinite loops as even if a malicious user tries to execute an infinite loop on the Ethereum EVM it would stop after the transaction runs out of gas. Any contract execution in Ethereum must be triggered by an externally owned account. The triggered transaction can in-turn trigger further transactions on other contract accounts. The triggered transactions are executed atomically, i.e., either the full transaction involving all the intended smart contracts is executed or all the changes made by the incomplete transaction are reversed. Ethereum also uses digital signature and hash functions like Bitcoin. However, Ethereum uses the KECCAK-256 algorithm and an Ethereum address is a 42-character hexadecimal address derived from the last 20 bytes of the public key. The signed transactions are sent to a mempool, thereafter the nodes verify and propagate the transactions using flooding.

## 7.1 Proof-of-Stake based consensus in Ethereum

Ethereum initially started as a blockchain based on the proof-of-work consensus mechanism but in order to reduce the negative externalities of proof-of-work, it switched to a proof-of-stake based consensus mechanism similar to one described in the section 2.3.9. The Ethereum proof-of-stake mechanism is based on each validator staking 32 ETH to be a part of the set of validators which propose and attest blocks. Besides this the validators have various other duties assigned to them, for which they get paid in the form of rewards at regular intervals. Any malicious behaviour by a validator is penalized through slashing – a mechanism in which the staked Ether is slashed for violating the proof-of-stake rules, it also results in removal of the validator from the network.

The validators deposit 32 ETH in a smart contract on Ethereum and must wait for certain period in the activation queue before they can take part in proposing and validating blocks. The 32 ETH deposit is made to a smart contract that keeps track of all the staking validators. The validators specify a withdrawal address in the smart contract to which the payouts are made, as shown in Fig. 29 The depositor sends 32 ETH along with its public address and a withdrawal address to the smart contract. It is important to note that the staked ETH becomes a part of the consensus layer and is accounted for separately from Ether in execution layer Ethereum accounts and contracts. No Ethereum execution layer transactions can be run on staked ETH and no transfers can be made between validator accounts.

The validators can top-up their ETH balance on the smart contract later if it goes below 32 ETH. The validators are rewarded for proposing as well as validating proposed blocks with a part of the user fee and newly issued ETH by the Ethereum consensus layer known as 'Issuance.' The validators are also penalized if they do not complete their assigned duties of block proposal and validation. However, these penalties are much less severe than the slashing penalties which are levied for malicious behaviour. The base fee in each block is burnt in accordance with EIP 1559[74] which may be smaller or larger than Issuance for the block.

[73] https://www.cryptokitties.co/about
[74] https://eips.ethereum.org/EIPS/eip-1559

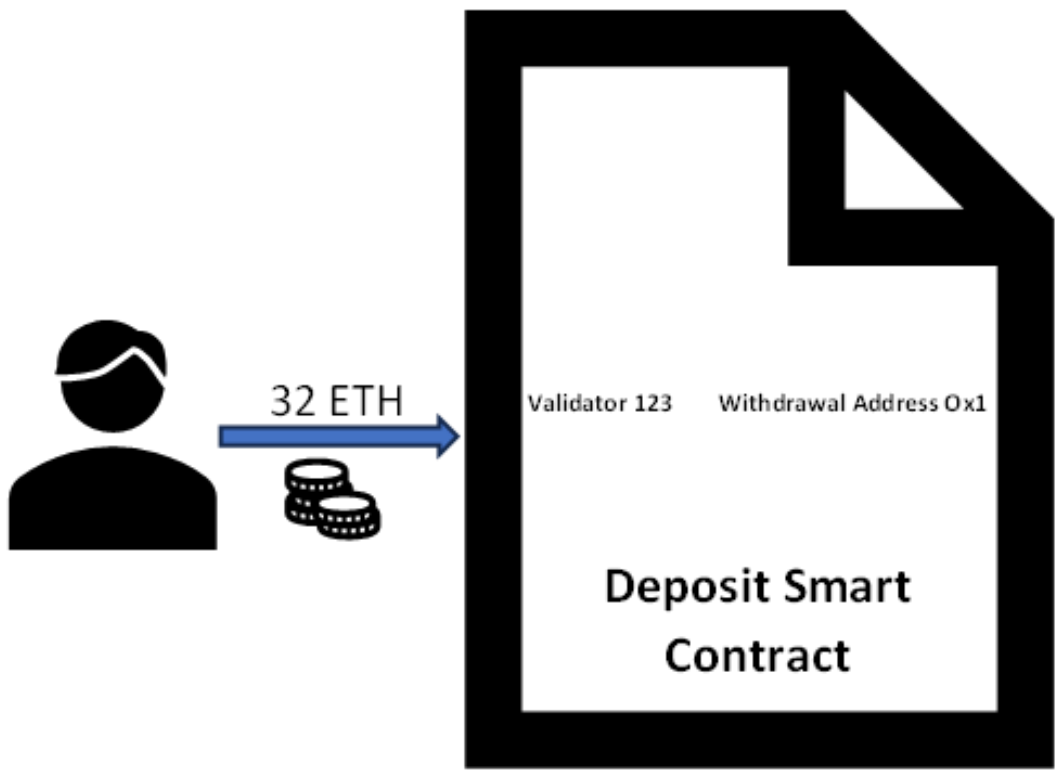

Fig. 29 Ethereum deposit smart contract

Validators are paid rewards at regular intervals of time. At any point if a validator is found violating proof-of-stake rules, its deposit is slashed and it exits the system. The remaining deposit of a slashed validator can only be withdrawn after certain time has elapsed. If a validator voluntarily decides to withdraw its stake and cease to be a part of the Ethereum consensus mechanism, it can initiate a withdrawal and if no breaches of the proof-of-stake rules are found, the entire staked ETH is returned to the validator which can be withdrawn after a certain interval. No gas is required for this transaction. The flowchart depicting deposits and withdrawals in the Ethereum proof-of-stake smart contract is given in Fig. 30 [75] The system of entry and exit of validators is not instantaneous to maintain a stable pool of validators over a certain period.

The proof-of-stake based consensus mechanism in Ethereum is like a voting-based mechanism in which the validators carry out the functions proposing the next blocks of the blockchain as well as attesting the blocks proposed by others. The attestation of a block is essentially a vote on the validity of the block and its inclusion in the blockchain. The validators maintain the integrity of the blockchain by honestly carrying out the duties assigned to them, with the other validators keeping an eye over their conduct. The mechanism is designed such that it incentivises honest behaviour by the validators and places heavy penalties on dishonest behaviour by the validators.

In the proof-of-work consensus mechanism, the high capital required for establishing a mining farm along with its high running costs creates barriers to entry in the mining ecosystem and plays a major role in incentivising honest behaviour by the validators. Similarly, in the proof-of-stake based consensus mechanism of Ethereum, staking of 32 ETH by each validator creates a high capital requirement. Ethereum tries to achieve consensus amongst the set of validators through a combination of two different consensus mechanisms – the Latest Message Driven Greedy Heaviest-Observed Sub-Tree (LMD GHOST) and the Casper Friendly Finality Gadget (Casper FFG). The LMD GHOST is different from the longest chain rule in the Bitcoin Blockchain and the proof-of-work based Ethereum as it counts every block in a fork as a vote, even if they are conflicting blocks and not a part of the longest chain. This means that the fork with the highest number of validator votes is the one to which new blocks are added.

[75] https://notes.ethereum.org/@hww/lifecycle

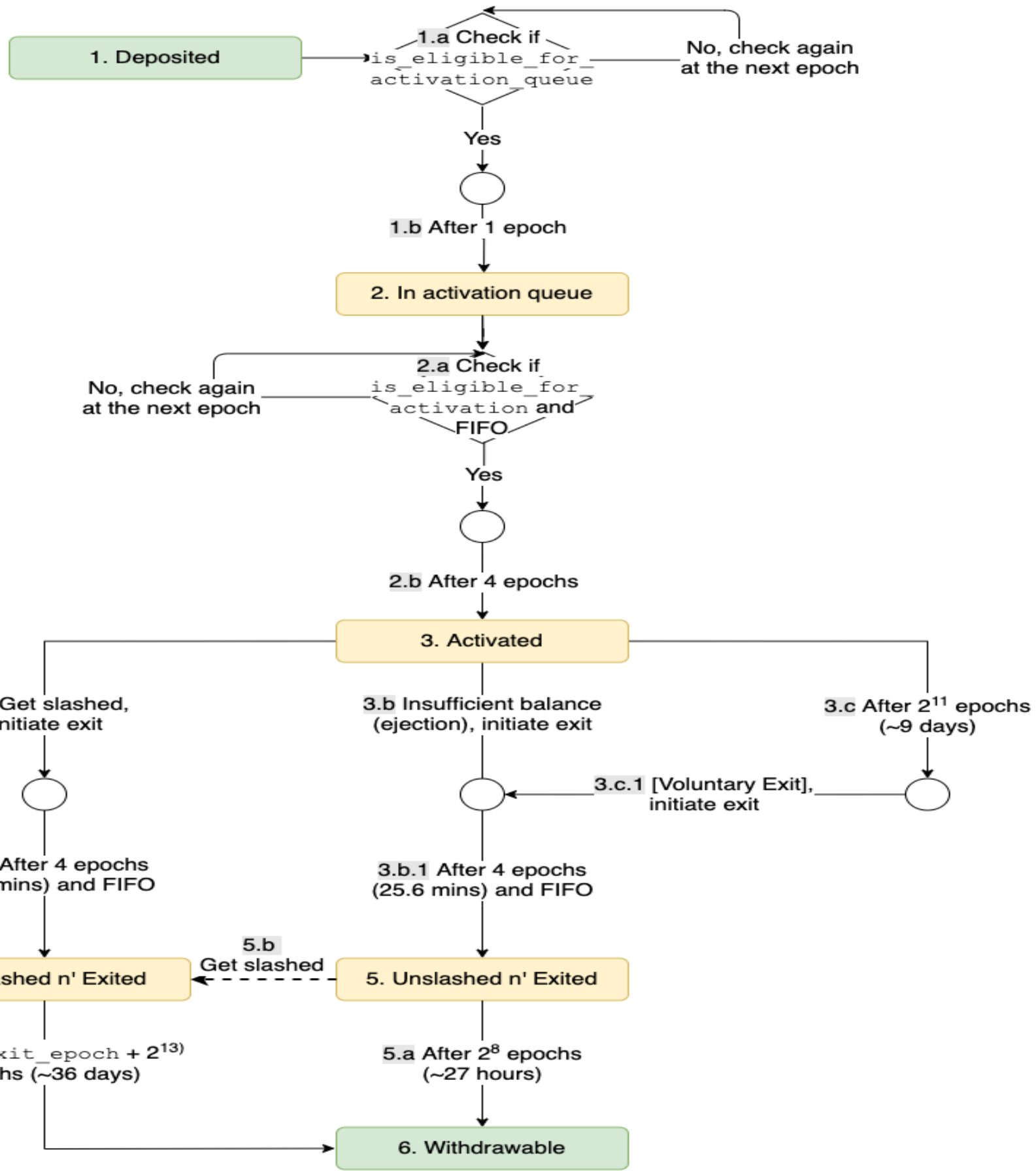

Fig. 30 Deposits and withdrawals in the Ethereum proof-of-stake smart contract[76]

### 7.1.1 Slots and Epochs

Time periods on the Ethereum Blockchain are divided into slots and epochs. One slot which results into a block proposal and attestations is 12 seconds long and 32 slots together form an 'epoch' as shown in Fig.31. By design, in Ethereum every 12 seconds a random validator is chosen to propose a block to be added to the Ethereum Blockchain. The block proposed by the validator needs to be verified by other validators in the validator set through an 'attestation.' These validators, other than the proposer have a responsibility for verifying the proposed block if it is correct. Attestations include information about the current block the validators are attesting to, as well as the previous block they are building upon. However, not all the validators are required to attest every block and, in each epoch, the set of validators is distributed into 32 randomly chosen committees, with each committee being responsible for one slot which is 12 seconds long.

At the start of each slot, the first validator of the committee proposes a block to be added to the Ethereum Blockchain. The rest of the members of the committee are supposed to attest the proposed block. Validators attest to the validity of proposed blocks by signing and broadcasting attestations. A validator can only attest one block per slot in an epoch and any violation of this results into slashing. Each proposer validator is required to propose a block in the first 4 seconds of the slot, failing which the validators in the committee are required to attest the previous block.

[76] https://ethos.dev/beacon-chain#

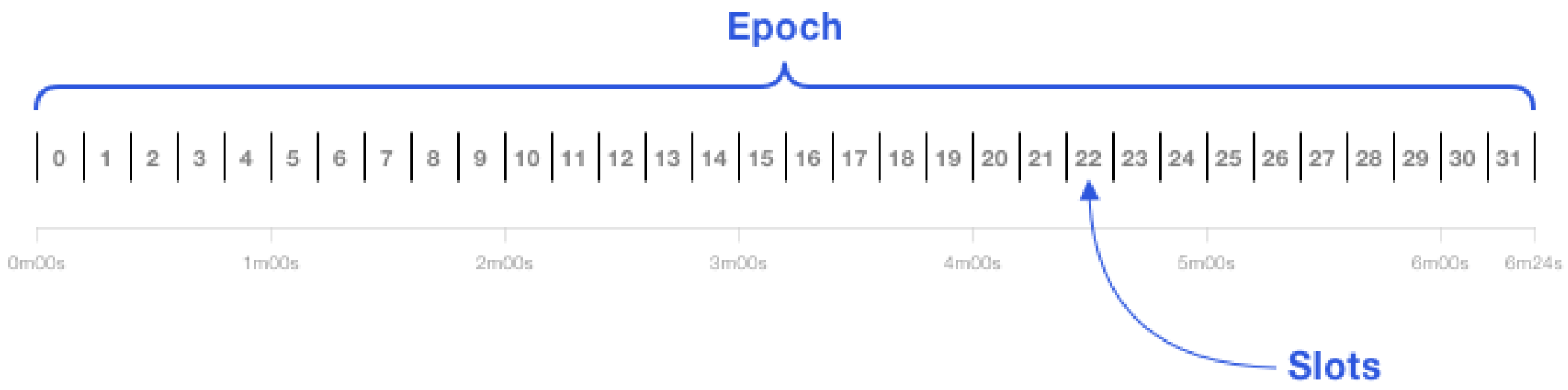

Fig. 31 Slots and epoch in Ethereum consensus mechanism[77]

Considering the current number of validators is >800,000 each committee is likely to consist of >25000 validators. It is an uphill task to collect signatures of so many validators for each slot. Consequently, validators in each committee are further divided into 128 subnets, as shown in Fig. 32, and validators produce an aggregate BLS signature in each subnet. Validators in the committee collect all the individual BLS signatures they receive and aggregate them into a single BLS signature using a process called "signature aggregation." This aggregation is possible because of the unique additive property of BLS signatures, allowing multiple signatures to be combined into one without losing any security properties. By collecting attestations from an ample number of validators, the network can finalize blocks, guaranteeing their inclusion in the Ethereum Blockchain and making them resistant to reversal unless a substantial portion of the validator set collaborates to do so.

The choice of the block proposal and assignment of validators into committees and subnets is dependent on an Ethereum Blockchain process that generates a pseudo-random number. The random number is generated on Ethereum using an algorithm called RANDAO[78]. The validator selection is fixed two epochs in advance. This implies that validators know a few minutes in advance about their upcoming proposer or attestation duties. This process pre determines a small set of 64 validators who are entrusted to propose blocks for each slot in the two epochs (12 mins 48 seconds).

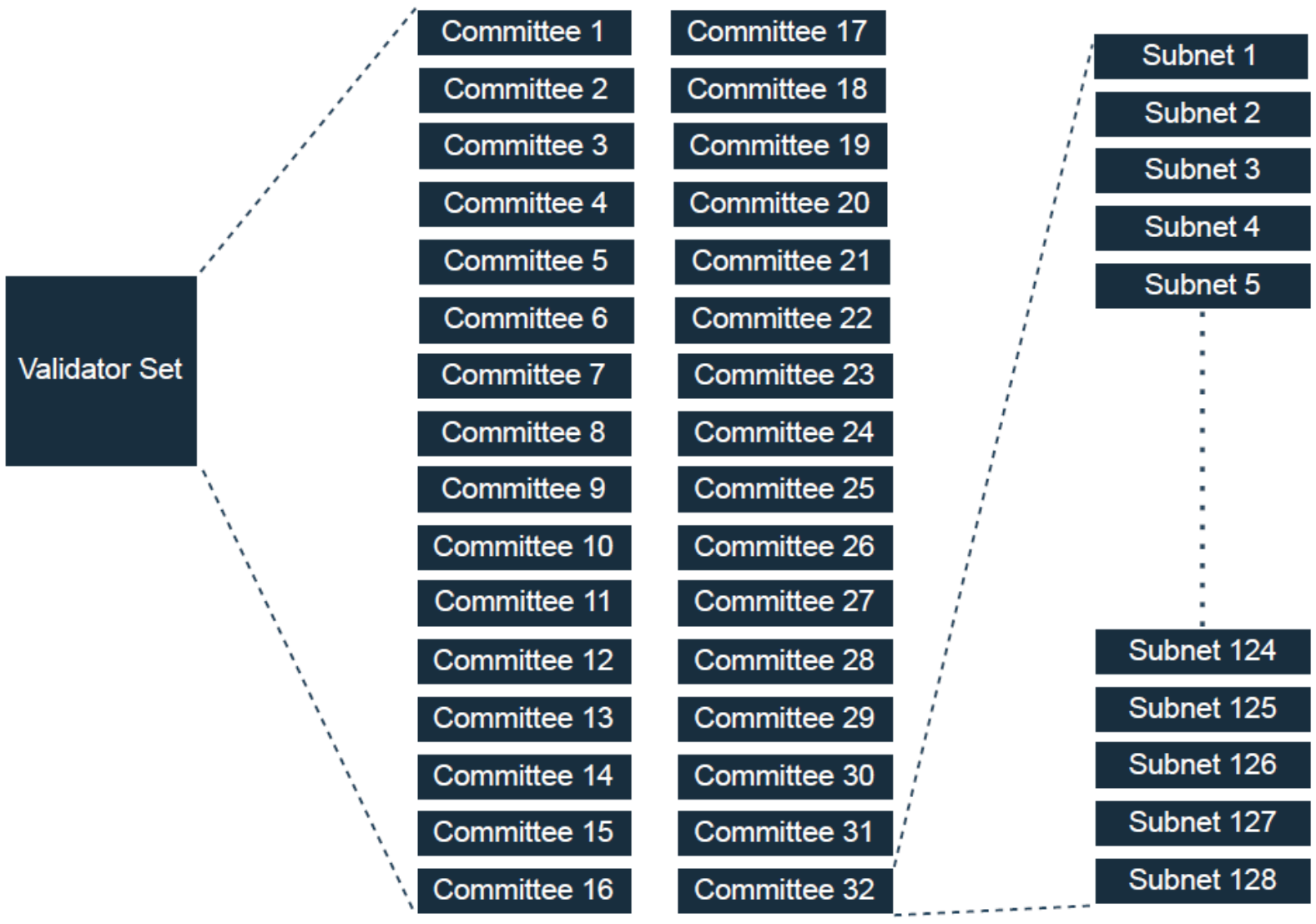

Fig. 32 Committees and Subnets in Ethereum consensus mechanism

[77] https://ethos.dev/assets/images/posts/beacon-chain/Beacon-Chain-Slots-and-Epochs.png.webp
[78] https://ethereum.org/gu/developers/docs/consensus-mechanisms/pos/block-proposal/

Besides the LMD GHOST which provides the fork choice rule[79], the Ethereum Blockchain also aims to achieve finality in the blockchain through the Casper FFG which is a Practical Byzantine Fault Tolerance (PBFT) inspired and improved consensus protocol. Finality refers to the guarantee that a block cannot be altered or removed from the blockchain without burning at least 33% of the total staked Ether. Finality in Casper FFG is achieved through "checkpoint" blocks, which are always the first blocks in an epoch.

Validators agree on the state of a block at checkpoint blocks, and if two-thirds of the validators agree, the block is finalized. It usually takes two epochs for the Ethereum Blockchain to attain finality as by the end on two epochs >2/3$^{rd}$ of the validators likely vote for two 'checkpoint' blocks. If a block is not able to exceed the two-thirds threshold, finality is not achieved, the fork choice rule would kick-in to determine which chain to follow and finality will be achieved when the 2/3$^{rd}$ majority is met. If finality is not achieved for more than four epochs, 'inactivity leak' mechanism gets activated.

The inactivity leak is a feature of the Ethereum proof-of-stake consensus mechanism that is activated when the network fails to finalize a checkpoint for more than four epochs (25.6 minutes). The inactivity leak is designed to restore finality in the event of the permanent failure of large numbers of validators. It does so by gradually reducing the stakes of validators who are not making attestations until the remaining validators control two-thirds of the stake and can resume finalizing checkpoints. The inactivity leak also prevents validators from receiving attestation rewards during this period, to discourage attacks that might deliberately cause the network to lose finality.

### 7.1.2 Rewards and Penalties

To incentivise the validators, the Ethereum Blockchain offers rewards to the participants in the network. Also, to penalize behaviour that is detrimental or outrightly malicious, the Ethereum Blockchain also has a system of penalties and slashing which act as a deterrence against such behaviour.

#### 7.1.2.1 Rewards

The main rewards which validators receive in the Ethereum ecosystem are the newly issued Ether created by the protocol and the transaction fee paid by the users transacting on the Ethereum Blockchain. The validators are assigned various duties like proposing and attesting blocks in a slot as well as participation in sync committees and they receive rewards for correct and timely performance of these duties. 'Correct' attestation implies that the attestation of the block by the validator agrees with the fork choice of the block proposer. This ensures that only participants in the winning fork (in case of an available fork choice) receive the rewards. Failure in performing duties in the specified timeframe results in missed rewards. The block proposers are also rewarded for reporting any violations of the slashing rules, which may not happen so often. In a block, majority of rewards of the validators come from attestations. An attestation contains three votes and each vote is eligible for a reward if it satisfies the conditions given in Table 5:

| S. No. | Validity | Timeliness |
|---|---|---|
| 1 | Correct source | Within 5 slots |
| 2 | Correct source and target | Within 32 slots |
| 3 | Correct source, target and head | Within 1 slot |

Table. 5 Timelines for attestation reward eligibility[80]

[79] The Fork Choice Rule (FCR) is a crucial mechanism in blockchain networks, determining which branch of a forked chain to accept as the canonical or main chain.
[80] https://eth2book.info/capella/part2/incentives/rewards/

Even if the above duties are performed in a timely manner by the validator, it may not receive the rewards due to various extraneous factors like the block proposer missing to propose a block within 4 seconds in a slot, or the block proposer proposing a block on a minority fork which is discarded after a few blocks. Thus, the rewards accrued to the validator due to randomness in allocation of duties as well as other extraneous factors stated above have a variance and vary over time. The breakdown of expected rewards for proposers and validators is given in Fig. 33

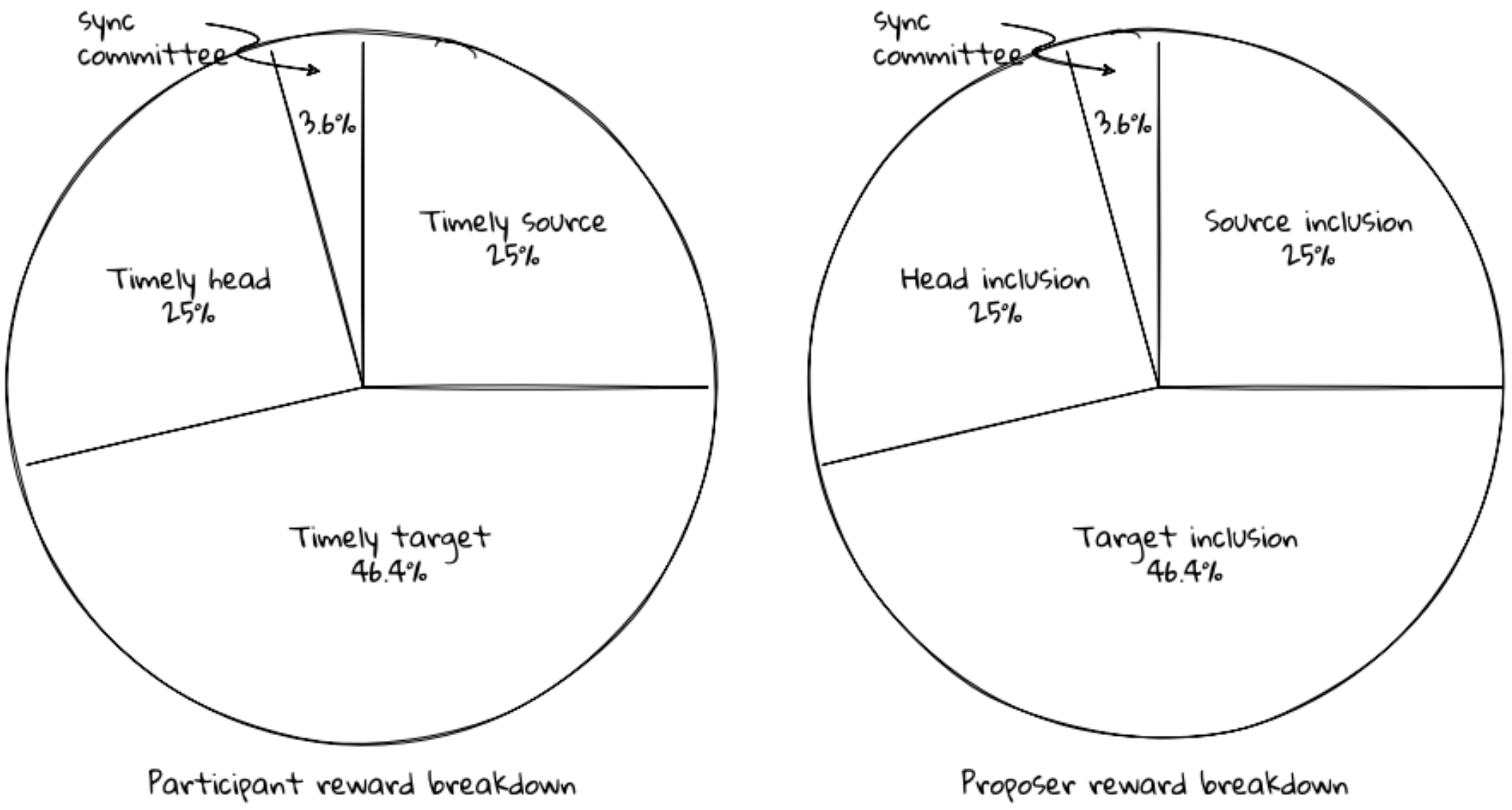

Fig. 33 Breakdown of expected rewards for proposers and validators[81]

Sync committees allow light Ethereum clients to keep track of the chain of beacon block headers. Light clients are nodes that do not download the entire blockchain, but only rely on the block headers and some cryptographic proofs to verify the state of the blockchain. Sync committees are groups of 512 validators that are randomly selected every 256 epochs (about 27 hours). They are responsible for signing the block headers that are the new head of the chain at each slot. The signatures of the sync committee members are broadcast to the network and are used by light clients to authenticate the block headers without downloading the full blocks. Validators are rewarded for participating in sync committees.

The maximum amount of newly issued Ether per year – Annual issuance, is proportional to the square root of the number of validators in the network. However, the annual returns of the validators are inversely proportional to the square root of the number of validators. This results in a design like the proof-of-work based consensus where the difficulty level of the target hash is related to the overall hashing power of the network. Similarly, in Ethereum as the number of validators increases the annual return on staked Ether goes down and vice versa, resulting in an optimal number of validators being present in the Ethereum consensus protocol. Mathematically

$$Annual\ Issuance \propto \sqrt{N}$$
$$Annual\ Return \propto \frac{1}{\sqrt{N}}$$

Here $N$ is the number of validators on the Ethereum consensus layer. This mechanism helps the consensus layer in reaching the equilibrium number of validators at a given time.

[81] https://eth2book.info/capella/part2/incentives/rewards/

### 7.1.2.2 Penalties and Slashing

To create negative incentives for validators who fail to contribute in a desired manner to the Ethereum Blockchain, penalties are levied. The validators are penalized for incorrect, late, or missing attestations(votes). The validators are penalized for incorrect, late, or missing source[82] and target votes[83] but there is no penalty for a missed head vote[84]. Besides this, validators who fail to participate in sync committees receive a penalty equal to the reward they would have earned had they participated in the committee correctly. These penalties act as 'sticks' in the consensus protocol to motivate the validators to perform their duties diligently. However, these are not penalties for malicious behaviour or potential attacks on the protocol.

The mechanism to deal with malicious activity on the Ethereum consensus layer is Slashing. Upon detection of violation of rules or any dishonest behaviour by a validator the Stake of such validator is slashed and it is removed from the network. Slashing protects the protocol against any attacks. For example, it prevents a validator from voting for two blocks in the same slot. Also, the incentive to the block proposers for reporting any violations of slashing rules acts as a protection against such malicious behaviour.

The penalties and slashing mechanisms in the Ethereum consensus layer have important tax implications which are related to the admissibility of the such penalties and slashing to be allowed as an admissible expense. While penalties can be allowed, as they might be caused due to bona fide reasons, the stake of a validator slashed due to malicious behaviour may not be allowed as an expense for the purpose of taxation.

### 7.1.3 Staking Pools

When an individual/entity runs an Ethereum validator node on its own and makes the entire 32 ETH deposit it is said to be involved in 'Solo' staking. Solo staking also involves a reasonable degree of technical knowledge and the validators retain the full control of the keys of their deposited Ether. This form of staking also involves incurring the hardware and operational costs only, without paying any service fee to staking pools, resulting in higher profits. It also helps to prevent the accumulation of majority stake with one central entity on the consensus layer.

For the validators who have access to the required capital but do not possess the required knowhow or do not want to run a validator node, usually delegate the technically difficult tasks to a service provider for a fee. The validators retain the control of the keys of their deposited Ether and do not need to purchase any hardware or software to run validator nodes. However, this method of staking might involve increased third party risks due to potential downtime or bugs with the software of the service provider.

---

[82] Source vote: This is a vote for the lower checkpoint of a link between two checkpoints from different heights. A checkpoint is a block that is divisible by 50, and a link is a connection between two checkpoints that represents a validator's attestation. A source vote is used to determine the justified and finalized checkpoints, which are the blocks that have received enough votes from validators to be considered final and irreversible

[83] Target vote: This is a vote for the higher checkpoint of a link between two checkpoints from different heights. A target vote is used to measure the participation rate of validators and to reward them for voting on the correct chain. A target vote also contributes to the finality of checkpoints, as a checkpoint can only be finalized if its previous checkpoint is justified

[84] Head vote: This is a vote for the most recent block that the validator sees as the head of the chain. A head vote is used to implement the LMD-GHOST fork choice rule, which selects the chain with the most votes from validators as the canonical chain. A head vote also helps to prevent stale blocks from being proposed and included in the chain

The 32 ETHs required to be deposited for being a validator on the Ethereum consensus layer is a significant barrier to entry for individuals/entities which want to participate in the consensus mechanism and earn rewards. Those inclined to participate in staking who do not possess the required capital or those who want to earn rewards over and above the staking rewards collaborate and stake Ether through a staking pool. Various stakers who possess smaller amounts of ETH collaborate and pool their assets to participate in staking on the Ethereum consensus layer and earn rewards.

In many staking pools the depositors are issued tokens that represent a claim on the staked ETH amount and its associated rewards. For example, on depositing ETH on a popular staking platform LIDO, the depositor gets and equivalent number of stETH tokens which are pegged 1:1 with Ether and can be traded or used like any other ERC token on Ethereum to earn additional income, over and above the staking rewards by transacting on various DeFi platforms as shown in Fig. 34 Consequently, more than 30% of the ETH currently deposited on the Ethereum consensus layer Deposit Smart Contract is staked through LIDO[85]. The stakers on such platforms have a choice to run a pool node and get additional rewards for it. The stakers pay a commission/fee (For example LIDO charges 10%) which is split between the pool node operators and the Decentralized Autonomous Organization (DAO) of LIDO.

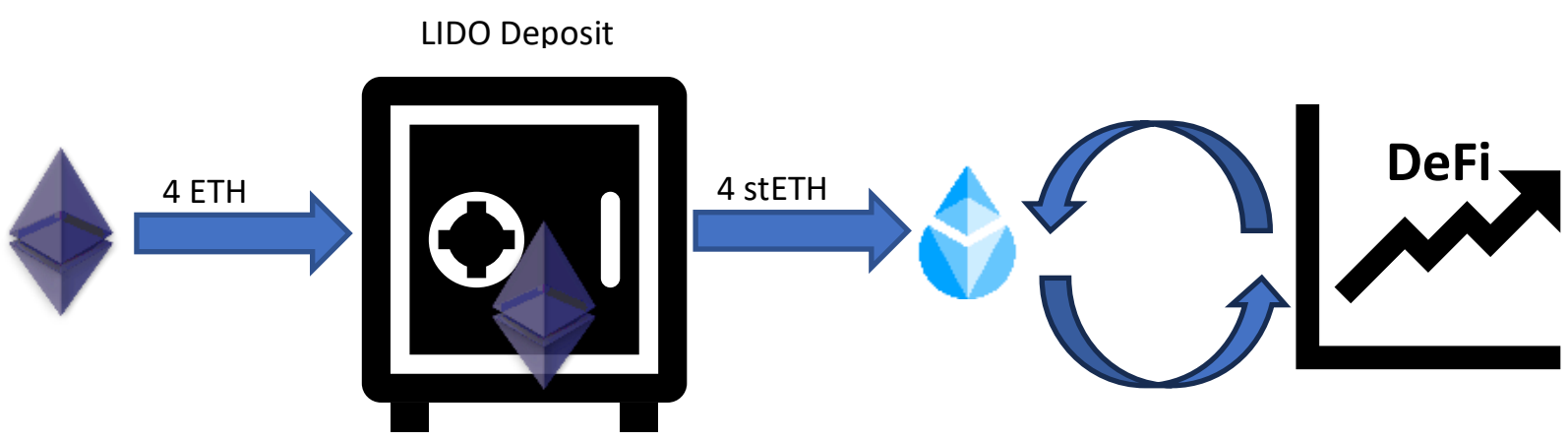

Fig. 34 Staking pool token issuance and trading

### 7.1.4 Taxation of Proof-of-Stake based consensus in Ethereum

The activities of the validators in the proof-of-stake based consensus in Ethereum and the rewards received by them can have direct and indirect tax implications as they result in accrual of income and involve providing certain services to the users of the Ethereum Blockchain. The main issues concerning the direct tax treatment of rewards accrued by staking revolve around the treatment of the reward as an active or passive income for the validator. For the levy of indirect taxes, the classification of nature of service provided by the validator, the classification of Ethereum as a negotiable instrument, property, asset etc. and the tax residency of the validation service recipient for ascertaining the place of supply would be important to determine the incidence of taxes.

#### 7.1.4.1 Direct Taxes related to Proof-of-Stake based consensus in Ethereum

As discussed above, the validators in the Ethereum consensus layer receive rewards for block proposals and attestation along with participation in sync committees and reporting any violation of the proof-of-stake rules. The rewards received by the validator above 32 ETH does not increase the weight of the validator in the consensus layer and is withdrawn automatically every few days as reward payment. The rewards are credited to the validator's payout address at regular intervals and as these are initiated at the consensus layer, no gas is required for such payout transactions.

The receipt of a staking reward by a validator would be a taxable event in most tax jurisdictions as it results in accession of wealth over which the validators have complete dominion. However, their tax treatment depends largely upon whether a jurisdiction considers the reward as a passive income for the stake deposited in the deposit contract by the validators or as an active self-employment income. The treatment would also be contingent upon the rewards being accrued because of solo staking or staking through a staking pool.

[85] https://dune.com/hildobby/eth2-staking

To classify the income from staking as active or passive income it is important to understand the nature of relationship between:

i) The Ethereum consensus layer and the Solo staking validator
ii) The Ethereum consensus layer and the staking pool validator
iii) The staking pool and person/entity contributing ETH to the staking pool and/or running a pool node.

The solo and staking pool validators are assigned with specific duties and tasks of proposing and attesting blocks on the blockchain. They agree to the rules of the consensus layer protocol and receive regular rewards for performance of assigned duties. Moreover, they are also subject to penalties for downtime and malicious behaviour. Also, the rewards received cannot be considered a compensation for the use or forbearance of money. Thus, the solo validators can be subject to self-employment taxes in some jurisdictions and the associated social security contributions. Some jurisdictions may consider the fact that as staking involves little or no effort and can be earned without the active involvement of the person or entity, and tax it as a passive income akin to interest. The staking pool validators, being involved in the business of staking would be treated correspondingly.

The taxpayers earning income by contributing ETH to a staking pool cannot be considered to be self-employed as instead of the contributors, the staking pool/pool node operator is assigned with specific duties and tasks of proposing and attesting blocks on the blockchain and is also subject to penalties for downtime and malicious behaviour. The rewards accrue to them on the Ethereum execution layer through the staking pool instead of the consensus layer. Thus, they may not be subject to self-employment taxes and the corresponding social security contributions.

The accrual and/or 'disposal' of the rewards would be a taxable event in most tax jurisdictions and attract income tax or a capital gains tax. If the validators are engaged in the business of providing validation services on a commercial scale, the income might be taxed as business income. The fair market value of the rewards at the time of accrual would be the basis for calculation of the capital gains tax. In this ecosystem, the staking pools are also likely to accrue income as an entity, which usually exists as a Decentralized Autonomous Organization. For example, LIDO which is the leading staking pool on Ethereum operates as a Decentralized Autonomous Organization. The tax issues related to such entities are discussed in subsequent sections.

#### 7.1.4.2 Indirect Taxes related to Proof-of-Stake based consensus in Ethereum

The validators, staking pools, the individuals/entities that contribute to the staking pools and Ethereum users involved in providing or receiving services may be subject to indirect taxes depending upon:

i) The nature of service provided
ii) The place of supply of services
iii) The recipient of the services

The solo validators and staking pools are involved in providing validation services to the users carrying out transactions on the Ethereum Blockchain and may be considered to be self-employed. They essentially provide two kinds of services, those to the users carrying out transactions on the Ethereum Blockchain (for which they get rewarded in the form of transaction fee) and to the Ethereum consensus layer for finalizing blocks (for which they get rewarded in the form of newly issued ETH).

For the transaction validation services to the user, the validators might be subject to GST if the validator is a resident of the jurisdiction same as that of the validator. Services provided by the validator to users in other jurisdictions would constitute an export of services and would be zero rated in most jurisdictions. In case of Ethereum, taxation of services provided by the staking pools or solo stakers to the Ethereum Blockchain itself, for which they receive the issuance reward programmed in the Ethereum Blockchain, would be a complicated issue. In this case also, the place of supply of such service cannot be determined and the payment for the service may amount to billions of dollars each year. Mechanism like those described in section 2.3.8.1 for collecting indirect taxes on miners and mining pools in Bitcoin can be applied for services of staking pools and solo stakers.

In case of other models of staking like using Staking-as-a-Service the services provided by the staking service provider might also be subject to GST depending upon the place of supply and the residency of the staking service provider and the individual/entity staking the ETH. In the case of staking pools, the commission charged by the pools would also be subject to GST as it is in lieu of a service provided to the depositors. As discussed in the case of mining rewards, only the direct validation services provided to the Ethereum users and the consensus protocol may be considered exempt services as other services would largely qualify as input to validation services. Similar to the treatment of miners providing services to users in other jurisdictions, the validators providing zero-rated services might be entitled to refunds for the input taxes paid for exporting the service.

#### 7.1.4.3 Taxes on MEV rewards on Ethereum

As described above the users broadcast their transactions on the Ethereum Blockchain into a ‘mempool’ and the validators which are supposed to propose blocks in a slot bundle the transactions into a ‘Block’. As all the mempool transactions are visible to everyone, the validators can also come to know about potential arbitrage opportunities on various DeFi applications. This can enable the validators to ‘front run’ such transactions to extract a profit by altering the order of transactions. This is known as Maximal Extractable Value on the Ethereum Blockchain and acts as an ‘invisible tax’ on Ethereum users and leads to imperfect markets. The role and taxation of MEV Ecosystem on the Ethereum Blockchain is described subsequent to the section on DeFi as the readers can better appreciate the mechanism through which value is extracted by validators and other actors by reordering DeFi transactions.

## 7.2 Smart Contracts

Smart contracts are agreements written in code which are automatically executed by the Ethereum Blockchain when certain conditions are met. For example, few individuals might pool Ether in a smart contract for organizing a fair lottery and making a pay-out to the winner without relying on any central authority for trust and fairness. Another analogy is that of a vending machine which delivers the desired product and the residual amount, if any. It follows a specific algorithm for delivering the item automatically.

These contracts try to eliminate third party risk as once crypto assets are kept in the custody of the smart contract, they can only be withdrawn or released when the contract conditions are met. The smart contracts contain code which has well defined functions which execute the contract. Invocation of functions can alter the balance of the contract by depositing or releasing Ether to or from the contract, change the internal contract state like assign a token like Tether (USDT) from one account to another and alter the data stored in the contract. Smart contracts are used to perform important functions in the Ethereum Blockchain like

i) Authenticating the identity of the user invoking the transaction with the smart contract
ii) Store and update the data intended to be stored by users. For example, ownership information of fungible and non-fungible tokens, maximum supply of a token etc.

iii) Provide trust as a third party governed solely by terms of an agreement subjected to public scrutiny by users.
iv) Expose functions to be invoked by other contracts

The transaction that creates a smart contract is special transaction as it has a special destination address called a *zero address*. The smart contracts once created cannot be altered ever in the Ethereum Blockchain. Only if a smart contract has a self-destruction function, it could be destroyed earlier, leaving a blank Ethereum account and destruction of the storage and state of that smart contract, but this function has been deprecated in the later releases of Ethereum.

From the discussion above it can be seen that a smart contract is essentially code running on the Ethereum Virtual Machine which executes when certain conditions are met. There are multiple viewpoints on the legality and enforceability of a smart contract and many argue that smart contracts are neither smart nor contracts (Durovic, 2021). Levi & Lipton (2018) argue that smart contracts can be categorized into code-only smart contracts and ancillary smart contracts which are essentially vehicles to effectuate the provisions of the traditional text-based contracts. In common law, a contract is a "*an agreement between private parties creating mutual obligations enforceable by law. The basic elements required for the agreement to be a legally enforceable contract are: mutual assent, expressed by a valid offer* and *acceptance; adequate consideration; capacity; and legality*"[86]. This definition can be satisfied by ancillary smart contracts. As agreements need not always be in writing to be enforced[87], code-only smart contracts might also be enforceable in some jurisdictions. Just like a user using the vending machine sees the price displayed for each item in the display and acts according to the instructions written on the vending machine, he/she gains certain implied rights and a contract is formed without any written terms, conditions and obligations.

Levi & Lipton (2018) argue that the Uniform Electronic Transactions Act (UETA) in the USA provides for records created by computer programs, and electronic signatures the same legal effect as written documents. UETA states that an Electronic Agent is "capable within the parameters of its programming, of initiating, responding or interacting with other parties or their electronic agents once it has been activated by a party, without further attention of that party,". The federal Electronic Signatures Recording Act (E-Sign Act) in the US provides that a contract or other record relating to a transaction "may not be denied legal effect, validity, or enforceability solely because its formation, creation, or delivery involved the action of one or more electronic agents so long as the action of any such electronic agent is legally attributable to the person to be bound."[88].

Article 9 of the EU Directive on Electronic Commerce 2000 states that "*Member States shall ensure that their legal system allows contracts to be concluded by electronic means. Member States shall in particular ensure that the legal requirements applicable to the contractual process neither create obstacles for the use of electronic contracts nor result in such contracts being deprived of legal effectiveness and validity on account of their having been made by electronic means*". Sates like Arizona[89] and Nevada[90] have amended their UETA to include blockchain and smart contracts which can have far reaching implications on their legal status and enforceability. However, most of the users who interact with a smart contract would have no means to verify what the code actually does and pre-emptively find out the vulnerabilities if any, this may be treated as a case where the user was unaware of the terms of agreement of the contract and the smart contract might not be enforceable.

---

[86] https://www.law.cornell.edu/wex/contract
[87] *Lumhoo v. Home Depot USA, Inc.*, 229 F. Supp. 2d 121, 160 (E.D.N.Y. 2002)
[88] 15 U.S.C. § 7001(h).
[89] https://www.azleg.gov/legtext/53leg/1r/bills/hb2417p.pdf
[90] https://www.leg.state.nv.us/nrs/nrs-719.html#NRS719Sec245

### 7.2.1 Issue with code-only smart contracts

Code-only smart contracts might also face challenges due to a fundamental issue in Computer Science called the 'halting problem.' The halting problem refers to the challenge of determining whether a given computer program, when provided with input, will eventually terminate, or run indefinitely. It is proven to be undecidable, indicating that there is no universal algorithm capable of solving the halting problem for all combinations of programs and inputs. For code-only smart contracts the undecidability of the halting problem poses challenges, as the user of the smart contract has no means to verify the deterministic execution of the smart contract code. If a party engages with a smart contract without understanding its code or implications, it may call into question whether true consent was given. The legality and other aspects of smart contracts are an active area of research and more judicial amendments and court rulings are expected to bring more clarity on the legality and enforceability of smart contracts. This may also have profound implications for tax treatment of various transactions involving smart contracts.

To prove that the halting problem is undecidable, we can use a proof by contradiction. We assume that there exists a Turing machine $H$ that can decide the halting problem, and then we derive a contradiction.

The Halting Problem

Let $H$ be a Turing machine that decides the halting problem. This means that for any pair of input strings $\langle M, w \rangle$, where $M$ is a description of a Turing machine and $w$ is an input string for $M$, $H$ halts and outputs "Accept" if $M$ halts on input $w$, and it halts and outputs "Reject" otherwise.

Now, we construct a new Turing machine $D$ with the following behaviour:
Given an input string $M$ (a description of a Turing machine):

1. Run $H$ with input $\langle M, M \rangle$.
2. If $H$ outputs "Accept," $D$ enters an infinite loop.
3. If $H$ outputs "Reject," $D$ halts.

Now, let us consider what happens when we feed $D$ its own description $\langle D \rangle$:

1. If $D$ halts on input $\langle D \rangle$, according to its construction, it should enter an infinite loop (step 2).
2. If $D$ enters an infinite loop on input $\langle D \rangle$, according to its construction, it should halt (step 3).

This creates a contradiction. If $D$ halts on input $\langle D \rangle$, then it should enter an infinite loop according to its construction. If $D$ enters an infinite loop on input $\langle D \rangle$, then it should halt according to its construction.

Thus, our assumption that there exists a Turing machine $H$ that decides the halting problem leads to a contradiction. Therefore, the halting problem is undecidable.

## 7.3 Tokens

The Collins dictionary defines token as a countable noun. It is defined as "A token is a piece of paper or card that can be exchanged for goods, either in a particular shop or as part of a special offer"[91]. In the context of a blockchain a token represents something similar. But, instead of a piece of paper or card, it is an abstraction created on the blockchain, often through smart contracts to represent assets, equity, collectible, currency etc. It is often confused and used interchangeably with the term coin in the parlance of crypto assets. Coins are the native assets of blockchains whereas tokens are abstractions created on top of blockchains which may or may not represent any underlying asset. For example, Ether and Bitcoin are the native assets of the Ethereum and Bitcoin Blockchain respectively whereas Tether (USDT) is a token created through a smart contract which regulates its supply and pegs it to the US Dollar. In order to interact with the USDT smart contract the user still has to pay gas in Ether denominated in gwei($10^{-18}$ Ether).

Smart contracts enable creation of tokens which may represent a wide variety of tangibles and intangibles. There can be different types of tokens like:

i) Payment Tokens: These are native assets of the blockchain, also known as coins, which are used for making payments. For example, Ether, Bitcoin and Monero

ii) Utility Tokens: These tokens provide access to a specific product or service on the blockchain. For example, the WRX or the WazirX token is used to make payments for transaction fee on the WazirX crypto exchange and sometimes enables users to get a discount on the transaction fee.

iii) Security Token: These tokens derive their value from external assets like stocks and bonds. For example, the tZero token which represents tokenized assets like stocks and bonds.

iv) Governance Token: These tokens provide voting rights to the holders regarding various proposals for changes in a Decentralized Application or a Decentralized Autonomous Organization. For example, The Maker token (MKR) allows holders to vote on decisions regarding the decentralized lending platform Maker.

v) Non-Fungible Tokens: These tokens represent a collectible or something unique which is not interchangeable with any other token. For example, the Bored Ape Yacht Club[92] NFTs with each ape being unique.

The tokens mentioned in examples above cannot be classified strictly into the categories. There are tokens which have multiple characteristics and can be categorized into more than one category. For example, the MKR token is a utility token in the Maker ecosystem as well as its governance token which can be used for initiating and voting on proposals. It can also be exchanged for other crypto assets and purchased on crypto exchanges. This makes the regulation of these tokens complicated. As security tokens are essentially tokenized assets like securities and bonds, they are regulated by securities market regulators like the SEC. In some instances, like in case of Ripple (XRP), the regulators may treat the issued/minted tokens as securities, giving rise to litigation[93].

From the discussion above it can be observed that there are certain tokens which purportedly represent real underlying assets like real estate, securities, US Dollars and Gold. The transfer of these assets on the blockchain may not mitigate the counterparty risk in a real-world environment. For example, a user on Ethereum Blockchain may have no means to authenticate the ownership of a corporation's equity for a particular Ethereum address without involving the real-world custodian of listed securities, posing a counter-party risk to the transaction. On the other hand, ownership of tokens like WRX, MKR can be cryptographically proven on the blockchain and their transfer on the blockchain does not involve any counterparty risk.

---

91 https://www.collinsdictionary.com/dictionary/english/token

92 https://etherscan.io/address/0xbc4ca0eda7647a8ab7c2061c2e118a18a936f13d

93 https://www.sec.gov/news/press-release/2020-338

### 7.3.1 ERC-20 Token Standard

The ERC-20 is an Ethereum Blockchain standard for fungible tokens. Many tokens are based on the ERC-20 standard. An ERC-20 token smart contract should have some mandatory functions like those for defining the total supply of tokens, transfer of tokens from one account to another, functions authorizing certain accounts to transfer tokens from authorizing accounts etc. Some examples of popular ERC-20 tokens are USDT, DAI and MKR which are discussed in detail in later sections. Anyone can create their own ERC-20 token. However, to make the token popular and widely acceptable, bootstrapping through incentives like ICOs and airdrops is required, which may be taxable events for the users as well as creators.

The USDT is a so-called stablecoin that is told to be pegged to the US Dollar. It is an asset backed ERC-20 token which enables the owner to exchange 1 USDT for 1 USD. The total supply of the USDT tokens on Ethereum is controlled using the USDT smart contract on the Ethereum Blockchain (Fig. 35). Whenever a request for exchanging a Dollar with USDT is made, the smart contract issues new USDT tokens to the Ethereum account with corresponding addition on the asset side. On the other hand, when a user converts USDT to USD or an equivalent asset, USDT tokens are burnt and equivalent assets held are released. As the so-called stablecoin is pegged to USD it can be used to buy or sell other crypto assets using a dollar-pegged stablecoin.

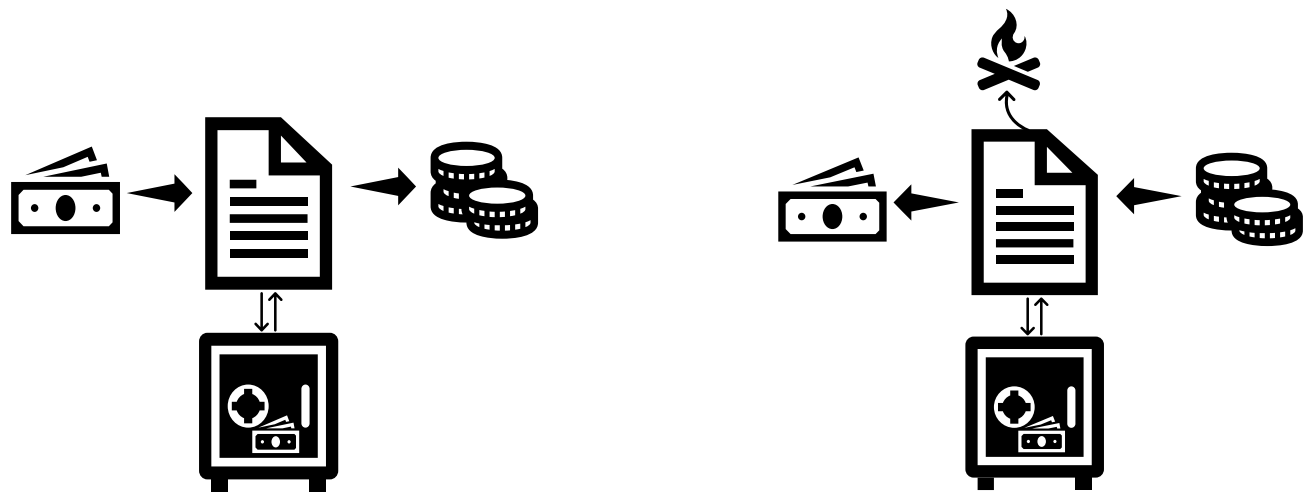

Fig. 35 Issuance and burning of tokens

## 7.4 Non-Fungible Tokens

Non-Fungible Tokens represent unique and non-interchangeable assets on a blockchain. The ownership of an NFT is usually established on the blockchain through a smart contract. The purported underlying asset in an NFT can be anything ranging from a piece of art, an antique piece, real-estate, music etc. Some of the famous NFT sales like the NFT of the first tweet by Jack Dorsey and the "Everydays - The First 5000 Days" NFT by Beeple created lot of buzz and increased the interest in NFTs. NFTs are enabling content creators and sellers of unique non-fungible items or assets to connect directly with the users. NFTs are an integral component of various metaverses where a lot of e-Commerce takes place through NFTs. For example, unique land parcels in metaverses like Decentraland are essentially NFTs which can be bought or sold in the metaverse using crypto assets. The non-fungible nature of NFTs gives them value, in case of crypto assets like Ether, every Ether denomination is same and completely fungible. On the other hand, NFTs like the NFT of the first tweet by Jack Dorsey are unique.

ERC-721 is the token standard for Non-Fungible Tokens on Ethereum. Although, there are other blockchains like Solana which also have a rapidly developing NFT and smart contract ecosystem. In an ERC-721 smart contract the unique NFTs are assigned Token IDs. The process of establishing the ownership of representation of a unique item on the blockchain is known as Minting. The unique attributes of the item are recorded on the blockchain in the smart contract, which assigns a unique tokenID to the item. As the collectibles/artworks are often very large files, the metadata describing the unique collectible and the URL of the actual artwork is provided in the URI (Uniform Resource Identifier) field in the smart contract. This highlights an important fact regarding many NFTs which represent digital artwork that the NFTs usually only store a pointer URI to the URL of the artwork and no art itself on the blockchain. Users can write their own smart contract to mint NFTs, however, many platforms like OpenSea and Rarible provide services to create smart contracts and mint NFTs for a fee.

Let us try to understand the ERC-721 framework using the famous Bored Ape Yacht Club collection on the Ethereum Blockchain. On visiting the website of Bored Ape Yacht Club on OpenSea[94] we find many unique images of bored apes. To understand how the ownership of these NFTs is recorded and transferred on the blockchain we can click any NFT to find out its current owner's account address, the address of the smart contract and the TokenID of the NFT.

The smart contract holds the information regarding the NFTs and maintains the record of its ownership for each token ID (Unique collectible). The smart contract of Bored Ape Yacht Club on EtherScan can be used to find owner account addresses of each NFT, their URIs as well as ownership history. The transaction history of the smart contract can be seen and if a user goes to the first few transactions, he/she can see the transactions that minted the BAYC NFTs. The transaction for minting a TokenID associates the TokenID and other information like the URI (containing the description of the artwork along with the URL of the artwork), with an Ethereum account in the BAYC smart contract.

To find where the metadata of the NFT image is stored on the Ethereum Blockchain one can query the smart contract on Etherscan by using the Read Contract Tab on the Contract stub[95]. One can find the following important functions in the smart contract which can be queried to read data without paying any gas on the Ethereum Blockchain. They are listed as below:

A) MAX_APES: Defines the maximum number of tokens (Unique Apes) that can be minted in this contract. 10000 for BAYC.
B) balanceOf: Returns the number of BAYC NFTs owned by an Ethereum address
C) ownerOf: It returns the Ethereum address that owns a specific BAYC TokenID
D) tokenURI: It takes token ID as the input and returns the Uniform Resource Indicator which is usually a link containing the metadata of the NFT

Thus, an ERC721 smart contract essentially keeps a track of the minted NFTs and the accounts that own them. Every time a new NFT is minted, gas must be paid to record the URI, ownership, and other data on the blockchain. While changing the ownership also, gas must be paid to record the change in the smart contract, along with some consideration if it is a sale or without consideration if it is merely a transfer.

As discussed earlier, the token URI usually contains the metadata regarding the NFT which might in-turn contain the URL where the actual digital image is stored. Querying the tokenURI field for any token ID provides the address:
ipfs://QmeSjSinHpPnmXmspMjwiXyN6zS4E9zccariGR3jxcaWtq/17XX
as shown in Fig. 36 which is address of data stored in Inter Planetary File System - a peer-ot-peer decentralized storage which stores data based on the hash identifier of the item[96]. Accessing the metadata present at the given URI we find a json document which contains the metadata of the NFT as shown in Fig. 37. This is the core of the NFT which lists its characteristics as well as provides the link to the URL of the digital image of the artwork. This URL is a source of potential risk for the

[94] https://opensea.io/collection/boredapeyachtclub
[95] https://etherscan.io/address/0xbc4ca0eda7647a8ab7c2061c2e118a18a936f13d#readContract
[96] https://ipfs.tech/

NFT as it makes it essentially a pointer to a URL. If the image URL is changed from the unique artwork in a token ID to any random picture, the NFT and the ownership of the image can be altered by a third party hosting the artwork. To mitigate that risk, BAYC uses the IPFS to store the image and metadata. As IPFS is a hash identifier-based peer-to-peer network which ensures that the URI would not change or go offline.

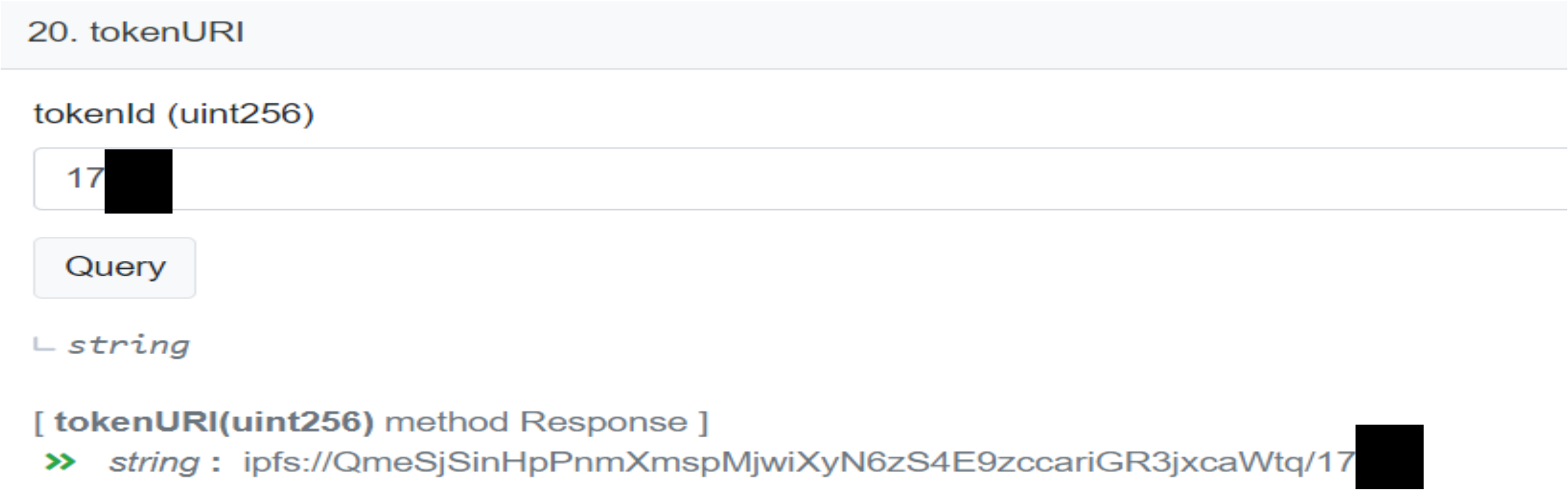

Fig. 36 tokenURI of an NFT

{"image":"ipfs://QmW2uHxzxDfxGDUroqYkDQRaNuXSb2Vc9zl[redacted]","attributes":[{"trait_type":"Clothes","value":"[redacted]"},{"trait_type":"Fur","value":"Golden Brown"},{"trait_type":"Hat","value":"Sea Captain's Hat"},{"trait_type":"Eyes","value":"Coins"},{"trait_type":"Mouth","value": [redacted]"},{"trait_type":"Background","value":"Yellow"}]} Angry"},{"trait_type":"Background","value":"Yellow"}]}

Fig. 37 NFT Metadata

Blockchain explorers also provide the transaction history of a particular TokenID since it was minted. For any NFT Token ID the sale and transfer transactions can be seen on the blockchain. The consideration paid in various NFT transactions is also given. However, a relook at the smart contract functions reveals that NFT smart contracts usually have no field for recording the price at which the sale takes place. NFT smart contracts typically do not directly record the price of sale of the NFT within the contract itself. However, they often include functionalities to facilitate the transfer of ownership and may emit events or logs that contain information about the sale, including the price. Also, the owners and intermediate holders of the NFT are merely Ethereum Account addresses which cannot be easily mapped to a real-world identity.

When purchasing an NFT, the transaction typically involves two separate components:

1. **Transfer of the NFT**: This is the actual transfer of ownership of the NFT from the seller's address to the buyer's address. This transfer is recorded on the blockchain and is immutable. Once the transfer is complete, the buyer becomes the new owner of the NFT.
2. **Consideration or Payment**: This is the payment made by the buyer to the seller in exchange for the NFT. While the transfer of the NFT itself happens on the blockchain, the payment for the NFT usually occurs off-chain. Buyers and sellers typically arrange payment separately from the blockchain transaction using traditional payment methods such as credit cards, bank transfers, or crypto asset transfers.

This hinders a taxman to find out the income and capital gains accrued to various holders of the NFT (Fig. 38). However, NFT marketplaces like OpenSea in their privacy policy[97] state that they capture the email address, first name, last name as well as the blockchain address and IP address along with mobile device IDs. These can be used by tax administrations and law enforcement to map NFT owner account addresses to natural or juridical persons in tax fraud and money laundering cases. The valuation of NFTs just like any other piece of art can be complex and subjective. NFTs can be wash traded to artificially enhance their prices and laundering money.

[97] https://opensea.io/privacy

Various popular NFT marketplaces presently do not capture any KYC information for trading and minting NFTs. This enhances the risk of tax evasion and money laundering in NFTs. The Royal United Services Institute (RUSI) in the UK has published a report[98] that assesses the money laundering risk in NFT marketplaces. The U.S. Department of the Treasury has also published a study[99] on the facilitation of money laundering and the financing of terrorism through the trade in works of high-value art. It finds that emerging digital art forms like NFTs present new money laundering and terror financing risks. The blockchain analytics firm Chainalysis identified 262 users who had sold an NFT to a self-financed address more than 25 times[100].

Another important feature of NFTs is that the creators can earn royalty income from the subsequent sale of the NFT. While minting the NFTs on platforms like OpenSea, the creators can specify the percentage of royalty to be paid for every subsequent sale of the NFT. However, as can be seen in the most NFT smart contracts, there is no field for specifying the royalty percentage. Thus, this specification of royalty is enforceable only when the sale of the NFT takes place on the specific platform. If the NFT is sold on some other platform or through a direct smart contract interaction, the NFT may be transferred without any royalty payment as the royalty specification and payment is not in-built into the NFT smart contract.

As multiple marketplaces have multiple royalty payment implementations which may not be compatible, developers have proposed an NFT Royalty Standard - EIP 2981[101] which enables all marketplaces to retrieve royalty payment information for a given NFT and enables marketplace agnostic royalty payments for secondary sales. However, the proposal is not yet implemented and if a marketplace chooses not to implement the proposal, no royalty would be paid automatically for secondary sales.

The increase in trading volumes of NFTs has created an ecosystem of Decentralized Finance applications which involve NFT related commercial activity. They enable users to earn passive income besides the royalty income on sales. As NFTs are often unique collectibles, they can be rented out for use by others. For example, a popular and expensive sword or a weapon can be rented by users in an online game or a metaverse. Platforms like reNFT[102] provide NFT rental infrastructure for the metaverse. It allows NFT owners to rent their NFTs for a set amount and time after taking certain amount as collateral. There are also protocols which do not transfer the original NFT but issues an expiring wrapped version of the NFT to the borrower. NFTs can also be staked or deposited into smart contracts in lieu of tokens issued by the platform, which can lend the NFT to generate passive income which is shared with the owners of the NFT. The users can also deposit NFTs to get tokens that are composable and fungible at a ratio of 1:1 and can be used to acquire other NFTs or their fractions, on the platform or staked further. Many services pool multiple NFTs and provide a liquid market for their sale and purchase with a floor price for each NFT.

---

98 https://rusi.org/explore-our-research/publications/commentary/nfts-new-frontier-money-laundering
99 https://home.treasury.gov/system/files/136/Treasury_Study_WoA.pdf
100 https://www.financialexpress.com/digital-currency/wash-trading-and-money-laundering-observed-in-nfts-report/2497965/
101 https://eips.ethereum.org/EIPS/eip-2981
102 https://www.renft.io/

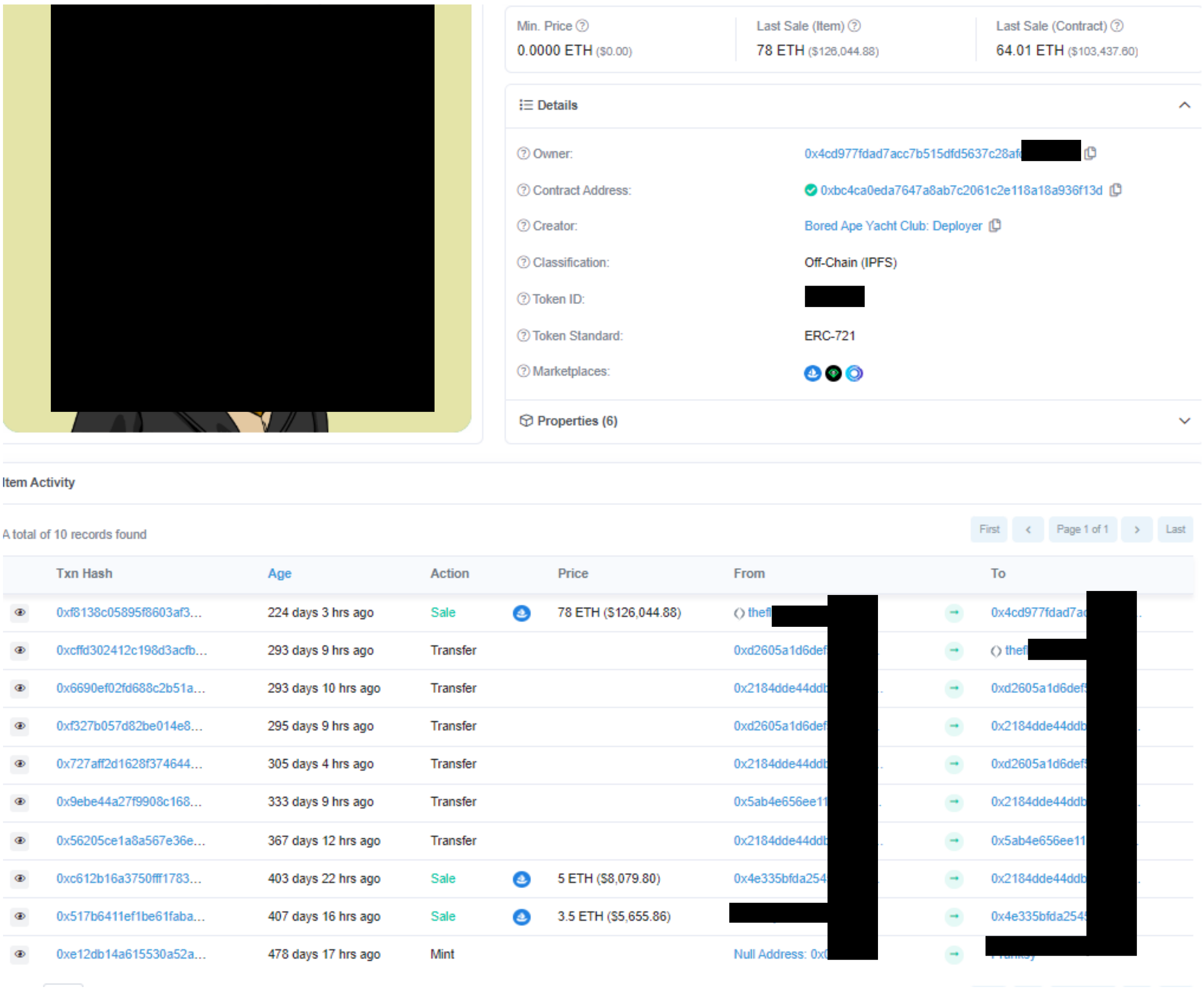

Fig. 38 NFT transaction history on Ethereum Blockchain explorer

## 7.4.1 Legal Status of NFTs

The basic idea behind NFTs purportedly representing art work, collectibles, real estate etc. is that by owning the NFT on the blockchain one can have property rights in another underlying asset like art or real estate. But what property rights of the underlying asset does the ownership of the NFT automatically transfer to the owner is a matter of debate. The fundamental question is about the property rights that the process of tokenization of a tangible or intangible asset on the blockchain brings with it. Moringiello & Odinet (2022) analysed the legal aspects of tokenization and examined how the token is connected to the underlying asset, if at all, and what does the current legal framework say about it. They argue that NFTs do not have attributes of other tokens like negotiable instruments and bills of lading. They also survey dataset of terms of service from the most prominent NFT platforms to explore their legal effects as compared to the claims.

tokenization is not a new concept in law and economics. Various negotiable instruments, securities and deeds are based on the principle that the tokens represent the ownership rights of some other assets. When a user buys an NFT and the NFT is assigned to the blockchain address controlled by the user on the blockchain, it is believed by many buyers that they acquire the ownership and other associated rights in the underlying artwork or collectible just by virtue of acquiring the NFT. The large amounts involved in NFT purchases might prompt a connoisseur of art to assume that the digital work itself has been transferred, whereas the NFT bought by the users essentially contains metadata and not the image itself.

Moreover, anything that can be digitized can be minted into an NFT as the process of minting essentially creates a unique TokenID for that digital work of art and records it into a smart contract in the blockchain. Thus, it is technologically possible to mint a copyrighted item for which the user does not have ownership or the copyright to be minted into an NFT purportedly "owned" by the user. The ownership, reproduction and use of an artwork or collectible in most jurisdictions is governed by the copyright law, which may or may not recognize NFTs as a valid deed of transferring the copyright or other rights in the artwork. Thus, it is argued that from a copyright perspective that minting an NFT describing an artwork without authorization might not be clearly a copyright infringement as it can be considered generation of a string of numbers in relation to a work but not reproduction or adaptation of the work itself[103].

Guadamuz[104] (2022) argues that most NFTs do not involve a transfer of rights and in some instances where a user must check a box for transferring rights while minting an NFT, it might not fulfil the requirements of the copyright law of the jurisdiction of the creator of the NFT. He gives the example of Copyright Designs and Patents Act 1988 of UK which requires the copyright agreement *"in writing signed by or on behalf of the assignor"*. For example, in case of real estate in India the transfer of ownership is governed and effectuated through the Transfer of Property Act. The sale or purchase of any real-world real-estate merely through the transfer of an NFT representing the real-estate would not be recognized as a transfer under the Transfer of Property Act.

Moringiello & Odinet (2022) surveyed the terms of service of various leading NFT platforms and found that none of the terms of service provided any clear link between ownership of an NFT and rights in the underlying creative work. One of the terms stated that buyer obtains ownership of "a cryptographic token representing the artist's creative Work as a piece of property, but" obtains no ownership of the "creative Work itself." They argue that the only right the owner of the NFT gets by making the purchase is the right to display the NFT.

Moringiello & Odinet (2022) reviewed the terms of service of a site SuperWorld which maps the Earth to create virtual land assets which participants can "buy". The users can buy or sell real estate tokens. The terms of service state that "purchase" of tokens conveys no rights in the underlying art. The users can monetize the property in SuperWorld. Marinotti[105] (2022) argues that although the NFTs used in metaverses are on the blockchain but the constituents that make the metaverse like land and goods exist on private servers which are stored on secured databases running proprietary code. He argues that the metaverse platforms effectively grant the users only access to the digital assets. As the metaverse NFTs have functional use in the metaverse, blocking access or deleting the NFT from the metaverse for violation of terms of service can significantly affect the utility and financial worth of the NFT. A more detailed and comprehensive discussion about the copyright issues of NFTs can be found in an article by James Grimmelmann, Yan Ji, and Tyler Kell[106] which also mentions the case of Spice DAO which involved a dispute over ownership and control of a digital art collection on the Ethereum blockchain.

Various tax administrations in recent rulings have also clarified the nature of NFTs. The Spanish tax administration in its ruling V0486-22 on 10th March 2022[107] stated the fact that "*the object of the sale is not the illustrations themselves but NFTs, that is, non-fungible tokens that grant the buyer rights of use but in no case the underlying rights to ownership of the work.*". The Norwegian tax administration in its guidance[108] also states that *"An NFT will not normally be considered home contents/movable property in this connection, as the NFT in itself is a digital code and not the actual object it is linked to."* These rulings and clarifications can have consequences for direct and indirect tax treatment of NFTs.

---

[103] https://www.wipo.int/wipo_magazine/en/2021/04/article_0007.html
[104] https://www.weforum.org/agenda/2022/02/non-fungible-tokens-nfts-and-copyright/
[105] https://theconversation.com/can-you-truly-own-anything-in-the-metaverse-a-law-professor-explains-how-blockchains-and-nfts-dont-protect-virtual-property-179067
[106] https://www.theverge.com/23139793/nft-crypto-copyright-ownership-primer-cornell-ic3
[107] https://petete.tributos.hacienda.gob.es/consultas/?num_consulta=V0486-22

One interesting perspective on NFTs is to consider them as vouchers, which can be either single-purpose or multipurpose in nature[109]. In the context of NFTs, single-purpose vouchers refer to tokens that are designed for a specific utility or use case. Once redeemed or utilized, their purpose is fulfilled, and they may lose their value or become non-transferrable. For example, an NFT representing a ticket to a virtual event provides access to that specific event only. Single-purpose NFTs provide a direct link between the digital asset and its owner, facilitating transparent ownership and provenance tracking. On the other hand, multipurpose vouchers in the realm of NFTs refer to tokens that possess versatile utility across various contexts or platforms. These NFTs retain their value and transferability even after redemption, allowing owners to utilize them in different ways or trade them in secondary markets.

Multipurpose NFTs may represent a range of assets or rights whose value often depends on factors such as scarcity, demand, and interoperability with different ecosystems. For instance, an NFT representing a virtual land parcel in a metaverse could be utilized for various purposes, such as building virtual structures, hosting events, or generating revenue through in-world activities. Similarly, an NFT representing membership in a decentralized governance platform might grant voting rights, access to exclusive content, or dividends from platform activities. Treating NFTs as single-purpose or multi-purpose vouchers can have consequences for taxation of NFTs from the perspective of VAT/GST.

### 7.4.2 ERC 1155 Token Standard

A closer look at the ERC-721 smart contracts reveals that an ERC-721 smart contract can be used only to map unique TokenIDs to Ethereum addresses. It cannot hold ERC-20 tokens along with NFTs (ERC-721 tokens). Also, there are no functions to transfer the NFTs in bulk. All these shortcomings result in payment of huge amount in gas fees while creating and transferring NFTs. Moreover, there are use cases where more than one copy of a collectible like a weapon or outfit in an online game might be required along with the native ERC-20 token of the game.

The ERC 1155 token standard solves this problem and enables the smart contract to hold fungible as well non-fungible tokens. ERC-1155 solves these problems by assigning TokenIDs to each kind of token(s) held by the smart contract, which also keeps a track of their ownership by mapping them to Ethereum addresses. In case of Non-Fungible Tokens, the respective TokenID is mapped to only one Ethereum address, whereas for an ERC-20 token the TokenID can have many tokens issued whose balances are maintained in the smart contract. It also enables the creation of multiple copies of a single artwork or collectible which can be minted in limited quantities and assigned to different users (Ethereum Accounts). For example, an online game can have an ERC 1155 smart contract which can hold the native ERC-20 token of the game as well as collectibles like weapons or armour which are assigned different TokenIDs and available in one or more than one quantity. This can be pictorially represented in Fig. 39

[108] https://www.skatteetaten.no/en/person/taxes/get-the-taxes-right/shares-and-securities/about-shares-and-securities/digital-currency/nft/#:~:text=The%20taxable%20value%20of%20NFTs,the%20value%20assessment%20upon%20request.

[109] VALUE ADDED TAX COMMITTEE (ARTICLE 398 OF DIRECTIVE 2006/112/EC) WORKING PAPER NO 1060

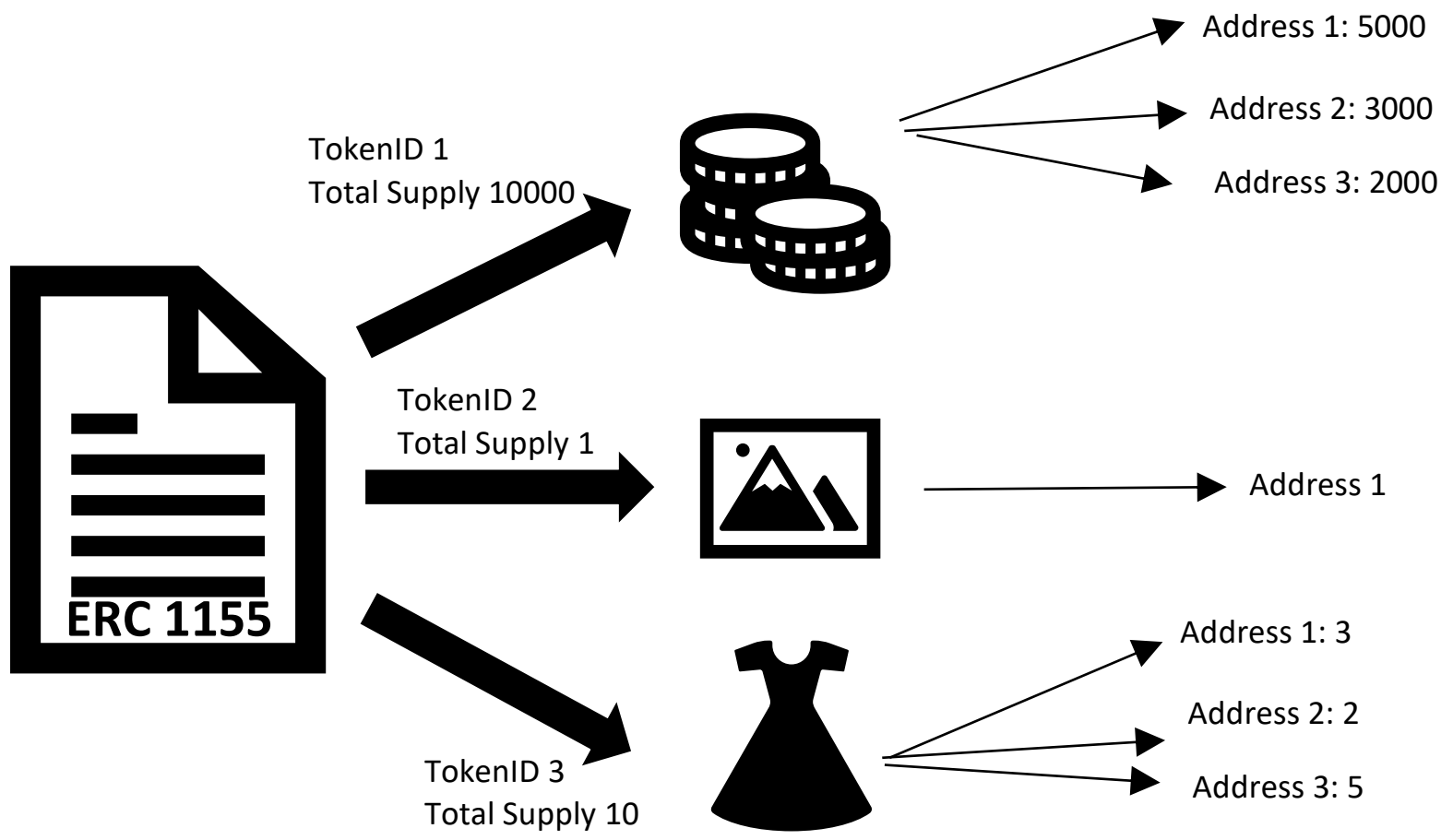

Fig. 39 ERC 1155 Token Standard (illustration)

### 7.4.3 Taxation of NFTs

NFTs are collectibles and can be considered as blockchain representation of various art forms like paintings, music, videos etc. They may also be tokens created purely for acting as tokens for a specific purpose like entry into a concert or represent the position of a liquidity provider in a liquidity pool. The first potential taxable event is the creation or 'Minting' of the NFT. The creation of the NFT usually involves recording the metadata and URI of the artwork on the blockchain. As this does not lead to accrual of any income and is merely a process for recording information about the artwork on the blockchain it might not be taxed in most jurisdictions. However, only a few jurisdictions have issued guidance related to taxation of NFTs. The Norwegian tax administration in its guideline has clarified that creating an NFT does not trigger taxation[110]. Also, the transfer of NFT from one account to the other owned by the same person/entity would also not be taxable.

The tax treatment of the transactions related to the NFT after minting depend upon the Intellectual Property Rights that get transferred to the purchaser. As discussed above, as buying an NFT does not necessarily transfer all the Intellectual Property Rights associated with the artwork, the first sale of the primary transfer of the NFT may be treated as a license for the purpose of taxation. The secondary and subsequent transactions of the NFT may be treated as a sale as the recipient of the primary transfer would sell all its rights to the recipient of the secondary or subsequent transfer. However, the secondary sale might also require payment of royalty to the owner of the underlying artwork and might also involve issues related to amortization of the basis by the taxpayer.

For the seller making the secondary sale it might attract income tax or capital gains tax depending upon the classification of NFTs by the jurisdictional tax administration and the prescribed rates as well as the nature of activity and the period of holding. For example, in the US NFTs are treated as collectibles and might be subject to higher tax rates. However, the tax treatment would depend on the facts and circumstances of each transfer and the rights relinquished and acquired by the transferor and the transferee respectively. The royalty payments on secondary sales might be subject to withholding taxes in many jurisdictions, depending upon the tax residency of the transferor. The lack of sale consideration data on the blockchain as well as the limited ability to map Ethereum addresses to Tax Identification Numbers can make it difficult for tax administrations to collect due taxes. However, as many NFT owners publicly mention their social media handles/IDs it might be possible for tax administrations to obtain the ownership information from such platforms.

[110] https://www.skatteetaten.no/en/person/taxes/get-the-taxes-right/shares-and-securities/about-shares-and-securities/digital-currency/nft/

NFT sales can also be subject to GST/VAT in many jurisdictions as it may be considered a taxable supply of services. For example, NFTs would be considered as digital services as per Council Implementing Regulation (EU) No 282/2011. The individuals or entities involved in trade of NFTs can might be considered 'taxable person' for GST/VAT as they carry out an 'economic activity' with direct nexus between the supply of service and the consideration. Even NFTs minted for functional purposes like for entry into concerts and restaurants might also be subject to GST/VAT in many jurisdictions depending upon their classification as a voucher, with possible exemptions up to a certain limit. The services offered by various NFT marketplaces for minting and sale of NFTs may also be subject to GST/VAT depending upon the tax residency of the clients. The royalty payments related to NFTs may also be subject to GST/VAT in various jurisdictions as royalties are considered as licensing services in many jurisdictions. GST/VAT liability on royalties received may depend on whether they pertain to the right of using the NFT or the right to resell the NFT. The issues and tax treatment related to place of supply, claiming input tax credit and exemption of supplies made to residents in foreign jurisdictions would be similar to the issues related to taxation of Bitcoin Mining activities.

Some tax administrations have already issued rulings and guidance that clarify the tax treatment of NFTs. The Spanish tax administration has clarified that the NFT sales are electronically supplied services and will be subject to VAT, depending on the place of supply[111]. The Belgian tax administration has also clarified that the NFT sale is to be regarded as a service provided by electronic means. If the supply of NFTs is deemed to take place in Belgium standard VAT rates would apply[112]. A striking feature of such rulings and guidance is the presumption of place of supply to exist somewhere in the real world, where a destination tax administration can collect the VAT/GST due. However, the supply of such services in virtual worlds like a metaverse raise fundamental questions regarding the taxability and tax treatment of such transactions. One such example is the judgement of the German Federal Finance Court in a case involving the online platform 'Second Life' where the court overruled the judgement of the Cologne Finance Court and concluded that such transactions in virtual world do not constitute participation in economic activity in the real world and are outside the scope of VAT/GST[113]. Despite these uncertainties and issues, it is worth mentioning that the market cap of NFTs is approximately 4 billion USD with all time sales volumes of around 137 billion USD[114] which makes it difficult for tax administrations to overlook the revenues associated with NFTs

## 8. Decentralized Finance

The financial system performs a critical job of financial intermediation and enables the flow of capital from lenders to borrowers. Similarly, in the domain of crypto assets using platforms like centralized crypto exchanges or decentralized platforms using smart contracts, such financial intermediation is possible. The smart contract ecosystem enables the automation of deposit, disbursal, lending and repayment of funds. This has created an entire ecosystem of financial services which operate on the blockchain, called DeFi or Decentralized Finance. DeFi enables enormous opportunities of financial innovation using crypto assets and smart contracts. At the same time, it also poses new challenges to tax administrations and law enforcement agencies as lack of KYC norms and regulatory oversight can enable tax evasion and money laundering. The wide variety of applications and products offered by DeFi warrant a separate paper to describe the mechanics as well as the taxation aspects of this ecosystem, which are not very clear and well understood today. However, this paper discusses two prominent DeFi platforms MakerDAO and UniSwap which account for a significant portion of assets locked up in DeFi applications.

---

[111] https://petete.tributos.hacienda.gob.es/consultas/?num_consulta=V0486-22
[112] https://www.dekamer.be/QRVA/pdf/55/55K0073.pdf - page 114
[113] https://www.justiz.nrw.de/nrwe/fgs/koeln/j2019/8_K_1565_18_Urteil_20190813.html
[114] https://coinmarketcap.com/nft/

## 8.1 MakerDAO

MakerDAO is one of the most popular DeFi projects. It is a Decentralized Autonomous Organization that issues and manages two ERC-20 tokens. The first token is known as DAI, which is a crypto-collateralized so called stablecoin soft pegged to the US Dollar, the second token is Maker (MKR) which is used as a governance token of the Maker protocol as well as a utility token for auctioning DAI for MKR and vice versa. In unforeseen situations it can also be used as a resource for recapitalization and repayment of Maker protocol debt. The MakerDAO is essentially a system of smart contracts running on the Ethereum Blockchain that manage the supply of DAI and MKR. A system of incentives and other mechanisms enable creation and destruction of DAI as well as its backing with appropriate collateral to maintain its soft peg to the US Dollar using smart contracts. Unlike the so called centralized stablecoins, the collateralization and assets in the MakerDAO ecosystem are publicly available and do not require any proof of reserve or audit. A detailed overview of the MakerDAO is available in the whitepaper[115].

The MakerDAO can be considered a digital vault system which gives a DAI denominated loan to users which lock-up their Ether or other authorized crypto assets into a smart contract, which is called a Collateralized Debt Position (CDP). This is the mechanism that generates DAI as shown in Fig. 40. The DAI loan accrues interest denominated in DAI which is called the 'stability fees' in the Maker protocol. When the DAI loan is returned along with the stability fee by the borrower, the retuned DAI is burnt, the stability fees go to the surplus smart contract account and the collateralized assets are released (Fig. 41).

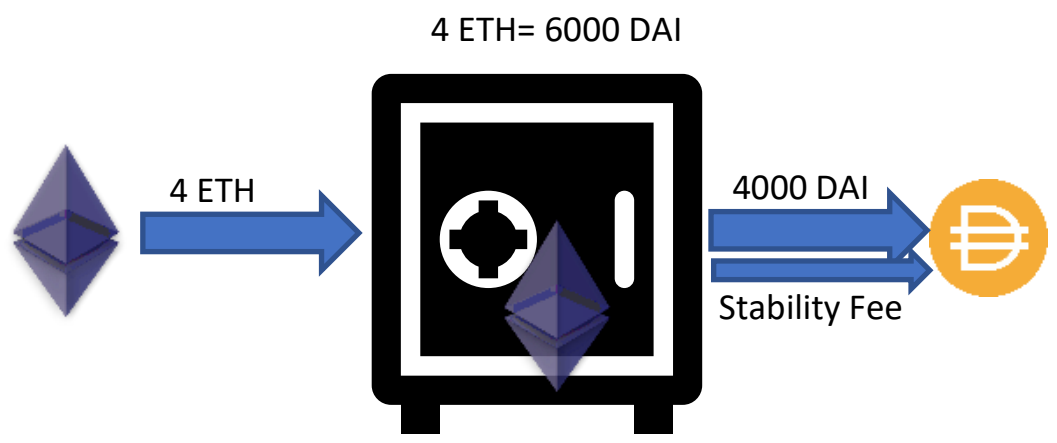

Fig. 40 Collateralized Debt Position

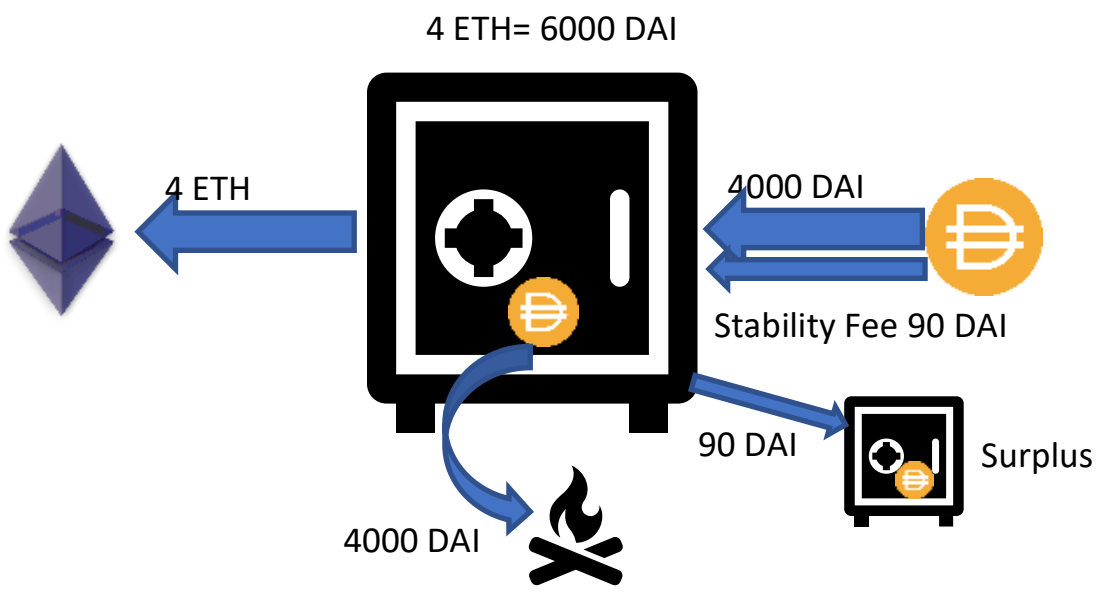

Fig. 41 Repayment of DAI loan

This system is analogous to a mortgage in the traditional financial system. One of the key differences is that the collateralized crypto assets like ETH and Basic Attention Token (BAT) are much more volatile than the assets mortgaged in traditional finance. Thus, MakerDAO users need to overcollateralize the DAI loans by at least 150% for ETH which can be higher (>175%) for some other crypto assets like BAT. This facilitates the stability of the protocol as well as the peg of DAI to the USD, as even if the value of CDP reduces, the Maker protocol would still be able to discharge the liability of issued DAI. If the CDP value falls below the threshold, the protocol enables three scenarios:

i) The borrower deposits more collateral to the CDP.
ii) The borrower repays some of the DAI debt.

[115] https://makerdao.com/en/whitepaper/

iii) The protocol opens the CDP for auction by keepers which liquidate the CDP.

The MKR token is the governance token of the Maker protocol which enables the holders to vote on proposals as well as initiate proposals for changes in the protocol. The parameters like stability fee, the crypto assets to be taken as collateral, as well as their collateralization ratios are decided through voting, with each MKR token holder entitled to one vote. If due to an extreme event the protocol accrues a debt, the governance mechanism decides through voting to issue new MKR tokens for recapitalization. This is very similar to a situation in traditional finance where new equity is issued to infuse capital in a sick enterprise.

It is also noteworthy that apart from the Maker protocol DAI and MKR can also be bought and sold on various crypto exchanges. In order to incentivise people to hold DAI rather than selling it and to stabilize its value in the open market, the Maker protocol also enables the DAI borrowers to deposit the borrowed DAI into smart contracts that yield returns at the rate of the DAI savings rate. The DAI savings rate is also determined by the governance ecosystem of MakerDAO. Just like in ordinary finance where the lending rate is higher than the savings rate, the stability fee is always higher than the DAI savings rate. Besides this, the borrowers of DAI can use it to earn passive income on other DeFi applications or create further leveraged positions. The DAI stability fee, savings rate and the collateralization ratio are like instruments in the armour of a central bank which can affect the demand and supply of DAI resulting in movement of price of DAI in the desired direction as compared to the US Dollar.

Various external actors also play an important role in the Maker protocol operations. They play a critical role in maintaining the peg against the US Dollar. The main external actors in the Maker protocol are:

i) Keepers: They are independent actors which take advantage of arbitrage opportunities which help DAI to move towards the target price of 1 USD. They participate in Surplus Auctions, Debt Auctions, and Collateral Auctions which help to maintain sufficient reserves against the issued DAI in the protocol. Keepers are usually bots operated by natural or legal persons.

ii) Oracles: As the blockchains cannot access any off-chain information like asset prices of assets locked in CDPs or the price of DAI, oracles deliver price information about the assets locked in CDPs in order to enable the protocol to know when to trigger a liquidation.

iii) Emergency Oracles: They are the oracles selected by Maker governance as last line of defence against Maker governance or oracles. They are authorized by Maker governance to freeze oracles and trigger an emergency shutdown in an unforeseen situation.

iv) DAO Teams: These are individuals and service providers who are authorized by Maker governance to provide services to MakerDAO.

Keepers play an important role in maintaining the peg of DAI to the US Dollar as they participate in various auctions like the Surplus Auctions, Debt Auctions, and Collateral Auctions. The various auctions in the MakerDAO ecosystem and their role in the protocol is as follows:

i) Surplus Auctions: The DAI loans with accrued stability fee denominated in DAI is repaid back to the CDP smart contract. Once the DAI principal along with the stability fee is repaid, the collateralized assets are released. The excess DAI obtained from the stability fee goes to a smart contract which accumulates it to a level determined by the Maker governance, after which an auction for the surplus DAI to be converted to MKR can be triggered by a keeper and the MKR tokens obtained consequent to the auction are burnt.

ii) Debt Auctions: It is possible in certain cases that in situations where the value of the collateralized asset collapses rapidly, the liquidations might not be able to repay the entire collateralized assets and the protocol may incur debt. When this debt crosses a certain threshold determined by the Maker governance, an auction can be triggered by a keeper to recapitalize the protocol by auctioning off MKR for a fixed number of DAI.

iii) Collateral Auctions: These auctions are triggered to recover debt in liquidated vaults. When the collateral position in a CDP vault drops below the specified liquidation ratio the vault becomes available for liquidation. The keeper bots liquidate the vaults by sending transactions which trigger an auction of the assets locked in the vault. It is important to note that as the liquidation transactions can only originate from an externally owned account on Ethereum, the Maker protocol cannot automatically trigger the collateral auctions and relies on keepers for the same. The amount of DAI recovered by the auction is used to repay the debt and the remaining collateral is returned to the user after deduction of a penalty fee. The keeper gets a fee for the transaction as an incentive to keep a watch on vaults that can be liquidated.

The MakerDAO users and owners of MKR and DAI do not undergo any KYC procedure to determine their tax residency or for identification of the beneficial ownership of the assets, this gives rise to a complex scenario where a number of individuals or entities come together to transact on a decentralized lending platform which also issues a so-called stable coin and maintains its peg to the US Dollar. The individuals or entities also participate in the governance of the Maker Decentralized Autonomous Organization and take decisions which have financial implications for the protocol and the holders of MKR token themselves.

The tax implications of transactions in such ecosystems can be very complex and due to unforeseen and trans-national nature of such organizations there is virtually no guidance available for taxation of the truly Decentralized Autonomous Organizations which do not have a well-defined legal identity. Although, the individual transactions are still governed by the tax guidelines and laws of the respective jurisdictions of the members and users of the Decentralized Autonomous Organization's platforms/protocols. The taxation of such Decentralized Autonomous Organizations is discussed in the subsequent sections.

## 8.2 Uniswap

Uniswap is one of the most popular Decentralized Finance projects and one of the biggest decentralized exchanges built on the Ethereum Blockchain which is censorship resistant, non-custodial and trustless. It enables users to swap one ERC-20 token for another for a fee, without presence of market makers[116] who are individuals or entities. Unlike the traditional order-book based exchanges there is no single authority like WazirX and Binance which matches buy and sell orders on Uniswap. The trades are executed against on-chain liquidity pools which pool the crypto assets provided by all the liquidity providers, unlike individual orders by individual market makers in a traditional centralized exchange.

In Uniswap, unlike a traditional centralized crypto or securities exchange, a counter-party is not required to execute a trade. Instead, an automated market maker facilitates every trade according to the supply and demand of the ERC-20 tokens. This can be considered to be analogous to barter trading in crypto tokens. There are two possible models in crypto exchanges, the first one is a centralized model which enables users to buy and sell their crypto assets for fiat currency and vice versa after undergoing a KYC process. The second model is that of an entirely decentralized exchange which can deal only in crypto assets and facilitates the exchange of one crypto asset for another purely on the basis of barter, without the involvement of any fiat currency or KYC procedures.

---

[116] https://en.wikipedia.org/wiki/Market_maker

Analogous to the first case if an individual who has only Rubies wants to have some Emeralds which he/she does not possess, he/she would either have to find someone who is willing to exchange his/her Rubies for Emerald(s) which would be extremely difficult. It would also be difficult to find out how many Rubies are a fair deal for one Emerald and vice versa. The individual would instead try to find an exchange where he/she can exchange Rubies for a common medium of exchange like the Indian Rupee or the US Dollar and then exchange it for Emeralds, as depicted in Fig. 42(a). However, exchanging the Ruby for money would require the individual to have a bank account and identification proofs for KYC before the individual can undertake the transaction. Also, the exchange would charge the individual a fee for providing this intermediation and liquidity facility both when converting Rubies to fiat currency and subsequently using the fiat currency for buying Emeralds. This is analogous to the mechanism of trading in centralized exchanges with the added disadvantage of crypto assets being in the custody of the exchange rather than the users.

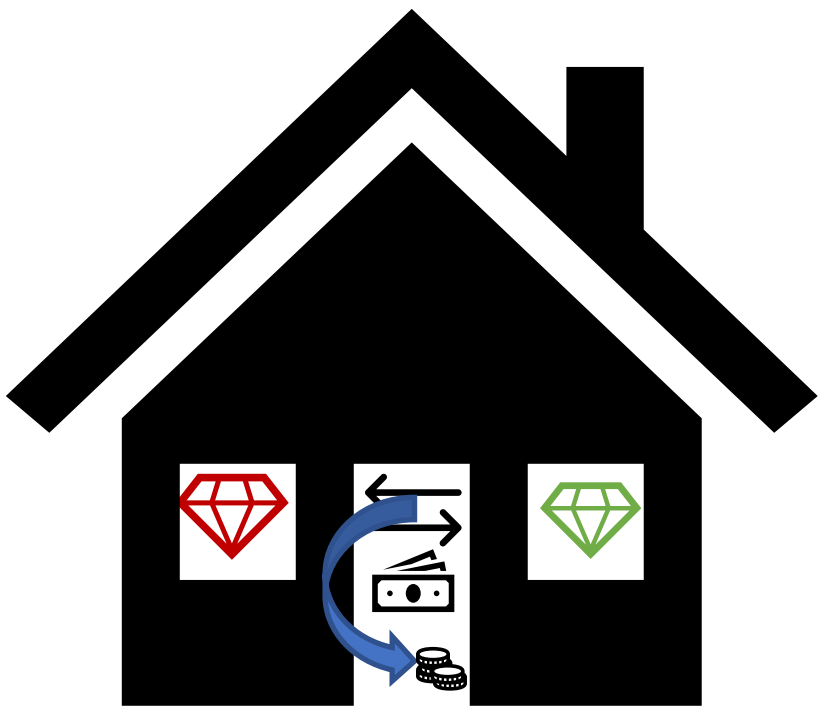

Fig. 42(a) Exchange based on fiat currency

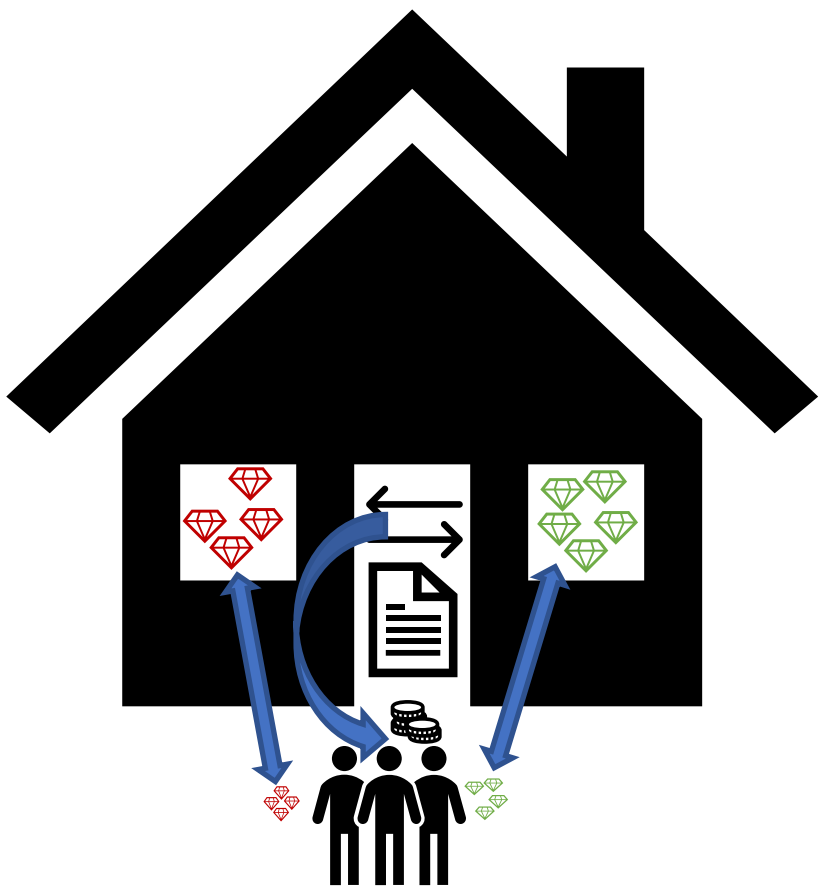

Fig. 42(b) Barter based on Automated Market Maker

Analogous to the second case, there can be an automated store like Amazon Go named Unistore which does not deal in fiat currency at all and carries out purely automated barter trades of precious stone pairs like Ruby/Emerald, Emerald/Sapphire, Sapphire/Diamond etc. as depicted in Fig. 42(b). If we consider the Ruby/Emerald pair in the store, the store has a pool of individuals who own Rubies and Emeralds and wish to earn passive income on their assets. The owners of the Rubies and Emeralds are given an offer by Unistore to deposit equal values of Rubies and Emeralds into the baskets in the store in order to earn a fixed fee on every Ruby/Emerald swap executed in Unistore. The owners of the Rubies and Emeralds can withdraw their precious stones at their will along with the accumulated fee. The liquidity providers are required to deposit Rubies and Emeralds of equal value (not equal number) in order to maintain the stability of the protocol which is based on an automated market maker (AMM).

Unistore facilitates its users to anonymously exchange Rubies for Emeralds and vice versa without converting them to fiat currency, in lieu of a small fee which is used to pay the owners of Rubies and Emeralds who decide to pool them into Unistore to provide the liquidity. This enables a user to anonymously walk into the store and exchange Rubies for Emeralds and vice versa through an Automated Market Maker, without any human intervention and questions about identity and jurisdiction. This is achieved through smart contracts which run an Automated Market Maker that decides the price for each Emerald exchanged in lieu of a Ruby and vice versa.

Uniswap uses a constant product market maker which can be mathematically represented as (1) where $Q_{Ruby}$ is the quantity of Rubies in the pool and $Q_{Emerald}$ is the quantity of Emeralds in the pool and K is a constant.

$$Q_{Ruby} \times Q_{Emerald} = K \quad (1)$$

When K=1600 this would result in a curve shown in Fig. 43

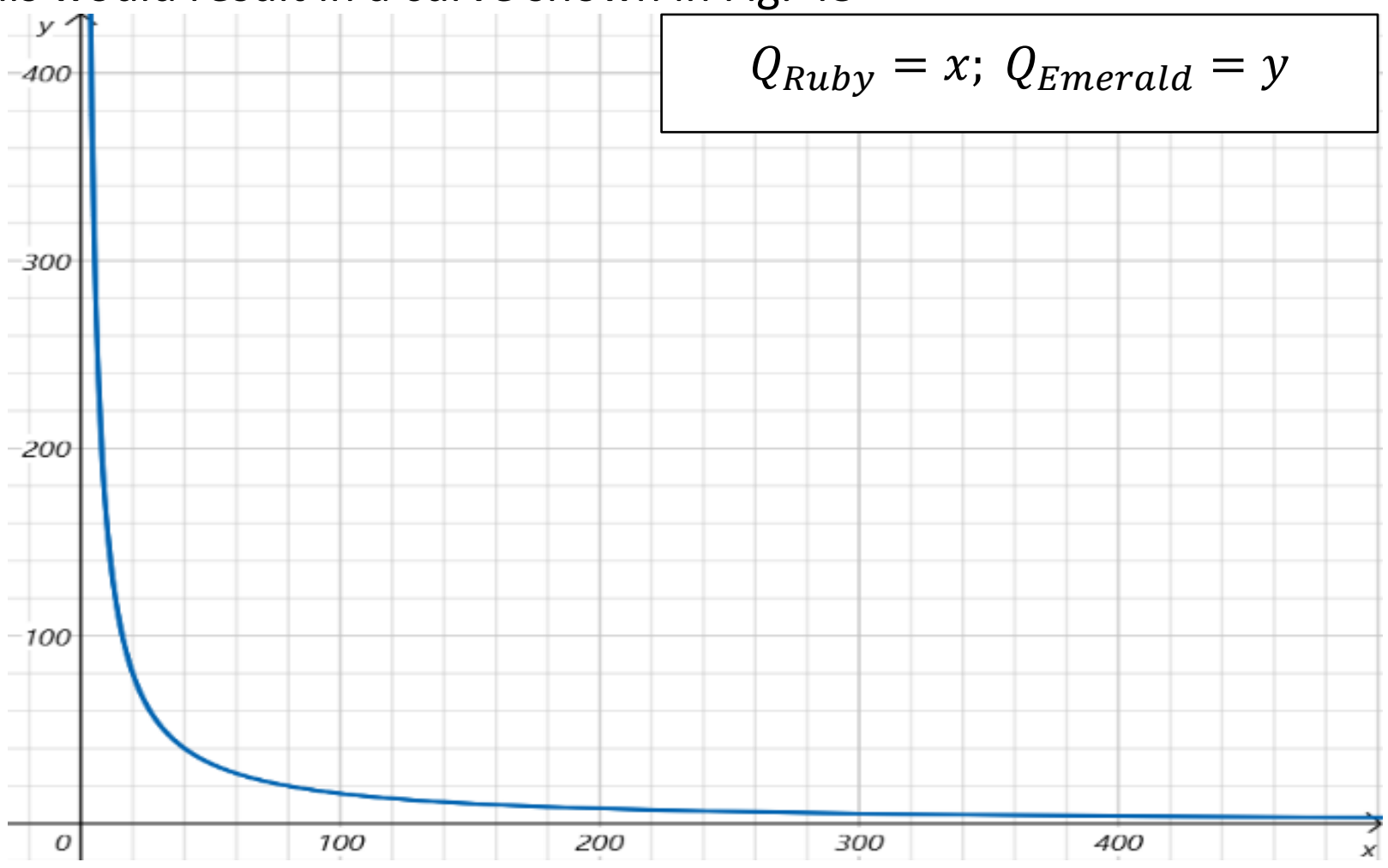

Fig. 43 Constant product market maker of Unistore where K=1600

For example, if a user swaps their Ruby for an Emerald when there are 40 Rubies and 40 Emeralds in the pool already, the supply of Ruby would increase and that of Emerald would decrease, thus the price of Ruby would go down and the user would get a smaller number of Emeralds for every subsequent Ruby swapped for an Emerald. Such fluctuations give rise to arbitrage opportunities which are used by rational actors in the DeFi ecosystem to carry out trades which move the exchange rate towards equilibrium levels. This also gives rise to opportunities for validators to earn income over and above the transaction fee and issuance and is known as Maximal Extractible Value (MEV).

The operation of the Automated Market Maker is purely algorithmic without the need for any counter-party's presence to facilitate individual transactions. Theoretically it can provide infinite liquidity with exchange rates of tokens rising asymptotically. A transaction would be available even for the last available Emerald for an infinite number of Rubies. Uniswap is identical to a Unistore trading in ERC-20 token pairs instead of precious stone pairs.

However, in Uniswap V3 the liquidity providers can choose the range on the curve for which they wish to provide liquidity, resulting in much greater market depth. This results in each liquidity provider providing a unique position on the liquidity curve. Therefore, these positions cannot be represented by ERC-20 LP tokens and are instead minted in the form of LP NFTs which specify the ownership of the corresponding liquidity position on the bonding curve of the Automated Market Maker as shown in Fig. 44.

It can also be seen from the curve in Fig. 43 that every trade would result in a change in price for every unit of the ERC-20 token bought, this is known as slippage. Hence, large trades executed in pools with low liquidity can result in significant price fluctuations when the trade is actually executed on chain. This is one of the main reasons why validators who have visibility of such transactions might reorder such transactions to benefit from such variations. The value derived by such reordering is called Maximal Extractible Value (MEV) and provides additional revenue to the validators. The taxable events triggered in the MEV ecosystem are discussed in the subsequent sections.

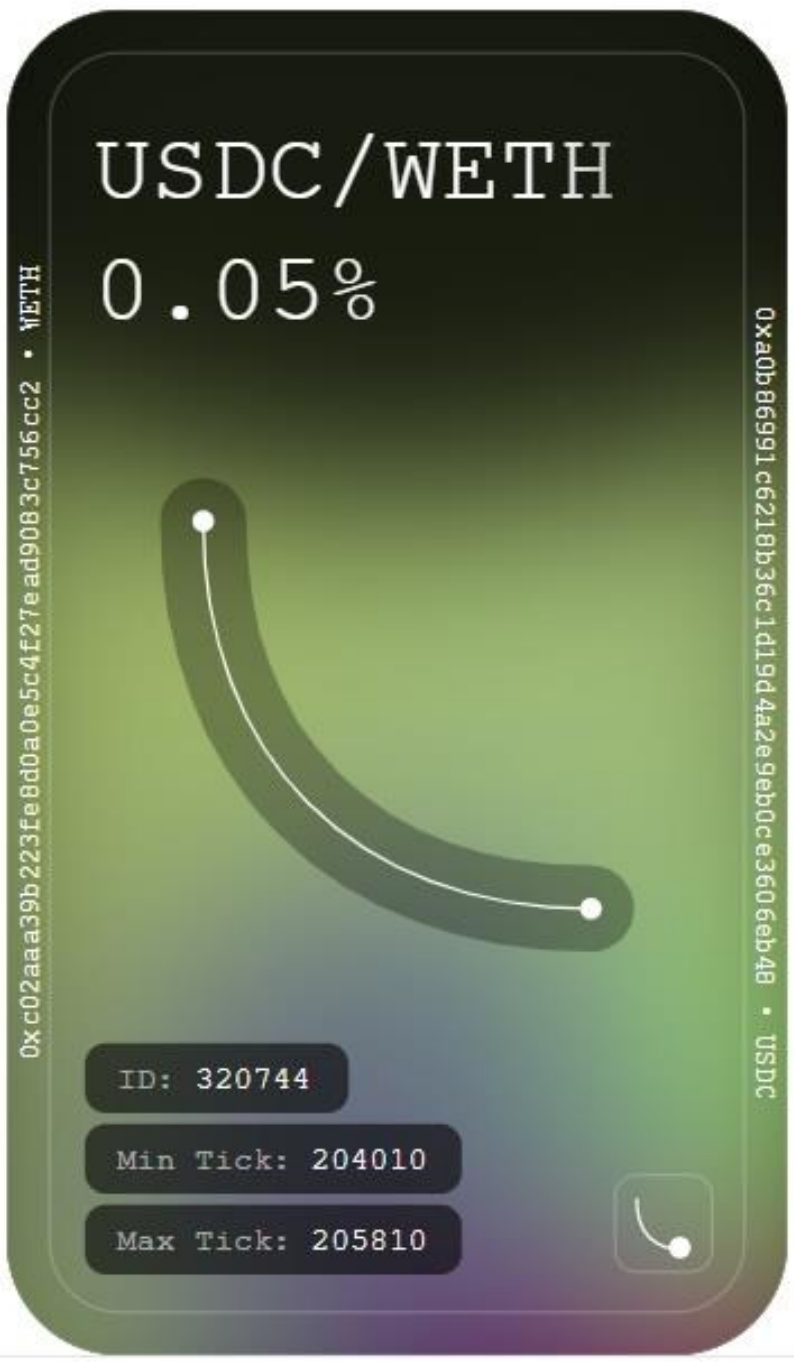

Fig. 44 A Uniswap LP NFT

In order to maintain and incentivize liquidity and pooling of tokens, Uniswap provides the users who pool their tokens (known as liquidity providers) with LP tokens which can be used by them to earn passive income in the form of share of the transaction fee of charge by Uniswap. Uniswap V2 charged a flat fee of 0.3% for all the swap transactions[117]. However, Uniswap V3 has a fee structure of 0.05%, 0.3% and 1% depending on the token being exchanged[118]. The users pooling their assets are given a percentage of LP tokens or LP NFTs that represent the liquidity reserve which accumulates the liquidity fee, when the users decide to stop providing their assets for liquidity, they redeem the LP tokens or LP NFTs to get their share of the liquidity reserve. The LP tokens or NFTs are burnt consequently. This mechanism is the main source of liquidity for decentralized exchanges like Uniswap. However, the conversion of the crypto asset pair into LP tokens/LP NFTs and vice versa might be considered a 'disposal' in many tax jurisdictions.

In this process, the project itself does not receive any income, as all the fee is given to the liquidity providers. It is noteworthy that LP tokens are also ERC-20 tokens which can also be used by liquidity providers to earn passive income on other DeFi platforms. However, providing liquidity has its own risks with a possibility of divergence loss which is caused due to the change in price of assets locked in the Liquidity Pool smart contract. A divergence loss is also known as an impermanent loss, it is said to have occurred when the current price of the tokens pooled into the liquidity pool changes relative to the price when the tokens were deposited. A detailed explanation can be found here[119]. Both upward and downward movement of relative price of assets causes divergence loss. Mathematically deriving from (1) it can be expressed in terms of the price ratio $p$ as given below and the plot of (2) is shown in Fig. 45.

$$Divergence\ Loss = \frac{2\sqrt{p}}{1+p} - 1 \quad (2)$$

[117] https://docs.uniswap.org/contracts/v2/concepts/advanced-topics/fees
[118] https://docs.uniswap.org/concepts/protocol/fees
[119] https://pintail.medium.com/uniswap-a-good-deal-for-liquidity-providers-104c0b6816f2

Where $p$ is the ratio of the price of assets currently to the price at the time of depositing. However, despite the possibility of a divergence loss users pool their assets in liquidity pools as fees from liquidity mining offsets the divergence losses.

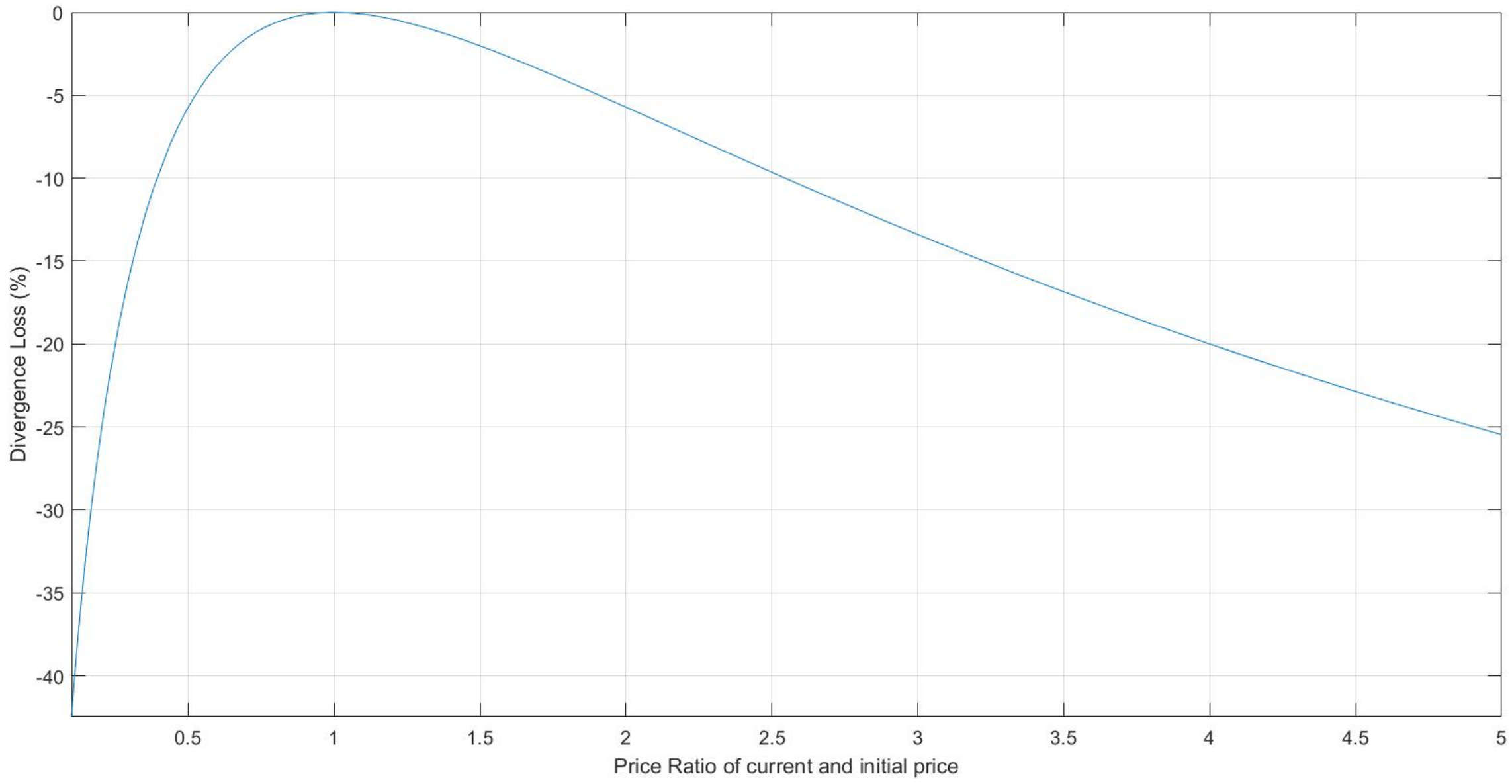

Fig. 45 Divergence loss

The users can list any ERC-20 token pair for swapping on the exchange, As on March 2024 there were 2056 token pairs listed on Uniswap[120]. The percentage of transaction fee to be charged is determined using the governance mechanism of Uniswap wherein each holder of the governance token UNI can initiate proposals for change and cast their votes on proposed changes. This makes Uniswap a Decentralized Autonomous Organization like the MakerDAO where no single 'person' owns the entity and it exists in a truly decentralized manner. Although, Uniswap has recently established an NPO – the Uniswap Foundation[121], a real-world legal entity which supports decentralized growth and long-term sustainability of Uniswap. Such Decentralized Autonomous Organizations pose various unique tax and legal challenges which are discussed in the subsequent sections.

The market '*depth*' in decentralized exchanges like Uniswap is obtained by having large pools of asset pairs which are liquidity pools and this facilitation of swap transactions in lieu of fee is known as liquidity mining. In Uniswap V 2.0 the liquidity providers had to deposit equal value of both the ERC-20 tokens in the liquidity pool, a condition which has been relaxed in Uniswap V3. For example, if the value of 3 BAT is same as 1 DAI then the liquidity miner would have to deposit 300 BAT and 100 DAI in the BAT/DAI liquidity pool. It is worth noting that the crypto assets being swapped in Uniswap are also available for open market trading on other platforms. Just as it would be possible for users of Unistore to utilize an arbitrage opportunity where the swapping ratio of the Ruby/Emerald pair is higher or lower than their open market price ratio, the crypto asset pairs like DAI/BAT or ETH/DAI having different price ratios than those prevailing in other exchanges would give arbitrageurs an opportunity to earn profit. This facilitates the maintenance of liquidity and price ratios close to the market price ratios in Uniswap liquidity pools.

120 https://www.coingecko.com/en/exchanges/uniswap#:~:text=Uniswap%20(v3)%20is%20a%20decentralized,pairs%20available%20on%20the%20exchange.

121 https://www.uniswapfoundation.org/

## 8.3 Taxable events in DeFi ecosystem

The market cap of DeFi crypto market is ~138 billion USD[122]. The number of DeFi users had increased to ~7.5 million in late 2021 and has declined since[123]. Many DeFi transactions result in accrual or realization of income to the depositors, borrowers, DeFi protocols and other actors and service providers in the DeFi ecosystem. As the income received or accrued along with the associated services might be taxable, it is important to understand the tax implications of various DeFi transactions. Some of the direct and indirect tax events in the DeFi ecosystem and their potential tax treatments are given below. This list is not exhaustive and the treatment of the events below mentioned might be significantly different in different jurisdictions[124].

### 8.3.1 Direct Taxes in DeFi ecosystem

The tax treatment of depositing crypto assets into a DeFi protocol smart contract and their locking up into a Collateralized Debt Position (CDP) would largely depend on the treatment of locking up of crypto assets into a Collateralized Debt Position (CDP) as 'disposal.' As the protocol can allow the CDP to be auctioned by keepers to liquidate the CDP, some tax administrations might take a view that this is tantamount to the transfer of beneficial ownership, thus making the creation of a CDP a 'disposal' of the underlying crypto asset. Some tax administrations might treat the CDP as a kind of escrow account which does not lead to transfer of beneficial ownership and hence may not be considered a taxable event. To understand the instance of transfer of beneficial ownership of the crypto assets locked in the CDP it would be worthwhile to delve deeper into the mechanism of a Collateral Auction in the MakerDAO protocol. A closer look at the documentation of the liquidation module of MakerDAO[125] reveals that a liquidation is triggered when a keeper detects a CDP that is below the liquidation ratio and triggers a liquidation by calling the Dog.bark function[126] . Thus, it can be inferred that the beneficial ownership of the CDP remains with the borrower till the value of assets in the CDP falls below the liquidation ratio and a liquidation is triggered by a keeper.

The issuance and disbursal of the loan amount in any other crypto asset in lieu of a CDP might also be taxable for the DeFi platform, but without any tax implications for the DeFi platform user. This might lead to issues of tax neutrality and taxation of such activities being inconsistent with their economic rationale and may also increase the administrative and compliance burden on users. Such transactions are very similar to repo transactions as they do not lead to transfer of all economic rights. Thus, some jurisdictions might consider including such transactions in the repo rules or create new rules for treating such sale and repurchase transactions as loans. This makes their tax treatment in sync with their economic rationale.

Some liquidity pools issue tokens to liquidity providers which give them a right to exchange the issued tokens for the original crypto asset pair deposited by the liquidity providers at the time of repayment. This arrangement may be considered as disposal by some tax administrations and subject to capital gains. However, such transactions are also like repo transactions and some jurisdictions might consider including such transactions in the repo rules or create new rules for treating such sale and repurchase transactions in line with their economic rationale. Only the HMRC has issued guidance on such liquidity pools and has also done a public consultation to try to sync the tax treatment of such transactions with their economic rationale.

[122] Top DeFi Tokens by Market Capitalization | CoinMarketCap

[123] https://www.statista.com/statistics/1297745/defi-user-number/

[124] Currently as there are no generally accepted definitions of DeFi which are used or accepted by tax administrations, the terms used below might not have the same meaning as used in regulatory or statutory parlance.

[125] https://docs.makerdao.com/smart-contract-modules/dog-and-clipper-detailed-documentation

[126] https://github.com/makerdao/dss/blob/liq-2.0/src/dog.sol

The interest paid by the borrower to the DeFi application, like the stability fee in case of MakerDAO may constitute taxable income in the form of interest for the application (after deducting the interest paid by the application to the borrowers) and might require withholding taxes by the borrower depending upon the legal residency of the DeFi application. The stability fee might be deductible as an expense for the borrower if the DAI is used as an investment or for trading. Actions like auctioning of the accumulated stability fee might also give rise to income for the DeFi application due to changes in the value of DAI with respect to MKR, even though the MKR acquired because of the surplus auction is eventually burnt.

The deposits made by borrowers in the MakerDAO smart contract for earning income at the DAI savings rate might also constitute ordinary income and taxable at the fair market value of the return in DAI when it is received. In this case, the DeFi application might be required to withhold taxes based on the tax residency of the borrower. Besides this, any further leveraging of the borrowed DAI or MKR and any income therefrom or any capital gains arising on disposal may also be taxable as income and capital gains respectively and may also have associated withholding requirements. However, some jurisdictions to reduce the administrative and compliance burden might consider taxing the net income or capital gains accruing due to such leveraging.

In case the governance of the application decides to recapitalize the protocol through a mechanism like Debt Auction in MakerDAO, the conversion of the issued MKR into DAI might also be taxable for the application. A CDP liquidation by a keeper might also be a taxable event for the borrower as the CDP of the borrower would be 'disposed' to repay the DAI loan and the excess ETH will be returned to the borrower after deduction of a penalty. The borrower might be subject to capital gains tax, the loss or gains would be determined by the basis of the liquidated crypto assets. The penalty might be allowed by some jurisdictions as a deductible expense if the borrowed DAI was used as an investment or for trading.

Any governance tokens issued to the participants of the DeFi application in the form of airdrops or otherwise would be taxed in most jurisdictions as income at the fair market value of the governance token at the time of receipt. Spending, trading or selling the acquired governance tokens might attract capital gains or taxed as business profit depending upon the nature and scale of the activity. The fee charged by decentralized exchanges (DEXs) for swap transactions might also constitute their income. Also, the returns paid by the DEXs to the liquidity providers might also constitute the income of liquidity providers and allowed as an expense for the DeFi application. Any income or capital gains obtained by further leveraging or selling tokens like UNI, LP or NFTs of a DEXs Liquidity Pool might also attract taxes on income and Capital gains respectively. In most of the liquidity pools like Uniswap, the liquidity providers get a fixed number of LP tokens or an LP NFT which represents their share in the liquidity pool which increases in value over time. The LP tokens or the NFT can be exchanged for the original tokens along with the financial return on the liquidity provided. This realized gain at the time of withdrawal or sale might be subject to capital gains in most jurisdictions.

DeFi applications might also hire teams/individuals to develop certain functionalities or provide services to the DAO like assistance in establishing a foundation for the protocol or developing or fixing software of the application. Such income might be characterized as self-employment income in some jurisdictions and may require deduction of social security contributions and other taxes. Depending upon the tax residency of the individuals hired by the DAO taxes might be required to be withheld.

Also, the fee charged or incentives given to various other actors or entities like keepers and oracles providing services to the DeFi application, might also be chargeable as income.

### 8.3.2 Indirect Taxes in DeFi ecosystem

Various stakeholders in DeFi ecosystem provide services to the users and other entities for a consideration. For example, Uniswap users need to pay a fee for swapping one crypto asset for another. Keepers also provide the services of monitoring the collateralization ratios of the CDPs and initiate Dutch auctions when they fall below the liquidation ratio. The keeper triggering the liquidation receives an incentive in the form of a percentage of the collateral auctioned. Oracles perform the service of providing external data to the smart contracts of DeFi platforms and charge a fee for the same. The fee charged by various entities in the DeFi ecosystem may also be subject to GST/VAT depending upon the tax residency, registration requirements, thresholds, and place of supply. The LP NFTs minted on UniswapV3 may also be subject to VAT/GST in many jurisdictions.

## 8.4 Maximal Extractable Value (MEV) on Ethereum

Maximal Extractable Value refers to "*the maximum value that can be extracted from block production in excess of the standard block reward and gas fees by including, excluding, and changing the order of transactions in a block.*"[127]. The nodes on Ethereum have visibility of the transactions in the mempool as transactions are not encrypted. This gives various actors on the Ethereum Blockchain the ability to front-run some transactions and capture lucrative arbitrage and other opportunities like liquidations. For example, at a certain price ratio in a liquidity pool, if such actors are aware of a transaction that is likely to cause the relative prices of tokens in that liquidity pool to move significantly, they might try to place their own transaction(s) before and/or after the transaction, taking advantage of the prior knowledge about the likely change in relative prices of the tokens in the liquidity pool. In one such transaction[128] due to difference in prices of the Ether and DAI between Uniswap and Sushiswap, a user was able to make a profit of 45 ETH in a single transaction.

As it is known in advance that a large transaction might cause prices on the decentralized exchanges to move, the user trying to extract MEV can pay much higher gas fee to place the buy and sell transactions before and after the target transaction. Such strategies can cause a 'tax' to be levied on the blockchain users due to slippage caused by sandwich trading. As users who want to extract MEV are ready to pay very high gas fees to include transactions extracting MEV, common users have to pay much higher gas fee with increased network congestion and gas prices. However, not all MEV on blockchain has such undesirable effects. MEV acts as an incentive for actors on the blockchain to keep the prices in various DeFi applications at the general equilibrium price and makes the markets more efficient.

As validators decide which transactions are included in the block and in what order, one might conclude that the entire MEV would accrue to the validators. However, exploiting arbitrage and liquidation opportunities requires running specialized algorithms and gaining faster access to the transactions transmitted on the Ethereum network. This requirement is fulfilled by certain specialized actors like builders, searchers, and specialized infrastructure like relays in the Ethereum ecosystem, which play a significant role in MEV extraction. Thus, unlike the traditional method of transactions being sent by users to the mempool over the Ethereum network, which are then subsequently included in blocks proposed by validators, is hardly executed in practice. A cursory look at the most recent blocks of the Ethereum blockchain[129] highlights that in most of the blocks the fee recipient is not the validator but a 'builder.' As the block proposer specifies the address which receives the fee, it can be inferred that the blocks were created by the 'builders' which received the fees.

---

[127] Maximal extractable value (MEV) | ethereum.org
[128] https://etherscan.io/tx/0x5e1657ef0e9be9bc72efefe59a2528d0d730d478cfc9e6cdd09af9f997bb3ef4
[129] https://etherscan.io/blocks

To prevent their transactions from being front run, the actors in the MEV ecosystem join a mechanism which communicates the transaction bundles prepared by a set of specialized bots called searchers which continuously look out for MEV opportunities on the Ethereum Blockchain. The searchers submit transaction bundles to specialized actors called 'builders' which construct most profitable blocks from these transaction bundles and transmit these blocks to validators through secure communication channels called relays. Relays receive blocks from builders and forward them to validators, however they can reorder or censor bundles based on their own policies. They also provide an available, reliable, efficient, and fast channel of communication between builders and validators and provides a layer of abstraction and anonymity between them.

Flashbots, a popular block space auction platform used by many searchers uses the first-price sealed-bid auction or blind auction mechanism, wherein users can privately communicate their bid transaction order preference without paying for failed bids[130]. The transactions bundle flow between a searcher and builder is shown in Fig. 46.

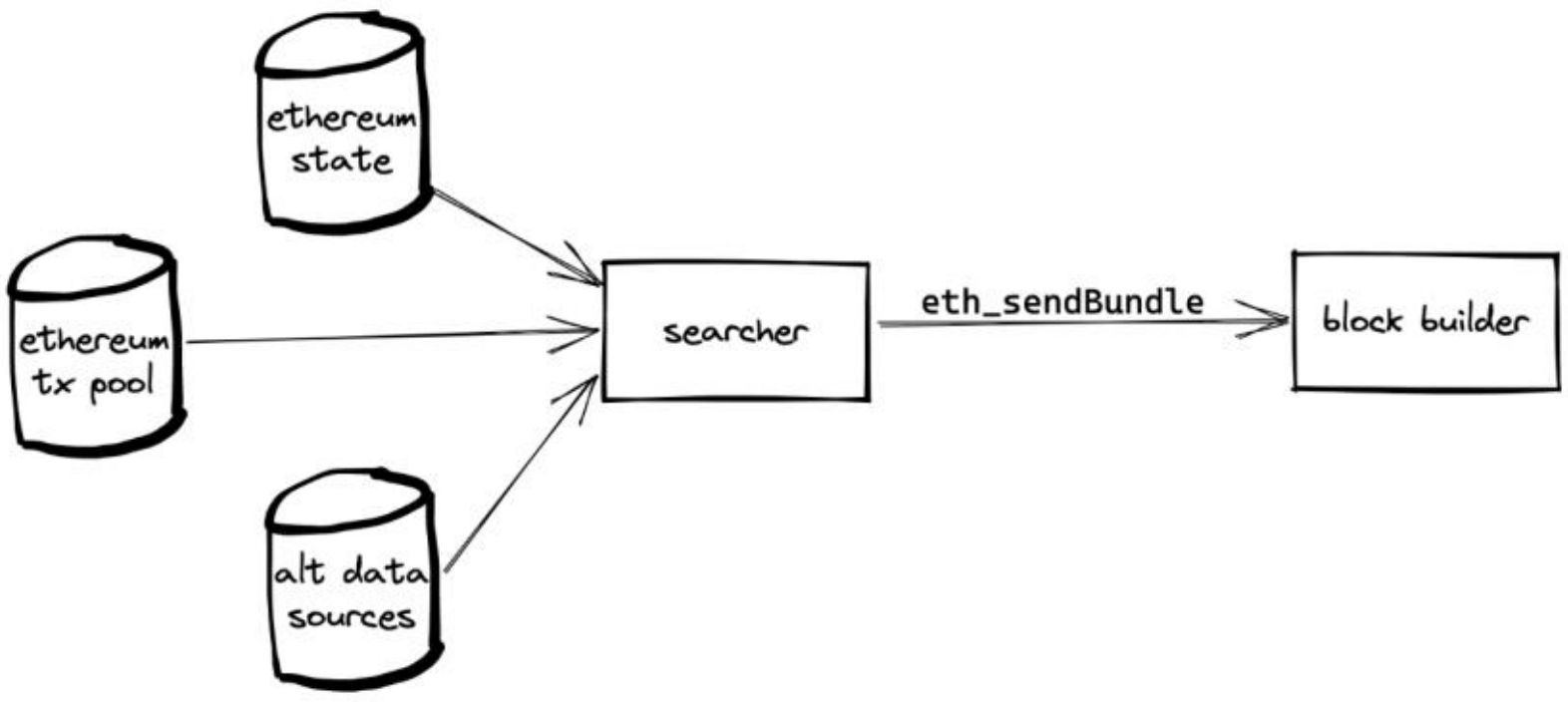

Fig. 46 Transactions bundle flow between a searcher and builder[131]

In the above scenario, the transaction bundles will be sent to the private transaction pool of Flashbots instead of the common mempool of Ethereum nodes, which protects the transactions from front-running. The searchers need to trust block builders as they have full visibility of the transactions in the bundle and can maliciously front-run the searchers as the builders can also operate searchers themselves. As shown in Fig. 47 a searcher can submit transaction bundles to multiple builders and the builders in-turn can submit the blocks to various relays which connect them to validators.

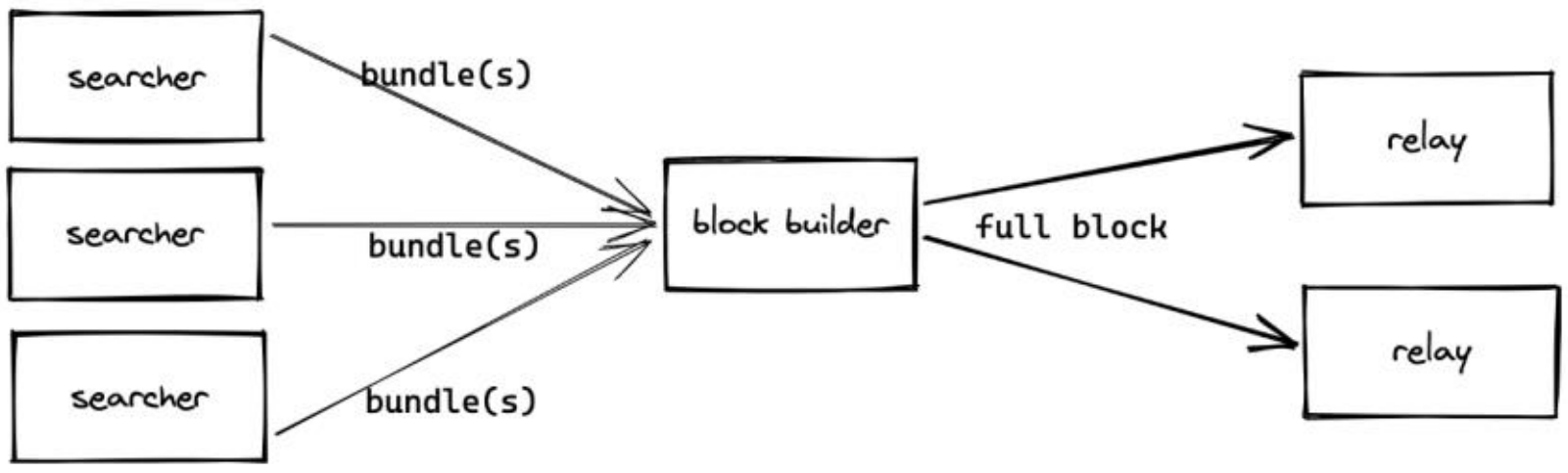

Fig. 47 Searchers submitting bundles to builders who submit full blocks to relays[132]

The mechanism of creation of blocks by builders and their transmission securely to the validators through relays creates centralization within the Ethereum ecosystem which can be used for censoring certain transactions by builders and relays. For example, on Nov 21 2022 it can be seen in Fig. 48 that

---

130 https://docs.flashbots.net/flashbots-auction/overview
131 https://docs.flashbots.net/assets/images/searcher-architecture-d9a0bd137035304fc54067ce243c32ce.png
132 https://docs.flashbots.net/assets/images/block-builder-flow-0c01103143daeac8b79cc377ff248630.png

79% of all Ethereum Blocks were OFAC compliant as MEV Boost, the then dominant off-chain marketplace used by validators for selling block space to builders is OFAC compliant.

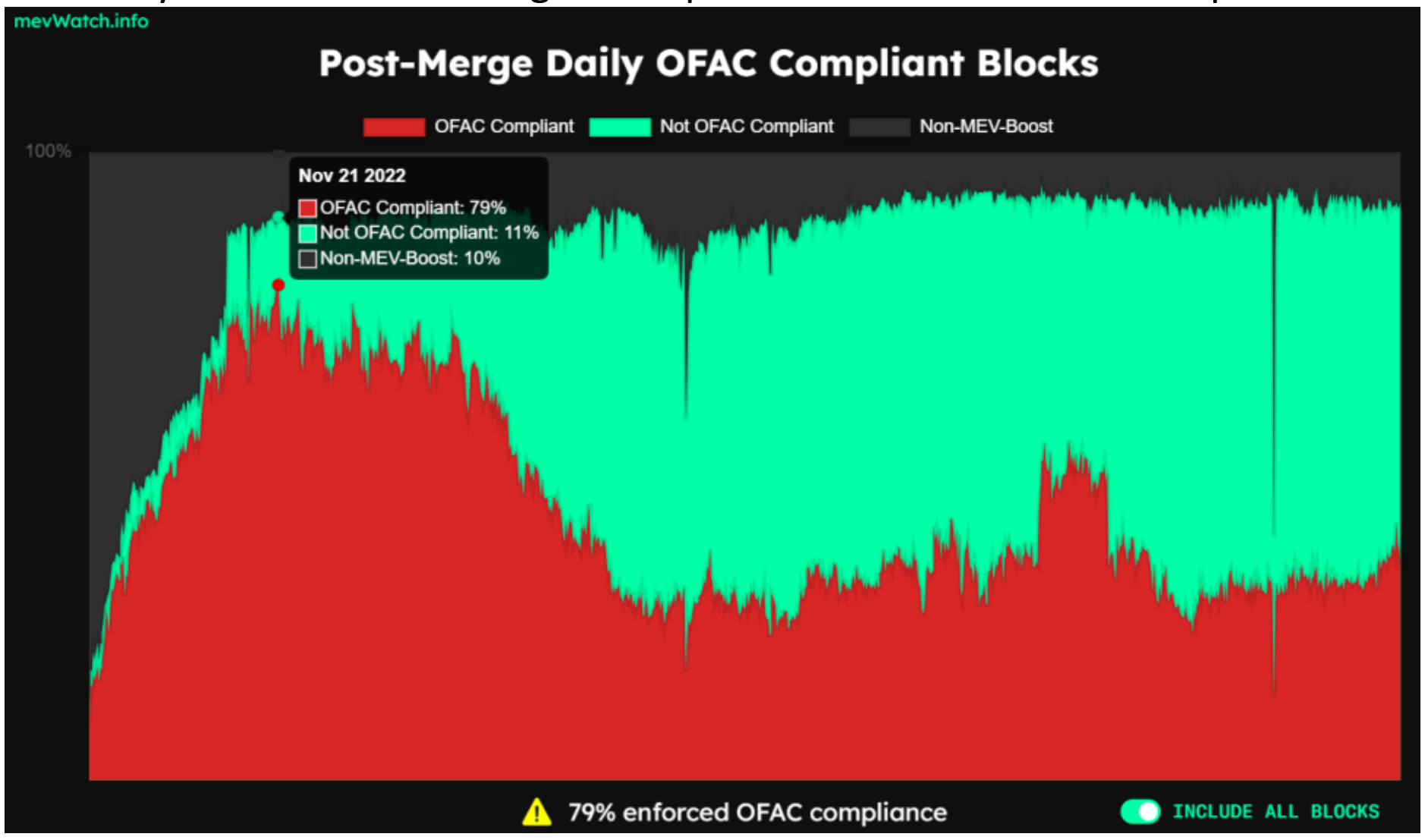

Fig. 48 Daily OFAC Compliant Blocks on Ethereum (source: https://www.mevwatch.info)

As relays also have full visibility of the transactions they need to be trusted by the builders. Relays do not provide the full block to the validators on request. Instead, the header of the most profitable block is provided to the validator which receives the full block from the relay only upon committing to propose the block by sending the signed block header to the relay, this is depicted in Fig. 49.

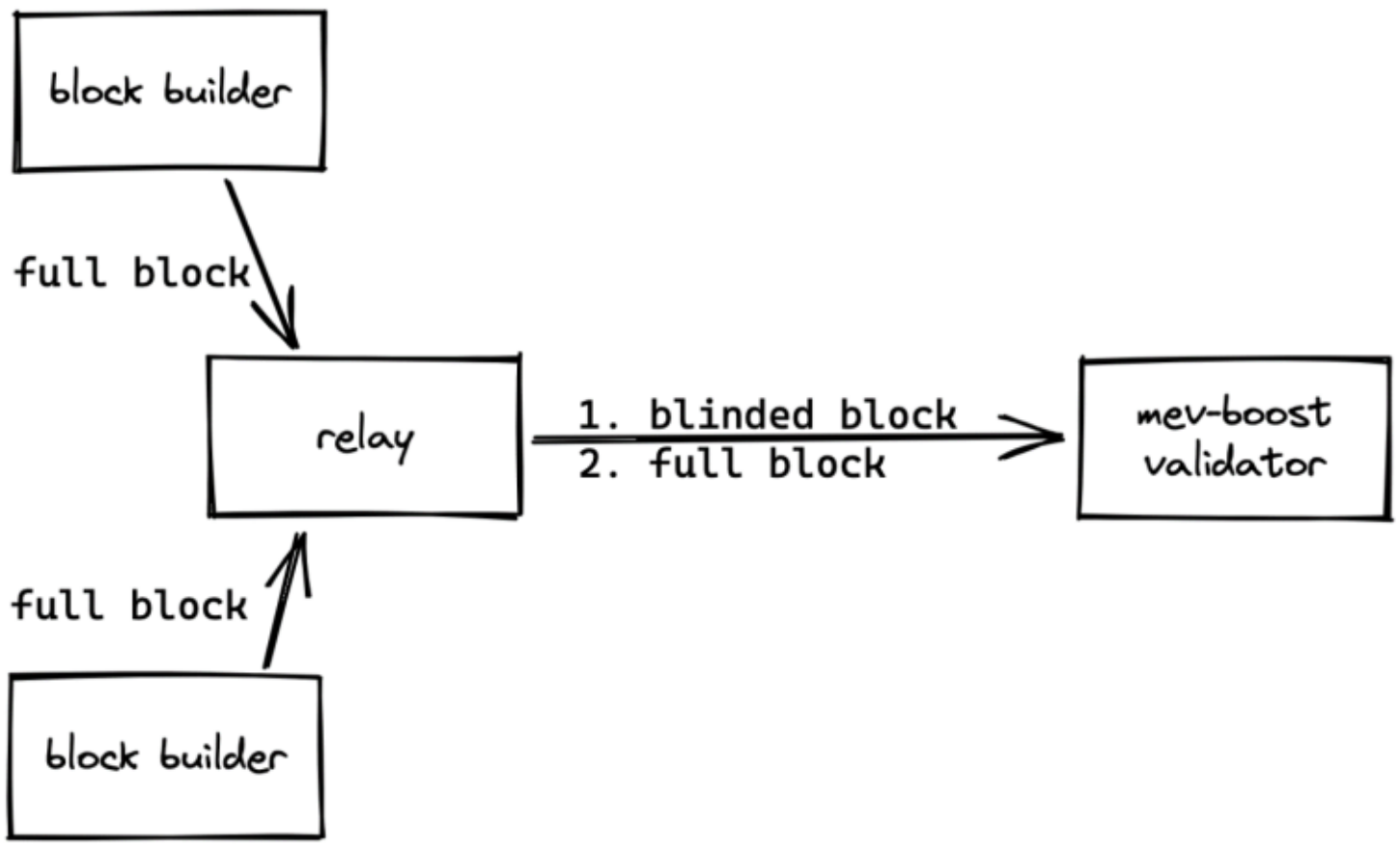

Fig. 49 Relay selecting the most profitable block and providing validator with the block header followed by the full block[133]

The validator can be connected to multiple relays and upon receipt of multiple block headers from multiple relays, proposes the most profitable block on the Ethereum Blockchain as shown in the Fig. 50

[133] https://docs.flashbots.net/assets/images/relay-flow-8f9aca183eaf4b8213220bc5bd71eb3a.png

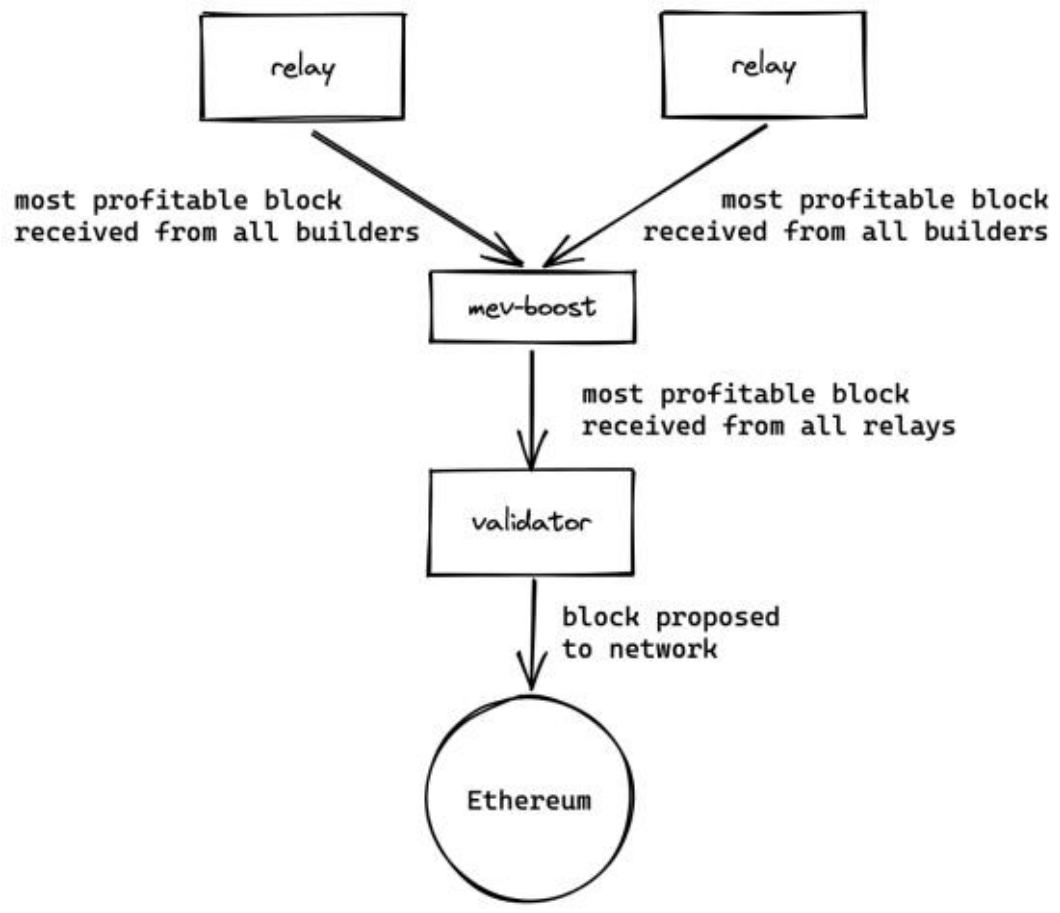

Fig. 50 Validator proposing the most profitable block on the consensus layer[134]

The overall communication and auction architecture for block space is shown in Fig. 51

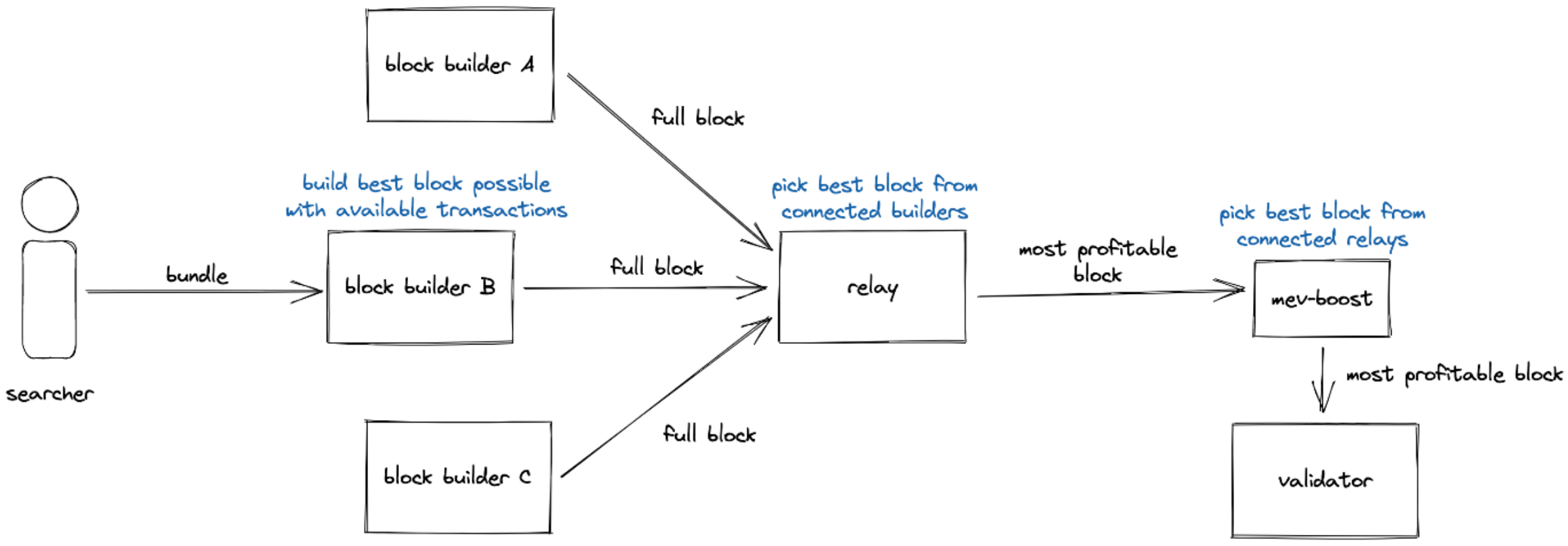

Fig. 51 Overall communication and auction architecture for Ethereum block space[135]

## 8.4.1 MEV Supply Chain and its taxation

The MEV ecosystem described above facilitates the extraction of MEV on the Ethereum Blockchain with the help of various specialized actors and service providers which play crucial role in the entire process. The income accruing to multiple actors in this ecosystem would be subject to tax in most jurisdictions with most jurisdictions classifying it as a business activity, allowing deduction of allowable expenses. However, as on date, there does not exist any specific guidance on fee or rewards received through MEV. The MEV related services might also be subject to indirect taxes as actors like builders, validators get paid for their service of creating and proposing blocks containing specific transaction bundles through the mechanism described above.

To understand the tax events in the process of MEV extraction starting from the creation of bundles by the searcher to the final proposal of the block by the validator, it is important to track the payments made to/by searchers, builders, and validators in a block on the Ethereum Blockchain. For example, in the Ethereum Block No. 18840841 it can be observed that the block reward (transaction fee – base fee burnt) was transferred to the address 0x1f9090aaE28b8a3dCeaDf281B0F12828e676c326 (rsync-builder.eth) as shown in Fig 52.

[134] https://docs.flashbots.net/assets/images/validator-flow-f3a8249b600db2b2b8d0a0344f336f95.png
[135] https://docs.flashbots.net/assets/images/mevboost-searcher-bundle-flow-bae4ba67a9d8d928efe337f36defa14a.png

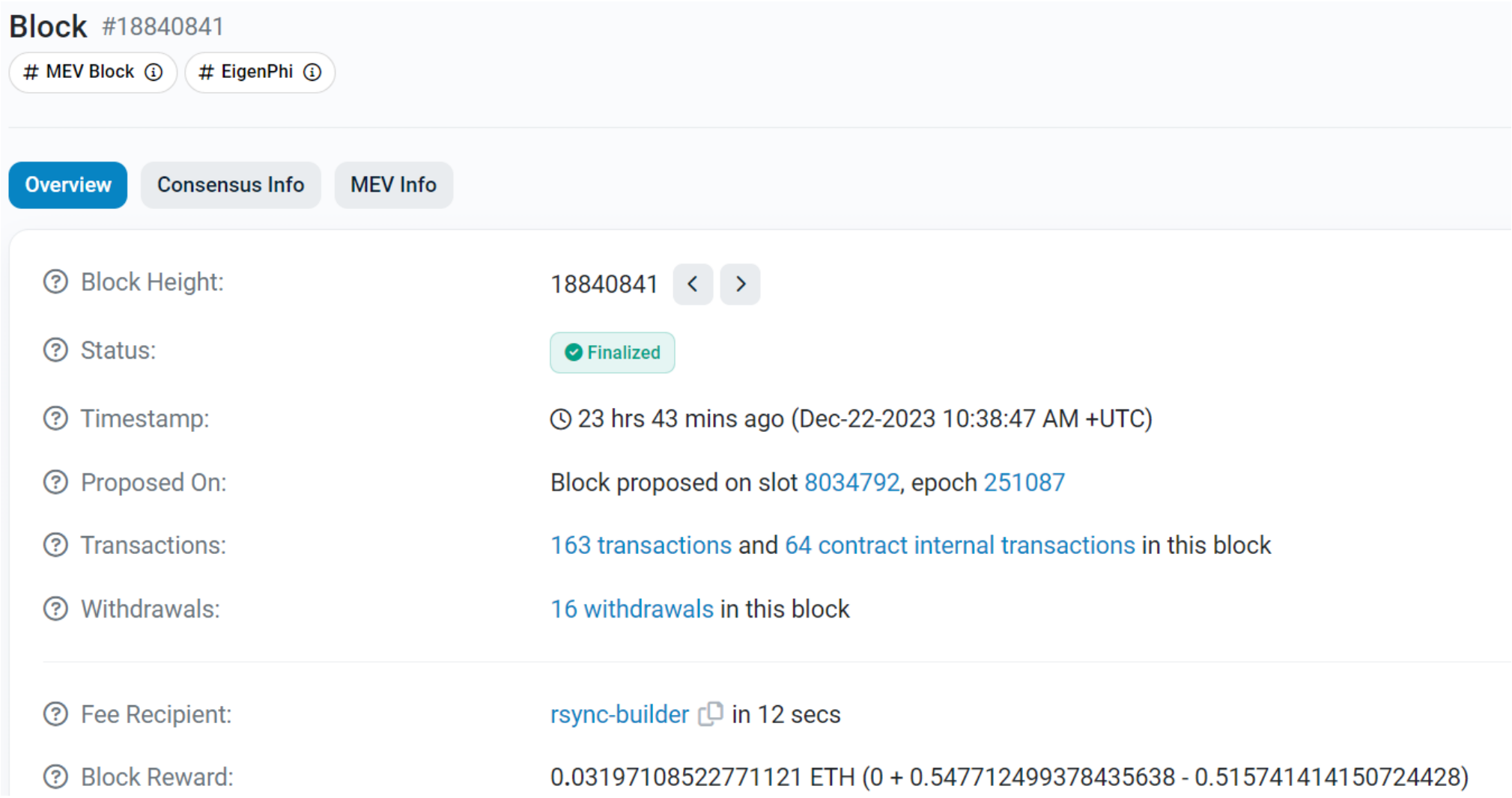

Fig. 52 Reward in Ethereum Block No. 18840841 accrued to rsync-builder.eth

A look at the MEV Info of the same block as shown in Fig. 53 indicates that the recipient of the Proposer fee for the block is the Lido: Execution Layer Rewards Vault (validator) with the smart contract address 0x388C818CA8B9251b393131C08a736A67ccB19297

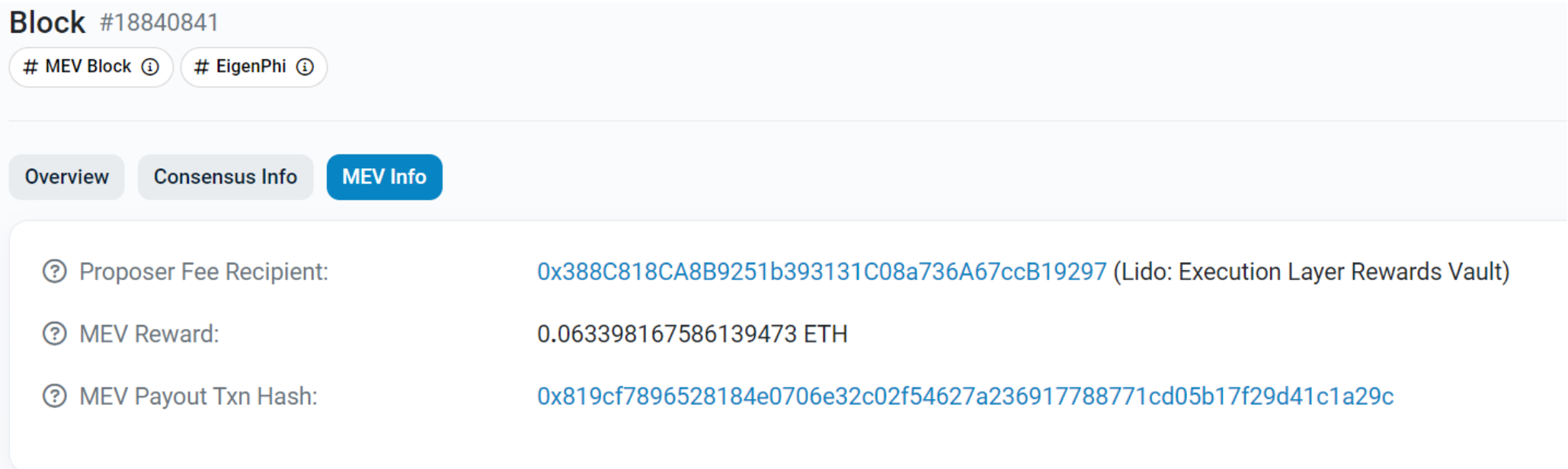

Fig. 53 MEV Info of the Block No. 18840841 of Ethereum

This transaction can also be seen in the list of transactions in the block 18840841. As shown inf Fig. 54 this is the reward received by the validator (Lido) over and above the issuance and other rewards received on the consensus layer. This is also the fee paid by the builder to the block proposer for proposing the block sent by the builder through the relay as described above.

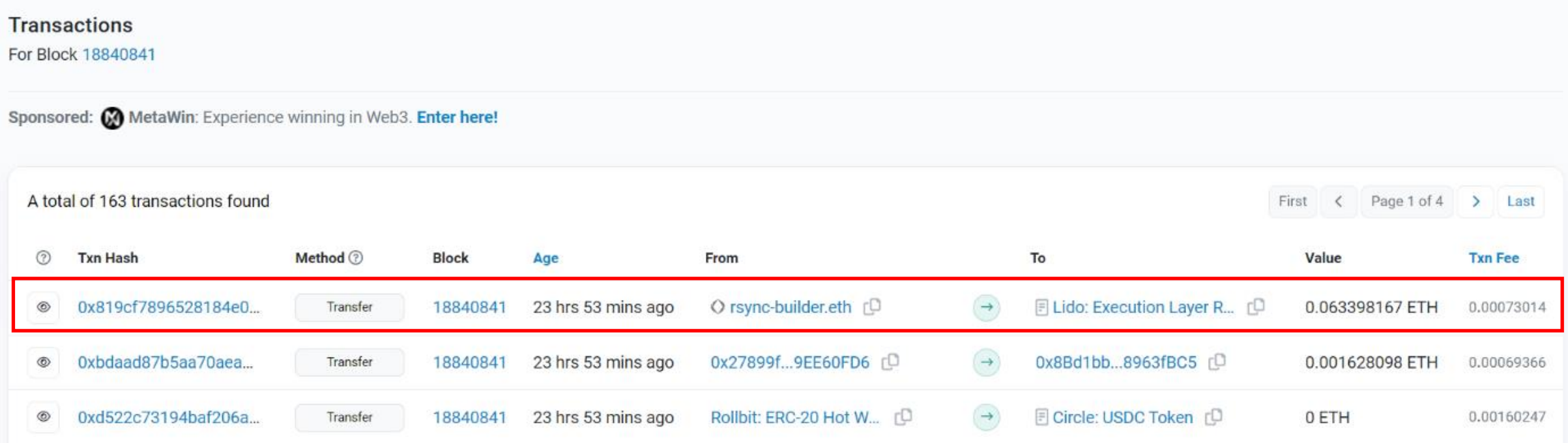

Fig. 54 list of transactions in the block 18840841

This might lead to a conclusion that for this block the builder paid more fee to the validator than the reward received by it, thereby incurring a loss. However, a closer look at the internal transactions in this block (Fig. 55) highlights two transactions where amounts of 0.013180 ETH and 0.032166 ETH were sent to the builder. This is the fee paid by the searchers or other users to the builder for inclusion of their transaction bundle in the block. Thus, the net fee received by the builder for the block is 0.031971 + 0.013180 + 0.032166 - 0.063398 = **0.013919 ETH**

| | | | | | |
|---|---|---|---|---|---|
| 0x43125de689c4cad0... | call | Wrapped Ether | → | Maestro: Router 2 | 2.034193282378778175 ETH |
| 0x7453abfd272396725... | call | 0x97C1A2...7Fe55447 | → | rsync-builder | 0.01318040562286369 ETH |
| 0xc524b89285930a30... | call | 0x74de5d...94016631 | → | MetaMask: DS Proxy | 0.007 ETH |
| 0xc524b89285930a30... | call | 0xE37e79...1457BD09 | → | Wrapped Ether | 0.6344 ETH |
| 0xc524b89285930a30... | call | 0xE37e79...1457BD09 | → | Wrapped Ether | 0.1586 ETH |
| 0xc524b89285930a30... | call | Metamask: Swap Router | → | 0x74de5d...94016631 | 0.8 ETH |
| 0x1b75b8c0ecc94434... | call | Banana Gun: Router 2 | → | Wrapped Ether | 0.19900497512437812 ETH |
| 0x95b23086aa5cdeb7... | call | 0x51C728...84502a7F | → | rsync-builder | 0.03216687894283185 ETH |

Fig. 55 Internal transactions in block number 18840841

It can be observed that the searcher paid the builder through a transaction on the blockchain rather than high gas fee for inclusion in the block, this method conditions the searcher's bid on the inclusion of their transaction in the block and obviates the need to pay for unsuccessful bids. However, to gain reputation and remain competitive some builders might discount some blocks and incur a short-term loss.

The payments made by the searchers and other actors to the builder and by the builder to the validator constitute payments in lieu of services of including the transaction bundles in the blocks formed by the builder and proposing the blocks formed by builder on the Ethereum Blockchain by the validator respectively. Thus, depending upon the tax residency of the builders and validators the services provided by them to searchers and builders may be subject to VAT/GST. As relays are not currently monetized, they would not be subject to VAT/GST. The total value of MEV rewards extracted before 'the Merge' of Ethereum is $675,623,114[136]. After 'the Merge' of Ethereum the total MEV extracted is 464,201 ETH[137] which is ~800 million USD. Although these figures are not huge, but with increasing transaction volumes it might not be possible for tax administrations to ignore this aspect of the crypto assets ecosystem.

## 9. Decentralized Autonomous Organizations - DAOs

Both the DeFi applications discussed above rely on smart contracts for their execution without the need for any centralized entity or agency. The parameters like savings rate, token pairs to be exchanged on the platform, collateralization ratio for various tokens etc. are decided by members holding voting rights through governance tokens of the DeFi applications. Such DeFi applications are classic examples of Decentralized Autonomous Organizations, which run on permissionless public blockchains without any centralized institutional structures, usually not having offices or addresses where operations are carried out or decisions are made. The rules that govern such organizations are encoded into and enforced through smart contracts on the blockchain, unlike conventional organizations where board of directors and personnel with clearly defined powers and responsibilities take decisions on behalf of the shareholders or participants.

DAOs attempt to address the 'agency problem' faced by many traditional organizations by making decision-making more participative. Each member of the DAO's governance contributes virtual assets to the DAO and obtains rights to vote through governance tokens. Holders of governance tokens can make proposals and vote on proposals through various voting models like Token-Based Quorum Voting, Quadratic Voting etc.[138]. The wide variety of activities carried out by participants is based on foundations of Cryptography and properties of the transactions undertaken on the blockchain. This provides pseudonymity to the DAO's participants who can use the applications to carry out transactions without any KYC requirements or disclosure of tax residency or beneficial ownership information. Also, this renders the DAO technologically unable to gather the required information for filing and compliance purposes, if any. As the application is essentially smart contract code running on a public blockchain, it is possible to create a DAO without any formal registration with a State, any government entity, or a regulator. This makes it possible for a DAO to have an organizational form which does not formally associate with any legal entity with the location of its operations and users not being known with certainty.

[136] https://explore.flashbots.net/
[137] https://transparency.flashbots.net/

The World Economic Forum's report Decentralized Autonomous Organization Toolkit defines DAOs as *organizational structures that use blockchains, digital assets and related technologies to allocate resources, coordinate activities and make decisions*. In simple terms, Decentralized Autonomous Organizations (DAOs) are like digital clubs or teams where people work together, but there is no boss or leader. Instead, decisions are made by everyone involved. Imagine a school club where all members vote on what activities to do, what snacks to bring, and how to spend the club's money. That is analogous to how DAOs work, but they use blockchains and smart contracts instead. DAOs are not just confined to financial applications like decentralized exchanges but can have a wide gamut of for-profit and non-profit objectives. Some examples of different kind of DAOs are:

i. **Protocol DAOs**: These are like the "rule-makers" for online applications. They decide how the application works, what features to add, and when to update it.
   example: Uniswap is a popular decentralized exchange (DEX) protocol. Its governance decisions are made by UNI token holders.
ii. **Grant DAOs**: These are akin to charity clubs. People pool their money to help others, like donating to a cause they care about.
   example: Gitcoin Grants allows contributors to fund open-source projects and public goods in the crypto space.
iii. **Social DAOs**: These are like online communities. Members collaborate on projects, share ideas, and organize events, undertaking tasks that require a big team effort.
   example: Friends With Benefits (FWB) is a social DAO where members participate in events, discussions, and projects.
iv. **Collector DAOs**: It is like a group of people who collect rare collectibles. They decide which collectibles to buy and how to take care of them.
   example: Flamingo DAO acquires rare NFTs (non-fungible tokens) and digital art.
v. **Investment DAOs**: These are like investment clubs. People put their money together to invest stocks, new technologies, startups etc.
   example: MetaCartel Ventures invests in early-stage blockchain projects.
vi. **Media DAOs**: Akin to a group of content creators (like YouTubers or bloggers) working together. They decide what videos to make or articles to write.
   example: Forefront is a media DAO where creators share revenue and collaborate on content.

Many of the different kinds of DAOs listed above have profit as one of their primary motives. Many DAOs may also involve contributions from its members as well as other economic transactions like acquisition of assets by the DAO, changes in value of assets or distribution of profits to the members. The DAO's governance tokens might also be tradable on secondary markets. All such transactions are likely to involve transfer of economic value and may be taxable events, resulting in tax liability for the

[138] DAO Voting Mechanisms Explained - LimeChain

DAO at the entity level or for the participant. However, in the absence of a legal entity structure and limited or no information about the location of its operations and users, the DAOs are likely to suffer from uncertainty about their legal structure, tax liability, residency, and jurisdiction as well as the applicable laws, rules, and regulations. It is likely that due to this uncertainty, till date no tax administration has issued any guidance for taxing DAOs.

Despite the uncertainties associated with DAOs, the total value of assets locked in DAO treasuries as of February 2024 is around 33 billion USD with around 10 million governance token holders, out of which around 3 million are active voters and proposal makers[139] . These statistics as shown in Fig. 56 are showing an increasing trend and are difficult for tax administrations across the world to ignore.

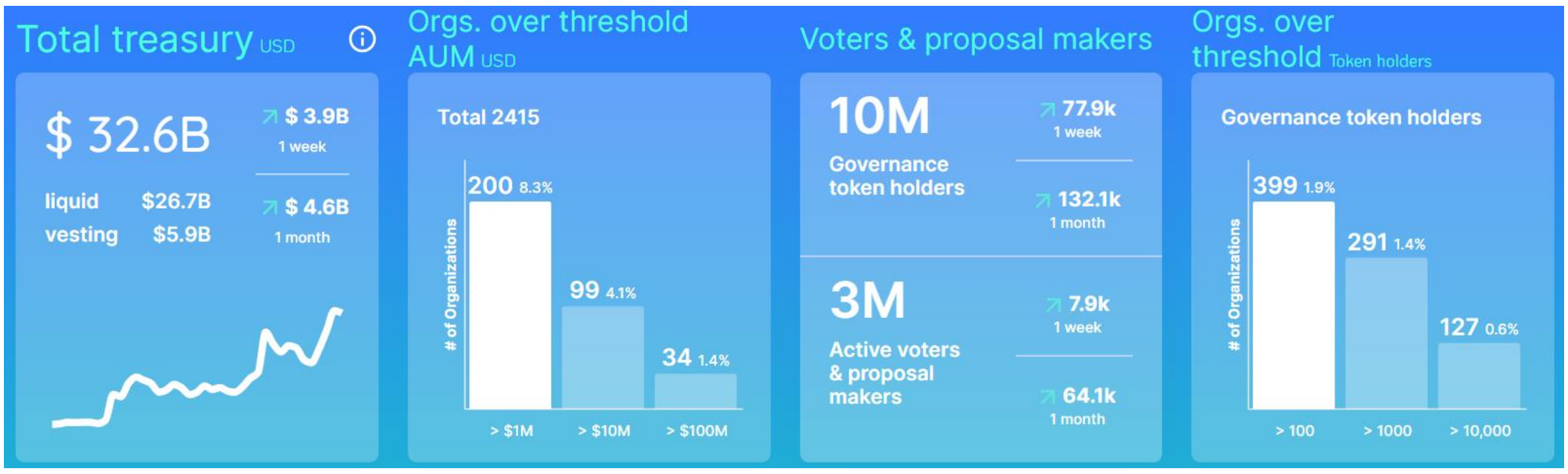

Fig. 56 DAO treasury values and number of active voters

Some pertinent questions related to DAOs which regulators in general and tax administrations in particular would want to seek answers to are:

a) How do DAOs classify under existing laws? Are existing legal structures and frameworks sufficient to provide legal personhood to DAOs or new legal technology is required for DAOs which pose challenges to any known classification?
b) How does the place of operation of DAO get determined and which laws, rules and regulations apply to it?
c) How do the organizational structure and nature of the tokens/crypto assets determine the regulatory framework of the DAO?
d) Is the DAO required to identify or collect information about its members? If yes, then how it does so to fulfil its legal and regulatory obligations?
e) Who has the power to act legally on behalf of the DAO and how does the DAO confer such powers?
f) How does a DAO carry out its legal affairs like filing of taxes and other compliances like social security contributions, withholding taxes etc.?
g) How the profits of DAOs from their multiple revenue streams are determined and who is obligated to pay the corresponding taxes?
h) How do investors/participants and the DAO are prevented from No Taxation as well as Double Taxation?
i) Do DAO participants have an unlimited or limited liability for debts, judgements, obligations, and liabilities of the DAO?

Though this is not an exhaustive list of issues being faced by stakeholders in the DAO ecosystem, but the issues highlight the unique challenges posed by DAOs. Most of these questions arise because of Blockchain based operations of DAOs and the pseudonymity of their members/users. These questions might be difficult or impossible to answer due to the technological bottom-lines of crypto assets. To

139 https://deepdao.io/organizations

answer these questions standardized technological solutions which address these issues on a global scale would be required. It would also require global policy action to formulate bespoke laws, rules and regulations that address the technological framework of DAOs, the blockchains on which they operate and their underlying crypto assets.

Due to such uncertainties, there have been multiple instances where DAOs have faced actions from regulators and other stakeholders, some of which are given below:

a) CFTC action on Ooki DAO[140]: Ooki DAO faced regulatory action by the U.S. Commodity Futures Trading Commission (CFTC) for allegedly operating an illegal trading platform without proper registration as a futures commission merchant. The CFTC filed a legal action against Ooki DAO, alleging violations of the Commodity Exchange Act. Ooki DAO failed to respond, leading to a default judgment. The court held that '*Ooki DAO is a 'person' under the Commodity Exchange Act and thus can be held liable for violations of the law*.' The court also ordered Ooki DAO to pay a penalty and permanently cease operations.

b) SEC action on The DAO[141]: In 2017, the SEC investigated digital assets associated with The DAO, a virtual organization which was created in 2016 with more than 11,000[142] anonymous investors who collectively raised more than 150 million USD[143]. The SEC applied the Howey test, which is a legal framework used to determine whether an arrangement constitutes an investment contract (and thus a security) and found that The DAO tokens met the criteria for investment contracts. The SEC concluded that these tokens were securities, subject to federal securities laws as The DAO did not meet the criteria for Regulation Crowdfunding exemption. It issued a warning in the report that the digital assets offered by virtual organizations like The DAO could be subject to federal securities laws.

c) MakerDAO arbitration[144]: On 12th March 2020, the crypto assets market witnessed a crash and the event later came to be known as Black Thursday. On the Black Thursday, the value of Ether plummeted dramatically. As explained earlier, this led to under-collateralization of the DAI loans in the Collateralized Debt Position vaults, triggering liquidations. Due to multiple issues related to the crash in price of Ether, several CDP vaults could be liquidated at a bid price of $ 0 resulting in heavy losses to the borrowers. Subsequently, a class-action lawsuit was filed against the Maker Foundation which governed the Maker protocol and DAI, claiming loss and damages for misleading investors about the risks associated with MakerDAO. The suit culminated in an arbitration and subsequent dissolution of the Maker Foundation.

There is a general lack of certainty about the tax and regulatory treatment of DAOs which results in instances of actions as given above. Many legal experts are of the opinion that DAOs would be classified as general partnerships (Shenk, Van Kerckhoven, & Weinberger, 2023). Some experts also argue that DAOs have certain characteristics of corporations and they should be given the ability to choose their legal entity status for tax and other regulatory purposes (Brunson, 2022). However, the truly Decentralized Autonomous Organization is an entity which lives and operates on the blockchain and suffers from real impediments in its interaction with the physical world. Such interaction requires agents who are natural persons acting on behalf of the entity for important tasks like regulatory compliances, filing and payment of taxes and compliance and cooperation with law enforcement.

[140] https://www.cftc.gov/media/8736/enfookidaoorder060923/download
[141] https://www.sec.gov/news/press-release/2017-131
[142] The DAO Attack: Understanding What Happened – CoinDesk
[143] https://www.bitstamp.net/learn/crypto-101/ethereum-dao-hack/
[144] https://www.coindesk.com/policy/2020/09/29/28m-makerdao-black-thursday-lawsuit-moves-to-arbitration/

Thus, it would be important for a DAO to have a real-world legal identity to avail benefits of legal personhood like access to banking, more certain tax and regulatory treatment and limited liability. This would have to be augmented by technological solutions which help to ameliorate issues related to pseudonymity on the blockchain.

## 9.1 Legal Entity Status of DAOs

DAOs can be defined as "*organizational structures that use blockchains, digital assets and related technologies to allocate resources, coordinate activities and make decisions*.[145]" As discussed above, such organizations can commence their operations without any registration due to the pseudonymous nature of its members and permissionless nature of the public blockchains on which DAOs function. Although this might have advantages of enabling decentralized decision making without a management structure with clearly defined roles, it makes it difficult for DAOs to avail benefits of legal personhood like being able to sue as an entity, having a bank account and being able to enter into contracts, limited liability of its members, and simplified compliances. The legal entity status of a DAO has profound consequences related to applicability of tax laws, AML/CFT/CPF and other regulatory obligations of the DAO and its members. This makes the issue of a DAOs legal entity status crucial. Classification of DAOs as domestic or foreign general partnerships, unincorporated associations, domestic or foreign corporations, foundations, or limited liability corporations has bearing on various important issues like:

A) Liability of the members for debts, judgements, obligations, and liabilities of the DAO
B) Collection and filing of information relating to taxes and other regulations in multiple jurisdictions
C) Determination of tax liability of the DAO and its members and identification of taxable events

It is also possible that various sub organizations within the DAO may have different classifications as legal entities. For example, components of the DAO which manage the treasury and the component managing voting on the protocol might have different classifications.

Many countries have enacted provisions to recognize DAOs as a legal entity. The Innovative Technology Arrangements and Services (ITAS) Act of Malta[146] enables DAOs to register by applying to the Malta Digital Innovation Authority. In Switzerland, DAOs can be structured as ownerless foundations and jurisdictions like Cayman Islands enable creation of an LLC structure that acts as real-world interface of the DAO[147].

Many states like Wyoming and Vermont in the US have also enabled DAOs to register as LLCs[148]. However, many of these legal entities require filing of physical forms along with the relevant information or establishing and maintain a registered agent in the jurisdiction, which might be operationally difficult for truly decentralized DAOs and against their foundational principles. The following discussion about DAOs and their legal aspects is from the perspective of the laws in the US, other jurisdictions may have similar or different legal treatment and classification depending on their domestic law.

---

[145] Decentralized Autonomous Organization Toolkit
[146] https://legislation.mt/eli/cap/592/eng/pdf
[147] https://www.forbes.com/sites/irinaheaver/2023/08/14/the-ultimate-crypto-legal-guide-to-structuring-your-dao/?sh=777f38957b81
[148] https://sos.wyo.gov/Business/Docs/DAOs_FAQs.pdf

### 9.1.1 Unregistered DAOs

One of the most important issues concerning the legal entity status of DAOs is the legal identity of an unregistered Decentralized Autonomous Organization which operates on the blockchain with pseudonymous members without registering with any State, government entity, or a regulator. As mentioned earlier, in an important case related to the Ooki DAO in the US, the US District Court for Northern District of California has adjudicated that Ooki DAO was an unincorporated association with token holders as its members, who joined it voluntarily. It was observed that Ooki DAO was a 'person' under the Commodity Exchange Act and it could be sued as it was an unincorporated association. The notice was served by Commodity Futures Trading Commission (CFTC) through the online community forum of the DAO, which was upheld by the court. A civil monetary penalty of $643,542 was levied and its operations were shut down.

In California, an unincorporated association refers to a gathering of individuals united by a shared purpose or goal, but lacking formal incorporation status with the state. Despite this absence of incorporation, such associations maintain legal standing and can undertake activities such as contractual agreements, property ownership, and litigation. While possessing fewer formalities and regulatory obligations compared to corporations, they still hold recognition under the law. Examples encompass social clubs, community organizations, and informal business partnerships. There are no State filing requirements for unincorporated associations, however, if the association is operating a business, it may need to obtain relevant licenses or permits, which the Ooki DAO purportedly failed to obtain from the CFTC. This case is important as it highlighted that for unregistered DAOs, members of unincorporated association may have unlimited personal liability for the debts, obligations, and liabilities of the DAO.

It may also be possible to classify many DAOs as general partnerships. General partnerships are a type of business structure where two or more individuals (or entities) come together to operate a business for profit. In a general partnership, each partner shares in the management, profits, and liabilities of the business. They may be formed through a written or oral agreement and the members have unlimited personal liability for the debts, obligations, and liabilities of the business. Thus, it might be possible that all members holding the governance tokens of DAOs which have not otherwise been incorporated as a legal entity form a general partnership inadvertently, as the intention to form a partnership is not a pre-requisite for forming a general partnership. However, the pseudonymity of the members of the general partnership and lack of the concept of a 'principal office' is a major roadblock in determination of the jurisdiction that governs the DAO as a general partnership.

The classification as a general partnership might obligate a DAO to reporting requirements in jurisdictions like the US. Even if none of the members of the DAO happen to be US tax residents, but if the income of the DAO is attributable to the US, it might be required to report the same to the IRS. Partnerships may also be required to submit an information return, Form 1065, to the IRS, detailing the partnership's financial information. Additionally, each member receives a Schedule K-1, which outlines their share of income, deductions, and other tax items, to be reported on their personal tax returns. Due to pseudonymous nature of membership of the DAO, such information might be extremely difficult to gather and furnish. Also, considering the nature of DAOs operations it is not clear that who would have the power to act legally on behalf of the DAO to make such filings and how does the DAO confer such powers?

As many of the DAOs may not be domestic i.e. 'Organized in the US' it would be possible to classify them as foreign partnerships. If the DAOs governance tokens are tradable on the secondary market, it may be classified as a Foreign Publicly Traded Partnership. Foreign partnerships that conduct certain activities within the US are subject to filing obligations with the IRS. This typically includes filing Form 1065, "U.S. Return of Partnership Income," if the partnership earns income effectively connected with a U.S. trade or business, or if it receives U.S. source income subject to withholding. Additionally, the partnership may need to issue Schedule K-1 to each partner to report their share of income, deductions, and other tax items

### 9.1.2 DAOs registered as a legal entity

Some DAOs are already registered in various jurisdictions under the existing legal frameworks. This provides the DAOs with the benefits of legal personhood, but might not eliminate the legal and regulatory uncertainty completely. DAOs choose multiple legal wrappers based on legal and regulatory considerations as well as tax implications. Some DAOs are constituted as foundations under the laws of jurisdictions like the Cayman Islands and Switzerland. Such structures have benefit of limited liability to identify members and beneficial owners, with limited liability of members of the foundation for its debts and other obligations.

The LLC structure provides members of the DAO with benefit of limited liability and the flexibility to choose how it is taxed based on specific circumstances. Brunson (2022) argues that just like check-the-box regulations led to more widespread adoption of LLCs as a legal entity, DAOs can also be given a choice to opt their entity status for tax purposes. However, as Decentralized Autonomous Organization live and operates on the blockchain and the Internet, and due to the pseudonymous nature of their members, DAOs registered as entities might find it technologically infeasible to collect and report information related to tax residency, beneficial ownership etc. of its members. Technological solutions that address the pseudonymity problem can help to resolve many legal, regulatory and tax related issues related to both registered and unregistered DAOs.

## 9.2 Taxation issues of DAOs

One of the fundamental aspects related to taxation of DAOs and their members is the classification of the tokens of the DAO. Some tokens might be classified as securities on application of the Howey Test or any other similar test for securities in a jurisdiction. In such a scenario the tokens will be taxed as securities. However, the tokens might serve dual purpose as utility tokens of the DAO's protocol/application. Unless jurisdictions prescribe specific tax treatment for such tokens, it is likely that actions like those taken by SEC in case of The DAO and by the CFTC in case of Ooki DAO may continue.

The individuals or entities may undertake many transactions with the DAO which might be taxable. The members may either receive governance tokens of the DAO or NFTs in the form of airdrops or they may pay a consideration for buying the tokens. In case of airdrops, in most jurisdictions, an income tax is chargeable upon the receipt of the airdrop. However, in some jurisdictions the receipt of an airdrop is not taxed but is considered acquisition of a crypto asset with zero basis. Also, the subsequent sale, swap, spend or gift transactions are also subject to capital gains. However, in some jurisdictions, depending upon the domestic tax laws and guidance, gifts might not be taxed.

In case the member pays a consideration for buying the governance tokens of a DAO which qualifies as a general partnership, in some jurisdictions like the US, no gain or loss would be recognizable (26 U.S. Code § 721[149]), making the initial contribution non-taxable. For DAOs which may be treated by default as general partnerships unless they register as any other entity, they might subject to pass-through taxation where the DAO does not pay taxes as a business entity and the individual partners pay taxes on their share of profits. The portfolios of the DAOs in the form of investments or NFTs etc. may appreciate and taxed when realized, DAOs might also have income from staking on DeFi platforms and the Ethereum Beacon Chain. DAO's sale of governance tokens for other crypto assets or fiat currency for funding the treasury, software development or any other purpose might also be a taxable event. The individual partners of the DAO would have to account for profits from all the activities of the DAO to pay their due taxes, which might be a challenge.

[149] Section 721 of the Internal Revenue Code (IRC) of the United States deals with the tax treatment of contributions to partnerships in exchange for partnership interests. Specifically, it states that no gain or loss shall be recognized to a partnership or to any of its partners in the case of a contribution of property to the partnership in exchange for an interest in the partnership.

The partners in a jurisdiction might pay taxes in their jurisdiction of residence as well as taxes in foreign jurisdictions for the DAOs profits attributable to those jurisdictions. This multi-jurisdictional liability might make profit attributions and tax liability ascertainment extremely complex and possibly lead to double or no taxation. If the DAO is registered as a corporation or is treated as a corporation for tax purposes, its profits will be taxed at the level of corporation as well as at the time of distribution of profits. The distribution of profits by the DAOs or burning of governance tokens might be taxed as dividends or share buyback in some jurisdictions.

DAOs also engage in the procurement of goods and services from multiple individuals and entities which would be taxable for the recipients of the consideration. These transactions might also give rise to VAT/GST liability and if the consideration paid is in the form of crypto assets it might be considered disposal for the purpose of capital gains. For example, a DAO may assign responsibility for some of their operations to a specialized entity or another DAO which might be subject to VAT/GST. DAOs that employ people might have obligations to pay employer taxes, social security contributions etc. Revenue sharing arrangements or royalty payments might also be subject to taxes. Many of the transactions and taxable events mentioned above might have withholding tax obligations. However, in the absence of tax residency information, DAOs might be have to withhold taxes at a higher rate. The mechanism for payment and seeking credit for withheld taxes is also a practical challenge.

## 10. International Cooperation and Exchange of Information

Ever since the first Bitcoin was mined in 2009 the landscape and scale of crypto assets has changed significantly. Even with recent dips in value of crypto assets this asset class remains, and is likely to remain a challenge for tax administrations across the world due to its pseudonymous and extra-territorial nature. Many tax administrations have formulated laws and policies to tax this asset class, whereas some jurisdictions have put an outright ban or a partial ban on crypto assets, whose efficacy has not been very high (Chen & Liu, 2022). However, many jurisdictions do not have any tax or regulatory policies regarding crypto assets which creates a regulatory arbitrage.

Due to the unique characteristics of crypto assets, in absence of clearly defined tax treatment and guidance, lack of third-party-reporting and regulatory arbitrage, the collection of taxes is primarily based on voluntary compliance, which might result in significant tax gaps. The UNCTAD has observed that crypto assets may enable tax evasion or avoidance through offshore flows whose ownership is not easily identifiable, undermine capital controls, erode the tax base, and consequently hinder Domestic Resource Mobilization for achieving the Sustainable Development Goals[150].

The IMF Executive Board also discussed a board paper on key elements of an appropriate policy response to crypto assets[151] and considered that *"the growing adoption of crypto assets in some countries, the extra-territorial nature of crypto assets and its providers, as well as the increasing interlinkages with the financial system, motivate the need for a comprehensive, consistent, and coordinated response*" and it also agreed that the applicability of tax regimes should be clarified.

The UNCTAD also issued a policy brief that "*highlights the importance and urgency of international cooperation regarding cryptocurrency tax treatments, regulation and information sharing as well as*

[150] https://unctad.org/system/files/official-document/presspb2022d10_en.pdf
[151] https://www.imf.org/en/News/Articles/2023/02/23/pr2351-imf-executive-board-discusses-elements-of-effective-policies-for-crypto-assets

*of redesigning capital controls to take account of the decentralized, borderless and pseudonymous features of cryptocurrencies."*[152]

The G20 Roadmap on crypto assets also highlighted the significance of international collaboration in tackling regulatory hurdles associated with crypto assets[153]. It underscored the necessity to share information and coordinate actions to establish uniform regulatory frameworks across different jurisdictions. This concerted effort aims to effectively address the challenges arising from the global nature of crypto activities and promote regulatory consistency to ensure a cohesive approach worldwide.

Presently, the FATF standards on virtual assets and virtual asset service providers and the Crypto-Asset Reporting Framework proposed by the Organisation for Economic Co-operation and Development (OECD) aim at achieving the comprehensive, consistent, and coordinated approach to Tax and AML/CFT related aspects of crypto assets. They are discussed in the following sections

## 10.1 FATF Standards on VAs and VASPs

The FATF standards pertaining to virtual assets (VAs) establish a comprehensive framework aimed at regulating virtual asset service providers (VASPs) to mitigate the risks associated with money laundering, terrorist financing, proliferation financing and other illicit activities. These standards mandate that countries enforce licensing and registration protocols for VASPs, conduct thorough customer due diligence (CDD), maintain transaction records, and promptly report any suspicious activities to relevant authorities[154]. Furthermore, the FATF guidelines stress the importance of robust supervision, enforcement mechanisms, international collaboration, and the adoption of risk-based approaches to safeguard the integrity of the financial system while fostering innovation in the virtual asset realm.

The FATF recommendations obligate VASPs to implement the Travel Rule, which requires VASPs to share the information related to a transaction's sender and recipient with other VASPs. It mandates that VASPs must obtain, hold, and transmit required originator and beneficiary information, immediately and securely, when conducting virtual asset transfers, particularly for those surpassing a designated threshold. This regulation aims to boost transparency and curb the potential for money laundering, terrorist financing, and other unlawful activities within the virtual assets ecosystem by enabling authorities to track and monitor transactions more efficiently, mirroring the obligations imposed on conventional financial institutions. It characterizes some products like so called Stablecoins and P2P transactions as high risk and guides the jurisdictions to take measures limiting the ability of users to transact anonymously. However, the guidance does not quantify the size of the P2P transactions and its associated money laundering/terror financing/proliferation financing risks.

The FATF defines virtual assets as a "*digital representation of value that can be digitally traded or transferred and can be used for payment or investment purposes. Virtual assets do not include digital representations of fiat currencies, securities, and other financial assets that are already covered elsewhere in the FATF Recommendations*."[155] This definition covers almost all types of virtual assets except few like NFTs. A VASPs is defined as:

---

[152] https://unctad.org/publication/cost-doing-too-little-too-late-how-cryptocurrencies-can-undermine-domestic-resource

[153] https://www.mea.gov.in/Images/CPV/G20-New-Delhi-Leaders-Declaration.pdf

[154] https://www.fatf-gafi.org/en/publications/Fatfrecommendations/Fatf-recommendations.html

[155] https://www.fatf-gafi.org/en/publications/Fatfrecommendations/Guidance-rba-virtual-assets-2021.html

*any natural or legal person who is not covered elsewhere under the Recommendations and as a business conducts one or more of the following activities or operations for or on behalf of another natural or legal person:*

*i) Exchange between virtual assets and fiat currencies;*
*ii) Exchange between one or more forms of virtual assets;*
*iii) Transfer of virtual assets; and*
*iv) Safekeeping and/or administration of virtual assets or instruments enabling control over virtual assets;*
*v) Participation in and provision of financial services related to an issuer's offer and/or sale of a virtual asset.*

This definition does not cover miners, validators and ancillary service providers like hardware wallet manufacturers, cloud service providers, software developers etc. DeFi Applications are not VASPs as per this definition. However, DeFi applications which have strong centralized operations, colloquially known as DeFi in Name Only (DINO), as well as not yet decentralized nascent DAOs may qualify as VASPs as per the definition.

The Travel Rule requires the following information to be obtained and transferred before the VA transfer takes place

i) originator's name
ii) originator's account number (wallet address)
iii) originator's physical (geographical) address, or national identity number, or customer identification number, or date and place of birth
iv) beneficiary's name
v) beneficiary's account number (wallet address)

The information is not required to be a part of the blockchain transaction. In case of transfer to an unhosted wallet the VASPs are required to obtain the counterparty information from the customer. Moreover, VASPs are required to perform CDD for customers transacting above a threshold amount of virtual assets. The ordering VASP is required to perform due diligence on the beneficiary VASP about the Data Protection and Privacy (DPP) measures related to security of travel rule information. It also encourages VASPs to collect additional information like:

a. the purpose of transaction or payment;
b. details about the nature, end use or end user of the item;
c. proof of funds ownership;
d. parties to the transaction and the relationship between parties;
e. sources of wealth and/or funds;
f. the identity and the beneficial ownership of the counterparty; and
g. export control information, such as copies of export-control or other licenses issued by the national export control authorities, and end-user certification.

### 10.1.1 Challenges in Travel Rule implementation

The FATF Travel Rule requires VASPs to obtain, hold, and transmit required originator and beneficiary information, immediately and securely, when conducting virtual asset transfers. However, due to the pseudonymous nature of crypto assets it might not be easy to ascertain the identity of the beneficiary VASP and might require reliance on blockchain analytics and specialized networks/protocols for travel rule solutions like TRISA[156], OpenVASP[157] and Notabene[158]. The FATF acknowledges that "*To date, the FATF is not aware of any technically proven means of identifying the VASP that manages the*

[156] https://trisa.io/
[157] https://www.openvasp.org/
[158] https://notabene.id/

*beneficiary wallet exhaustively, precisely, and accurately in all circumstances and from the VA address alone*"[159]. The technology neutral stand of FATF recommendations makes them agnostic to the products, services, solutions, or technologies provided by various vendors if they comply with the AML/CFT obligations.

There are many commercial solutions which provide travel rule services, but some of suffer from various shortcomings like:

a) Lack of interoperability – this remains the most important challenge in Travel Rule implementation
b) Lack of consensus on counterparty VASP due diligence
c) Inability to transfer information of both originator and beneficiary VASP before the transaction on blockchain
d) Issues in recordkeeping and retrieval of Travel Rule information
e) Transmission of transaction IDs instead of wallet addresses by some Travel Rule solution providers

Besides these, there are jurisdiction specific issues like different approaches to transaction thresholds, data privacy laws and approach on transactions to unhosted wallets. There are also issues related to jurisdictions where the FATF recommendations related to VASPs have not been applied yet – The Sunrise Issue. Moreover, as the FATF recommendations might not be complied by some jurisdictions and tax administrations might find it difficult to access information that is mainly intended for AML use, they may face challenges related to real-time use of such information for tax administration purposes. Also, there are many technological challenges in compliance to the Travel Rule. In the Targeted Update on Implementation of the FATF Standards on virtual assets and virtual asset service providers, FATF has called on industry to accelerate efforts to strengthen solutions that are global, and can accommodate nuances in requirements across jurisdictions, in line with the expectations of the FATF Standards.

## 10.2 Crypto-Asset Reporting Framework

The Crypto Assets Reporting Framework introduced by the OECD addresses the challenges posed by crypto assets in the realm of taxation, specifically addressing the lack of taxpayer information. This framework establishes guidelines and standards for the reporting and automatic exchange of information regarding crypto asset transactions among tax authorities globally. With the rapid proliferation of crypto assets and increase in their market capitalization, the OECD recognized the imperative to ensure transparency and compliance within this evolving landscape.

The CARF seeks to enable tax authorities to effectively monitor and regulate crypto asset transactions, to help them combat tax evasion. By establishing clear reporting requirements for crypto asset transactions, the framework aims to enhance tax compliance. Moreover, by facilitating the exchange of information among tax authorities across jurisdictions, it fosters international cooperation in addressing the unique tax challenges posed by crypto assets. Besides providing a framework for multilateral and bilateral agreements for exchange of information, it also provides an XML schema for information exchange.

The CARF[160] is similar to the Common Reporting Standard (CRS) developed by the OECD which helps jurisdictions to obtain offshore accounts information through annual automatic exchange of information. It defines the term "*Relevant Crypto-Asset*" which means "*any Crypto-Asset that is not a Central Bank Digital Currency, a Specified Electronic Money Product or any Crypto-Asset for which the*

[159] https://www.fatf-gafi.org/content/dam/fatf-gafi/guidance/Updated-Guidance-VA-VASP.pdf page 39
[160] https://www.oecd.org/tax/exchange-of-tax-information/crypto-asset-reporting-framework-and-amendments-to-the-common-reporting-standard.pdf

*Reporting Crypto Asset Service Provider has adequately determined that it cannot be used for payment or investment purposes.*" This definition includes crypto assets like NFTs and excludes specific categories of crypto assets like CBDCs. It makes the following three types of transactions reportable to tax authorities:

- exchanges between Relevant Crypto-Assets and Fiat Currencies;
- exchanges between one or more forms of Relevant Crypto-Assets; and
- Transfers (including Reportable Retail Payment Transactions) of Relevant Crypto-Assets.

The amount and details of units of crypto assets transferred to unhosted wallets are also required to be collected and retained for 5 years. The individuals or entities carrying out transactions related to crypto assets are known as "Reporting Crypto-Asset Service Providers. A Reporting Crypto-Asset Service Provider is defined as '*any individual or Entity that, as a business, provides a service effectuating Exchange Transactions for or on behalf of customers, including by acting as a counterparty, or as an intermediary, to such Exchange Transactions, or by making available a trading platform*.'

The commentary to the CARF clarifies the applicability of the definition to individuals and entities like software developers and decentralised exchanges. It states "*An individual or Entity that is making available a platform that solely includes a bulletin board functionality for posting buy, sell or conversion prices of Relevant Crypto-Assets would not be a Reporting Crypto-Asset Service Provider as it would not provide a service allowing users to effectuate Exchange Transactions. For the same reason, an individual or Entity that solely creates or sells software or an application is not a Reporting Crypto-Asset Service Provider, as long as it is not using such software or application for the provision of a service effectuating Exchange Transactions for or on behalf of customers.*"

The Reporting Crypto-Asset Service Provider are required to submit reports on Reportable Users[161] and also undertake due diligence on the customers, both individuals and entities. They are required to report the following information regarding Crypto-Asset Users that are Reportable Users or that have Controlling Persons that are Reportable Persons:

- Name
- Address
- Jurisdiction(s) of Residence
- Tax Identification Number(s)
- Date and Place of Birth

The name, address and identifying information of the Reporting Crypto-Asset Service Provider along with details of 'Relevant Transactions' need to be provided. Tax Identification Number (TIN) and Place of Birth may not be required to be reported if the domestic law does not require collecting the information or TIN has not been issued by the reportable jurisdiction. The information is to be obtained on a self-certification basis and its reasonableness needs to be confirmed relying on other information collected by the Reporting Crypto-Asset Service Provider, like AML/KYC related information.

The CARF is an amendment to CRS and based on a separate legal framework from the CRS, which makes it possible for countries to implement CARF without signing up for CRS (Falcao & Michel, 2023). The definitions of Relevant Crypto-Assets and Reporting Crypto-Asset Service Provider are wider than the corresponding FATF definitions. This can enable tax authorities to obtain tax related information for a wider gamut of crypto assets and from a larger set of exchanges/intermediaries. However, the determination of place of residence of a taxpayer might be a challenge in case of decentralized exchanges which would likely qualify as Reporting Crypto-Asset Service Provider.

[161] D Section IV: Defined Terms

Although the XML Scheme standardizes the information exchange but absence "*of any technically proven means of identifying the VASP that manages the beneficiary wallet exhaustively, precisely, and accurately in all circumstances and from the VA address alone*" as acknowledged by FATF, would also be a challenge for CARF as self-declaration based due diligence and blockchain analytics based attribution of crypto asset addresses to natural and legal persons might not provide reliable and accurate information for levy of taxes. Also, the information is provided annually which might lead to flight of assets in tax fraud cases.

## 10.3 Need for Global Public Digital Infrastructure

Pseudonymity and extra-territoriality are two key aspects of crypto assets which are a major hurdle for tax administrations across the world to realize the full potential of taxes due on crypto assets. Most crypto asset service providers and law enforcement agencies rely on blockchain analytics solutions which attribute crypto asset addresses to a natural or legal person. Such attribution would not be required for off-chain transactions undertaken by an individual or entity with a Reporting Crypto-Asset Service Provider as the information can be obtained through CRS. The accurate determination of such information for on-chain transactions i.e. P2P transactions is essential for levying taxes as the ownership of private keys corresponding to the wallet would be required to be established before the levy of tax. As the blockchain analytics-based attribution is mainly obtained through clustering techniques and transaction behaviours, it might not establish legal ownership or control of the private keys beyond doubt[162].

One example of such attribute of a crypto asset address is its jurisdictional tax administration. Accurate information about the originator and beneficiary tax jurisdiction is essential for addressing multiple issues related to taxation in a transaction, like:

a) Percentage of amount to be withheld, if at all
b) Attribution of profits to a jurisdiction
c) Preparation and submission of information returns of an entity for a jurisdiction
d) Application of Double Taxation Avoidance Agreements
e) Determining tax treatment of a DAO and its members

The information related to off-chain transactions would be collected and retained for the prescribed periods by the Reporting Crypto-Asset Service Provider. Information related to the transfer of crypto assets from a Reporting Crypto-Asset Service Provider needs to be obtained from the user, which might require a cryptographic process similar to the Address Ownership Proof Protocol[163]. As the transaction information is already stored on the blockchain it might be more prudent for tax administrations to attribute crypto asset addresses to TINs using cryptographic methods and then use the blockchain information about P2P transactions to determine tax liability of the taxpayer.

Presently the CRS information is exchanged through secure networks established by the participating countries' tax authorities or the Common Transmission System (CTS) provided by the OECD which are designed to ensure the confidentiality, integrity, and privacy of the exchanged data. However, this data is exchanged annually after a lag and is not designed to provide real-time information which might be the requirement for many tax jurisdictions to attribute the crypto asset address to a tax jurisdiction and TIN. To meet this requirement tax administrations across the globe would require a global public digital infrastructure for obtaining, maintaining, and facilitating the access to such attribution data ensuring the confidentiality, integrity, and privacy of the data, where exchange of data can take place in accordance to policies based on Exchange of Information Agreements between tax jurisdictions. The tax administration might use measures like prescribing higher withholding tax rates for transactions with wallets for which such attribution information could not be obtained from any tax jurisdiction in real time.

[162] https://finance.yahoo.com/news/crypto-tracing-revolutionizing-crime-fighting-124050120.html
[163] https://www.21analytics.ch/what-is-aopp/

One probable design of such a system may use digital signature certificates issued to every TIN holder from its respective tax authority. Using the digital signature certificate issued by the tax authority and digital signatures from their wallet addresses the tax jurisdiction of the TIN can obtain a cryptographic proof of the ownership of the crypto asset private keys by the TIN holder. The tax jurisdiction would be required to maintain a record of all public addresses of its TIN holders for which it holds the cryptographic proof, and in real time, respond to queries from other jurisdictions about the ownership of a Crypto address by a TIN holder in the jurisdiction. As the number of unique Ethereum addresses which have been used for transactions till date is in tens of Millions and the number of unique Bitcoin addresses that have been used for transactions till date is in few billions, using appropriate data structures and search methods, the tax administrations can respond to requests from other jurisdictions about a TIN holder in real-time.

Whenever a new transaction is to be carried out, the wallet software can broadcast a message signed with the digital signature issued by the originating jurisdiction enquiring about the tax jurisdiction of the beneficiary crypto address as depicted in Fig. 57. Depending upon the EOI agreements some or no tax jurisdiction may respond to the broadcast about the presence of the TIN of the beneficiary crypto asset address owner in the affirmative. In case a response is received the transaction can be carried out as usual on the blockchain, as the jurisdictions, by virtue of processing requests based on cryptographic proofs, would automatically obtain the visibility of the transactions, and may also exchange the personal information related to the TINs. In case no response is received broadcast about the presence of the TIN of the beneficiary crypto asset address from any tax jurisdiction, the originator may be obligated by its tax jurisdiction to withhold taxes at a higher rate to safeguard the interest of the tax administration of the transaction originator. This design may also enable tax administrations to perform their own analytics on on-chain transaction data and the beneficial ownership and controlled entity information captured in the tax returns filed by their taxpayers.

Another such secure information sharing mechanism that can be used by virtual asset service providers can be XMTP Protocol[164] based messaging. This messaging protocol provides a wallet-to-wallet messaging functionality provided both the wallets on-board to an XMTP protocol-based application. This service can provide a VASP with technically proven means of identifying the VASP that manages the beneficiary wallet exhaustively, precisely, and accurately in all circumstances and from the VA address alone, if the beneficiary VASP also uses an XMTP protocol-based application. The onward and backward communication between two wallets is depicted in Fig. 58 and 59. Multiple applications built on the protocol would be interoperable. An application can be developed as a global public digital infrastructure like the SWIFT network for exchanging Travel Rule information between VASPs, ensuring the adherence to Data Privacy and Protection. As most VASPs follow the interVASP Messaging Standard 101 (IVMS 101)[165], which is a data format for exchanging travel rule information, like the XML scheme in CARF, a standard protocol for communication can solve the problem of interoperability. Just as the protocols approved by the Internet Engineering Task Force like the Hypertext Transfer Protocol[166] (HTTP) and Transport Layer Security[167] (TLS) Protocol provide the broad framework which needs to be adhered to by developers, similar adoption of standards would be required to overcome the problem of interoperability. The SWIFT network used by financial institutions for the rapid, precise, and secure transmission of transaction information is another such example.

[164] https://xmtp.org/ .
[165] https://www.intervasp.org/
[166] https://www.ietf.org/rfc/rfc2616.txt
[167] https://www.ietf.org/rfc/rfc5246.txt

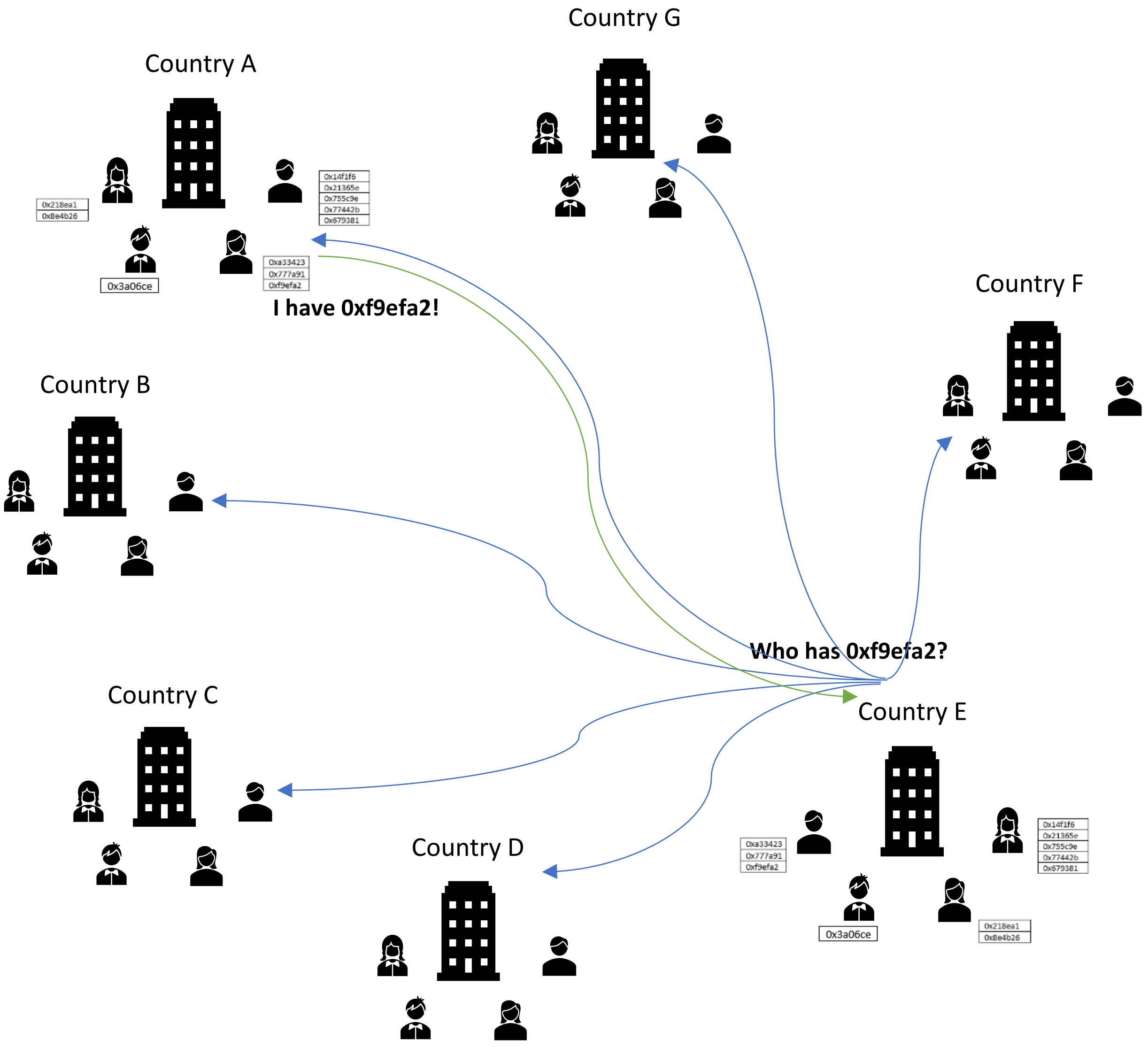

Fig. 57 A Digital Signature based solution to find tax jurisdiction of a crypto asset address

## 10.4 The Challenge of Anonymity Enhancing Crypto Assets

Tracing anonymity-enhancing crypto assets, particularly those like Monero, presents multifaceted challenges due to the sophisticated technological features embedded within these cryptocurrencies. Monero leverages cutting-edge cryptographic techniques such as ring signatures, stealth addresses, and confidential transactions to provide users with unparalleled privacy and fungibility. One of the primary hurdles in tracing Monero transactions lies in its use of ring signatures, which obfuscate the sender's identity by mixing the sender's transaction with those of other users, making it virtually impossible to determine the true source of a transaction.

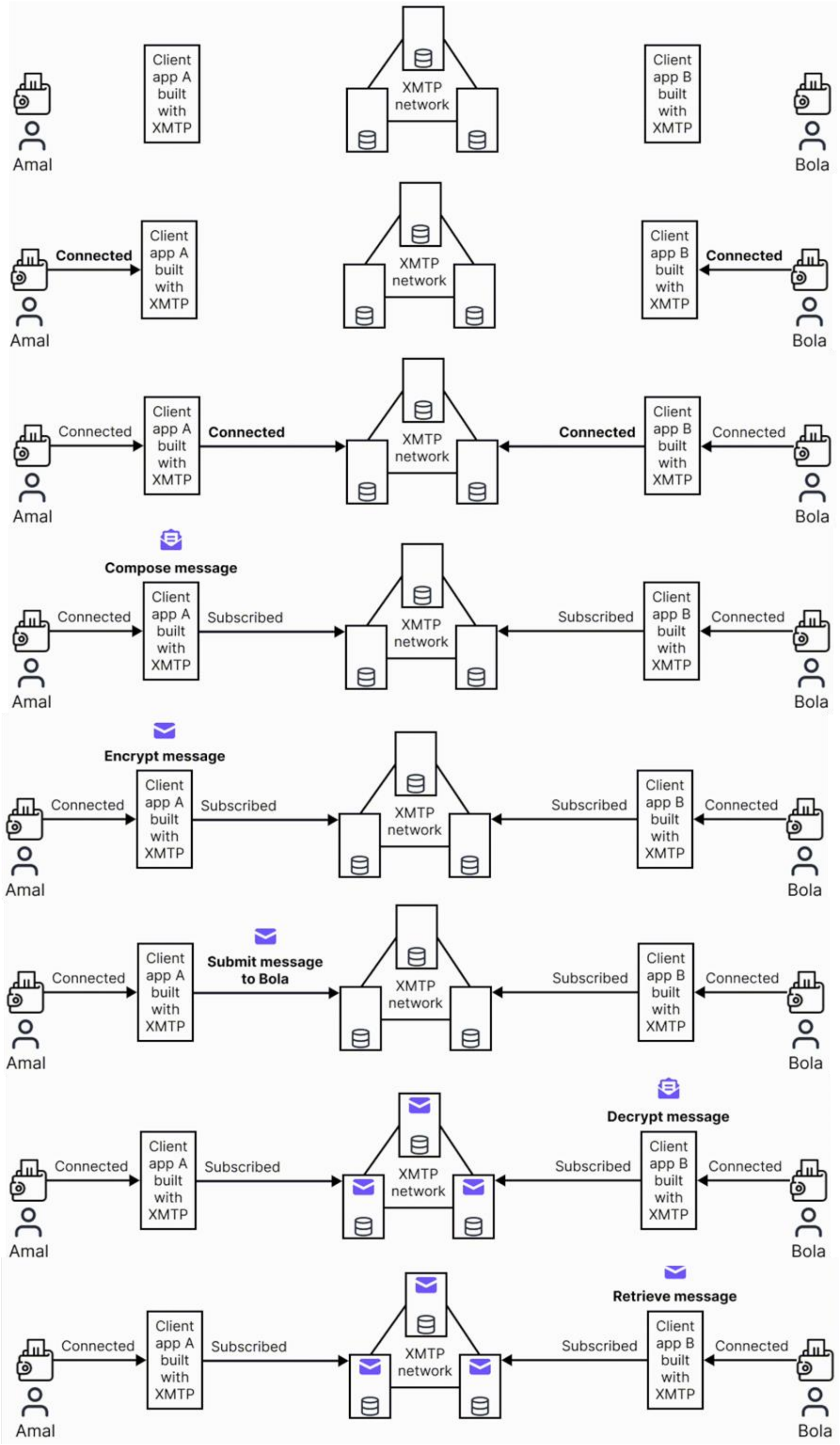

Fig. 58 XMTP: Onward communication between two wallets[168]

[168] https://xmtp.org/ Portions of this page are modifications based on work created and shared by XMTP and used according to terms described in the Creative Commons 4.0 Attribution License

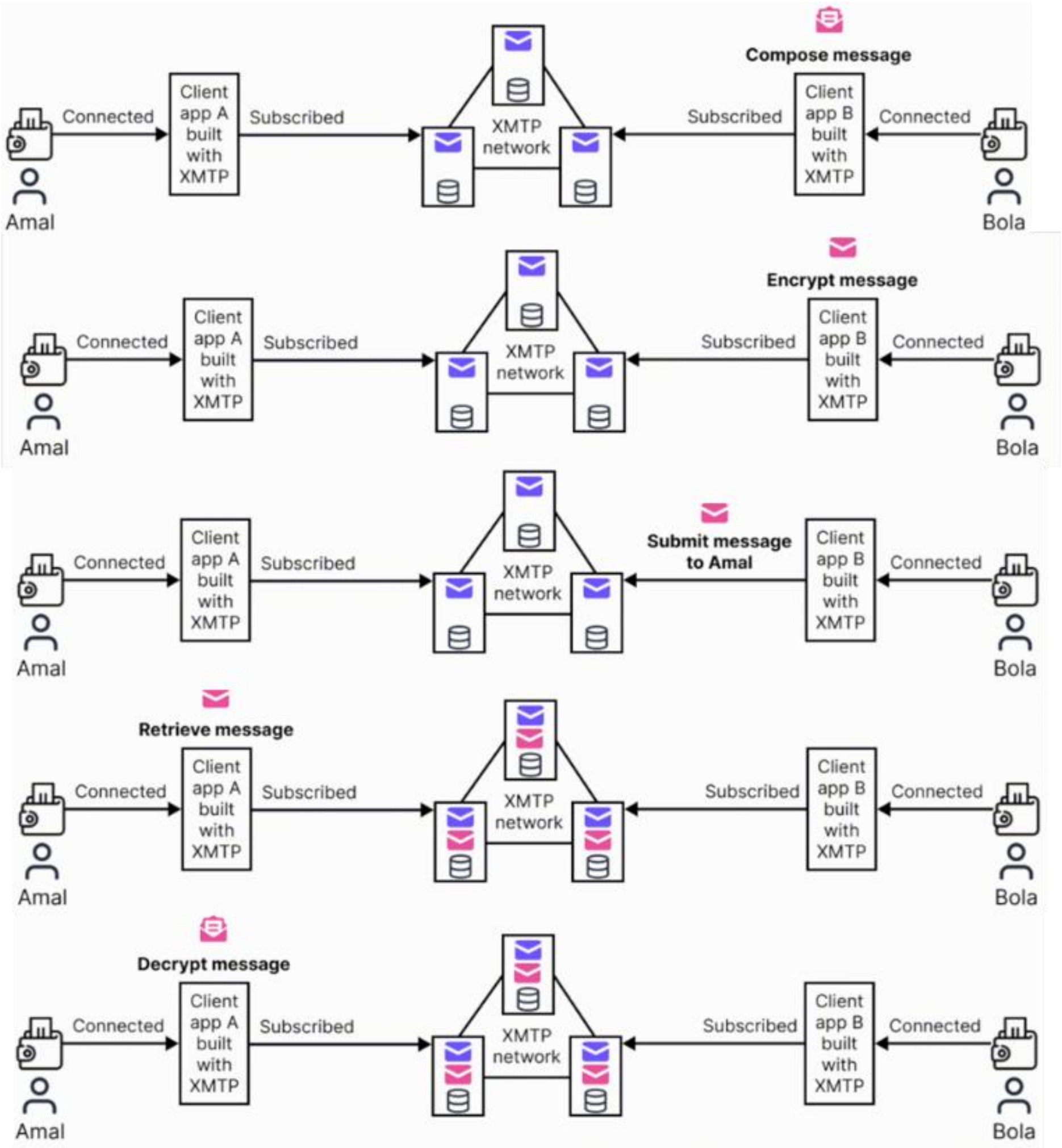

Fig. 59 XMTP: Backward communication between two wallets[169]

Unlike Bitcoin, where transactions can be traced through a transparent ledger, Monero's blockchain obscures transaction details, including sender, recipient, and transaction amount, thereby thwarting traditional tracing methods. Additionally, Monero employs stealth addresses to enhance privacy by generating unique, one-time addresses for each transaction, making it challenging for external observers to link transactions to specific recipients. Furthermore, Monero's integration of confidential transactions encrypts transaction amounts, adding another layer of complexity to tracing efforts. These technological innovations collectively render Monero transactions highly opaque and resistant to external scrutiny, posing significant challenges for tax authorities and law enforcement agencies seeking to enforce taxation laws and tracing illicit activities. Blockchain mixers and tumblers also pose similar challenges.

Addressing these challenges necessitates the development of novel investigative techniques and regulatory frameworks tailored to the unique characteristics of privacy-centric cryptocurrencies like Monero, as well as collaboration between public and private sectors as well as governments on a global scale. Various blockchain intelligence and analytics solution providers claim the ability to trace Monero. However, it remains extremely difficult, if not impossible, to trace Monero purely through conventional methods used for other crypto assets like Bitcoin and some other conventional investigative inputs might be required.

[169] https://xmtp.org/ Portions of this page are modifications based on work created and shared by XMTP and used according to terms described in the Creative Commons 4.0 Attribution License

## 11. Conclusion

Crypto assets have posed unique challenges to tax administrations throughout the world. On one hand they need to appropriately classify crypto assets as property/asset or means of payment/negotiable instrument and on the other hand address challenges due to their pseudonymous and extra-territorial nature. Tax administrations are innately designed to tax conventional means of holding and transmitting value as well as economic activity. This technologically challenging realm of economic activity creates tax design issues related to neutrality, efficiency and equity as well as their practical implementation. Policymakers would need to address the complex issues related to direct and indirect taxation of crypto assets, including issues related to the environmental costs of crypto assets, to provide clarity about their tax treatment without impeding progress or inhibiting innovation in this domain.

With ‘Crypto Winter’ slowly turning into a ‘Crypto Spring’ the revenues involved would be significant, especially for developing countries due to the large policy and compliance gaps that exist. The significant amounts involved in GST/VAT related to sales of goods and services in lieu of crypto assets as well as GST/VAT on cross border services like mining services, MEV and NFTs need urgent attention of tax administrations. This would require a policy response which is coherent and compatible with the technological bottom-lines of crypto assets and provides certainty to facilitate compliance. At the same time, they need to be conscious about the tax policy arbitrage opportunities that exist in this ecosystem, which may result in non-compliance and revenue loss. The swift pace of innovation in this field presents a formidable challenge for policymakers to effectively stay abreast to formulate prudent and practical policies. The migration of commercial activity from digital commerce in real world to virtual worlds like metaverses can fundamentally challenge the ideas on which our present tax systems are designed. In near future, tax administrations might be compelled to have a presence on various blockchains to collect the due taxes.

Capacity development, especially in the context of developing countries, must be at the core of the strategy of tax administrations to deal with this asset class. Having policymakers well versed with the technology related aspects of crypto assets, which have profound tax policy implications, can formulate appropriate policies to help realize the revenues associated with asset class. Each auditor who can accurately determine the tax liability of crypto asset owners in his/her jurisdiction can collect the fair share of taxes as well as prevent the misuse of this technology for tax evasion, bypassing capital controls and money laundering. Just as issues related to international taxation and transfer pricing remain focus areas for capacity development in many jurisdictions, similar focus needs to be adopted for the taxation related issues of crypto assets.

The inherent nature of crypto assets transcends national borders, rendering traditional tax frameworks inadequate in capturing and regulating these assets effectively. The decentralized and pseudonymous nature of blockchain technology poses formidable challenges in tracking and reporting taxable transactions. Effective exchange of taxpayer information among countries and implementation of the travel rule are indispensable for combating tax evasion and money laundering as well as ensuring compliance within the crypto ecosystem. It would also help to avoid non-taxation and double taxation of crypto assets. Without attribution of transactions to natural and legal persons even high-capacity jurisdictions might be unbale to effectively tax this asset class. As this ecosystem continues to evolve, collaborative efforts are essential to address emerging issues such as DeFi, NFTs and the metaverse. The creation of global public digital infrastructure for exchange of such information in real time like the SWIFT network, based on standardized protocols can help to mitigate many practical challenges associated with pseudonymity. Also, the issues related to anonymity enhancing crypto assets as well as services like tumblers and mixers need to be addressed as those intending to evade taxes may resort to them as avenues for tax regulatory arbitrage shrink.